\documentclass{article}

\PassOptionsToPackage{numbers, compress}{natbib}
\usepackage[numbib]{tocbibind}
\usepackage{amsmath,amssymb,amsthm,fullpage,mathrsfs,pgf,tikz,caption,subcaption,mathtools,cancel}
\usepackage[ruled,vlined,linesnumbered]{algorithm2e}
\usepackage{natbib}
\usepackage{amsmath,amssymb,amsthm,mathtools}
\usepackage[T1]{fontenc}    
\usepackage{hyperref}       
\usepackage{url}            
\usepackage{booktabs}       
\usepackage{amsfonts}       
\usepackage{nicefrac}       
\usepackage{microtype}      
\usepackage[margin=1in]{geometry}
\usepackage{bbm}
\usepackage{enumitem}
\usepackage{xcolor}
\usepackage{cleveref}
\usepackage[most]{tcolorbox}

\let\oldnl\nl

\newif\ifmynotes
\mynotesfalse
\def\mynote#1{\ifmynotes{#1}\else\fi}

\newcommand{\nonl}{\renewcommand{\nl}{\let\nl\oldnl}}
\newcommand{\maxh}[1]{\mynote{\textcolor{red}{[MH: #1]}}}
\newcommand{\yotamd}[1]{\mynote{\textcolor{blue}{[YD: #1]}}}
\newcommand{\TP}[1]{\mynote{\textcolor{purple}{[TP: #1]}}}

\newcommand{\Enc}{\textsf{Enc}}

\newcommand{\Ball}{{\cal B}}
\newcommand{\N}{{N}}
\newcommand{\M}{{N'}}
\newcommand{\Time}{{\rm Time}}
\newcommand{\Size}{{\rm Size}}
\newcommand{\Depth}{{\rm Depth}}

\newtheorem{theorem}{Theorem}[section]
\newtheorem{corollary}[theorem]{Corollary}

\newtheorem{proposition}[theorem]{Proposition}
\newtheorem{lemma}[theorem]{Lemma}
\newtheorem{definition}[theorem]{Definition}

\newtheorem{example}[theorem]{Example}
\newtheorem{remark}[theorem]{Remark}
\newtheorem{observation}[theorem]{Observation}
\newtheorem{claim}[theorem]{Claim}

\newcommand{\F}{\mathbb{F}}

\DeclareMathOperator{\poly}{poly}
\DeclareMathOperator{\polylog}{polylog}

\DeclareMathOperator*{\dist}{dist}

\newcommand{\set}[1]{\left \{ {#1} \right \}}
\newcommand{\sett}[2]{\left \{ {#1} : {#2} \right \}}
\newcommand{\abs}[1]{\left | {#1} \right |}
\newcommand{\Ex}[2]{\underset{#1}{\mathbb{E}} \left [ {#2} \right ]}
\newcommand{\Prob}[2]{\underset{#1}{\mathbb{P}} \left [ {#2} \right ]}

\newcommand{\iprod}[1]{\langle {#1} \rangle}
\newcommand{\one}{\mathrm{1}}
\newcommand{\fsize}{{\mathfrak{q}}}
\newcommand{\poldeg}{\kappa}
\newcommand{\gaussb}[2]{\bigl[\!\begin{smallmatrix} {#1} \\ {#2} \end{smallmatrix}\!\bigr]}
\newcommand{\kms}{\mathfrak{X}}

\newtheorem*{theorem*}{Theorem}
\newtheorem*{proposition*}{Proposition}
\newtheorem*{claim*}{Claim}

\newcommand{\MYstore}[2]{%
  \global\expandafter \def \csname MYMEMORY #1 \endcsname{#2}%
}

\newcommand{\MYload}[1]{%
  \csname MYMEMORY #1 \endcsname%
}

\newcommand{\MYnewlabel}[1]{%
  \newcommand\MYcurrentlabel{#1}%
  \MYoldlabel{#1}%
}

\newcommand{\MYdummylabel}[1]{}

\newcommand{\torestate}[1]{%
  \let\MYoldlabel\label%
  \let\label\MYnewlabel%
  #1%
  \MYstore{\MYcurrentlabel}{#1}%
  \let\label\MYoldlabel%
}

\newcommand{\restatetheorem}[1]{%
  \let\MYoldlabel\label
  \let\label\MYdummylabel
  \begin{theorem*}[Restatement of \cref{#1}]
    \MYload{#1}
  \end{theorem*}
  \let\label\MYoldlabel
}

\newcommand{\restateproposition}[1]{%
  \let\MYoldlabel\label
  \let\label\MYdummylabel
  \begin{proposition*}[Restatement of \cref{#1}]
    \MYload{#1}
  \end{proposition*}
  \let\label\MYoldlabel
}

\newcommand{\restateclaim}[1]{%
  \let\MYoldlabel\label
  \let\label\MYdummylabel
  \begin{claim*}[Restatement of \cref{#1}]
    \MYload{#1}
  \end{claim*}
  \let\label\MYoldlabel
}

\title{High Rate Efficient Local List Decoding from HDX}
\author{Yotam Dikstein\thanks{Institute for Advanced Study, Princeton. yotam.dikstein@gmail.com. Supported by the National Science Foundation 
under Grant No. DMS-2424441.}, Max Hopkins\thanks{Institute for Advanced Study, Princeton. nmhopkin@ias.edu. Supported by the National Science Foundation under Grant No. DMS-2424441.}, Russell Impagliazzo\thanks{University of California, San Diego. rimpagliazzo@ucsd.edu. Supported by the National Science Foundation under Grant No. CCF-2212136.}, and Toniann Pitassi\thanks{Columbia University. tonipitassi@gmail.com. Supported by the National Science Foundation 
under Grant No. CCF-2212136.}}
\begin{document}
\pagenumbering{Roman}
\maketitle
\yotamd{IF THIS TEXT IS VISIBLE WE DID NOT TURN OFF COMMENTS}
\begin{abstract}
We construct the first (locally computable, approximately) locally list decodable codes with rate, efficiency, and error tolerance approaching the information theoretic limit, a core regime of interest for the complexity theoretic task of \textit{hardness amplification}. Our algorithms run in \textit{polylogarithmic time} and \textit{sub-logarithmic depth}, which together with classic constructions in the unique decoding (low-noise) regime leads to the resolution of several long-standing problems in coding and complexity theory:

\begin{enumerate}
\item Near-optimally \textit{input-preserving} hardness amplification (and corresponding fast PRGs)
\item Constant rate codes with \textit{$\log(N)$-depth} list decoding (RNC$^1$)
\item Complexity-preserving distance amplification
\end{enumerate}

Our codes are built on the powerful theory of (local-spectral) \textit{high dimensional expanders} (HDX). At a technical level, we make two key contributions. First, we introduce a new framework for ($\polylog(N)$-round) \textit{belief propagation} on HDX that leverages a mix of local correction and global expansion to control error build-up while maintaining high rate. Second, we introduce the notion of \textit{strongly explicit local routing} on HDX, local algorithms that given any two target vertices, output a random path between them in only \textit{polylogarithmic} time (and, preferably, \textit{sub-logarithmic} depth). Constructing such schemes on certain coset HDX allows us to instantiate our otherwise combinatorial framework in polylogarithmic time and low depth, completing the result.

\end{abstract}



\newpage
\tableofcontents
\newpage
\setcounter{page}{1}
\pagenumbering{arabic}
\section{Introduction}
Error-correcting codes (ECCs) are a method of reliable communication over noisy channels that have long been recognized as fundamental objects in communication theory, information theory, and combinatorics. In the 1990's, developments in probabilistic interactive proof systems (culminating in the PCP theorem) made clear that error-correction is also central to TCS, a deep connection now ubiquitous throughout the field. In turn, applications in TCS have motivated the study of new ECCs with stronger properties. The development of \textit{list decoding} \cite{elias1957list,wozencraft1958list,sudan1997decoding} 
showed it is possible to decode from extreme noise (e.g., up to $\frac{1}{2}-\varepsilon$ errors in the binary case) if one allows the decoder to output a short list of possible messages. The advent of \textit{local decoding} showed it is possible to build highly efficient decoding algorithms which, given an index $i \in [N]$ and a corrupted encoding $w$ of a message $f$, recovers $f(i)$ while reading only a tiny \textit{sub-constant} fraction of $w$. Since their inception, these strengthened notions have played a key role in algorithms and complexity (e.g.\ in hardness amplification \cite{impagliazzo1995hard,o2002hardness}, hardness of approximation \cite{raz1997sub,arora1997improved,haastad2001some}, worst-case to average-case reductions \cite{cai1999hardness,hirahara2023hardness}), pseudorandomness (extractors \cite{trevisan2001extractors,guruswami2009unbalanced}, derandomization \cite{impagliazzo1997p,sudan1999pseudorandom}), cryptography (one way functions \cite{yao1982theory,levin1985one,goldreich1989hard}, private information retrieval \cite{chor1998private}), and machine learning \cite{carmosino2016learning,bogdanov2019xor,BlancHuangMalkinServedio2026}.

Despite their central importance, stark gaps remain in the literature between the best constructions of such codes and corresponding information-theoretic lower bounds (and similarly in downstream application). In this work, we develop the first \textit{approximate} locally list-decodable codes (aLLDCs) approaching the optimal trade-offs between rate, query complexity, efficiency, and error tolerance. Leveraging our codes, we resolve several long-standing problems in coding and complexity theory including \textbf{(1)} Input-preserving \textit{hardness amplification} and near-linear time \textit{pseudorandom generators}, and \textbf{(2)} Complexity-preserving distance amplification. The latter leads in particular to the first construction of good codes with \textit{low depth} (RNC$^1$) list-decoders, extending classic results of Zyablov and Pinsker \cite{zyablov1974decoding} and Sipser and Spielman \cite{sipser2002expander} to the high noise regime, and to new families of high rate LLDCs with modestly improved sub-polynomial decoding time.


Our construction is based on new local algorithmic machinery for the powerful class of spectral \textit{high dimensional expanders}, hypergraph analogs of expander graphs that have seen an impressive range of breakthrough applications in TCS \cite{kaufman2020high,anari2019log,panteleev2021asymptotically,anshu2023nlts,dinur2021,dikstein2024low,bafna2024constant,bafna2024quasi,hsieh2025explicit}. We introduce a new framework for \textit{belief propagation} on HDX that leverages a mix of local correction and global expansion to control error buildup over $\polylog(N)$-rounds of propagation while maintaining \textit{constant} encoding redundancy (rate). Core to our framework is the new notion of \textit{strongly explicit local routing}, a local algorithm that given any pair of vertices $v_1,v_2$ in a graph $G$, outputs a random path between $v_1$ and $v_2$ in \textit{polylog time} and \textit{sub-logarithmic depth}. Critically, we develop local routing schemes for certain HDX that are further \textit{tolerant to vertex-faults}, allowing polylog-time routing even when a constant fraction of vertices cannot communicate.

\subsection{Motivation and Background}
\subsubsection{Hardness Amplification}

Approximate local list decoding is a relaxation of local list decoding allowing the decoded message to be \textit{close to} (rather than exactly match) the original message. The notion arises naturally in the study of average-case complexity and hardness amplification \cite{trevisan2003list,impagliazzo2003hardness}: by encoding a weakly hard function $f: \{0,1\}^n \to \{0,1\}$ with an aLLDC correcting up to $\frac{1}{2}-\varepsilon$ errors, the encoded function $\Enc^f$ is automatically $\frac{1}{2}+\varepsilon$ hard (i.e., no algorithm in the class computes $\Enc^f$ with better than $\varepsilon$ advantage) against a weaker class of algorithms depending on the complexity of \(\Enc\)'s decoder. To ensure the worst-case complexity of $\Enc^f$ is nevertheless comparable to that of $f$ (allowing for quantitative hardness amplification within a fixed complexity class), it is further desirable the encoding function $\Enc$ be \textit{locally computable} (determined by querying just a few values of $f$), which is only possible under the relaxed approximate decoding model.

Up to $\varepsilon=\frac{1}{4}$ (the `unique decoding regime'), there are simple constructions of approximate locally decodable codes based on expander graphs \cite{guruswami2006hardness}, but amplifying \textit{beyond} $1/4$ (the `list-decoding regime’) expanders seem to break and every known method polynomially blows up the domain \cite{impagliazzo1995hard,impagliazzo1997p,yao1982theory,impagliazzo2008uniform,impagliazzo2009approximate}. Indeed even dropping the requirement of local computation, there are no constructions of higher rate approximate LLDCs any better than exact ones, and while expanders have seen greater success in this setting when combined with list-decodable algebraic codes \cite{kopparty2017high,gopi2018locally,kopparty2018improved}, the best known constructions still require quasipolynomially many queries in $n$, substantially impacting the quantitative hardness of $\Enc^f$.
As a result, core questions in hardness amplification questions have remained open: can we amplify a weakly hard function $f: \{0,1\}^n \to \{0,1\}$ to $\Enc^f: \{0,1\}^{n'} \to \{0,1\}$ where $n' \leq n+O(\log\frac{1}{\varepsilon})$? Can we do this in a way such that $\Enc^f$ is locally computable and remains hard against algorithms of roughly the same complexity as $f$? Such results are important both to our quantitative understanding of average-case complexity and in downstream applications such as the construction of efficient pseudorandom generators.

\subsubsection{Complexity Preserving Distance-Amplification}

A second strong motivation to study approximate local list decoding, also observed in Trevisan's \cite{trevisan2003list} seminal work on the topic, is \textit{distance amplification}. Distance amplification is a widespread tool in coding theory and complexity that amplifies the error correction capabilities of a base code $C$ while maintaining desirable properties such as rate. Expander-based distance amplification methods \cite{alon2002construction,alon1995linear} in particular have seen great success over the years in the construction of high rate codes with efficient list-decoding \cite{guruswami2003linear,hemenway2019local,jeronimo2025explicit,srivastava2025list}, including in the best known constructions of high rate locally list decodable codes \cite{kopparty2017high,kopparty2018improved}.

Similar to their failure in hardness amplification, expander-based methods do not generically preserve the \textit{complexity} of decoding in the high error regime (e.g.\ query, circuit complexity, etc).\footnote{Or at least are not known to. If the base code is additionally \textit{list-recoverable}, it is possible to preserve decoding complexity to an extent using \cite{alon1995linear} (see, e.g., \cite{kopparty2018improved}), but this is generally a much stronger condition on the base code more akin to list-decoding.} In contrast, Trevisan \cite{trevisan2003list} observed aLLDCs give a generic amplification method \textit{without} significantly increasing the base code's encoding or decoding complexity: by simply composing a base code $C$ with an aLLDC, we may fully (list)-decode the message by first running the aLLDC, then feeding each $99\%$-accurate output into $C$'s decoder. 

Unfortunately, since all known low complexity aLLDCs have vanishing rate, this connection has not lead to the construction of good error correcting codes with high noise tolerance and low decoding complexity, leaving open several core questions in the study of list decoding including:

\paragraph{List Decodable Codes with Low-depth Encoding and Decoding}

A classical question in coding theory is to construct families of good codes with low \textit{circuit complexity}, and in particular codes that are encodable and decodable \textit{in parallel}, that is by polynomial (or even near-linear) size, $O(\log N)$-depth circuits. 
Low-depth decoders have a long history in error correction dating back to the late 60s with Gallager's celebrated introduction of LDPC codes \cite{gallager2003low} and Zyablov and Pinsker's \cite{zyablov1974decoding} corresponding log-depth belief propagation decoders (famously made \textit{explicit} in the 90s by Sipser and Spielman \cite{sipser2002expander} using expanders). Surprisingly, despite some 30-50 intervening years, in the list decoding regime (even for $\varepsilon=1/4$), it is still not known even \textit{existentially} whether there are good codes with log depth decoders. While many codes, e.g.\ \cite{sudan1997decoding,guruswami2003linear}, have spectral or algebraic list decoding in (V)NC$^2$, none are known to break $\log^2(N)$-depth.

\paragraph{Locally List Decodable Codes}


Despite major progress over the last 30 years, local (list)-decoding with near-optimal parameters remains a central open problem. Unlike the easier parallel setting, local decoding is far from solved in the unique decoding regime, where the gap between the best known upper and lower bounds remains quasi-exponential. Nevertheless, it is interesting to ask whether constant rate LDCs in the unique decoding regime can be generically amplified. For instance, does any good LDC correcting $1\%$ errors in $q$ queries imply a good LLDC correcting $\frac{1}{2}-\varepsilon$ errors in $\poly(q)$ queries? Beyond providing new families of LLDCs, such a result would completely reduce the problem to solving the `easier' unique decoding case.

\bigskip

So far we have discussed several related open problems: near-optimal hardness amplification, complexity-preserving distance amplification, and local and parallel list decoding, all of which reduce to one core question and the real main focus of our work:

\begin{center}
    \begin{tcolorbox}[colframe=black, colback=gray!8, boxrule=.75pt, width=.85\textwidth, sharp corners, left=5pt, right=5pt, top=6.3pt, bottom=5.8pt]
    \centering
    \textbf{Main Q:} \emph{Do `high rate’ aLLDCs exist? With what query/circuit complexity?}
    \end{tcolorbox}
\end{center}

Achieving locality of this form while maintaining high rate and distance is a central challenge in complexity-theoretic coding; such algorithms must infer global codeword structure from just a few bits of local information, but achieving this local-to-global propagation typically requires destroying the code's rate. At a technical level, known sub-optimal constructions often leverage strong combinatorial expansion or algebraic structure, but these rarely combine cleanly to yield codes with efficient local algorithms. 

In this work, we essentially resolve this problem. In \Cref{sec:intro-intro-results} below, we give new constructions of aLLDCs with dramatically improved rate and locality approaching the information theoretic optimum. In \Cref{sec:intro-applications} we give applications of these codes in coding and complexity theory, breaking longstanding barriers in the area. In \Cref{sec:proof-overview} we discuss our techniques, a marriage of combinatorial and algebraic methods based on new codes and local algorithmic machinery for the powerful class of \textit{high dimensional expanders}.
\subsection{Main Results: High Rate aLLDCs}\label{sec:intro-intro-results}

 
 We construct new explicit codes in two main regimes. We first study the case where $\varepsilon$ is strongly sub-constant in the message length, i.e.\ when $\varepsilon \ll \frac{1}{\polylog(N)}$ (and can be taken as small as $1/\poly(N)$). Such codes inherently cannot have `constant rate', but are of core interest
in hardness amplification where the regime corresponds to amplifying $f:\{0,1\}^n \to \{0,1\}$ to $\Enc^f:\{0,1\}^{n'} \to \{0,1\}$ 
\textit{super-polynomially hard}.\footnote{Note $n=\log(N)$ here since we encode the entire truth table of $f$.} We give codes with essentially optimal parameters in this regime:\footnote{Below we assume approximate decoding to $0.01$ accuracy, roughly meaning the decoded outputs are within $0.01$-distance of the `true' valid messages. It is easy to amplify this value to any sufficiently small constant (or even subconstant at a proportional cost in rate and queries, see \Cref{lem:unique-amp}).}

\begin{theorem}[Polylog Rate aLLDCs (\Cref{cor:binary-polylog-aLLDCs})]\label{thm:intro-intro-polylog}
    For every $N \in \mathbb{N}$ and $\varepsilon \leq \frac{1}{\log(N)}$, there is a binary aLLDC $\mathcal{C}_N$ decodable from $\frac{1}{2}-\varepsilon$ errors with:
\begin{enumerate}
    \item \textbf{Rate:} $\poly(\frac{1}{\varepsilon})$
    \item \textbf{List Size:} $\tilde{O}(\frac{1}{\varepsilon^2})$
    \item \textbf{Queries:} $\poly(\frac{1}{\varepsilon})$
\end{enumerate}    
    Moreover $\mathcal{C}_N$ is decodable in $\poly(\frac{1}{\varepsilon})$-size and $\tilde{O}(\log^2\frac{1}{\varepsilon})$-depth, and encodable in time $\poly(\frac{1}{\varepsilon})$.
\end{theorem}
While \Cref{thm:intro-intro-polylog} requires the full power of high dimensional expanders, we also give a simpler `warmup' construction in \Cref{sec:subpoly} achieving $2^{-\sqrt{\log(N)}}$ rate with our framework just using subspaces. Even this is a substantial improvement over the prior state of the art at $\frac{1}{\poly(N)}$-rate. Finally, we remark through fairly standard concatenation analysis, it is possible to give a \textit{strongly local} variant of the above with only $\polylog(\frac{1}{\varepsilon})$ encoding time (though the rate suffers somewhat). We refer the interested reader to \Cref{thm:KO-locality-reduced} or \Cref{tab:aLLDC-info} and \Cref{tab:aLLDC-computation} for a summary of parameters.



When $\varepsilon$ is fixed and the message length $N$ tends to infinity, the codes in \Cref{thm:intro-intro-polylog} have vanishing rate $\frac{1}{\polylog(N)}$. Our second main result is an alternative HDX-based construction in this regime that achieves \textit{truly constant rate} and $\polylog(N)$ queries:
\begin{theorem}[Constant Rate aLLDCs (\Cref{cor:constant-rate-binary})]\label{thm:intro-intro-constant}
    $\forall \varepsilon>0$ and infinitely many $N \in \mathbb{N}$, there is a binary aLLDC $C_N$ decodable from $\frac{1}{2}-\varepsilon$ errors with:
\begin{enumerate}
    \item \textbf{Rate:} $\exp(-O(1/\varepsilon^3))$
    \item \textbf{List Size:} $\tilde{O}(\frac{1}{\varepsilon^2})$
    \item \textbf{Queries:} $\log(N)^{\exp(O(1/\varepsilon^3))}$
\end{enumerate}    
    Moreover $C_N$ is encodable and decodable in $\log(N)^{\exp(O(1/\varepsilon^3))}$-size and $\poly(\log\log N,\exp(O(1/\varepsilon^3)))$-depth.
\end{theorem}
Combined with known constructions of good local and parallel codes in the low noise regime, \Cref{thm:intro-intro-constant} essentially immediately implies good codes with sub-polynomial time list decoding, and good codes with RNC$^1$ list decoding (see \Cref{sec:apps-codes} below for formal statements).

To complete the picture, we prove a new lower bound showing any locally computable, rate $R$ aLLDC requires $\Omega(\frac{\log(N)}{\varepsilon\log \frac{1}{R}})$ queries (see \Cref{thm:lowerbound}), implying that our codes are within a polynomial gap of optimal (albeit a polynomial scaling poorly with $\varepsilon$ in the constant rate setting).\footnote{This is only an issue in the binary case --- for larger alphabets we build codes making $\polylog(N)\cdot\exp(O(1/\varepsilon))$ queries.} While the pure information-theoretic setting is not the main focus of our work, we remark we are able to match the lower bound up to low order factors in the polylog rate, large alphabet setting. Doing so with a corresponding computationally efficient decoding algorithm remains an interesting open problem.

\subsection{Applications in Complexity and Coding Theory}\label{sec:intro-applications}

Below we overview our main applications of \Cref{thm:intro-intro-polylog} (and its locality-reduced variant \Cref{thm:KO-locality-reduced}) to hardness amplification and PRGs (\Cref{sec:apps-PRG}), and of \Cref{thm:intro-intro-constant} to the construction of good locally list decodable and low-depth codes (\Cref{sec:apps-codes}). We start with the latter which require less explanation.

\subsubsection[Local and Parallel List-Decoding]{Local and Parallel List-Decoding}\label{sec:apps-codes}
Despite being only `approximately' list-decodable, our codes can easily be composed with good codes in the low noise regime to perform bona fide list decoding. In particular, due to the $\log(N)$ LDC query lower bound of Katz-Trevisan \cite{katz2000efficiency}, such aLLDCs indeed imply any good family of locally decodable codes correcting a constant fraction of errors in $q$ queries and $T$ time may be amplified to a family of LLDCs correcting up to $\frac{1}{2}-\varepsilon$ errors in $\poly(q)$ queries and $\poly(T)$ time.\footnote{Possibly over a larger alphabet. One must be slightly careful here if the base family does not have codes for every message length $N$. We handle this detail in the proof of \Cref{thm:intro-sub-Poly-LLDCs} below in \Cref{app:list-decoding}.} Applying this to the sub-polynomial time low noise LDCs of \cite{kopparty2017high} improves the best known LLDC decoding time from $\exp(\log(N)^{3/4})$ \cite{kopparty2018improved} down to $\exp(\log(N)^{1/2})$.\footnote{This comes at the cost of significantly worse rate in $\varepsilon$ compared to \cite{kopparty2017high,kopparty2018improved}, who focus mainly on optimizing rate rather than decoding time. We note it is plausible using their tools and aiming to optimize decoding time, one could instead derive similar guarantees to \Cref{thm:intro-sub-Poly-LLDCs}.}
\begin{corollary}[LLDCs in sub-Polynomial Time (\Cref{thm:sub-Poly-LLDCs})]\label{thm:intro-sub-Poly-LLDCs}
For every \(\varepsilon >0\) and infinitely many $N \in \mathbb{N}$ there is an explicit linear code $\mathcal{C}:\{0,1\}^N \to \{0,1\}^{N'}$ that is locally list-decodable from $\frac{1}{2}-\varepsilon$ errors with
\begin{enumerate}
  \item \textbf{Rate}:  \(\exp(-O(1/\varepsilon^3))\)
  \item \textbf{Query and Time Complexity:} $ \exp\left(O\left(\sqrt{\log N \cdot \log\log N}\right)\right)$
  \item \textbf{List Size:} $\leq \tilde{O}(\frac{1}{\varepsilon^2})$
\end{enumerate}
\end{corollary}

In fact we note these codes are locally \textit{correctable}, a stronger notion that corrects the codeword itself.

Moving to the parallel setting, composing our codes with Sipser and Spielman's \cite{sipser2002expander} famous expander codes gives the first good family of list decodable codes in RNC$^1$, breaking the $\log^2(N)$-depth barrier:

\begin{corollary}[List Decoding in RNC$^1$ (\Cref{thm:constant-rate-NC})]\label{thm:intro-constant-rate-NC}
For every $\varepsilon>0$ and infinitely many $N \in \mathbb{N}$ there is an explicit linear code $\mathcal{C}: \{0,1\}^N \to \{0,1\}^{N'}$ that is list decodable from $\frac{1}{2}-\varepsilon$ errors with
\begin{enumerate}
    \item \textbf{Rate:} $\exp(-O(1/\varepsilon^3))$
    \item \textbf{List Size:} $\tilde{O}(\frac{1}{\varepsilon^2})$
\end{enumerate}
Moreover $\mathcal{C}$ can be list decoded in RNC$^1$ by $O(N\log(N)^{\exp(O(1/\varepsilon^3)})$-size circuits.
\end{corollary}
These codes can be made either LDPC, or encodable in NC$^1$ depending on the base code used in the construction. We refer the reader to \Cref{app:list-decoding} for further discussion of this result and the model.
\subsubsection[Hardness Amplification and PRGs]{Hardness Amplification and Fast PRGs}\label{sec:apps-PRG}

For a complexity class \(\mathcal{C}\) we use the notation $f \notin {\rm Heur}_{\delta}$-$\mathcal{C}$  to mean $\forall g \in \mathcal{C}$, $\dist(f,g)>\delta$. Below, following \cite{karp1980some,trevisan2007pseudorandomness}, the class BPTIME$(T(n))/a(n)$ denotes the class of functions probabilistically computed by Time $T(n)$ probabilistic Turing machines given $a(n)$ bits of advice.

The standard locality-reduced variant\footnote{More accurately, both \Cref{thm:intro-intro-polylog} and this variant (\Cref{thm:KO-locality-reduced}) follow from concatenating the same large alphabet construction (\Cref{thm:polylog-aLLDCs}) with the appropriate binary inner codes. We do this in \Cref{sec:alphabet}.} of \Cref{thm:intro-intro-polylog} has the following immediate implications for semi-uniform hardness amplification:

\begin{theorem}[Uniform complexity-preserving hardness amplification (\Cref{thm:ucpha})]\label{thm:ucpha-intro}
Let $T(n)$ be a time complexity,   $1> \epsilon(n) \in 2^{-o(\sqrt{n})} , $ and $A$ a non-negative integer.  Let 
\begin{center}
$f: \{0,1\}^n \rightarrow \{0,1\}$ such that $f \not\in {\rm Heur}_{\frac{1}{n^A}}$-${\rm BPTIME}(T(n))/O(\log (n/\epsilon) )$
\end{center}
Then there is a
length $n' = n(1+o(1))$ and an $F: \{0,1\}^{n'} \rightarrow \{0,1\}$ so that 
\begin{center}
$F \in P^f $ and $F \not\in {\rm Heur}_{\frac{1}{2} - \epsilon(n)}$-${\rm BPTIME}(T(n) \poly (\frac{\epsilon(n)}{n}))$
\end{center}
\end{theorem}
Note that since $F \in P^f$, the above reduction is closed under polynomial time Turing reductions, and in particular allows for amplification within NP $\cap$ Co-NP, P$^{NP}$, etc (see \cite{impagliazzo2009approximate,impagliazzo2008uniform} for further discussion). \Cref{thm:ucpha-intro} is the first hardness amplification result that, even for constant $\varepsilon$, does not polynomially blow up the domain of the encoded function. We remark that the requirement $\epsilon(n) \in 2^{-o(\sqrt{n})}$ comes from the locality reduction step, and is present in all known complexity-preserving hardness amplification results. It would be very interesting to remove this limitation, but doing so is largely orthogonal to improving the domain size as a function of $n$, which is the focus of our work.

Using the same construction, we also get the following complexity-preserving hardness amplification for the standard non-uniform model:

\begin{theorem}[Non-uniform complexity preserving hardness amplification (\Cref{thm:nucpha})]\label{thm:nucpha-intro}
Let $S(n)$ be a circuit size function, $1> \epsilon(n) \in 2^{-o(\sqrt{n})}$, and $A$ a non-negative integer.  
\begin{center}
    $f: \{0,1\}^n \rightarrow \{0,1\}$ such that $f \not\in {\rm Heur}_{\frac{1}{n^A}}$-${\rm Size}(S(n))$.
\end{center}
Then there is a
length $n' = n(1+o(1))$ and an $F: \{0,1\}^{n'} \rightarrow \{0,1\}$ so that 
\begin{center}
    $F \in P^f $ and $F \not\in {\rm Heur}_{\frac{1}{2} - \epsilon(n)}$-${\rm Size}(S(n) \poly(\frac{\epsilon(n)}{n}))$
\end{center}
\end{theorem}
Again, \Cref{thm:nucpha-intro} is the first construction that avoids polynomial blowup in the domain \cite{impagliazzo1997p,impagliazzo2008uniform} or commensurate loss in the resulting circuit size \cite{sudan1999pseudorandom,chen2021simple}.

Besides its inherent interest, one of the most powerful applications of hardness amplification is to the construction of \textit{pseudorandom generators} (PRGs). PRGs are a central tool from cryptography and complexity theory that allow one to extend a small seed of true randomness to many `pseudorandom' bits that are indistinguishable from uniform by any efficient (non-uniform) algorithm. A function $G: \{0,1\}^r \to \{0,1\}^m$ is called a $(N,\varepsilon)$-PRG if for every circuit $C$ of size $N$:
    \[
    |\Pr[C(G(U_r))=1] - \Pr[C(U_m)=1]| < \varepsilon.
    \]
PRGs have several key parameters one may try to optimize. We focus on three central parameters in this work: the \textit{seed-length} $r$, number of \textit{output bits} $m$, and the \textit{running time} of $G$.

By plugging \Cref{thm:intro-intro-polylog} (or rather its direct non complexity-preserving hardness amplification translation) into the recent framework of Chen and Tell \cite{chen2021simple}, we immediately get the first construction of PRGs that is near-optimal in all three parameters under reasonable complexity theoretic and cryptographic assumptions:

\begin{theorem}[Near-Optimal Fast PRGs (\Cref{thm:PRG})]\label{thm:intro-PRG}
There exists an $\alpha_1>0$ so that for every $\alpha_0 < \alpha_1$ the following holds. Assume 
\begin{enumerate}
    \item There are one-way functions secure against polynomial sized circuits
    \item $f \in TIME(2^n)$ is not
in $i.o.$-$Size(2^{n(1- \alpha_0)}) $
\end{enumerate}
Then for all sufficiently large $N$ there is an $(N,1/4)$-pseudorandom generator 
\[
G: \{0,1\}^r \to \{0,1\}^N
\]
with seed length $r=(1+O(\alpha_0))\log N$ bits that is computable in time $N^{1+O(\alpha_0)}$
\end{theorem}

Taking $\alpha_0 \to 0$, we get PRGs on $N$ output bits approaching $\log(N)$ seed-length and linear time. In other words, under these assumptions, we prove any time $N$ randomized algorithm can be simulated using roughly $\log(N)$ truly random bits with almost no loss in running time, which is known to be optimal under NSETH \cite{williams2016strong}. Prior constructions of PRGs in this regime either suffered larger seed-length or \textit{quadratic} runtime.

\subsubsection{Summary of aLLDC Parameters}\label{sec:summary}
As discussed, we briefly summarize the main parameters of our aLLDC constructions in \Cref{tab:aLLDC-info} and \Cref{tab:aLLDC-computation} below without the assumption that $\varepsilon \leq \frac{1}{\log N}$, and including \Cref{thm:KO-locality-reduced}, a locality-reduced version of our code used for our complexity-preserving hardness amplification theorems above.

\begin{table}[ht!]
    \centering
    \caption{Summary of Information-Theoretic Parameters of our Binary aLLDCs}
    \label{tab:aLLDC-info}
    \begin{tabular}{@{}lccc@{}}
        \toprule
        \textbf{Result} & \textbf{Rate} & \textbf{Queries} & \textbf{List Size} \\ 
        \midrule
        Thm.~\ref{thm:intro-intro-polylog} (Cor.~\ref{cor:binary-polylog-aLLDCs}) 
        & $\poly(\tfrac{1}{\log N},\varepsilon)$ 
        & $\poly(\log N,\tfrac{1}{\varepsilon})$ 
        & $\tilde{O}(\tfrac{1}{\varepsilon^{2}})$ 
        \\[4pt]
        Cor.~\ref{thm:KO-locality-reduced} 
        & $\tfrac{1}{\polylog N}\cdot 2^{-O(\log^{2}(1/\varepsilon))}$ 
        & $\tfrac{\polylog(N,1/\varepsilon)}{\varepsilon^{2}}$ 
        & $\tilde{O}(\tfrac{1}{\varepsilon^{2}})$ 
        \\[4pt]
        Thm.~\ref{thm:intro-intro-constant} (Cor.~\ref{cor:constant-rate-binary}) 
        & $2^{-O(1/\varepsilon^3)}$ 
        & $\log(N)^{\exp(O(1/\varepsilon^3))}$ 
        & $\tilde{O}(\tfrac{1}{\varepsilon^{2}})$ 
        \\ 
        \bottomrule
    \end{tabular}
\end{table}
\begin{table}[ht!]
    \centering
    \caption{Summary of Computational Parameters of our Binary aLLDCs}
    \label{tab:aLLDC-computation}
    \begin{tabular}{@{}lccc@{}}
        \toprule
        \textbf{Result} & \textbf{Encoding Time} & \textbf{Decoding Size} & \textbf{Decoding Depth} \\ 
        \midrule
        Thm.~\ref{thm:intro-intro-polylog} (Cor.~\ref{cor:binary-polylog-aLLDCs}) 
        & $\poly(\log N,\tfrac{1}{\varepsilon})$ 
        & $\poly(\tfrac{1}{\varepsilon},\log N)$ 
        & $\tilde{O}\!\left(\log^{2}\!\left(\tfrac{\log N}{\varepsilon}\right)\right)$ 
        \\[4pt]
        Cor.~\ref{thm:KO-locality-reduced} 
        & $\polylog(N)\cdot \log^{2}\!\tfrac{1}{\varepsilon}$ 
        & $\poly(\tfrac{1}{\varepsilon},\log N)$ 
        & $O\!\left(\log\!\left(\tfrac{\log N}{\varepsilon}\right)\log\log N\right)$ 
        \\[4pt]
        Thm.~\ref{thm:intro-intro-constant} (Cor.~\ref{cor:constant-rate-binary}) 
        & $\log(N)^{\exp(O(1/\varepsilon^3))}$  & $\log(N)^{\exp(O(1/\varepsilon^3))}$  & $\exp(O(1/\varepsilon^3))\poly\log\log(N)$
        \\ 
        \bottomrule
    \end{tabular}

\end{table}

\subsection{Proof Overview}\label{sec:proof-overview}
In this section we give a proof overview of our two main results, \Cref{thm:intro-intro-polylog} and \Cref{thm:intro-intro-constant}.
The applications in hardness amplification, pseudorandom generators, and coding theory are all (in essence) derived from plugging these codes into known frameworks (plus some required bookkeeping). Although our main theorems are stated for binary alphabet, in this overview we describe codes with large alphabet. These can be reduced to binary alphabet via fairly standard concatenation techniques (see \Cref{sec:alphabet}).

\subsubsection[aLLDCs with Sub-Constant Error (Theorem 1.1)]{aLLDCs with Sub-Constant Error (\Cref{thm:intro-intro-polylog})}

All codes constructed in this work are based on the classic \textit{direct product} construction (sometimes called `ABNNR' in the coding literature after \cite{alon2002construction}). Let \((V,S)\) be an \(r\)-uniform hypergraph with vertices \(V=[N]\) and sets (hyperedges) \(S\) of size \(r\). The direct product code based on $(V,S)$ encodes strings of length \(N\) by viewing them as functions \(f:V \to \set{0,1}\), and writing down for each hyperedge $s \in S$ the value of $f$ on each vertex of $s$. In other words, the direct product code encodes \(f\) as a function \(\Enc^f:S \to \set{0,1}^r\) by
\[
\Enc^f(s)=f|_s,
\]
recording \(f(v)\) for all vertices \(v \in s\). Thus this code maps strings of length $|V|$ to strings of length $|S|r$ (more accurately, to strings of length $|S|$ with an $r$-bit alphabet), and therefore has rate $\frac{|V|}{|S|r}$.

In the context of direct product codes, we say a word \(w:S \to \set{0,1}^r\) is \((1-\varepsilon)\)-close to the codeword \(\Enc^f\) if there is a subset \(A \subseteq S\) of relative size \(\varepsilon\) such that \(w_s=f|_s\) for every \(s \in A\). Our goal is to locally decode such \(f\) while only querying \(\polylog (|V|)/\varepsilon\) symbols from \(w\). 

Fix $|V|=N$. Below, even though our theorem is general, for simplicity we recommend thinking of  \(\varepsilon = \frac{1}{2^{\sqrt{\log N}}}\) as quasi-polynomial, and \(r=\poly(\log N,\frac{1}{\varepsilon})\).

\paragraph{Reduction to List Recovery} Our aLLDCs in the subconstant regime start with a now standard reduction from high noise list-decoding to low-noise list-recovery of Guruswami and Indyk \cite{guruswami2001expander,guruswami2003linear}. This is followed by a new list-recovery decoder which is the main novel step in the subconstant error regime.

In the \emph{list recovery} problem, the decoder recieves a list \(\mathcal{L}_s\) of alphabet symbols for every index of the code \(s \in S\) (in function notation \(\mathcal{L}_s=\set{g_1,g_2\dots,g_{\ell}:s\to \set{0,1}}\). A function \(f:V\to \set{0,1}\) is \((1-\rho)\)-close to a list \(\set{\mathcal{L}_s}\), if \(f\) appears in a \(\rho\)-fraction of the lists. Like local list decoding, our goal in this problem is to list decode all functions that are \((1-\rho)\)-close to an input list, while querying the values of as few lists as possible. The key idea of the Guruswami-Indyk reduction is to reduce the problem of list decoding $(1-\varepsilon)$-close functions $f$ (that is those that are only computed by an $\varepsilon$ fraction of $S$) to a list recovery problem in which \textit{almost every list} contains $f$, i.e. list recovery of $o(1)$-close functions.


\paragraph{List recovery setup} Our main goal is now to perform list-recovery in the sub-constant noise regime. To be slightly more formal, in the local list recovery setup we need to build a (probabilistic) algorithm \(\mathcal{A}\) that outputs a list of circuits \(\set{C_1,C_2,\dots,C_\ell}\). These circuits take as input vertices \(v \in V\), query access to lists \(\mathcal{L}=\{\mathcal{L}_s\}_{s \in S}\) over $S$, and output \(b \in \set{0,1}\). Our goal is to show that for any \(f\) that is \((1-\rho)\)-close to \(\mathcal{L}\), one of the output circuits \(C_i\) computes a function \(0.01\)-close to \(f\) in normalized Hamming distance on \(V\). We remind the reader $\rho$ should be thought of as $1-o(1)$ in this setting, i.e.\ $f$ appears in almost all the lists.

Our generating algorithm for this task is extremely simple: we will sample a random \(s_0 \in S\) and output a circuit \(C_i^{s_0}\) for every function \(g_i \in \mathcal{L}_{s_0}\). We henceforth call \(C_i^{s_0}\) `the \(i\)-th decoder', or simply `the decoder'. The \(i\)-th decoder will implicitly assume that \(g_i\) is the restriction to $s_0$ of some global \(f:V\to \set{0,1}\) that is \((1-\rho)\)-close to \(\mathcal{L}\), and try to propagate this information to compute \(f\) across all of $V$. Since $\rho=1-o(1)$, any $f$ which is truly $(1-\rho)$-close to $\mathcal{L}$ will almost certainly be in the list of $s_0$, so if we can show the above type of decoder succeeds we will be done.




\paragraph{List Recovery via Belief Propagation} We have now reduced to the setting where we know the value of $f$ on $s_0$, and would like to determine the value of $f$ on a target vertex $v \in V$. We do this in two steps. First, we'll sample a random \(s_m \ni v\). By the same reasoning as before, we know almost surely that \(f \in \mathcal{L}_{s_m}\), so our real task is to identify \textit{which} of the functions in \(\mathcal{L}_{s_m}\) is (close to) \(f\) using its value on \(s_0\). Once we have identified such a \(g \in \mathcal{L}_{s_m}\) at least morally it is clear we may simply output \(g(v)\) as our decoding and be done.


Towards this end, we'd like to take advantage of the following basic agreement heuristic that has appeared in many prior works in approximate local list decoding, e.g.\ \cite{impagliazzo1997p,impagliazzo2009approximate,impagliazzo2008uniform}: if $s_0$ and $s_m$ happen to intersect and some function \(g \in \mathcal{L}_{s_m}\) is close to $f$ on their intersection (which can be checked, as we know the value of $f$ on $s_0$ by assumption), it is reasonable to suspect this local agreement on $s_0 \cap s_m$ stems from \textit{global} agreement across all of $s_m$. In other words, we might hope that if $g$ is close to $f$ on $s_m \cap s_0$, then $f \approx g$ on all of $s_m$. If this is indeed the case, we'd have \(g(v)=f(v)\) with high probability as desired.

Roughly speaking, to make this heuristic work formally, we must address two core problems. First, the heuristic could simply be false. More accurately, for a specific intersection \(s_0 \cap s_m\), it could be the case that \(f=g|_{s_0 \cap s_m}\) but \(f\ne g\) on all \(v \in s_m \setminus s_0\), in which case our decoding would clearly err. The second issue is that, especially in a high rate code, the intersection \(s_0 \cap s_m\) is almost certainly empty, so we have no direct way to check what function in $\mathcal{L}_{s_m}$ is close to $f$ based on $s_0$.

To deal with the first problem, we require our set system has a strong \emph{local sampling} property. To understand, imagine first the intersection \(s_0 \cap s_m\) is distributed as a uniformly random subset of \(s_m\). Then by a Chernoff bound, if \(f\) is far from \(g\) on $s_m$, then \(f\) is also far from \(g|_{s_0 \cap s_m}\) (i.e.\ on the random subset) with extremely high probability. Local sampling generalizes this requirement. We say \(s_m \in S\) is locally sampled if for any pair of functions \(f,g:s_m \to \set{0,1}\), the Hamming distance between \(f,g\) in \(s_m\) is close to the distance between \(f,g\) \emph{in most non-empty intersections \(s_0 \cap s_m\)} (over the randomness of $s_0$). This requirement is important already in \cite{impagliazzo2009approximate,impagliazzo2008uniform}, where it is used to build direct product codes with inverse-polynomial rate. However, our high rate constrains us to use very sparse set systems (where \(|S| \approx O(|V|)\)), where it is highly non-trivial to construct systems with this property. Thankfully, this is exactly the type of behavior one expects for systems arising from high dimensional expanders. We elaborate below.

\begin{figure}
    \centering
    \includegraphics[width=0.5\linewidth]{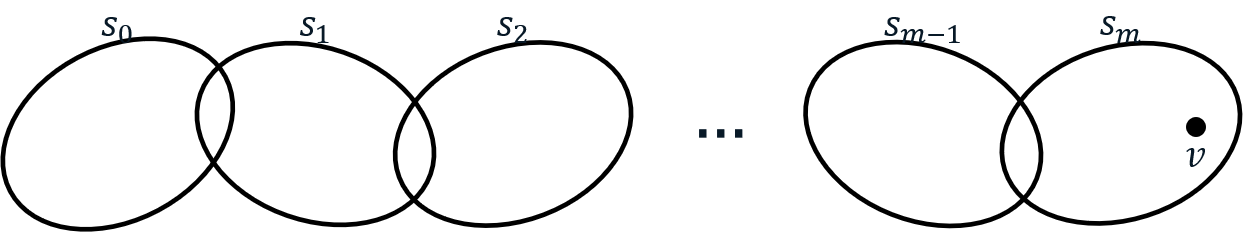}
    \caption{A Decoding Path}
    \label{fig:decoding-path}
\end{figure}

It is left to deal with the problem that $s_0$ and $s_m$ may not intersect. An elementary degree calculation shows one cannot simultaneously have good rate and expect many $v$ to have \textit{any} hyperedge $s_m \ni v$ that intersects $s_0$ (indeed most $v$ will be `$\log(N)$-far' from $s_0$ in the appropriate sense). To overcome this, we will work with hypergraph systems for which there always exist \emph{short intersection paths} of hyperedges \((s_0,s_1,\dots,s_m)\) of length at most $m \leq \polylog(N)$ such that every \(s_j\) intersects significantly with \(s_{j-1}\) (see \Cref{fig:decoding-path}). We may then propagate the value of $f$ on $s_0$ along the path inductively by finding for each $j \in [m]$ a function \(g_j \in \mathcal{L}_{s_j}\) that is close to \(f\) by comparing it the close function $g_{j-1}$ from the previous set over the intersection $s_{j-1} \cap s_j$, i.e.
\begin{enumerate}
    \item We start with \(f=g_{0} \in \mathcal{L}_{s_0}\).
    \item Given \(g_j:s_j \to \set{0,1}\) we find some \(g_{j+1} \in \mathcal{L}_{s_{j+1}}\) such that \(g_j \approx g_{j+1}\) on the intersection \(s_j \cap s_{j+1}\).\footnote{This algorithm is presented sequentially but can be performed in low-depth using Savitch's algorithm \cite{Savitch1970}.}
\end{enumerate} 
Here \(g_j \approx g_{j+1}\) just means the Hamming distance between the two functions is under some threshold. By the Triangle inequality and a basic induction, we have that if \(f \approx g_j\) and \(g_j \approx g_{j+1}\) on the \(s_j \cap s_{j+1}\), then \(f \approx g_{j+1}\) on \(s_j \cap s_{j+1}\). Finally, the local sampling property for \(s_{j+1}\) implies that \(f \approx g_{j+1}\) on all of \(s_{j+1}\) (with high probability), continuing propagation down the chain.

We note that even if \(g_0 = f\) on all of \(s_0\), we can still only hope for the approximate guarantee \(f \approx g_j\) in later steps of the algorithm (as local sampling will never imply perfect global agreement). Handling this error is a bit delicate, and the simple strategy described above actually doesn't work (identifying a unique $g_j$ in each round forces error to \textit{double} at each step\footnote{The reason is that because you may not select the closest function to $f$, in each step you must double the error threshold you allow to ensure you can actually find a close function in the next step.}). Thankfully, this can be fixed by maintaining a \textit{sub-list} of functions at each $s_j$ close to $f$, and recording at step $j+1$ the subset of all functions in $\mathcal{L}_{s_{j+1}}$ close to any function in the prior sub-list (starting again just at $g_0=f$). A more careful analysis shows that error in the sub-list only grows \textit{linearly} (see \Cref{sec:subpoly-list}). Thus, as long as we can keep our local sampling error below $\frac{1}{\polylog(N)}$, our final decoding after $m \leq \polylog(N)$ steps will almost surely succeed.

\paragraph{Strongly Explicit Low Congestion Routing}

While this strategy sounds appealing, a hidden key technical difficulty remains. In order to decode efficiently, we need not only that there \textit{exist} short paths between any pair of hyperedges $s_0,s_m$, but moreover that they can be generated efficiently in \textit{polylogarithmic time}. Furthermore, the paths need to be short (since error grows linearly with the path length), and must have `\textit{low congestion}' in the following sense. Even under our sampling guarantees there is still some probability that an intersection \(s_j \cap s_{j+1}\) will not measure well the distance between \(f\) and \(g_j\). We have no control over this set of bad edges $(s_j,s_{j+1})$ other than the fact that it is small (due to the assumed local sampling). As such, we need to ensure the paths we output do not hit any small set of edges with large probability. 

We formalize this through a new algorithmic primitive we call strongly explicit (or `local') low congestion routing (\Cref{def:robust-routing-scheme}). Roughly, a graph $G=(V,E)$ admits a strongly explicit low congestion routing scheme if there exists a (randomized) algorithm $\mathcal{O}$ which on input any vertex pair $v,v'$, outputs a path $P=(v=v_1,\ldots,v_m=v')$ of length at most $m \leq \polylog(|V|)$ in time $\polylog(|V|)$ such that for any $B \subset E$
\[
\Pr_{v,v'}[\mathcal{O}(v,v')~\text{hits $B$}] \leq \poly(m,\Pr[B])
\]
where $v,v'$ are i.i.d random vertices from $G$. In other words, the set of paths output by $\mathcal{O}$ does not concentrate on any small set of edges (hence the name `low congestion' which is commonly used to refer to variants of this property throughout the literature).\footnote{In reality, we give local routing schemes with the stronger guarantee $\Pr_{v,v'}[\mathcal{O}(v,v')~\text{hits $B$}] \leq m\Pr[B]$, but this is not necessary for list-recovery.}

Our requirement above then corresponds to constructing a local routing scheme for the graph $G$ whose vertices are the hyperedges $S$, and whose edges are given by hyperedges with non-trivial intersection.\footnote{In reality, we will only include a natural subset of these edges based on the hypergraph structure.}

\paragraph{Summary of Key Properties} Summarizing all the above, we need a set system \((V,S)\) with three key properties for this decoding strategy to work:
\begin{enumerate}
    \item Local sampling properties to infer closeness to \(f\) from closeness on the intersection.
    \item A polylog-time low congestion routing scheme for intersecting hyperedges of $(V,S)$
    \item For high rate and good locality:
    \begin{itemize}
        \item $|S| \leq N\poly(\log(N,\frac{1}{\varepsilon}))$, and
        \item Each $s \in S$ should have at most \(\poly(\log N,\frac{1}{\varepsilon})\) vertices.
    \end{itemize}
\end{enumerate}
How can one construct such a set system?

\paragraph{HDX to the Rescue} We finally turn to the theory of high dimensional expanders (HDX), powerful hypergraph analogs of their low dimensional cousins that have played a major role in recent advances and coding theoretic breakthroughs including high rate (non-local) list decodable codes \cite{dinur2021list}, locally testable direct product codes \cite{dinur2017high, dikstein2019agreement,dikstein2024low, bafna2024constant}, and even efficient PCPs \cite{bafna2024quasi}. In a nutshell, HDX are hypergraphs where neighborhoods of vertices and hyperedges are themselves small expander graphs. 

As sparse, well-connected hypergraphs, HDX are a natural choice when aiming for a high-rate local codes. Indeed, it was already observed in early work of \cite{dinur2017high} that variants of the local sampling property we need arise naturally in HDX. Interestingly, a variant of low congestion routing also appeared in the recent work of \cite{bafna2024constant} on efficient PCPs, but the setting is quite different (see \Cref{sec:related}). In particular \cite{bafna2024constant}'s schemes spend polynomial time and processes the entire HDX. This is useless for our setting, where the entire point is to decode in sub-linear (indeed even polylogarithmic time). Our local routing scheme uses entirely different techniques and may be of independent interest to the HDX and algorithms communities.

The set system we derive from HDX for our codes differ from the one in \cite{dinur2017high} and other prior works, where $(V,S)$ consist of vertices and hyperedges of the HDX respectively. Instead, in the subconstant error setting we rely on \(2\)-dimensional HDX (i.e.\ high dimensional expanders whose hyperedges are \(3\)-uniform), and the sets $S$ of our system are not the hyperedges of the HDX themselves, but rather \textit{vertex neighborhoods}. As a result, the intersection path \(s_0,s_1,\dots,s_m\) actually corresponds to a path in the underlying graph of the HDX rather than a high dimensional random walk. This seemingly minor change is critical, as the natural hyperedge encoding used in prior works cannot support the $\frac{1}{\polylog(N)}$-strength local sampling we need for list recovery.

Finally, we construct local low congestion routing schemes by leveraging the strong algebraic symmetries of the recent spectral HDX construction of \cite{dePeraltaVB2025high} (a variant on the `elementary' construction of \cite{kaufman2018construction}).\footnote{We remark the construction of \cite{kaufman2018construction} works just as well. We work with \cite{dePeraltaVB2025high} since their local properties are simpler to describe.} In particular, we reduce the routing problem on these HDX to giving short solutions to the group theoretic word problem (i.e.\ constructing arbitrary group elements $g$ from a specified set of simple generators) on $SL_d(\mathbb{F}_{\fsize^{\poldeg}})$, and give a $\polylog(N)$-step solution. En route, we also show these HDX (and their corresponding routing) can be implemented by small, low-depth circuits.

\subsubsection[aLLDCs with Constant Rate (Theorem 1.2)]{aLLDCs with Constant Rate (\Cref{thm:intro-intro-constant})}

Our constant rate construction also uses belief propagation over an appropriate direct product code, but a central issue arises: since our list-recovery algorithm inherently requires $\frac{1}{\poly\log(N)}$-strength sampling, we must instantiate our framework with a $\polylog(N)$ degree HDX causing a $\polylog(N)$ loss in the rate of the encoding.\footnote{A secondary issue is that the simple reduction to list-recovery we use in the previous section also loses $\polylog(N)$ factors in rate, but this can likely be side-stepped using a more complicated AEL-based reduction as in \cite{gopi2018locally}. Since we will already perform a `lossless' reduction to list-recovery in our solution to the core sampling problem, there is no need for this approach.} Can we propagate information for $\polylog(N)$-steps on a \textit{bounded degree} object where every vertex sits in only $O_{\varepsilon}(1)$ hyperedges?

\paragraph{List Recovery via Local Codes} Inspired by the `double sampler' method of \cite{dinur2017high,dinur2019list} (c.f.\ \Cref{sec:related}), we answer this question by introducing a \textit{third layer} $U$ of hyperedges (the `local codes') in such a way that we may reduce list-decoding on $(V,S)$ (which is bounded degree) to list-recovery over $(V,U)$ (which can be $\polylog(N)$-degree without adversely affecting the rate). In particular, we will find a bounded degree system $(V,S)$ and a set of hyperedges $U$ such that the `local view' of every $u \in U$ in $(V,S)$, i.e.\ the intersections \(\set{u \cap s}_{s \in S}\), have a similar structure to a complete hypergraph. Using ideas in \cite{impagliazzo2008uniform}, it is relatively straightforward to approximately list decode each such local code, allowing us to simulate lists \(\mathcal{L}_u\) for each $u \in U$. If we can ensure that $U$ `globally samples' $S$ in the sense that almost all inner codes $u$ see many sets $s$ that computing $f$, we will end up with a sub-constant noise list-recovery problem on the $\polylog(N)$-degree system $(V,U)$, over which we may reasonably hope to apply our original propagation algorithm.

Even putting aside the existence of a triple system $(V,S,U)$ satisfying the above, however, there is a problem with this approach: due to information theoretic barriers, we cannot guarantee the lists \(\mathcal{L}_u\) contain our global function \(f\) exactly, only a function \(f'\) that is \(\delta\)-close to \(f\) for some \(\delta\) polynomially related to \(1/\text{rate}\). Since we are interested in maintaining \textit{constant} rate, plugging in these lists in the previous propagation algorithm results in \(m\delta \gg 1\) normalized distance between $f$ and the final selected function \(g_m\) after our $m$-step propagation, rendering the result completely useless since $m \geq \Omega(\log(N))$.

To overcome this barrier, we appeal to a variant of a list-processing procedure by \cite{dinur2019list} that ensures functions in our local lists $\mathcal{L}_u$ are \textit{well-separated}, while still containing a function close to $f$. This preprocessing step is unfortunately expensive, and may result in the internal error \(\delta\) growing to  \(\delta' = \delta\exp(O(\frac{1}{\varepsilon}))\).\footnote{Note this blowup means we cannot use such pre-processing in the proof of \Cref{thm:intro-intro-polylog} where \(\varepsilon\leq \frac{1}{\log N}\), as we'd end up back at $\poly(\delta'^{-1}) \leq \frac{1}{\poly(N)}$-rate, similar to prior constructions.} However, after this initial preprocessing, we will have the property that if \(g_k\) is \(2\delta'\)-close to \(g_{k+1}\) it will be \(6\delta'\)-far from every other function in the list. As a result, in the propagation algorithm we may \textit{uniquely} match $g_k$ to the next closest $g_{k+1}$ to $f$, and don't incur any additional error at all. The argument is similar to that in \cite{dinur2019list}, so we omit it in this overview.

\paragraph{Fault Tolerant Strongly Explicit Routing}

Even now, there is a final core difficulty brushed under the rug. In reality, for technical reasons we cannot pre-process all sets in our system in this way, only a large constant fraction (say $99\%$). This leaves $1\%$ of local codes which, if used, would still incur untenable error in our propagation. Furthermore, if one were to just route from \(u_0\) to some random \(u_m\), we will almost certainly encounter a (indeed many) $u_i$ where pre-processing failed, killing our propagation. To overcome this, we significantly strengthen our routing scheme to be \emph{tolerant to vertex faults}. In particular, given a (large enough) arbitrary subset $T \subset V$, a vertex fault-tolerant (`adaptive') routing scheme must output paths \(u_0,u_1,\dots,u_m\) such that each internal node $u_i \in T$ and maintain low congestion. Here, $T$ should be thought of as the $99\%$ of nodes on which our pre-processing succeeds, so routing entirely through $T$ allows us to perform the `lossless' list-recovery described above. Due to a notational collision with prior work on a different notion of fault-tolerant routing (e.g.\ \cite{bafna2024quasi,bafna2025constant}), we will call the above `subset internal routing' in the main body of the paper. Our subset internal routing scheme is very general, and can efficiently route inside any big enough set of vertices or hyperedges in \cite{dePeraltaVB2025high}'s HDX (up to some sub-constant failure probability).

\paragraph{Summary of Key Properties}

In summary, to construct these codes we now need a set system which (roughly speaking) has all of the previous properties and additionally:
\begin{enumerate}
    \item An extra layer \(U\) such that
    \begin{itemize}
        \item The intersections of $U$ with $(V,S)$ are (similar to) complete hypergraphs
        \item $U$ `samples' $S$ except with probability $\frac{1}{\polylog(N)}$ (to ensure $f$ is in most lists)
    \end{itemize}
    \item A \emph{fault-tolerant} polylog-time routing scheme with low-congestion.
    \item To ensure constant rate and $\polylog(N)$ log-alphabet size:
    \begin{itemize}
        \item Each $v \in V$ sits in at most $O_{\varepsilon}(1)$ vertices
        \item Every $s$ contains at most $\polylog(N)$ vertices
    \end{itemize}
\end{enumerate}
At outset, it is not clear whether such a system where $(V,S)$ is bounded degree in the above sense should exist. Indeed it is impossible to construct such a 3-layer system based on hyperedges of increasing arity used in prior works, as super constant strength sampling implies super constant degree (nor can we directly use the vertex neighborhood encoding of the prior section which suffers a similar issue). 

We sidestep this barrier by using HDX whose dimension $d$ (i.e.\ maximal set size) grows exponentially with \(\frac{1}{\varepsilon}\), and, somewhat perversely, encoding message bits on \textit{hyperedges} of the complex while sets in the system correspond to \textit{vertices} (namely each vertex records the bit value of every hyperedge containing it). This encoding ensures the rate scales like $\frac{1}{d}$, a constant. We then pick the `third layer' sets $u \in U$ to be $\polylog(|V|)$-size overlapping neighborhoods in the HDX, similar to the neighborhood encoding used in \Cref{thm:intro-intro-polylog}, in such a way that strong sampling is implied directly by high dimensional expansion, while local `complete' structure follows from the strong local independence properties of \cite{dePeraltaVB2025high}'s HDX construction. The latter structure, discussed in \Cref{subsec:defining-KMS}, is also key to making local routing fault-tolerant.

It is an interesting question whether our framework can be instantiated on \(\poly(1/\varepsilon)\)-dimensional HDX to improve the corresponding rate.

\subsection{Related Work}\label{sec:related}

\paragraph{Codes, Local Testing, and List Decoding on HDX}

There has been a great deal of work in the past decade developing a theory of error correcting codes on high dimensional expanders \cite{dinur2017high,gotlib2022list,dikstein2023agreement,bafna2024characterizing,kaufman2011edge,kaufman2014high,dikstein2019agreement,dikstein2020locally,dinur2023new,breuckmann2021balanced,first2022good,dikstein2024chernoff,dikstein2024low,bafna2024constant}, mostly focused on the topic of \textit{local testability},\footnote{Also, relatedly, on quantum LDPC codes (see \cite{EvraKZ20,KaufmanT21quantum,panteleev2021asymptotically,leverrier2022quantum,gulshen2025quantum} among others).} eventually leading to the resolution of the c$^3$-LTC conjecture \cite{dinur2021,panteleev2021asymptotically} and construction of quasi-linear PCPs with low soundness \cite{bafna2024quasi}. These works, especially those in the `low soundness' testing regime, are related to our problem in the sense that their local tests often implicitly rely on list-decoding. However, despite the tests in these works being local, the corresponding list-decoding algorithms are not. In particular, the typical strategy is to reduce the problem to solving an expanding unique games instance across the entire encoding. Since we do not know how to solve unique games locally, even on expanders, there is no clear path towards making these algorithms local.

Global list decoding on HDX is more explicitly addressed in \cite{dinur2019list}, where the authors analyze the natural direct product construction where message bits sit on the vertices of a $k$-dimensional HDX and one takes the direct product over hyperedges $X(i)$ for some $i \ll k$. Similar to our hypergraph triples, the authors' list-decoding algorithm then takes advantage of this `three-layer' system $(V,X(i),X(k))$ which forms a so-called \textit{double sampler} $(V,S,U)$ such that the `local view' of every $u \in U$ is \textit{complete}, which also allows them to build a list of candidates on each $u \in U$, then use the global expansion/overlap structure of $X(k)$ to list-decode via solving a corresponding unique game.\footnote{We note this itself is a variant of an idea from \cite{dinur2017high}, which uses the same local view structure for agreement testing.} Later works \cite{alev2020list,jeronimo2021near,dikstein2024chernoff} substantially improved on this method using SDPs and regularity lemmas, removing the need for the double sampler.

Our hypergraph triple systems in the constant rate regime, and in particular the idea of moving to an object with built in `local codes', is obviously inspired by the double-sampler method though we require substantially more delicate properties of our system (due, e.g., to the need for keeping $(V,S)$ constant degree while ensuring various other local and global sampling graphs related to $U$ have \textit{sub-constant} failure). This precludes us from using the `natural' hyperedge encoding on HDX, which even using the rest of our new machinery, seems to achieve at best rate $\exp(-\poly\log\log(N)\exp(O(1/\varepsilon)))$, a far cry from the parameters needed to resolve any of the major questions in this work.

\paragraph{Routing and PCPs} As alluded to above, recent work \cite{bafna2024quasi}, building on the development of local testability (`agreement testing') of the hyperedge direct product encoding on certain special HDX \cite{gotlib2022list,dikstein2023agreement,bafna2024characterizing,dikstein2024low,bafna2024constant}, gave the first low-soundness quasi-linear size PCPs. This also corresponds to a sort of size-efficient `hardness amplification', but the challenges in the setting are quite different than the ones we address. For instance, we either work with \textit{extremely sub-constant} $\varepsilon$, or require \textit{constant degree} encodings, while \cite{bafna2024quasi} works with constant $\varepsilon$ and are happy to lose polylog factors in $N$ which changes the problem quite a bit. More obviously, \cite{bafna2024quasi} need local testing, while we need local decoding. These turn out to be inherently quite different: for instance local testing actually requires co-boundary expansion \cite{dikstein2023agreement,bafna2024characterizing} (a very strong `topological' notion of expansion). We do not require this, but need our complexes to satisfy various other properties like the existence of local routing schemes.

On that note, we remark \cite{bafna2024quasi} also use a related notion of routing they call a `fault-tolerant routing protocol'. Again the challenges here are fairly orthogonal, even though the constructions share some of the same HDX tools. Roughly speaking, the goal in fault tolerant routing is to build a protocol in which every vertex sends a message to every other vertex even when some fraction of the edges are corrupted with no individual vertex doing too much computation. The only other constraint is the protocol should be computable in $\poly(N)$ time. This is fairly different from our setting in which we only compute a set of paths, but we must do so in $\polylog(N)$ time. While we do need a version that is tolerant to vertex faults, our scheme \textit{knows} the corruptions, which makes the problem fairly different from \cite{bafna2024quasi}.\yotamd{Maybe add an open research direction if we have an interesting one?}

\paragraph{High Rate Local List Decoding}

The last decade has seen exciting progress on the construction of locally (list) decodable codes with high rate \cite{kopparty2010local,kopparty2014high,guo2013new, guo2015high,guo2016list, hemenway2015local}, culminating in the sub-polynomial time codes of \cite{kopparty2017high} in the low noise regime, and later improvements and extensions to the list decoding regime \cite{hemenway2019local,kopparty2018improved,gopi2018locally} (see \Cref{app:LLCC} for further discussion). These latter works also rely on combinatorial distance amplification, in particular the so-called AEL construction \cite{alon1995linear}, relying on the fact that their base code (a sub-constant distance multiplicity code) is \textit{list-recoverable} to show local list-decodability of the final code. Our methods remove this constraint and improve the query complexity and running time, but come at the cost of significantly worse rate in terms of $\varepsilon$. Indeed in the large alphabet setting, \cite{kopparty2018improved} achieve sub-polynomial time local list decoding up to information-theoretic capacity (the maximum rate with respect to the error correction radius). It would be very interesting if our techniques could be adapted to hold in this stronger setting.



Finally, it is worth noting that some instantiations of expander codes themselves are known to be locally list correctable in the low noise regime in time $N^{\alpha}$ (i.e.\ sublinear but not sub-polynomial time). In some moral sense our result `upgrades' this to sub-polynomial time decoding in the \textit{high noise} regime using \textit{high dimensional} expanders, though while there are some similarities (use of inner codes, paths) it is not clear the methods are much related beyond this.

\paragraph{Pseudorandom Generators} There is a great deal of work on the construction of pseudorandom generators from various average and worst case complexity theoretic or cryptographic assumptions \cite{yao1982theory,nisan1994hardness,impagliazzo1997p,blum2019generate,aszl1993bpp,haastad1999pseudorandom,umans2002pseudo,shaltiel2005simple}. Quite recently two exciting works \cite{doron2022nearly} (under stronger non-deterministic hardness assumptions) and \cite{chen2021simple} (assuming one way functions) gave the first such PRGs with near-optimal $(1+\alpha)\log(N)$ seedlength, and as a result showed how to de-randomize BPP under these assumptions with only a quadratic blowup in runtime.\footnote{Under slightly stronger assumptions \cite{chen2021simple} even give a finer-grained result scaling with how $N$ depends on the input size that can be better than quadratic.}

Our pseudorandom generators are strongly inspired by \cite{chen2021simple}. Indeed the construction is exactly the same trading out the use of (concatenated) Reed-Muller codes with our aLLDC families. By avoiding the use of Reed-Muller, we avoid having to compute the entire truth table of our initial hard function to compute even a single input to $G$, which is what caused prior constuctions to run in quadratic time. Instead, our aLLDC only queries a small fraction of $f$'s values, and therefore can run in near-linear time. We note that while one might think as a result we'd have to start from average case hardness, we avoid this due to the starting worst-case assumption being for very large size circuits (the same assumption as in \cite{chen2021simple}), in which regime strong enough average-case and worst-case hardness are essentially equivalent. We note this improvement is most relevant in settings where one wants to use a PRG simply to reduce the number of required truly random bits rather than to fully derandomize. The latter requires running over all $(1+\alpha)\log(N)$ seeds and therefore requires near-quadratic total runtime, which is necessary but can also be achieved using Reed-Muller codes.
\subsection*{Roadmap}
We provide a brief roadmap of the main body of the paper. In \Cref{sec:preliminaries} we cover general preliminaries. In \Cref{sec:subpoly}, we introduce propagation decoding and local routing schemes, and give a simple sub-polynomial rate construction using subspaces. In \Cref{sec:polylog}, we instantiate this framework assuming sparser hypergraph systems to prove \Cref{thm:intro-intro-polylog}. In \Cref{sec:constant}, we introduce our hypergraph triple framework for constant rate codes and prove \Cref{thm:intro-intro-constant} assuming appropriate hypergraph systems. In \Cref{sec:instantiations} we construct the assumed hypergraph systems and routing schemes for both cases using high dimensional expanders. In \Cref{sec:lower}, we give query lower bounds for aLLDCs showing variants of our codes are near the information-theoretic optimum.
Finally, direct applications of our aLLDCs to hardness amplification, PRGs, and coding theory (along with further discussion on these problems) are given in \Cref{app:hardness} and \Cref{app:list-decoding} respectively.

\section{Preliminaries} \label{sec:preliminaries}

\subsection{Approximate Locally List Decodable Codes}\label{sec:apx-local-list}
\begin{definition}
    An error correcting code ${\cal C}: \Sigma_1^{\N} \rightarrow \Sigma_2^{\M}$ is a mapping from messages $f \in \Sigma_1^{\N}$ to codewords $w \in \Sigma_2^{\M}$.
    \begin{itemize}
    \item A locally computable code ${\cal C}$ is equipped with a  local encoding function
    $\Enc^f: [\M] \times [\log |\Sigma_2|] \rightarrow \{0,1\}$, that has oracle access to a message $f:[N] \to \Sigma_1$. On input $s \in [\M]$ and an index $i \in [\log |\Sigma_2|]$, $\Enc^f(s,i) \in \{0,1\}$ outputs the $i$-th bit of $\mathcal{C}(f)_s$, the value of the  $s^{th}$ coordinate of the codeword corresponding to message $f$.
    \end{itemize} 
\end{definition}
Unless otherwise specified, we will always think of the input alphabet $\Sigma_1$ as constant size. The \textit{distance} between two strings, $dist(w,w')$ is the normalized hamming distance between them, and the \textit{distance} of $\mathcal{C}$ is the minimum distance between any two codewords.


In this work we will be interested in the ``high noise'' decoding regime, where we can only guarantee list-decoding; 
additionally, we will further relax list-decoding to approximate list-decoding:


\begin{definition}[Approximate Locally List-Decodable Codes (aLLDCs)]\label{def:aLLDC}
For code ${\cal C}: \Sigma_1^{\N} \to \Sigma_2^{\M}$, let $\Ball_{1-\varepsilon}(w)$ denote the set of codewords 
within distance $1-\varepsilon$ of $w$.
    $\mathcal{C}$  
    is $\delta$-approximately, $(\varepsilon,\ell,q)$-locally list decodable if there exists a probabilistic TM $\mathcal{A}$ which on input $1^N$ outputs (probabilistic) circuits $C_1,\ldots,C_\ell$ such that
    \[
    \forall w \in \Sigma_2^{N'}, \forall f \in \Ball_{1-\varepsilon}(w): \Pr_{{\cal A}}\left[\exists i: \Pr_{j \in [\N], r}[{C_i}^{w,r}(j) = f(j)] \geq 1-\delta \right] \geq \frac{3}{4}.
    \]
    Moreover, every \(C_i\) queries \(w\) on at most \(q\) locations.
\end{definition}
We remark that in the main body, the reader will often see an extra parameter $\delta_{in}$ appearing. This is a strengthened notion that allows for only approximate agreement within alphabet symbols, and we define it only for direct product codes. See \Cref{def:doubly-apx} for this full definition.

Note we have also, for simplicity, relaxed the requirement that the output circuit list cover \textit{every} $f$ such that $\Enc^f$ is $(1-\epsilon)$-close to $w$ simultaneously, to
 just covering each \textit{fixed} element $f$ such that $\Enc^f$ is $(1-\epsilon)$-close to $w$, with high probability.
 This change is just for convenience; in our constructions, one can always move between this and the usual `stronger' definition covering \textit{all} functions just by running the decoder enough times to amplify the success probability such that we can union bound over the small combinatorial list size. In this vein, we also note it is sometimes useful to have a slightly stronger confidence guarantee over the internal randomness of the circuits $C$ and the generating algorithm $A$, namely that
    \[
    \forall w \in \Sigma_2^{n}, \forall f \in \Ball_{1-\varepsilon}(w): \Pr_{\cal A}\left[\exists i: \Pr_{r}[{C_i}^{w,r}\text{ $\delta$-computes $f$}] \geq 1-\tau' \right] \geq 1-\tau.
    \]
In other words, $C_i$ computes $f$ on almost every setting of the randomness, rather than in expectation over $r$, and with high (rather than $\frac{3}{4}$) probability. It is easy to move to this notion in two steps: 1) define new circuits $C'_i$ which takes the majority output over $O(\log \frac{1}{\tau'})$ independent runs of $C_i(\cdot,r)$ (a standard Markov+Chernoff argument ensures that the result $O(\delta)$ computes $f$ except with probability $(1-\tau'$), and 2) repeat the algorithm outputting these circuits $O(\log \frac{1}{\tau})$ times to ensure one of them indeed computes $f$ with high probability.

\bigskip

\noindent{\bf Locality of Encoding and Decoding.}
As discussed in the introduction, crucial to this work is the circuit complexity of encoding and decoding, with respect to both circuit size and depth, as well as the encoding an decoding locality. We denote by $q$ the decoding locality (the number of queries to decode) and by $r$ the encoding locality (the number of message coordinates each code coordinate depends on). We remark that in most circumstances, it will be simpler to just write the encoding \textit{time}. On top of the locality $r$, this also includes the time to compute which $r$ message bits to query (then computing the corresponding output based on this). When necessary we will record both the encoding time and the encoding locality, but when possible only record the former for simplicity which upper bounds the latter.

\subsection{Direct Products and Circuit Complexity}\label{sec:direct-products}

Essentially all constructions of approximately locally list decodable codes are built via the so-called \textit{direct product} construction. Direct product codes are based on an underlying \textit{hypergraph} $H=(V,S)$ where $V=[N]$ is the vertex set and $S \subset \binom{[N]}{r}$ is an ($r$-uniform) set system (family of hyperedges). A hypergraph is \textit{regular} if every vertex $v$ appears in the same number of hyperedges in $S$. All hypergraphs used in this work for encoding will be regular.

The direct product code corresponding to $(V,S)$, 
denoted $\mathcal{C}_{(V,S)}$, encodes a message $f: V \to \Sigma_1$ to a codeword $\Enc^f: S \to (\Sigma_1)^r$ by concatenating the symbols along every hyperedge, that is:
\[
\Enc^f((v_1,\ldots,v_r)) = (f(v_1),\ldots,f(v_r))
\]

where we have fixed some arbitrary ordering of the hyperedge $s=\{v_1,\ldots,v_r\}$. Equivalently, one can think of $\Enc^f$ as a family of assignments $\{\Enc^f_s: s \to \Sigma_1\}_{s \in S}$ where $\Enc^f_s(v_i)=f(v_i)$.

In this work, we consider direct product codes that are equipped an ancillary \textit{decoding graph} $G$, typically given by some natural intersection structure over hyperedges. We call the collection of a hypergraph and corresponding decoding graph a `hypergraph system':
\begin{definition}[Hypergraph System]
    A decoding system $(V,S,G)$ consists of a hypergraph $(V,S)$, and a `decoding graph' on hyperedges $G=(S,E)$
\end{definition}
Hypergraph systems also come with a natural family of \textit{intersection graphs} for each $s \in S$ that will be critical to our decoding procedure. Given $(V,S,G)$ and $s \in S$, the intersection graph \(G_s = (L,R,E)\) is the bipartite graph where $R=s$, and $L$ is indexed by the set of possible intersections of $s$ with neighboring $s' \in G$, that is
\[
L \coloneqq \{s \cap s': (s,s') \in E\}
\]
$G_s$ is the inclusion graph of this system. More formally, we generate edges of the graph via
the following probabilistic experiment:
\begin{enumerate}
    \item Sample a neighboring $s' \sim \pi_E(s)$ and write $t=s \cap s' \in L$
    \item Sample \(v \in t\) uniformly at random and output the edge \(\set{v,t}\).
\end{enumerate}
We say that \(G_s\) is `uniform' if the resulting marginal distribution over $R=s$ is uniform. We call a hypergraph system \textit{regular} if $(V,S)$ and $G$ are regular, and all $G_s$ are uniform. All hypergraph systems considered in this work are regular.

Since the above definitions are a bit abstract, let's consider a basic example for concreteness. 
\begin{example}[Complete Complex]
    Let $V = [\N]$ and  $S=\binom{[N]}{r}$ be the complete hypergraph. A `typical' choice of decoding graph would be to connect hyperedges that intersect in exactly half their vertices (the resulting graph is usually refered to as a `Johnson graph'). Given a fixed $s \in \binom{[N]}{r}$, the intersection graph $G_s$ is just the `inclusion graph' with $L=[s]$, $R=\binom{[s]}{r/2}$ (i.e.\ each $r/2$-set in $R$ is connected to each vertex in $L$ it contains). This is because we move from $s$ to a neighboring $s'$ in the decoding graph by picking a random half of $s$ to intersect on, then $r/2$ new vertices to include in $s'$ at random, so the intersection with a random neighbor is just $r/2$ random vertices from $s$.
\end{example}

\paragraph{Circuit Implementation of $(V,S)$.} 
The complexity of codes based on a hypergraph system scale with the complexity of constructing the hypergraph and decoding graph itself. In particular, we will typically need circuit implementations of the adjacency operators of the hypergraph (more accurately, adjacency circuits of the bipartite inclusion graph of $(V,S)$). More explicitly, we will need circuits
\begin{enumerate}
    \item $C_{V\to S}(v,i)$: outputs the $i$-th neighbor in $S$
    \item $C_{S\to V}(s,i)$: outputs the $i$-th neighbor in $V$
\end{enumerate}
for some instance-specific ordering of the neighbors. Note here that $v$ and $s$ are given by a corresponding bit-representation in $\{0,1\}^{\log V}$ and $\{0,1\}^{\log S}$ respectively.

Somewhat less obviously, we will also need circuits which given neighboring $(v,s)$ in our system, output \textit{which} neighbor $v$ is with respect to $s$ and vice-versa. We call these the `reverse indexing' circuits:
\begin{enumerate}
    \item $C_{V\to S}^{-1}(v,s)$: outputs $i$ such that $C_{V\to S}(v,i)=s$
    \item $C_{S \to V}^{-1}(s,v)$: outputs $i$ such that $C_{S\to V}(s,i)=v$
\end{enumerate}

Finally, our decoding algorithm will also need a circuit which takes in two hyperedges $s,s'$ and computes an element of their intersection. 
\begin{enumerate}
    \item $C_{S \cap S \to V}(s,s',i)$: outputs the \(i\)-th element $v \in s \cap s'$, or `FAIL' if $s \cap s' = \emptyset$
\end{enumerate}


\paragraph{Encoding and Decoding Complexity:} Finally, given the above, we briefly formalize encoding and decoding direct product codes in the circuit model. An \textit{encoding circuit} for $C_{(V,S)}$ is a circuit $C_{Enc}(s,i)$ that takes as input the bit name of $s \in S$ and the index $i$, and (given oracle access to a message $f:V \to \{0,1\}$) computes $f(C_{S\to V}(s,i))$, i.e.\ the value of $f$ on the $i$-th neighbor of $s$.

Similarly, the decoding circuits of our codes discussed in the prior section are given access to the word $w$ as an oracle (typically thought of as a circuit $C_w$) which takes in the bit representation $(s,i)$ and outputs the $w$'s claimed value for the $i$-th neighbor of $s$. Note that this allows us to give decoding circuits whose size depends only \textit{logarithmically} on the alphabet size of the code (i.e.\ on the index length). In other words, these decoding circuits may choose not to read the entire alphabet symbol encoded at $s \in S$, and may query specific bits of the symbol. In general the decoding circuits will be randomized unless otherwise noted.

\subsection{List Recovery}
Our list decoders critically rely on an intermediate decoding task called \textit{list recovery}. A list-recoverable code $\mathcal C: \Sigma_1^N \to \Sigma_2^{N'}$ takes as input to its decoder a \textit{list} of $\ell_{in}$ symbols in $\Sigma_2$ for each $i \in [N']$, denoted $\mathcal{L}=\{\mathcal{L}_s\}_{s \in [N']}$. Note if $\ell_{in}=1$, then $\mathcal{L}$ is just a word $w \in \Sigma_2^{N'}$ as in the list decoding setting. Unlike the list-decoding case, however, we will mainly focus on list recovery in the \textit{low noise} regime, where we think of the global function $f$ we'd like to decode as sitting in \textit{almost every list} input to the decoder.

We now formalize the setup. An $\ell_{in}$-input list-recoverable code $\mathcal{C}:\Sigma_1^N \to \Sigma_2^{N'}$ is given oracle access to an input list $\mathcal{L}=\{\mathcal{L}_s\}_{s \in [N']}$ where each $\mathcal{L}_s \subset \Sigma_2$ is a list of symbols of size at most $\ell_{in}$. We say a message $f \in \Sigma_1^N$ is $(1-\eta)$-computed by $\mathcal{L}$ if all but an $\eta$ fraction of lists contain $\Enc^f(s)$. We write the set of such $f$ as the $\eta$-ball:
\[
\Ball_{\eta}(\mathcal{L}) \coloneqq \{f:\mathcal{L} \text{ $(1-\eta)$-computes $\mathcal{C}(f)$}\}
\]
A code is called list-recoverable if it can recover any such $f$ in this ball:
\begin{definition}[Approximate Locally List-Recoverable Codes (aLLRCs)]\label{def:aLLRC}
A code ${\cal C}: \Sigma_1^{\N} \to \Sigma_2^{\M}$
    is $\delta_{out}$-approximately, $\ell_{in}$-input $(1-\eta,\ell_{out},q)$-locally list recoverable if there exists a probabilistic TM $\mathcal{A}$ which on input $1^N$ outputs (probabilistic) circuits $C_1,\ldots,C_{\ell_{out}}$ such that
    \[
    \forall \mathcal{L} \in (\Sigma_{2}^{\ell_{in}})^{\M}, \forall f \in \Ball_{\eta}(w): \Pr_{\cal A}\left[\exists i \in [\ell_{out}]: \Pr_{j \in [\N], r}[{C_i}^{\mathcal{L},r}(j) = f(j)] \geq 1-\delta_{out} \right] \geq \frac{3}{4}.
    \]
    Moreover, every \(C_i\) queries \(w\) on at most \(q\) locations.
\end{definition}
We have written the above to emphasize the low noise regime (think of $\eta$ as small, say $1\%$ or even $o_N(1)$), however it is worth noting that when $\eta=1-\varepsilon$ is large (the `high noise' regime), this definition directly generalizes $(\varepsilon,\ell_{out},q)$-list decoding, which is just the $\ell_{in}=1$ case. While our main focus is the low noise regime, we will also use high noise list-recovery in our concatenation analysis when moving to the binary and low locality settings, in which case we write $(\varepsilon,\ell_{out},q)$-LLRC to highlight that we are in the high noise setting.

\subsection{Code Concatenation and Doubly Approximate DP Codes}

Code concatenation is a simple paradigm in coding theory where one replaces a large symbol in an `outer' code with a short `inner code'. We will critically use code concatenation in several places in our construction, mostly to move from large alphabet to binary codes needed for hardness amplification and PRGs.


More formally, given 
an ``outer code''
${\cal C}_{\rm out}: \Sigma_{\rm out}^k \rightarrow \Sigma_{\rm out,2}^{n_1}$ 
with encoding map $\text{Enc}_{\rm out}$ and an 
``inner code'' 
${\cal C}_{\rm in}: \{0,1\}^{\log |\Sigma_{\rm out,2}|} \rightarrow \Sigma_{\rm in}^{n_2}$ (we assume the input is binary for simplicity, but this need not be the case)
with encoding map $\text{Enc}_{\rm in}$, the concatenated code  $\mathcal{C}_{\rm out}  \circ \mathcal{C}_{\rm in}: \Sigma_{\rm out}^k \rightarrow \Sigma_{\rm in}^{n_1 n_2}$ is given by encoding each symbol of the outer code $\mathcal{C}_{\rm out}$ with the smaller alphabet of the inner code  $\mathcal{C}_{\rm in}$. That is, for $(i,j) \in [n_1] \times [n_2]$ and a message $f \in \Sigma_{\rm out}^k$:

\[
\text{Enc}_{\mathcal{C}_{\rm out} \circ \mathcal{C}_{\rm in}}(f)_{i,j} = Enc_{\rm in}(Enc_{\rm out}(f)_i)_j
\]
As we will discuss throughout the work, code concatenation typically preserves most nice properties of the code (possibly up to some small loss), like locality and query complexity.
\subsubsection{Doubly Approximate LLRCs}
In the world of approximate decoding, concatenation has a slight problem. When the inner code in the concatenation is only approximately decoded, the outer code will typically receive symbols that are only \textit{close} to correct, rather than entirely correct. This is a standard issue in the literature \cite{impagliazzo2008uniform}, and leads to our final core definition which ensures that the LLRC/LLDC decodes even when an $\varepsilon$ fraction of the lists have a function that \textit{approximately} computes the encoded message. For simplicity, we will only work with this definition for direct product codes. Given a DP code $\mathcal{C}_{(V,S)}: \Sigma_1^N \to (\Sigma_1^r)^{N'}$, we say an input list $\mathcal{L}$ $(1-\eta,\delta_{in})$-computes a message $f$ if there is an $(1-\eta)$-fraction of blocks $s \in [N']$ on which some function $g \in \mathcal{L}_s$ is $\delta_{in}$-close to $f$ in the sense that $\Pr_{v \sim s}[g(v) \neq f(v)] \leq \delta_{in}$, i.e.\ they agree on all but a $\delta_{in}$. We define the set of such $f$ as the (approximate) $(\eta,\delta_{in})$-ball around $\mathcal{L}$:
\[
\Ball_{\eta,\delta_{in}}(\mathcal{L}) \coloneqq \{f \in \Sigma_{in}^{\N}: \Pr_{s \in \Sigma_{out}^{\M}}[\exists g \in \mathcal{L}_s: \text{$g$ is $\delta_{in}$-close to $f|_s$}] \geq 1-\eta\}
\]

Abusing notation, we will also simply call DP codes that can recover such $f$ `approximate locally list decodable (recoverable) codes', but include the additional parameter $\delta_{in}$.
\begin{definition}[(Doubly) Approximate Locally List-Recoverable Codes]\label{def:doubly-apx}
    A code $\mathcal{C}: \Sigma_{1}^{\N} \to \Sigma_{2}^{\M}$ is a $(\delta_{in},\delta_{out})$-approximately, $\ell_{in}$-input, $(1-\eta, \ell_{out},q)$-LLRC if there exists a probabilistic TM $\mathcal{A}$ which on input $1^{\N}$ outputs (probabilistic) circuits $C_1,\ldots,C_{\ell_{out}}$ such that
    \[
    \forall \mathcal{L} \in (\Sigma_{2}^{\ell_{in}})^{\M}, \forall f \in \Ball_{\eta,\delta_{in}}(\mathcal{L}): \Pr_{\mathcal{A}}\left[ \exists i \in [\ell_{out}]: \Pr_{j \in [\N], r}[{C_i}^{w,r}(j) = f(j)] \geq 1-\delta_{out} \right] \geq \frac{3}{4}.
    \]
    Moreover, every \(C_i\) queries \(\mathcal{L}\) on at most \(q\) locations.
\end{definition}
When $\ell_{in}=1$ and $\eta=(1-\varepsilon)$ (i.e.\ the high noise regime), the above definition specializes to (doubly) approximate local list decoding. We include the definition explicitly for convenience.
\begin{definition}[(Doubly) Approximate Locally List-Recoverable Codes]\label{def:doubly-apx-list}
    A code $\mathcal{C}: \Sigma_{1}^{\N} \to \Sigma_{2}^{\M}$ is a $(\delta_{in},\delta_{out})$-approximately, $(\varepsilon, \ell_{out},q)$-LLDC if there exists a probabilistic TM $\mathcal{A}$ which on input $1^{\N}$ outputs (probabilistic) circuits $C_1,\ldots,C_{\ell_{out}}$ such that
    \[
    \forall w \in \Sigma_{2}^{\M}, \forall f \in \Ball_{1-\varepsilon,\delta_{in}}(\mathcal{L}): \Pr_{\mathcal{A}}\left[ \exists i \in [\ell_{out}]: \Pr_{j \in [\N], r}[{C_i}^{w,r}(j) = f(j)] \geq 1-\delta_{out} \right] \geq \frac{3}{4}.
    \]
    Moreover, every \(C_i\) queries $w$ on at most \(q\) locations.
\end{definition}

When $\delta_{in}=0$, we recover the standard notion of approximate decoding defined above. In what follows, we will simply call these aLLRC or aLLDCs, and distinguish between this and the original definition just by whether or not we include the parameter $\delta_{in}$.

\subsection{Graphs and samplers}
Let \(G=(L,R,E)\) be a bipartite bi-regular graph. We denote by the probability of \(A \subseteq R\)
\[\Prob{v \in R}{A} \coloneqq \frac{|A|}{|R|}.\]
(resp. \(A \subseteq L\) replacing \(R\) with \(L\) and vice versa). Sometimes we omit \(v \in R\) for brevity. for \(A \in R\) and \(v \in L\) we denote by 
\[\Prob{u \sim v}{A} \coloneqq \frac{|A \cap N(v)|}{D},\]
where \(N(v)\) is the set of vertices adjacent to \(v\), and \(D=|N(v)|\) is the left-degree.

The following graph property is crucial for our analysis.
\begin{definition}[Sampler graph]
    Let \(\beta \geq 0\) and let \(G=(L,R,E)\) be a bipartite graph. We say that \(G\) is a \(\beta\)-sampler if for every \(A \subseteq R\) and every \(\delta \in [0,1]\),
    \[\Prob{v\in L}{\abs{\Prob{u \sim v}{u \in A} - \Prob{u \in R}{A}} \geq \delta} < \frac{\beta}{\delta^2}.\]
\end{definition}

The following standard lemma states that any bipartite expander graph is automatically a good sampler (see e.g.\ \cite{dinur2017high}). Let \(\lambda \geq 0\). Recall that a bipartite graph is a \(\lambda\)-spectral expander if its normalized adjacency operator has second largest eigenvalue \(\leq \lambda\).\footnote{The adjacency operator is the \(V\times V\) real-valued matrix where \(A(v,u)=\frac{1}{deg(v)}\) if and only if \(\set{v,u}\) is an edge, and \(0\) otherwise. Spectral expansion is a standard notion, see e.g.\ \cite{hoory2006expander} for an extensive survey. } 
\begin{claim}[{\cite{dinur2017high}}]\label{claim:expander-sampler}
            Let \(G=(L,R,E)\) be a bipartite graph that is a \(\lambda\)-spectral expander. Then \(G\) is a \(\lambda^2\)-sampler. 
\end{claim}

\subsection{Complexity classes and average-case complexity}
In this subsection, we introduce notation concerning complexity classes and average case complexity that will
be used to state the consequences of our work for hardness amplification and pseudo-randomness.

We assume familiarity with the basic complexity classes:
$$P, NP, BPP, PSPACE, E=TIME(2^{O(n)}) $$.  \TP{Some of below definitions were already defined in intro.}\maxh{I think it's fine to repeat, as long as they are the same.} \TP{Actually now they appear twice in prelims -- once in very beginning and once here so we have to merge.} We use ${\rm Size}(S,n)$ to denote the set of Boolean functions on $n$  bit inputs that can be computed by circuits with at most $S$ gates (or just ${\rm Size}(S)$ when clear from context), and ${\rm SizeDepth}(n,S,d)$ those computable by circuits of depth $d$ with at most $S$ gates.  For a complexity class $C$ and a Boolean function $f$, we use $C^f$ to denote the class of functions computable by algorithms of the type defining $C$ but with an oracle for $f$.  

For simplicity, we will only consider average-case complexity with respect to the uniform distribution on $n$ bit strings, but most results will easily carry over to other distributions.  For a class $C$, we let ${\rm Heur}_{\delta(n)}[C]$ represent the class of functions $f$ so that there is a function $g \in C$ so that for all $n$, $Prob_{x \in \{0,1\}^n} [f(x) \neq g(x)] \leq \delta(n)$; in other words, there is a heuristic in $C$ that can solve all but a $\delta(n)$ fraction of instances. If $C$ is a class defined by randomized algorithms, we take this probability over both the input $x$ and the randomness used by the algorithm.  $i.o[C]$ represents the class of functions that are equal to some function in $C$ on infinitely many input sizes.  If $f \not\in i.o[{\rm Heur}_{\delta(n)}[C]]$ we say that $f$ is {\em $\delta(n)$-hard } for $C$.  

Boolean circuits are a non-uniform model of computation, one where each input size is allowed to have a separate and independent algorithm.  Complexity classes such as $P$ insist on a single algorithm for all input sizes.  Following \cite{trevisan2007pseudorandomness,karp1980some}, we define $C/a(n)$ to be the ``semi-uniform'' class of functions $f$ so that there is a $g \in C$ so that for all $n$, there is a $z, |z|=a(n)$ so that $f(x)= g(x, z)$ for all $x$ with $|x|=n$.   If $C$ is a class determined by randomized algorithms, we allow $z$ to be a function of the random coins used by the algorithm, so that there is a $g$ of the corresponding type using randomness $r$ and a function $z(r), |z(r)|=a(n)$ so that, for all $x$, with high probability over $r$, $f(x)=g(x,r,z(r))$.  For most classes $C$, we can find a corresponding type of circuits $Circ$ and algorithms producing circuits, so that $f \in C/a(n)$ if and only if there is a randomized algorithm that given $n$, with probability $2^{-a(n)}$ returns a circuit $Circ$ computing $f$.  

Hardness amplification seeks to show that the existence of somewhat hard problems implies that of more reliably difficult problems.  A  $\delta(n), \delta'(n), C_1, C_2, C_3$ hardness amplification construction is a functional that maps $f$ to $F \in C_1^f $ so that if $f \not\in {\rm Heur}_{\delta (n)}[C_2]$ then $F \not\in {\rm Heur}_{\delta'(n)}[C_3]$.  This is interesting if $C_1$ is a ``small'' class compared to $C_2$ and $C_3$, if $C_3$ is not much smaller than $C_2$, and if $\delta'(n)$ is much larger than $\delta(n)$.

\section{Sub-Polynomial Rate aLLDCs}\label{sec:subpoly}
In this section, we give the following `warmup' construction of aLLDCs with sub-polynomial rate that introduces core parts of our framework while avoiding much of the technical difficulty.
\begin{theorem}[Sub-Polynomial Rate aLLDCs]\label{thm:sub-poly-aLLDCs}
    For all large enough $N \in \mathbb{N}$ and \(\varepsilon \leq \frac{1}{\log^2(N)}\),
    there exists $\M \geq \N$ and a strongly explicit 
    direct product code $\mathcal{C}: \Sigma^{\N} \rightarrow \Sigma^{\M}$ that is a $(\frac{1}{10^6\log N},0.01)$-approximate $(\varepsilon,\ell_{out},q)$-LLDC with\footnote{Below, \(\tilde{O}\) hides \(\polylog N\) factors.}
    \begin{enumerate}
        \item \textbf{Rate:} \(R \geq 2^{-16\sqrt{\log N \log \frac{1}{\varepsilon}}}\)
        \item \textbf{Query Complexity and List Size:} $q,\ell_{out} \leq \tilde{O}(\frac{1}{\varepsilon})$
    \end{enumerate}
    In particular, the rate is \textbf{sub-polynomial} whenever \(\varepsilon > N^{-o(1)}\).
    Moreover, $\mathcal{C}$ is locally encodable in $\Time(\polylog(N)\cdot \log(1/\varepsilon))$ and locally decodable in size $\poly(\frac{1}{\varepsilon})$ and depth $O\left(\log \frac{1}{\varepsilon}\cdot \log^2(\log(N))\right)$.
\end{theorem}
\Cref{thm:sub-poly-aLLDCs} is stated to highlight the $\varepsilon \ll \frac{1}{\log(N)^{\omega(1)}}$ regime (in the sense we have suppressed $\log(N)$ factors, and focused only on subconstant $\varepsilon$). This corresponds to amplifying a function $f:\{0,1\}^n \to \{0,1\}$ that cannot be computed by Size($S$) circuits on more than $99\%$ of its inputs\footnote{We note we can also handle generic $\delta_{out}$ for small enough $\delta_{in}$ up to a poly($\delta_{out}$) loss in the rate and query complexity.} to $\Enc^f:\{0,1\}^n \to \Sigma$ that is \textit{super-polynomially} hard, which is the more interesting regime in hardness amplification.

However, from a coding-theoretic standpoint, the construction underlying \Cref{thm:sub-poly-aLLDCs} is also interesting when $\varepsilon$ is constant (or more slowly decaying with $N$). Indeed the code actually achieves \textit{nearly asymptotically optimal rate/query/list-size trade-off} in this setting, with rate $2^{-\tilde{O}(\sqrt{\log(N)})}$, list size $O(\frac{1}{\varepsilon})$, and query complexity $\tilde{O}\left(\frac{\sqrt{\log(N)}}{\varepsilon}\right)$. This is tight by our general lower bound for (locally encodable) codes in \Cref{sec:lower}.

While \Cref{thm:sub-poly-aLLDCs} achieves sub-polynomial rate, it has somewhat large alphabet and locality which is problematic for applications to PRGs and uniform hardness amplification. The first of these problems is quite easy to resolve simply by concatenating with a sufficiently good polynomial rate binary inner code. In particular, concatenating \Cref{thm:sub-poly-aLLDCs} with \cite{impagliazzo2008uniform}'s polynomial rate aLLRC to reduce the log-alphabet size to $\poly(\log N, \frac{1}{\varepsilon})$, then with any sufficiently good binary inner code (we use concatenated Reed-Solomon codes \cite{guruswami2000list}, see \Cref{sec:alphabet} for details) gives the following corollary:\footnote{Note in the query complexity below, it is important \Cref{thm:sub-poly-aLLDCs} reads only $\poly(\log N, \frac{1}{\varepsilon})$ bits of each symbol it queries. Our algorithm achieves this essentially `for free'. See \Cref{rem:fine-grained} for further discussion.}
\begin{corollary}[Binary Sub-Polynomial Rate aLLDCs]\label{cor:sub-poly-aLLDCs-binary}
    For all large enough $N \in \mathbb{N}$ and \(\varepsilon \leq \frac{1}{\log^2(N)}\), there exists $N' > N$ and a strongly explicit binary linear code $\mathcal{C}: \{0,1\}^{\N} \to \{0,1\}^{\M}$ that is a $0.01$-approximate $(\frac{1}{2}+\varepsilon,\ell_{out},q)$-aLLDCs with
    \begin{enumerate}
        \item \textbf{Rate:} $r \leq 2^{-O(\sqrt{\log N\log\frac{1}{\varepsilon}})}$
        \item \textbf{Query Complexity and List Size:} $q,\ell_{out} \leq \poly(\frac{1}{\varepsilon})$ 
        \item \textbf{Locality:} $\poly(\frac{1}{\varepsilon})$
    \end{enumerate}
    Moreover, $\mathcal{C}$ can be locally encoded in $\Time(\poly(\frac{1}{\varepsilon}))$ and locally decoded in logspace uniform size $\poly(\frac{1}{\varepsilon},\log N)$ and depth $\tilde{O}\left(\log^2\left(\frac{\log(N)}{\varepsilon}\right)\right)$.
\end{corollary}
In some applications, e.g.\ to hardness amplification within complexity classes like NP $\cap$ Co-NP \cite{impagliazzo2008uniform} or to applications in learning \cite{carmosino2016learning}, $\poly(\frac{1}{\varepsilon})$ encoding time is insufficient when $\varepsilon \ll \frac{1}{\polylog(N)}$ as the resulting encoded function is too complex (e.g.\ requires computing $f$ in super-polynomially many locations). This is the reason we need our codes above allow for some $\delta_{in}$ symbol corruption. Given this property, it is not too hard to improve the locality to only $\polylog(N)\cdot \log^2(\frac{1}{\varepsilon})$ via concatenation (albeit at the cost of worse rate). Since this does not quite follow black-box from results in the literature in our regime, we state and prove the existence of these codes carefully in \Cref{sec:alphabet} for our improved $\polylog(N)$-rate construction.


\paragraph{Construction Overview:} We give a brief overview of the construction before moving on. At a high level, we use the hypergraph systems framework covered in \Cref{sec:preliminaries}. We recall it briefly. A hypergraph system consists of a vertex set $V$, hyperedges $S$, and a `decoding graph' $G$ on $S$ of overlapping hyperedges used for advice propagation. The system $(V,S,G)$ automatically induces `intersection graphs' $G_s$, corresponding to what vertices in $s$ are sub-sampled in the intersection with a random neighbor $s'$ of $s$ in $G$.

The construction for \Cref{thm:sub-poly-aLLDCs} is split into three components, each of which has a corresponding subsection below. The first part, which is completely generic, is a variant of the now standard trick of Guruswami and Indyk \cite{guruswami2003linear,guruswami2001expander} that reduces the problem of high noise list-decoding to low-noise list \textit{recovery}, i.e.\ the setting where every hyperedge $s \in S$ sees a small \textit{list} of symbols, and we are promised there is a global function $f:V \to \{0,1\}$ where \textit{almost all} hyperedges have $f|_s$ in their list. We give a version of their reduction in the aLLDC setting, reducing the problem to building good approximate locally list-recoverable codes in the low-noise regime.

The second and \emph{main} component of the construction is then a new framework for local list recovery on hypergraphs systems $(V,S,G)$ and its subsequent instantiation. Namely, we build a simple propagation-style decoder for any system with the following two key properties:
\begin{enumerate}
    \item All intersection graphs $G_s$ are (very) good samplers, and
    \item $G$ admits a `low congestion local routing scheme'. 
\end{enumerate}
The first property is self-explanatory, but the second requires some discussion. A \textit{routing scheme} is a small low-depth circuit which, given as input $s_{in},s_{out} \in S$, outputs a short path $P$ in \(G\) between $s_{in}$ and $s_{out}$. A routing scheme has \textit{low congestion} roughly when no edges in \(G\) are reused in too many paths or, equivalently, when no small set of edges is hit by too many paths in the scheme (see \Cref{def:robust-routing-scheme} for the exact notion we use). Morally, we use this routing to locally propagate advice from a random starting $s_{in}$ to a random $s_{out} \ni v$ for a vertex $v$ we'd like to decode. If $G_s$ is a sufficiently good sampler, very few edges in the graph are `bad' for propagation decoding, and low congestion promises our routing usually misses this bad set and output the correct $f(v)$ with high probability.

The final component is then to instantiate the above framework. In particular, we need to find a sparse explicit family of hypergraphs with structured overlapping hyperedges (to ensure intersection sampling), good `global' expansion (to ensure the existence of short paths), and a strongly explicit algorithm that locally routes on $G$ between any two hyperedges in the system in a sufficiently symmetric way to ensure low congestion. Somewhat amazingly, we show there is a completely elementary way to do this with sub-polynomial degree using a hypergraph system based on subspaces. Later, we will see how this fits into a broader set of constructions based on neighborhood encodings on high dimensional expanders,\footnote{In particular, the construction in this section arises from taking neighborhoods on a very simple HDX called the `Flags Complex' or `Spherical Building'.} and how to modify the overall framework to achieve polylog or even constant rate.

\subsection{List Recovery and Routing}\label{sec:outer}
While Guruswami and Indyk's reduction to list-recovery is conceptually simple and the `first step' of our construction, it has a variety of unwieldy parameters that need to be set to match the underlying list-recoverable code. Thus, it will be easier for us to first present our new list-recovery procedure, then discuss the reduction from list-decoding to list recovery with the correct corresponding parameters. 

We split this section on list-recovery into two sub-parts. We first formally describe our notion of low congestion routing schemes, then prove any hypergraph system with such a scheme (and sufficiently good sampling) has a simple propagation-based list-recovery algorithm.
\subsubsection{Low Congestion Routing}\label{sec:route}
As we briefly described above, a core requirement of our list recovery algorithm is the ability to locally compute paths in the decoding graph $G_s$ along which we propagate advice. We will critically need these paths to exhibit two main properties: 1) they should be short (this corresponds to the query complexity), and 2) for any small set of edges $B$ in $G$, the probability our path hits $B$ should remain small (this ensures we miss any small subset of `bad edges'). We abstract these guarantees as the notion of a `low congestion  routing scheme':
\begin{definition}[Low Congestion Routing Scheme] \label{def:robust-routing-scheme}
    For any $t \in \mathbb{N}$ and graph $G=(V,E)$, a length-$t$ routing scheme on $G$ is a (possibly randomized) algorithm $\mathcal{O}$ that on input $v,v' \in G$, outputs a path $\mathcal{O}(v,v')$ between $v$ and $v'$ of length at most $t$. $\mathcal{O}$ has \textbf{unit congestion} if for every subset $B \subset E$, the probability $\mathcal{O}$ hits $B$ is at most
    \[
    \Pr_{v,v',\mathcal{O}}[\exists e \in \mathcal{O}(v,v'): e \in B] \leq t\Pr[B]
    \]
\end{definition}
As defined, it is not particularly difficult to construct low congestion routing schemes for, e.g., expander graphs. Rather, the difficulty really lies in the third key criterion alluded to above: the scheme must have a \textit{local implementation}. In other words, we'd like $\mathcal{O}$ to be implementable by a $\polylog|V|$-size, low-depth (say $\poly\log\log |V|$), logspace uniform circuit class. 

We emphasize that even forgetting hardness amplification, building such local schemes is critically important in application, as otherwise there is no way to compute the query set needed to decode a particular vertex $v$. Indeed without such an implementation we cannot even achieve sub-linear time local decoding, much less, e.g., the sub-polynomial time constant rate LLDCs eventually built in this work.

In this and later sections, we will see how to exploit subspace (and subsequently group/polynomial) structure to build such efficient schemes even on highly sparse hypergraph systems. In fact, in later sections we will show something even stronger --- there exist local routing schemes which work even when $1\%$ of the graph is \textit{adversarially corrupted} (in a `testable' way), which is critical to achieve constant rate.\footnote{Note the length $t$ of the routing scheme must be at least the diameter of the graph, so $\polylog(N)$ in the settings we study. Handling a $1\%$ fraction of corruptions is highly non-trivial in this regime, even if the algorithm `knows' which vertices are bad.}

\subsubsection{List Recovery for Hypergraph Systems}\label{sec:subpoly-list}
We now show any hypergraph system with a low congestion routing scheme and whose intersection graphs are good samplers admits a simple list-recovery algorithm.

\begin{theorem} \label{thm:list-recovery}
Let \(\delta > 0\) and \(\ell_{in},t \in \mathbb{N}\).
Let \((V,S,G)\) be a regular hypergraph system such that
\begin{enumerate}
    \item For every \(s \in S\), \(G_s\) is a \(\frac{10^{-3}\delta^3}{t\ell_{in}}\)-sampler.
    \item $G$ admits a length-$t$ unit congestion routing scheme $\mathcal{O}$.
\end{enumerate}
Then $\mathcal{C}_{(V,S)}$ is a $(\delta,15t\delta)$-approximate, $\ell_{in}$-input, \((1-\frac{10^{-5}}{t},\ell_{in},t)\)-LLRC.
\end{theorem}
\Cref{thm:list-recovery} states only combinatorial properties of our code. As discussed, it is critical our decoding algorithm is also \textit{highly efficient}, i.e.\ that the output circuits should be small, low-depth, and generated efficiently. We show this holds assuming oracle access to a circuit implementation of the hypergraph and routing scheme (see \Cref{sec:direct-products} for the definition of circuit implementation):
\begin{lemma}[Circuit Complexity of \Cref{thm:list-recovery}]\label{lem:list-recovery-circuit}
    Given a circuit implementation $(C_V,C_S,C_{V^{-1}},C_{S^{-1}},C_I)$ of $(V,S)$, a circuit implementation of the input list $C_{\mathcal{L}}$, and a circuit implementation $C_{\mathcal{O}}$ of $\mathcal{O}$, the generated decoding circuits $\{C_1,\ldots,C_{\ell_{in}}\}$ are in logspace uniform
\begin{enumerate}
    \item \textbf{Size:} $\text{Size}(C_i) \leq \poly(t,\ell_{in},\delta^{-1})$
    \item \textbf{Depth:} $\text{Depth}(C_i) \leq O(\log (\ell_{in})\log(t))$
\end{enumerate}
with one oracle call to $C_\mathcal{O}$ and $C_V$, and $\tilde{O}(\frac{t\ell_{in}^2}{\delta})$ (parallel) oracle calls to $C_I$, $C_{S^{-1}}$, and $C_{\mathcal{L}}$
\end{lemma}

Our main decoding process is given by the propagation algorithm in \Cref{fig:list-rec-algorithm}. Roughly speaking, the process starts at a hyperedge $s_1$ and operates under the assumption that a fixed function $g \in \mathcal{L}_{s_1}$ in its list is close to the global $f$ we'd like to decode (in reality, we output one circuit for each $g$). Given the target $v$, the algorithm samples a random hyperedge $s_t \ni v$ (Step 1), then calls the routing scheme to find a path from $s_1$ to $s_t$ (Step 2). The goal is now to find, based on our initial advice $g$, a function $g_t$ in the list of $s_t$ which is close to $f$ with high probability (we may then decode by outputting $g_t(v)$). 

This is done by the following iterative procedure described in Steps (3)-(4) below, where we build a list of `good candidates' $L_i$ for every $s_i$ in the path. In particular, starting from our advice $L_1=\set{g}$, in each step we compute $L_{i+1}$ as all functions in the list of $s_{i+1}$ which are close to any function in $L_i$ from the previous round (as measured on the intersection of $s_i \cap s_{i+1}$). We argue that at the end of this process, $L_t$ will be non-empty, and \textit{every} function in $L_t$ will be close to $f$ with high probability, so we can pick one arbitrarily to decode $v$. We comment that this process can be performed in low depth using Savitch's algorithm \cite{Savitch1970}.

In \cref{app:distance-tester} we analyze a very simple circuit \((\frac{\log \frac{2t\ell_{in}^2}{\delta}}{\delta},4\delta)\)-list-distance-tester that is given oracle access to lists of functions \(L_1,L_2\) on a common domain (in this case \(s_i \cap s_{i+1}\)), and outputs a list \(L_2' \subseteq L_2\) such that with probability \(1-\frac{\delta}{2t\ell_{in}^2}\):
\begin{enumerate}
    \item If \(f \in L_2\) is \(2\delta\)-close to a function in \(L_1\) then \(f \in L_2'\).
    \item If \(f \in L_2'\) then there exists some function \(f' \in L_1\) such that \(\dist(f,f')\leq 6\delta\).
\end{enumerate}
The circuit samples \(O(\ell_{in}\frac{\log \frac{2t\ell_{in}^2}{\delta}}{\delta})\) random points in the domain, and queries all functions on these points. 
We use this procedure to avoid working with the entire symbol at each $s_i$ which may be quite large (namely this ensures our decoding circuit size stays independent of the alphabet, and that we make fewer queries after concatenation).

\begin{figure}[ht!]
\fbox{\parbox{\textwidth}{
\vspace{.1cm}
\underline{List-Recovery Algorithm \(Dec(s_1,g,v)\)}:
\begin{enumerate}
    \item[*] \textbf{Input:}
    \begin{itemize}
        \item A set \(s_1\) and a member of its list \(g \in \mathcal{L}_{s_1}\)
        \item A target vertex \(v \in V\)
    \end{itemize}
    \item[*] \textbf{Output:} a value in \(\Sigma\).
    \item Sample a random \(s_t \ni v\). 
    \item Sample a random path \(s_1,s_2,\dots,s_t\) from $\mathcal{O}$.
    \item Set \(L_1 = \set{g} \subseteq \mathcal{L}_{s_1}\).
    \item For \(i=2,3,\dots,t\) do: 
    \begin{enumerate}
        \item $L_i \gets $ \((\frac{\log \frac{2t\ell_{in}^2}{\delta}}{\delta},4\delta)\)-distance-tester($L_{i-1}$,$\mathcal{L}_{s_i}$)
        \item If \(L_i = \emptyset\) output `FAIL'.
    \end{enumerate} 
    \item Choose a random \(g_{t} \in L_t\) and output \(g_{t}(v)\).
\end{enumerate}}}
    \caption{List-recovery algorithm}
    \label{fig:list-rec-algorithm}
\end{figure}

\begin{proof}[Proof of \Cref{thm:list-recovery}]
    Fix an input list $\mathcal{L}$ and $f \in \Ball_{1-\varepsilon,\delta}(\mathcal{L})$. Our goal is to prove that most $s_1$ have some choice of list element $g_j$ such that $Dec(s_1,g_j,\cdot)$ approximately computes $f$:
        \[
        \Prob{s_1}{\exists g \in \mathcal{L}_{s_1} : \Prob{v,r}{Dec(s_1,g,v) = f(v)} \geq 1-15t\delta} \geq 0.99
        \]
    In particular, outputting the circuits $\{Dec(s_1,g,\cdot)\}_{g \in \mathcal{L}_{s_1}}$ then gives the desired list-recovery guarantee.
Toward this end, fix any input list \(\mathcal{L}\) and \(f \in \Ball_{1-\varepsilon,\delta}(\mathcal{L})\). We first define a few rare `bad events', then argue outside these cases decoding succeeds. In particular, let \(B\) denote the set of edges \(\set{s,s'}\) such that for either \(s'' \in \set{s,s'}\) one of the following conditions holds:
    \begin{enumerate}
        \item \textbf{Bad List:} There is no \(g'' \in \mathcal{L}_{s''}\) that is \(\delta\)-close to \(f\)
        \item \textbf{Bad Sampling:} There exists some function \(g'' \in \mathcal{L}_{s''}\) such that 
    \[
    \abs{\dist_{s''} (f,g'') -  \dist_{s \cap s'}(f,g'') } > \delta.
    \]
    \end{enumerate} 
We call a path $(s_1,\ldots,s_t)$ good if no edge in the path is in $B$. We claim it is enough to prove that most $s_1$ have mostly good oracle paths to a random endpoint $s_t$:
\begin{equation}\label{eq:good-paths-distance-1}
\Pr_{s_1}\left[\Pr_{s_t,\mathcal{O}}[\mathcal{O}(s_1,s_t)\text{ is good}] \geq 1-\frac{\delta}{2}\right] \geq 99\%
\end{equation}
Let's first complete the proof assuming this claim. Fix any $s_1$ such that the inner condition holds. By assumption, there exists some $g \in \mathcal{L}_{s_1}$ that's $\delta$-close to $f$. We claim that for any such fixed choice, the probability Dec$(s_1,g,v)$ succeeds in decoding $f(v)$ for a random $v$ is at least $1-15t\delta$.

To see this, first recall that on input $v$, Dec draws a random $s_t \ni v$, paths to it via $\mathcal{O}(s_1,s_t)$ and outputs $g_t(v)$ for some valid choice of $g_t \in L_t$. Thus our goal is to lower bound:
\[
\Pr_{r,v}[Dec(s_1,g,v) = f(v)] =\Pr_{v \in s_t,\mathcal{O},g_t}[g_t(v) = f(v)]
\]
where $g_t \in L_t$ is chosen uniformly at random and independent of \(v\).

Now, examining the right-hand probability, notice we have drawn $v \in V$ uniformly at random, then a random neighbor $s_t \ni v$. Since $(V,S)$ is regular, this is the same as \textit{first} drawing a random $s_t \in S$, then sub-sampling a random vertex $v \in s_t$. This means it is enough to argue that, over the randomness of $s_t,\mathcal{O}$, the final function $g_t$ on $s_t$ is $O(t\delta)$-close to $f$ with high probability:
\[
\Pr_{s_t,\mathcal{O}}\left[\forall g_t \in L_t: \text{dist}(g_t,f) \leq 14t\delta\right] \geq 1-\delta
\]
In particular, the probability we correctly output $f(v)$ over a random choice of $s_t \in S,g_t \in L_t$ and $v \in s_t$ is at least $(1-\delta)(1-14t\delta) \leq 1-15t\delta$ as desired.

To show this, we introduce one final bad event outside which our decoder succeeds: the event that our distance tester in Step 4(a) errs (recall we sub-sample points in the intersection to test distance rather than check the full symbol to ensure our decoding circuits remain small). By construction (and \Cref{claim:independent-distance-test}), this occurs with probability at most $\frac{\delta}{2}$. 

Now condition on the path $\mathcal{O}(s_1,s_t)$ being good \textit{and} the above event that our distance tester doesn't err on the path (which occurs with probability at least $1-\delta$ by a union bound). We claim under this conditioning \textit{every} $g_t \in L_t$ is $14t\delta$-close to $f$ when we start from the promised $g$ that is $\delta$-close to $f$. More formally, we will argue the following two claims by induction for every $L_{i}$ along our path:

    If these events do not occur then we argue that:
    \begin{enumerate}
        \item \(L_{i}\) contains a function that is \(\delta\)-close to \(f\) on \(s_i\).
        \item Every function in \(L_{i}\) is \(14i \delta\) close to \(f\).
    \end{enumerate}
    Given these assertions it is clear that any function in \(L_{s_t}'\) is \(14t\delta\)-close to \(f\) in which case it answers correctly on \(1-14t\delta\) of its vertices, which is what we need.

    We prove the two items by induction on \(i\). In the proof we will 
    use the fact that the distance tester never errs, which we recall implies: (1) any two functions that are \(4\delta\)-close are accepted in the test, and (2) if a pair of functions was accepted by the distance test, then their distance is at most \(12\delta\).

    Now, first note that the base case for \(L_{1}\) holds by assumption. Therefore assume the two items hold true for \(L_{i-1}\). We will prove them for \(L_{i}\). Starting with the first item, suppose there exists some \(g_{i-1} \in L_{i-1}\) that is \(\delta\)-close to \(f\). By assumption that \(\set{s_{i-1},s_i} \notin B\) (first item), there exists some \(f_i \in L_{i}\) that is \(\delta\)-close to \(f\). As the edge is not in \(B\) (second item),
    \[\dist_{s_{i-1} \cap s_i}(g_i,f) \leq \dist_{s_{i-1}}(g_{i-1},f) + \delta \leq 2\delta, \]
    \[\dist_{s_{i-1} \cap s_i}(g_{i},f) \leq \dist_{s_i}(g_i,f) + \delta \leq 2\delta. \]
    Therefore by the triangle inequality  \(\dist_{s_{i-1} \cap s_i}(g_{i-1},g_i) \leq 4\delta\). As the tester did not not err, \(g_i \in L_{i}\).

    Moving on to the second item, let us assume all functions in \(L_{i-1}\) are \(14(i-1)\delta\)-close to \(f\). As before, because \(\set{s_{i-1},s_i} \notin B\) then for every \(g_{i-1} \in L_{i}\) 
    \[
    \dist_{s_{i-1} \cap s_i}(g_{i-1},f) \leq \dist_{s_{i-1}}(g_{i-1},f) + \delta \leq (14(i-1)+1)\delta.
    \]
    As the tester did not err, for every \(g_i \in L_{i}\), we also have that
    \[\dist_{s_{i-1} \cap s_i}(g_{i-1},g_i) \leq 12\delta.\]
    Hence by the triangle inequality,
    \[
    \dist_{s_{i-1} \cap s_i}(g_{i-1},f) \leq \dist_{s_{i-1} \cap s_i}(g_{i-1},f_i) + \dist_{s_{i-1} \cap s_i}(g_{i-1},g_i) \leq (14(i-1)+1)\delta + 12\delta = (14(i-1)+13)\delta.
    \]

    Again using the fact the the edge \(\set{s_{i-1},s_i} \notin B\), we have 
     \[\dist_{s_i}(g_i,f) \leq \dist_{s_{i-1} \cap s_i}(g_{i},f) + \delta \leq 14i\delta.\]

     Finally, it is left to prove \Cref{eq:good-paths-distance-1}, that many $s_1$ have few bad paths. To see this, first observe that by Markov's inequality it is enough to prove a random choice of $s_1,s_t$, and path $\mathcal{O}(s_1,s_t)$, the probability $\mathcal{O}(s_1,s_t)$ hits $B$ is very low, say:
\[
\Pr_{s_1,s_t,\mathcal{O}}[\mathcal{O}(s_1,s_t)\text{ hits } B] \leq \frac{\delta}{200}.
\]
This is now exactly the regime that is bounded by low-congestion routing. In particular, since $\mathcal{O}$ is unit congestion, it is enough to prove the measure of $B$ is small, namely $\Pr[B] \leq \frac{\delta}{200t}$. This follows essentially immediately from sampling, and from the promise that all but a $10^{-3}\frac{\delta}{t}$ fraction of $s \in S$ contain a function $\delta$-close to $f$. The latter ensures the probability a random edge satisfies the `Bad List' condition is at most $2\frac{10^{-3}}{t}$ by a union bound. The former says that for any fixed $s$, the probability a random neighbor $s'$ mis-samples any fixed function in the list of $s$ is at most $\frac{10^{-3}\delta}{t\ell_{in}}$. Union bounding over the lists of both $s$ and $s'$ in a random edge, this event therefore occurs with probability at most $2\frac{10^{-3}\delta}{t}$, and a final union bound gives $\Pr[B] \leq \frac{\delta}{200t}$ as desired.
\end{proof}

We briefly remark that the proof above has the nice property that it only reads $O(\frac{1}{\delta}\log  \frac{t \ell_{in}^2}{\delta})$ bits per symbol it queries. We call the number of bits (rather than full symbols) read the `fine-grained query complexity' of the decoder.
\begin{remark}[Fine-grained query complexity]\label{rem:fine-grained}
    The fine-grained query complexity of \Cref{thm:list-recovery} is at most $O(\frac{t}{\delta}\log  \frac{t \ell_{in}^2}{\delta})$.
\end{remark}
This will be useful later in \Cref{sec:alphabet} when we want to build query-optimal concatenated codes and pay for each bit of the input word read. Now it is left to prove the claimed circuit complexity of our decoding algorithm $\text{Dec}(s,g,\cdot)$:
\begin{proof}[Proof of \Cref{lem:list-recovery-circuit}]
We recall each of the components used as a blackbox in $Dec(s_1,g,\cdot)$:
\begin{enumerate}
    \item The hypergraph circuit implementation $(C_V,C_S,C_{V^{-1}},C_{S^{-1}},C_I)$ described in \Cref{sec:direct-products}    
    \item A circuit \(C_\mathcal{O}\) implementing the routing scheme.
    \item A circuit \(C_{\mathcal{L}}\) that takes as input \(s \in S\) and a neighbor $v \sim s$ (formally, an index $i$ corresponding to the neighbor $v$), and outputs \(\{g^{(j)}(v)\}_{j \in [\ell_{in}]}\) for \(g^{(j)}\) the \(j\)-th member of \(\mathcal{L}_s\).
\end{enumerate}
We now describe the implementation of our $Dec(s_1,g,\cdot)$ using oracle calls to the above:
\begin{enumerate}
    \item Dec feeds the input $v \in V$ (and a random index $i_v$) into $C_V$ to generate \(s_t \ni v\).
    \item \(s_t\) is fed into \(C_{\mathcal{O}}\) (along with the hardcoded \(s_1\) and an additional random string \(r_{\mathcal{O}}\)) to produce the path \(s_1,s_1,\dots,s_t\).
    \item For every edge \(\set{s_{i-1},s_i}\), use \(d=O(\frac{\log \frac{t\ell_{in}}{\delta}}{\delta})\) copies of \(C_I\) to sample \(d\)-random points in \(s_i \cap s_{i-1}\). 
    \item Using $2d$ copies of $C_{S^{-1}}$, compute which neighbors of $s_i$ and $s_{i-1}$ these sampled vertices correspond to.
    \item Feed these into \(C_{\mathcal{L}}\) to compute \(g_{i-1}(v),g_i(v)\) for every pair of functions \(g_{i-1} \in \mathcal{L}_{s_{i-1}}, g_i \in \mathcal{L}_{s_i}\), and feed $(g_{i-1}(v_1),\ldots,g_{i-1}(v_d)) \oplus (g_{i}(v_1),\ldots,g_{i}(v_d))$ into a $4\delta$-threshold circuit to implement the distance tester.\footnote{Note we have slightly modified this step from \Cref{fig:list-rec-algorithm} for the circuit implementation by computing thresholds over the entire list rather than the current $L_{i-1}$.}
    Every threshold circuit can be implemented in depth \(O(\log (d) ) \) and size \(O(d\log d)\).
    \item We now implement a circuit which takes as input these values, and outputs the index of a random $g \in L_t$. Toward this end, we construct the following `adjacency graph' on the lists
    \begin{enumerate}
        \item The vertices are the functions in all the lists \(\mathcal{L}_{s_1},\mathcal{L}_{s_1},\dots,\mathcal{L}_{s_t}\).
        \item Edges are between consecutive \(\mathcal{L}_{s_{i-1}},\mathcal{L}_{s_i}\) where \(g_{i-1} \sim g_i\) if the corresponding threshold circuit output $1$ (i.e.\ they measured as $4\delta$-close)
    \end{enumerate}
    Now observe a function \(g'' \in \mathcal{L}_{s_t}\) is in \(L_{t}\) as described in \Cref{fig:list-rec-algorithm} if and only if there is a \(t\)-length path from \(g \in \mathcal{L}_{s_1}\) to \(g''\) in the above graph. Thus we can output the (indices of the) functions in \(L_{t}\) by a circuit \(C_{Sav}\) implementing Savitch's algorithm \cite{Savitch1970}. In our setting, where the graph has $t$ layers each of $\ell_{in}$ vertices, Savitch's algorithm can be implemented in logspace uniform depth \(O(\log t \log \ell_{in})\) and size \(\tilde{O}(\ell_{in}^3 t)\) by reusing many of the recursive computations (roughly, each of the $\log t$ recursion layers requires taking up to $\ell^2t$ parallel OR of $\ell$ 2-bit ANDs across computations from the previous round).
    \item Finally, we choose randomly some \(g'' \in L_{t}\) and output \(g''(v)\) by an additional query to \(C_{\mathcal{L}}\) (and to $C_{S^{-1}}$ to find the appropriate neighbor index of $s_t$ to query $C_{\mathcal{L}}$ on).
\end{enumerate} 
Thus, altogether we make the claimed number of oracle calls, and otherwise have
\[
\Depth(\text{Dec}(s_1,f,\cdot)) \leq \underbrace{O\left(\log \left(\frac{\log(t\ell)}{\delta}\right)\right)}_{\text{Distance Tester}} +  \underbrace{O(\log t\log \ell_{in})}_{\text{Savitch's Algorithm}}
\]
and
\[
\Size(\text{Dec}(s_1,f,\cdot)) \leq \poly(t,\ell_{in},\delta)
\]
\end{proof}

\subsection{From List-Decoding to List-Recovery}
We now give our simple local variant of Guruswami and Indyk's \cite{guruswami2003linear,guruswami2001expander} reduction to list-recovery. The key property of this reduction is that it moves from the noisy setup where only an \(\varepsilon\)-fraction of the sets contain information we want to compute, to a cleaner setup where \((1-o(1))\)-fraction of the sets contain information on the function we want to compute (although this information is hidden in a list of functions). This matches the list-recovery algorithm described in the previous subsection, which requires that all but an $O(\frac{1}{t})$ fraction of codeword blocks have a function in their list close to the global $f$ we'd like to decode.

The idea behind the reduction is remarkably simple. Suppose we have a direct product code on set system \((V,S)\), and a regular sampler graph \(G'=(L,R,E)\) where \(L=S\). By `composing' these, we construct a second direct product code \((V,T)\) where the hyperedges \(T = \set{t_r}_{r \in R}\) are indexed by vertices in $R$, and formally correspond to the \textit{multi-set}
\[
t_r \coloneqq \biguplus_{s \sim r} s
\]
where the multi-set union is over neighbors of $r$ in the sampler $G'$. In other words, if $(V,S)$ is $r_1$-uniform, and $G'$ is $r_2$ right-regular, then the composed code encodes a message $f: V \to \Sigma$ to a codeword $\Enc^f: T \to (\Sigma^{r_1})^{r_2}$, which on input $t_r \in t$, outputs the symbol $f(s) \in \Sigma^{r_1}$ for every neighboring $s \sim r$.

The point of the construction is the following. Say we now have a noisy input word which agrees with $\Enc^f$ on an $\varepsilon$-fraction of $t_r \in T$. By sampling, a random $s \in S$ in our original hypergraph system will neighbor at least an $\frac{\varepsilon}{2}$ fraction of these `good' symbols with high probability. This means we can decode a list of possible candidates for $f$ on the original hyperedge $s$ simply by writing down any symbol which appears many times across its neighbors $t_r$. This reduces us to the list-recovery case, where we can simply run our algorithm from the previous section for $(V,S)$.

The following lemma formalizes and extends the above to the setting of approximate agreement, specified to the parameters of our list-recovery algorithm from the previous section.
\begin{lemma} \label{lem:gi-list-decoding-to-list-recovery}
For any $\varepsilon>0$, $t \in \mathbb{N}$, and $\delta \leq \frac{10^{-5}}{16t}$, suppose $\mathcal{C}_{(V,S)}$ is a $(\delta,15t\delta)$-approximate, $\frac{100 \log t}{\varepsilon}$-input, \((1-\frac{10^{-5}}{t},\frac{100 \log t}{\varepsilon},t)\)-LLRC. Then for any $\frac{10^{-5}}{4t\varepsilon^2}$-sampler graph \(G'=(S,R,E)\), the corresponding direct product code \((V,T)\) is
    \((\delta^2,15t\delta)\)-approximately \((\varepsilon,\frac{200 \log t}{\varepsilon},\frac{100 t \log t}{\varepsilon})\)-locally list decodable. 
\end{lemma}
\begin{proof}
    Fix an input word  \(w=\set{w_{t_r}}\) and a global function $f \in \Ball_{1-\varepsilon,\delta^2}(w)$ we'd like to decode. We will construct from $w$ (implicitly) a randomized set of lists \(\mathcal{L}_S=\set{L_s}_{s \in S}\) of size $\frac{100 \log t}{\varepsilon}$ such that with $99\%$ probability, \(f \in C_{1-\frac{10^{-5}}{t},\delta}(\mathcal{L}_S)\). Given query access to these lists, the list-recoverable algorithm produces a list of $\ell_{out}$-circuits, one of which is promised to be $15t\delta$-close to $f$ with $\frac{3}{4}$ probability as long as the above high probability event holds. Thus running the list-recoverable algorithm twice ensures we output such a circuit with at least $\frac{3}{4}$ probability as desired.

    We now discuss how to construct our randomized family of lists \(\mathcal{L}_S=\set{L_s}_{s \in S}\), and how the list-recovery algorithm computes $L_s$ on demand from $w$. In particular, the list $L_s$ for $s \in S$ will be generated by picking $\frac{100\log t}{\varepsilon}$ random neighbors $r \sim s$, and outputting the value of $w_{t_r}$ on $s$. To ensure that this can be implemented by a small size circuit, we will use \textit{the same choice of randomness} for all $s \in S$. In other words, these indices are chosen ahead of time and hard-coded into the output circuits, which determines a particular family $\mathcal{L}_S$ that can then be deterministically computed by the decoding circuits with $\frac{100\log t}{\varepsilon}$ queries to $w$.

    We now need to argue that the above procedure for selecting $\mathcal{L}_S$ indeed creates a set of lists in which \(f \in C_{1-\frac{10^{-5}}{t},\delta}(\mathcal{L}_S)\) with at least $99\%$ probability. By Markov's inequality, it is enough to show that for all but $\frac{10^{-5}}{2t}$ fraction of $s$, $L_s$ contains a function $\delta$-close to $f$ with probability at least $1-\frac{10^{-6}}{2t}$.

    Toward this end, define the set of `bad' hyperedges $s$ who do not see enough good neighbors as
    \begin{equation}\label{eq:def-B-GI}
    B = \sett{s \in S}{\Prob{r \sim s}{\dist(w_{t_r}|_s,f|_s) < \delta} < \frac{\varepsilon}{4}}.
    \end{equation}
     We will prove that \(\Prob{s}{B} \leq \frac{10^{-5}}{2t}\). The fact that $L_s$ for any $s \notin B$ contains a function $\delta$-close to $f$ with probability at least $\frac{10^{-6}}{2t}$ is immediate from Chernoff.
     

    To bound $\Pr[B]$, we divide further into two `bad events' \(B_1,B_2\) such that \(B\subseteq B_1 \cup B_2\). Let \(E \subseteq R\) be the promised set of right-hand vertices such that \(w_{t_r}\) is \(\delta^2\)-close to \(f\). We define \(B_1\) as all sets \(s \in S\) such that \(\Prob{r \sim s}{E} < \frac{\Prob{}{E}}{2}\). Since $\Pr[E] \geq \varepsilon$ by assumption, sampling of $G'$ promises \(\Prob{s\in S}{B_1}<\frac{10^{-5}}{4t}\). Let \(B_2 \subseteq S \setminus B_1\) be all \(s \in S\) whose neighborhood contains at least an  \(\frac{\Pr[E]}{2}\) fraction of \(r \in R\) which \(\delta^2\)-compute \(f\), but for which at least half these good neighbors \(r \in E \cap N_G(s)\), \(w_{t_r}|_s\) is \(\delta\)-far from \(f\). It is clear that \(s \notin B_1 \cup B_2\) implies \(s \notin B\).

    To bound $\Pr[B_2]$, note that for a \textit{random} edge $(s,r)$ of $G'$
    \[
    \Prob{s \sim r}{r \in E, \dist(w_{t_r}|_s,f_s) > \delta} \leq \delta \Prob{r}{E}.
    \] 
    This follows immediately from viewing the edge $(s,r)$ by first drawing $r$ uniformly from $R$, then $s \sim r$ as a random neighbor. Namely, the probability our initial choice $r \in E$ is exactly $\Pr[E]$, and for \textit{any} $r \in E$, Markov's inequality promises at most a $\delta$-fraction of $s \sim r$ have $w_{t_r}|_s$ can be $\delta$-far from $f|_s$. On the other hand, every \(s \in B_2\) sees at least \(\frac{\Prob{}{E}}{4}\) such edges by definition, so no more than a $4\delta \leq \frac{10^{-5}}{4t}$ fraction of such $s$ can exist as desired.
\end{proof}

We note that assuming oracle access to a neighborhood circuit implementation of the graph $G'$, the output circuits of this procedure have the same asymptotic circuit complexity as the outputs of the original input list-recoverable code, since the only changes are 1) running the original code twice, and 2) simulating the input list by querying a hard-coded set of neighbors in $G'$. The `fine-grained' query complexity requires an additional multiplicative factor of $O(\frac{1}{\delta} \log \frac{t}{\delta \varepsilon})$ for distance testing at each step.

Finally, we make two simple remarks before moving on. The first is that \Cref{lem:gi-list-decoding-to-list-recovery} carries over easily to the \textit{high noise list-recovery} regime, where we write $\varepsilon=1-\eta$, which we will need for our concatenated codes:
\begin{corollary}[GI List Recovery]\label{cor:GI-recovery}
    The output code in \Cref{lem:gi-list-decoding-to-list-recovery} is an $\ell_{in}$-input aLLRC with $\ell_{out} \leq 200\frac{\ell_{in}\log t}{\varepsilon}$ and otherwise the same parameters so long as $\mathcal{C}_{(V,S)}$ is $\frac{100 \ell_{in} \log t}{\varepsilon}$-input list-recoverable.
\end{corollary}
The proof is exactly the same, with each $L_s$ being constructed by taking the union of the lists at its selected neighbors. In fact it is easy even to check the above is even a `single-circuit' aLLRC (see \Cref{def:single-circuit}), since it easy to check we can use the same query set for each of the output circuits. This allows us to achieve nearly optimal query complexity post concatenation.

Second, we remark that as written, \Cref{lem:gi-list-decoding-to-list-recovery} and \Cref{cor:GI-recovery} have the undesirable property that their list size grows with $t$, which itself grows with $N$. This is an artifact of the simplified proof we've presented above, and is not necessary. In \Cref{sec:polylog} we discuss a simple pruning procedure that improves the list size to the asymptotically optimal bound $O(\frac{\ell_{in}}{\varepsilon})$.

\subsection{The Subspaces aLLDC}
\renewcommand{\S}{\mathcal{SS}}

Until this point, our statements have been abstract in the sense that we only specify what properties of a hypergraph system are needed to achieve list-decoding. In this subsection we complete the picture with a simple set system based on subspaces that allows us to instantiate \Cref{lem:gi-list-decoding-to-list-recovery} and \Cref{thm:list-recovery}, and thereby prove \Cref{thm:sub-poly-aLLDCs}. While the systems presented in subsequent sections achieve substantially better parameters, the one presented here is simpler and requires no background on high dimensional expanders.

\subsubsection{The Subspace Set System}
Let \(\fsize\) be a prime power and let \(d > 1\) be an integer. 
Let \(\F_\fsize^{2d}\) be a vector space. Our code will simply be a direct product over the inclusion graph between $(d-1)$ and $(d+1)$ dimensional subspaces of \(\F_\fsize^{2d}\). Below, we state this formally in our language of hypergraph systems.

For subspace $v,u$ of $\mathbb{F}_\fsize^{2d}$, write \(v \leq u\) to indicate \(u\) is a linear subspace of \(v\). Let \(\S_{\fsize,d}=(V,S)\) be the hypergraph whose vertices
\[
V = \sett{v \leq \F_\fsize^{2d}}{\dim(v) = d-1}
\]
are the set of $d-1$ dimensional subspaces of $\F_\fsize^{2d}$. To describe the hyperedges (and later intersection graphs), for any subspace $u \leq \F_\fsize^{2d}$ of dimension greater than $d-1$, denote by $s_u$ the set of $d-1$ dimensional subspaces contained in $v$
\[
s_u \coloneqq  \sett{v \in V}{v \leq u}.
\]
We take our hyperedges to be given by all such $s_u$ at dimension $d+1$:
\[
S = \sett{s_u}{\dim(u) = d+1}
\]
In other words, our message bits sit on $(d-1)$-dimensional subspaces, and our encoding is indexed by $(d+1)$-dimensional subspaces which record every bit they see on a $(d-1)$-dimensional subspace below them.

\begin{figure}
    \centering
    \includegraphics[scale=0.5]{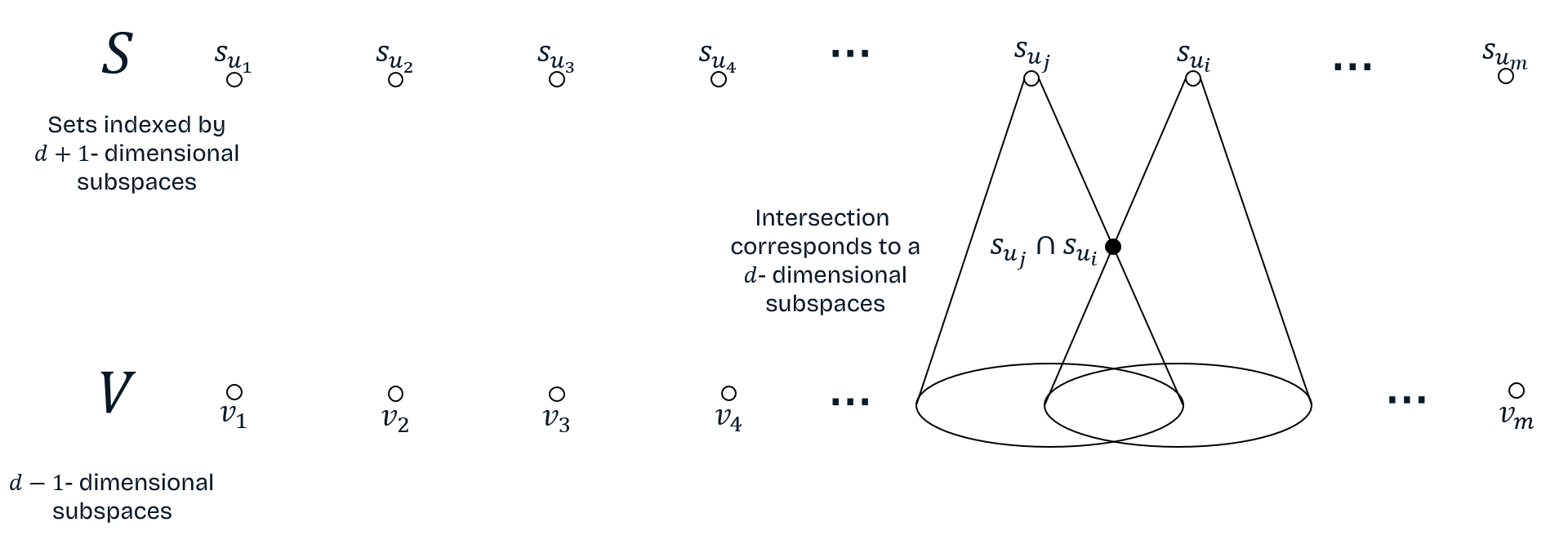}
    \caption{The Subspace Set System}
    \label{fig:subspaces}
\end{figure}

We briefly remark on the parameters of the direct product code on this system $(V,S)$. The length of the message is \(|V|=N=(1 \pm o(1))\fsize^{d^2-d}\). The length of the encoding is also \(|S|=|V|=N\) since $\mathbb{F}_{\fsize}^{2d}$ has the same number of $(d-1)$ and $(d+1)$ dimensional subspaces. The set size of \(s_u \in S\), which corresponds to the (log)-alphabet, is \(|s_u|=(1\pm o(1))\fsize^{2d-2}\). 
For a given target message length $N$ and $\varepsilon \leq \frac{1}{\polylog(N)}$, one should roughly think of taking $\fsize \sim \poly(\frac{1}{\varepsilon})$ and $d \sim O\left(\sqrt{\frac{\log N}{\log \frac{1}{\varepsilon}}}\right)$ (see \Cref{sec:proof-subpoly} for exact parameters). Thus the rate of the code is \(2^{\Theta(\sqrt{\log N\log \frac{1}{\varepsilon}})}\), and at worst \(2^{\Theta(\sqrt{\log N\log \log N})}\) for constant $\varepsilon$.

\paragraph{The Decoding Graph} The decoding graph \(G\) of our system, which we recall is a graph on vertex set $S=\{s_u:\text{dim}(u)=d+1\}$, has edges between \(s_u\) and \(s_{u'}\) iff \(\dim(u \cap u')=d\). We note that the intersection is equal to \(s_u \cap s_{u'} = s_{u \cap u'}\), all subspaces containing \(u \cap u'\). It is easy to verify this is a regular hypergraph system and the stationary distribution on $G$ is uniform.

\paragraph{The Intersection Graphs}
We now describe the intersection graphs \(G_{s_u}=(L,R,E)\) and argue they are good samplers. The right-hand side is the subspaces in \(s_u\) itself. The left-hand side is indexed by the possible intersections. We can partition the vertices in \(L\) to sets according to the intersection with \(s_u\), that is, \(t_{w} = \sett{s_{u'}}{u\cap u'=w}\) for all \(w \leq u\) of dimension \(d\). We note that all \(t_w\) have the same size and that every pair \(s_{u_1},s_{u_2} \in t_w\) have the same neighbor set, which is just \(N(t_w)=\sett{v \leq w}{\dim(u)=d-1}\). 
\begin{claim} \label{claim:subspace-local-graphs-expand}
    For every \(s_u \in S\), the graph \(G_{s_u}\) is a \(\frac{2}{\sqrt{\fsize}}\)-expander. Consequently, it is also a \(\frac{4}{\fsize}\)-sampler.
\end{claim}
This claim is fairly standard; a similar variant appears e.g.\ in \cite[Lemma 2.12]{impagliazzo2008uniform}.
\begin{proof}
    Instead of analyzing \(G_{s_u}\), it suffices to analyze the graph where we replace all vertices in \(t_w\) with a single vertex whose neighborhood is \(N(t_w)\). This graph is just the containment graph between subspaces of dimension \(d\) and subspaces of dimension \(d-1\) that lie in \(u\). This, in turn, is well known to be isomorphic to the lines vs.\ planes graph via mapping \(w \leq u\) to \(w^{\perp} = \sett{x \in u}{\forall y \in w, \; \iprod{x,y}=0}\) where \(\iprod{x,y}=\sum_{i=1}^{2d} x_i y_i\) is the standard bilinear form. Finally, it is a standard fact that the lines vs.\ planes graph is a \(\frac{2}{\sqrt{\fsize}}\)-expander, so \(G_{s_u}\) itself is a \(\frac{2}{\sqrt{\fsize}}\)-expander and the result follows from \Cref{claim:expander-sampler}. 
\end{proof}

\paragraph{Guruswami-Indyk Sampler}

The above construction describes the hypergraph system $(V,S,G)$ we will use for list recovery. We also need a corresponding bipartite sampler $G'=(S,R,E)$ whose left vertices correspond to $S$ with which to apply the Guruswami-Indyk reduction. Luckily, in this case we already have such an expander graph in hand: $(V,S)$ itself! More accurately, we will use the inclusion graph $(S,V)$ between $(d+1)$-dimensional subspaces on the left and $(d-1)$-dimensional subspaces on the right. This is a so-called (generalized) `Grassmann graph' and is well known to be a $\frac{1}{\fsize}$-expander (see e.g.\ \cite{dikstein2018boolean}) and therefore a \(\frac{2}{\fsize^2}\)-sampler as well. This choice of graph at most squares the log-alphabet (so the rate remains \(2^{\Theta(\sqrt{\log N\log \frac{1}{\varepsilon}})}\)), leaves the final blocklength unaffected since $|S|=|V|$, and is additionally convenient in that we will not need to implement a second class of adjacency circuits for a new choice of $G'$.

\subsubsection{Low Congestion Routing} \label{subsec:routing-subspaces}
Let \(s_u,s_{u'} \in S\). Informally our routing scheme is quite simple. We pick a random basis for the intersection $u \cap u'$, random extensions of this basis to each of $u$ and $u'$, then route by trading out basis vectors one by one to move from $u$ to $u'$. More formally, consider the following routing scheme:

\begin{enumerate}
    \item Denote by \(k \coloneqq \dim(v \cap v')\) and choose a random basis \(x_{d+1-k},x_{d+2-k},\dots,x_{d+1}\) to \(u \cap u'\).
    \item Choose \(d-k\) random vectors \(y_{1},y_{k+2},\dots,y_{d-k} \in v\) that extend \(x_{d+1-k},x_{d+2-k},\dots,x_{d+1}\) to a basis of $u$, and similarly random vectors \(z_{1},z_{2},\dots,z_{d-k} \in u'\) for \(u'\).
    \item Let $u_1=u$, $u_{d+1-k}=u'$, and for \(i \in [d-k]\):
    \[u_i = span(z_1,z_2,\dots,z_i,y_{i+1},y_{i+2},\dots,y_{d-k}) + (v \cap v').\]
    \item Output the path \(s_{u_1},s_{u_2},\dots,s_{u_{d+1-k}}\).
\end{enumerate}
\begin{claim}\label{claim:routing-on-subspaces}
    The routing scheme above has the following properties:
    \begin{enumerate}
        \item Length \(d\)
        \item Unit congestion
        \item It can be implemented by a circuit of size $\poly(d,\fsize)$ and depth $O(\log^2(d)\log(\fsize))$.
    \end{enumerate}
\end{claim}

\begin{proof}
    The first property is immediate from construction. The second property follows from the fact that our routing scheme is `spread uniformly' over $G$. In other words, if we draw random start and end points $s_u$ and $s_{u'}$, a random path from the scheme $\mathcal{O}(s_u,s_{u'})$, and finally a random edge $(s_{u_i},s_{u_{i+1}})$ from the path, it is easy to see by symmetry the resulting edge is distributed uniformly on \(G\).
    Thus 
    \[\Pr_{u,u',\mathcal{O}}[\exists e \in \mathcal{O}(u,u'): e \in B] \leq d\Pr_{u,u',\mathcal{O}, e \in \mathcal{O}(u,u')}[ e \in B] \leq d\Prob{}{B}.\]
    We defer the proof of the circuit complexity to the final subsection, where we construct circuits for all our operations in this section.
\end{proof}

\subsubsection[Proof of Sub-Polynomial Rate aLLDC]{Proof of \Cref{thm:sub-poly-aLLDCs} (Modulo Circuits)}\label{sec:proof-subpoly}

We are now ready to prove the main result of the section (though, again, we defer exact circuit implementations to \Cref{sec:circuit-subpoly}). First, we claim that whenever \(\fsize\) is large enough, \(\S_{\fsize,d}\) is list-recoverable.

\begin{lemma} \label{lem:subspaces-are-an-LLRC}
For any \(\varepsilon > 0\), $d \in \mathbb{N}$, and \(\fsize\) a prime power with \(\fsize > \frac{3 \cdot 10^{8} d^3 \log d}{\varepsilon^2}\), \(\S_{\fsize,d}\) is a \((\frac{1}{1500 d},0.01)\)-approximate, \(\frac{100\log d}{\varepsilon}\)-input, \((1-\frac{1}{10^5 d},\frac{100\log d}{\varepsilon}, d)\)-LLRC  with the properties specified in \Cref{thm:list-recovery}.
\end{lemma}
\begin{proof}
    \Cref{claim:routing-on-subspaces} states that \(\S_{\fsize,d}\) has a length \(\leq d\) unit congestion routing scheme and combining our assumption on the field size $\fsize$ with \Cref{claim:subspace-local-graphs-expand} promises all intersection graphs of \(\S_{\fsize,d}\) are \(\frac{10^{-12}\varepsilon}{d^3\log d}\)-samplers. The lemma then immediately follows from \Cref{thm:list-recovery} with $t=d$ and $\delta=\frac{1}{1500d}$.
\end{proof}

We are ready to prove \Cref{thm:sub-poly-aLLDCs}.
\begin{proof}[Proof of \Cref{thm:sub-poly-aLLDCs}]
Our goal is to build a code on $N$ message bits with \(\varepsilon < \frac{1}{\log^2 N}\). We will show there exists a valid setting of parameters for such a code on
\[
N' \in \left[N,N\cdot 2^{8\sqrt{\frac{\log(N)}{\log \frac{1}{\varepsilon}}}}\right]
\]
message bits for any $\varepsilon \leq \log^2(N)$. The desired code on $N$ bits then follows easily by encoding $\frac{N'}{N}$ copies of the original message with the slightly larger code above.\footnote{More formally, $\lfloor \frac{N'}{N} \rfloor$ copies plus a truncated final copy to reach exactly size $N'$.} We cover this in \Cref{lem:many-to-all} for completeness.

With this in mind, our code will be the (Guruswami-Indyk) composition of:
    \begin{enumerate}
        \item The list recoverable code on \(\S_{\fsize,d}=(V,S)\) promised in \Cref{lem:subspaces-are-an-LLRC} with parameters:        
        \begin{itemize}
            \item A prime power \(\fsize  \in [\frac{1}{\varepsilon^3},\frac{2}{\varepsilon^3}]\) (e.g.\ $\fsize=2^s$ for the first $s$ in this range).
            \item The first integer $d>1$ such that $\gaussb{2d}{d+1}_\fsize \geq N$.
        \end{itemize}
        \item The discussed sampler graph $G'=(S,V,E)$ of $(d+1)$ vs $(d-1)$ dimensional subspaces
    \end{enumerate}

    Let's now calculate the parameters more carefully. Recall our target number of message bits is $N$, and the true number of resulting message bits of the GI-composition above is $|V|=\gaussb{2d}{d+1}_\fsize =(1 \pm o(1))\fsize^{d^2-d}$. We therefore take the first $d>1$ such that $(1 \pm o(1))\fsize^{d^2-d} \geq N$. In particular, for large enough $N$, one can check the first such $d$ lies in:
    \[
    d \in \left[\sqrt{\frac{\log N}{\log \fsize}}, \sqrt{\frac{\log N}{\log \fsize}} +2\right]
    \]
    Thus plugging in the upper bound of $d=\sqrt{\frac{\log N}{\log q}} +2$, we get $|V| \leq N\fsize^{4d} \leq N2^{8\sqrt{\log(N)\log \frac{1}{\varepsilon}}}$ as claimed.

    Now, one can also see that for our choice of parameters we have
      \begin{enumerate}
          \item $\fsize > \frac{3 \cdot 10^{12} d^3 \log d}{\varepsilon^2}$, and
          \item $G'$ is a \(\frac{1}{10^7 d\varepsilon^2}\)-sampler graph
      \end{enumerate}
      for large enough $N$. Thus by \Cref{lem:subspaces-are-an-LLRC} and \Cref{lem:gi-list-decoding-to-list-recovery}, the GI-composition of the above is immediately a $(\frac{1}{1500^2d^2},0.01)$-approximate $(\varepsilon,\ell_{out},q)$-LLDC with 
      \[
      q=O(\frac{d\log  d}{\varepsilon})
      \]
    query complexity, and
       \[
        \ell_{out} \leq O(\frac{\log  d}{\varepsilon})
       \]
       list size. As discussed, the list size can be improved to just $O(\frac{1}{\varepsilon})$ (or $O(\frac{\ell_{in}}{\varepsilon})$ in the list-recovery case), but we defer discussion of this to the next section. Plugging in $d=\Theta\left(\sqrt{\frac{\log N}{\log \frac{1}{\varepsilon}}}\right)$ gives the claimed parameters.
      
      It is left to bound the rate of the code. The block length is $|V|$ after the GI-composition with $G'=(S,V,E)$, which contributes a $2^{-8\sqrt{\log N \log \frac{1}{\varepsilon}}}$ factor in the rate since $|V| \leq N2^{8\sqrt{\log(N)\log \frac{1}{\varepsilon}}}$. Since $(V,S)$ is balanced and regular, the log-alphabet is simply the degree of the graph $\gaussb{d+1}{2}_q$ squared, which is at most $(1+o(1))\fsize^{4d-4} \leq 2^{8\sqrt{\log N \log \frac{1}{\varepsilon}}}$ for large enough $N$. Thus together, the rate is at least $2^{-16\sqrt{\log N \log \frac{1}{\varepsilon}}}$ as claimed.
\end{proof}

Finally, we briefly remark on the parameters of the code for a general parameter $\varepsilon$, dropping the assumption that $\varepsilon \leq \frac{1}{\log^2(N)}$ which has only been used to simplify rate calculations and notation.

\begin{remark}[General $\varepsilon$]
    For any $\varepsilon>0$, the construction above may be instantiated to have rate $2^{O(\sqrt{\log(N)\log\log(N)})}$, makes $q \leq \tilde{O}\left(\sqrt{\frac{\log N}{\log \frac{1}{\varepsilon}}}\right)$ queries, and has list size $\ell_{out} \leq O(\frac{1}{\varepsilon})$ (after the list `pruning' described in \Cref{sec:subpoly}).
\end{remark}


\subsubsection{Circuit Implementations}\label{sec:circuit-subpoly}

We are left with the task of actually designing the circuit implementation for $(V,S)$ and our routing scheme. At a high level, our circuit implementations here and in the subsequent sections generally have two components. First, we will build a circuit that takes in the `name', i.e.\ the bit-representation of a vertex or hyperedge in the graph, and outputs a more reasonable `canonical representation' of the object that's easier to work with. In this case our graphs consist of subspaces, so we will build a circuit that inputs a bit string and outputs a nicer matrix representation of the corresponding subspace.  We will similarly need an `inverse circuit' that takes the canonical representation and maps it back to its corresponding name. Second, once we have implemented these, we will implement routing and various adjacency circuits operating over this nicer space of canonical representations. The final circuit implementations can always then be constructed by composing these with the `bit $\leftrightarrow$ canonical' representation circuits.

\paragraph{Bit Representation $\leftrightarrow$ Canonical Representation:}

We start with our basic circuit translating from bit names to subspace representations. Somewhat surprisingly, it is not immediately obvious how one should actually go about implementing this. For instance, it is tempting to simply read off the input bit-representation $\{0,1\}^T$ as a basis (for some appropriate choice of $T$), but this runs into the problem that the encoding is not unique. Indeed, since every subspace has $\fsize^{O(d^2)}$ bases, using such an encoding would blow up the rate of our code \textit{polynomially}, which we obviously cannot afford in the sub-polynomial regime.

Thus instead we should try to find some sort of canonical representation or basis for each subspace. There is a natural choice for this: we can put the basis (or rather the matrix whose rows are the basis vectors) into reduced row echelon form (RREF). This gives a convenient and bijective set of representations to work with (since two matrices have the same RREF if and only if their row span is the same). However, it is no longer obvious how to build a simple circuit that maps a particular input $\{0,1\}^T$ to a corresopnding RREF matrix.

While it is likely possible to construct such a circuit efficiently, we will instead solve the problem in a simpler way by cheating slightly. In particular, instead of using the full set of subspaces as claimed in our description earlier in the section, we will actually only use subspaces with the following very simple RREF form (for $k$-dimensional subspaces of $\F_{\fsize}^{2d}$):
\begin{equation}\label{eq:canonical-RREF}
A(M) \;=\; \begin{bmatrix} I_k & M \end{bmatrix}_{k\times 2d}
\end{equation}
for arbitrary $M \in \F_{\fsize}^{k \times 2d-k}$ and \(k \in \set{d-1,d+1}\). It is of course very easy to give a bit map to this representation (at least for $q$ a power of $2$) simply by reading off the matrix $M$ from the input (moving from the above form back to bit-representation is similarly easy). We'll call the set of subspaces with RREF form as in \Cref{eq:canonical-RREF} `\textit{valid}'. It is easy to see that almost all subspaces are valid, as the fraction of valid subspaces is equivalent to the fraction of random matrices in $\mathbb{F}_{\fsize}^{k \times 2d}$ whose initial $k \times k$ minor is full rank:
\begin{observation}
    All but an $O(1/\fsize)$-fraction of subspaces are valid.
\end{observation}
Since our routing scheme has unit congestion, as long as $\fsize$ is sufficiently larger than $d$, we can ensure our routing on the full subspace system only encounters an invalid subspace with vanishingly small probability (indeed we've set the parameters in the prior section such that this is true). This means our analysis of the full subspace code can be ported directly to the case where we only encode valid subspaces as above.

\paragraph{Hypergraph Adjacency Circuits:} Now that we have the above translation in hand, let's recall what circuits we actually need to implement in for our full construction. We will do the list-decoding case; the list-recovery case is no different. After GI-composition, we are given as input a boolean-valued circuit $\mathcal{C}_w(v,i,j)$ computing the word $w$, which takes a $(d-1)$-dimensional subspace $v \in V$, as well as two indices $i$, $j$ specifying a neighbor $s \sim v$, then a second neighbor $v' \sim s$. Note the code is thought of as having blocks indexed by $v$, and alphabet indexed by $i,j$. Of course we have not yet defined what subspace is the `$i$-th' neighbor of a particular $v \in V$ or conversely $s \in S$. We will do this carefully below via a similar RREF-type transform as above.

Putting this aside for the moment, let's now recall the process we need to implement using the above circuit. Given a target $v$, we first sample a random neighbor $s_t \ni v$, then call the routing oracle $C_\mathcal{O}$ (which we'll implement shortly) to compute a path $s_1,\ldots,s_t$ of $(d+1)$-dimensional subspaces. Our algorithm computes a list of local functions on each $s_\ell$ by picking several random $v' \sim s_\ell$, and querying the value of $\mathcal{C}_w(v',i', \cdot )$ where $i'$ is the index such that $s_{\ell}$ is the $i'$-th neighbor of $v'$. Finally given these lists, our list matching procedure draws several random $v'' \in s_{\ell} \cap s_{\ell+1}$, and checks across pairs $v'_1,v'_2$ (corresponding to decoded lists) whether $\mathcal{C}_w(v'_1,i'_1,j'_1) = \mathcal{C}_w(v'_2,i'_2,j'_2)$ where $j'_z$ is the index such that $v''$ is the $j'_z$-th neighbor of the $i'_z$-th neighbor of $v'_z$ (for $z \in \{1,2\}$). At the end of the matching process we pick a good list at $s_t$ (indexed by some neighbor $v_t$), and output $\mathcal{C}_w(v_t,i,j)$ where $i$ is the index of $s_t$ in $v_t$, and $j$ is the index of $v$.

Now with the above notational mess out of the way, we claim it is enough to implement the five `standard' circuits in the circuit implementation of $(V,S)$, which we recall here include the adjacency and intersection circuits (for some yet to be specified neighbor ordering):
\begin{enumerate}
    \item $C_{V\to S}(v,i)$ outputs the $i$th neighbor $s \sim v$ (in matrix form)
    \item $C_{S\to V}(s,i)$ outputs the $i$th neighbor $v \sim s$ (in matrix form)
    \item $C_{S \cap S \to V}(s,s',i)$ outputs a $(d-1)$-dim subspace $v \in s \cap s'$ (in matrix form),
\end{enumerate}
and the two `reverse index' circuits:
\begin{enumerate}
    \item $C^{-1}_{V\to S}(v,s)$ outputs the index $i$ such that $C_V(v,i)=s$
    \item $C^{-1}_{S\to V}(s,v)$ outputs the index $i$ such that $C_S(s,i)=v$
\end{enumerate}
Above, we think of $v$ and $s$ as being input to our circuits in canonical RREF form (and, in particular, we will only work with valid subspaces), and $i$ as the bit index of a valid neighbor. Note that given the above we can easily implement the `$(i,j)$-indexed' circuits for the GI-composition in the previous description: given $v' \in V$, a neighbor $s_\ell \sim v$, and a second $v'' \sim s_{\ell}$, we can compute the appropriate index of $\mathcal{C}_w(v',\cdot,\cdot)$ to query just by running $C^{-1}_{V \to S}(v',s_\ell)$ and $C^{-1}_{S\to V}(s_\ell,v'')$. In other words, once we have indexed $(V,S)$ successfully, indexing the two-step walk on $(V,S)$ (which corresponds to our full GI-encoding) is immediate.


Let's now implement our circuits (and in doing so define the appropriate neighborhood orderings). We will start with the circuits $C_{S\to V}$ and $C^{-1}_{S\to V}$.

\paragraph{Implementing $C_{S\to V}$ and $C^{-1}_{S \to V}$:}

Our strategy to build these circuits essentially follows a variant of the canonical RREF form method from before, but performed inside a fixed valid $(d+1)$-dimensional subspace $s$.

Let \(A_s\) be the canonical RREF form of \(s\). A subspace \(v\leq s\) is valid if and only if one can write it as a product of \([A]_s(M)\cdot A_s\) where
\begin{equation}\label{eq:canonical-RREF-2}
[A]_s(M) \;=\; \begin{bmatrix} I_{d-1} & M \end{bmatrix}_{d-1\times d+1}.
\end{equation}
To see why this is true, note that in one direction, every matrix of the form \([A]_s(M)\cdot A_s\) is a distinct RREF form of a valid \(d-1\)-subspace contained in \(s\).

As for the other direction, for every valid \(v \leq s\), there is a basis \(x_1,x_2,\dots,x_{d-1}\) such that projecting it to the first \(d-1\) coordinates gives back the standard basis. This means that \(x_i\) is equal to the \(i\)-th row of \(A_s\) plus some linear combination of the \(d\) and \(d+1\)-th row of \(A_s\) (i.e.\ the kernel of the projection inside \(s\)). Therefore, writing these linear combinations row by row will give us a matrix \([A]_s(M)\) such that \([A]_s(M) A_s\) is the RREF form of \(v\).

Therefore, the circuit \(C_{S\to V}\) receives \(s\) (represented as above), and the index \(i\) which is actually equal to the \(M\) in \([A]_s(M)\), and outputs \([A]_s(M)A_s\).


We note that this circuit only covers the valid subspaces inside \(s\), but similar to the above, the fraction of non-valid subspaces in \(s\) is \(O(\frac{1}{\fsize})\). To handle this modification in our approximate list-recovery framework, we can just treat the missing $O(1/\fsize)$ fraction of subspaces as part of the approximate input agreement $\delta_{in}$. In other words, these entries are simply assumed to disagree with the global $f$ we'd like to decode (more formally, we just treat them as always responding with some fixed symbol, e.g. $0$). This changes the parameters by at most a constant factor since we already assume $\delta_{in} > \Omega(\frac{1}{\fsize})$.

In the other direction, to compute \(C_{S\to V}^{-1}\), given the global names or canonical representations of a $v \subset s$ pair, we need only compute a basis for $v$ in terms of $s$ and pass to the RREF form in this basis, which immediately gives us the index as desired. These linear algebraic operations can be performed by logspace uniform circuits of size $\poly(d,\log \fsize)$ and depth $O(\log^2(d)\log \fsize)$ \cite{csanky1975fast}. We remark that since these circuits are produced uniformly, this also gives the claimed Time$(\polylog(N)\cdot\log\frac{1}{\varepsilon})$ local encoding.

\paragraph{Implementing $C_{V\to S}$ and $C^{-1}_{V\to S}$:} We now need to implement the `left-to-right' adjacency operator circuits from $(d-1)$ to $(d+1)$ dimensional subspaces. Thankfully, our graph is `self-dual' via mapping a subspaces $w$ on either side to their `orthogonal complement' $w^\perp$. For a subspace \(w\), the `orthogonal complement' 
\[w^{\perp} = \sett{x \in \F_{\fsize}^{2d}}{\forall y \in w, \iprod{x,y}=0}\]
where \(\iprod{x,y}\) is the standard bilinear form (we call it `orthogonal complement' even though there is no notion of orthogonality in finite fields).

It is standard that a the dual of a \(k\)-dimensional subspace is a \(2d-k\)-dimensional subspace and that the dual of the dual subspace is the original subspace. Moreover, a \(d\pm 1\)-subspace is valid if and only if the dual is valid with respect to the last \(d\mp 1\)-coordinates. That is, that we can write it as the row span of a matrix of the form
\[
    A'(M) \;=\; \begin{bmatrix} M & I_k \end{bmatrix}_{k\times 2d}
\]
This is true because a \(d \pm 1\)-subspace is valid if and only if \(w+E_{d \mp 1} = \F_\fsize^{2d}\) where \(E_{d \mp 1}\) is the subspace spanned by the last \(d \mp 1\) standard vectors. This is if and only if \(w^{\perp} \cap E_{d \mp 1}^{\perp} = \set{0}\). The subspace \(E_{d \mp 1}^{\perp}\) is the subspace spanned by the first \(d \pm 1\) standard vectors, so this is if and only if \(w^{\perp} + E_{d \mp 1}^{\perp} = \F_{\fsize}^{2d}\) if and only if \(w^{\perp}\) is valid with respect to the last \(d \mp 1\) coordinates.

Therefore, to implement \(C_{V \to S}\) we just compute the perpendicular space, find its canonical RREF with respect to the last coordinates, use \(C_{S\to V}\) (with respect to the last coordinates), and find the dual of the  subspace output by \(C_{S \to V}\). We do a similar operation for \(C_{V \to S}^{-1}\).


Computing these just requires computing orthogonal complements of the input (and running the above circuits), so remains in logspace uniform size $\poly(d,\log \fsize)$ and depth $O(\log^2(d)\log \fsize)$. We note this (of course) has the same caveat as the above that it fails to encode an $O(\frac{1}{\fsize})$ fraction of the neighbors of $v$, which we can treat as part of the $\delta_{in}$ error in our algorithms.

\paragraph{Implementing $C_{S\cap S \to V}$:} Recall $C_{S\cap S \to V}(s,s',i)$ takes in two matrices $\mathbb{F}_{\fsize}^{(d+1) \times 2d}$ in canonical RREF form, and must output the RREF form of a random valid $(d-1)$-dimensional subspace $v \in s \cap s'$. Unfortunately, an intersection of two valid \((d+1)\)-subspaces \(s \cap s'\) may not contain any valid \((d-1)\)-subspace \(v\). The circuit we actually implement takes the intersection of \(s \cap s'\), computes its RREF, \emph{and if it is a valid \(d\)-dimensional subspace}, it finds a valid subspace in it in the same manner that \(C_{S \to V}\) finds valid subspaces (of course, the intersection is \(d\)-dimensional but other than changing \(d+1\) to \(d\), the circuit operates the same way). We observe that there are only \(O(\frac{1}{\fsize})\)-edges \(\set{s,s'}\) for which the intersection is not valid, and therefore this circuit fails on a \(O(\frac{1}{\fsize})\)-fraction of the edges. For the rest of the edges, it only gives a mapping to a \(1-O(\frac{1}{\fsize})\)-fraction of the possible subspaces.

This suffices for our purposes since the analysis of the algorithm essentially stays the same even if we fail on an additional \(O(\frac{1}{\fsize})\)-fraction of the edges, and as before, even if on the non-failed edges we only cover a \(1-O(\frac{1}{\fsize})\)-fraction of the possible subspaces, this essentially just adds an additional \(O(\delta_{in})\)-error in each propagation step which we can essentially absorb.



Finally, standard parallel linear algebra allows us to compute such a form given matrix representations of $s$ and $s'$ with logspace uniform circuits of size $\poly(d,\log \fsize)$ and depth $O(\log^2(d)\log \fsize)$ as desired.

\paragraph{The Routing Circuit}

Finally, we are left to implement the routing circuit. Recall our routing circuit will take in the RREF form of two $d+1$ dimensional subspaces $s$ and $s'$ with $\text{dim}(s \cap s')=k$, and routes by
\begin{enumerate}
    \item Picking a random basis $x_{d-1-k},\ldots,x_{d-1}$ of $s \cap s'$
    \item Independently extending this to random bases of $s$ and $s'$, $(y_1,\ldots,y_{d-2-k})$ and $(z_1,z_2,\ldots,z_{d-2-k})$.
    \item For every $i \in [d-k-2]$ output the subspace
\[
v_i = span(z_1,z_2,\dots,z_i,y_{i+1},y_{i+2},\dots,y_{d-2-1}) + (v \cap v').
\]
\end{enumerate}
With all the above machinery in hand, this routing procedure is actually quite easy to implement. We have the RREF forms of $s$ and $s'$ and can compute a matrix form of $s \cap s'$ as discussed. Computing a random basis of $s \cap s'$ (and corresponding extensions) is then as simple as drawing the appropriate number of random combinations of the rows.\footnote{Again, this has some small failure probability of not resulting in a full rank system, but this occurs with probability at most $O(1/\fsize)$ and can easily be amplified by repeating a few times, so it can be safely ignored.} Once these basis vectors are generated, we can directly output the RREF representation of each $v_i$ defined above. Again, all procedures can be performed by logspace uniform circuits of size $\poly(d,\log \fsize)$ and depth $O(\log^2(d)\log \fsize)$. Finally, as described, this process has the chance of outputting $v_i$ not in our encoding (i.e. that do not have the appropriate RREF form), but this occurs with probability at most \(O(d/\fsize)\) by unit congestion so can be ignored similar to the above such bad events.

\paragraph{Circuit Complexity of \Cref{thm:sub-poly-aLLDCs}}

We can finally put everything together to prove the claimed circuit complexity bounds. By \Cref{lem:list-recovery-circuit}, \Cref{thm:list-recovery} uses at most $\poly(t,\frac{1}{\varepsilon},\delta^{-1})$ parallel calls to the circuit implementations of our routing oracle and hypergraph implementation (noting $\ell_{in}$ is set to $O(\frac{\log t}{\varepsilon})$, on top of at most $\poly(t,\frac{1}{\varepsilon},\delta^{-1})$ parallel calls to the GI-graph to simulate the input lists. Thus the total size is at most $\poly(t,\ell_{in},\delta^{-1})$ times the maximum size of the circuit implementations, which are all $\poly(\log(N),\log \fsize)$, and the depth is at most the worst-case depth of these implementations which is $O(\log(\frac{\log N}{\varepsilon})\log^2(\log(N)))$ as claimed.
\section{Polylog Rate aLLDCs}\label{sec:polylog}
In this short section, we formally state and prove our main polylog rate construction using the machinery developed in the last section along with the assumed existence of sparse hypergraph systems $(V,S,G)$ with unit congestion routing. Our main result is the following (large alphabet) aLLDC with polylog rate:
\begin{theorem}[Polylog Rate aLLDCs]\label{thm:polylog-aLLDCs}
    There is a universal constant $C>0$ such that for all large enough $N \in \mathbb{N}$ and \(\varepsilon >0\),
    there exists $\M \geq \N$ and a strongly explicit direct product code $\mathcal{C}: \Sigma_1^{\N} \rightarrow \Sigma_2^{\M}$ that is a $\left(\frac{1}{C\log^{10} N},0.01\right)$-approximate, $(\varepsilon,\ell_{out},q)$-LLDC with
    \begin{enumerate}
        \item \textbf{Rate:} \(R \geq \poly(\frac{1}{\log(N)},\varepsilon)\)
        \item \textbf{Query Complexity:} $q \leq \frac{\log^5(N)}{\varepsilon}\cdot \poly\left(\log\log N,\log \frac{1}{\varepsilon}\right)$
        \item \textbf{List Size:} $\ell_{out} \leq O(\frac{1}{\varepsilon})$
    \end{enumerate}
    Moreover, $\mathcal{C}$ can be locally encoded in $\Time(\polylog(N)\log\frac{1}{\varepsilon})$ and decoded in logspace uniform size $\poly(\frac{1}{\varepsilon},\log N)$ and depth $O\left(\log(\frac{log N}{\varepsilon})\cdot \log\log N\right)$.
\end{theorem}
We remark that if one is only interested in information-theoretic query bounds, the above can be improved to $q \leq \tilde{O}(\frac{\log N}{\varepsilon})$, simply by using completely random paths (computed via brute-force) rather than routing. This matches our query lower bound for such codes in \Cref{sec:lower}.

\Cref{thm:polylog-aLLDCs} also satisfies a slightly stronger high noise $\ell_{in}$-input, $(\delta_{in},\delta_{out})$-approximate, $(\varepsilon,\ell_{out},q)$-list-\textit{recovery} guarantee replacing $\varepsilon$ in the parameters above with $\frac{\varepsilon}{\ell_{in}}$.\footnote{This is a generic translation which simply follows from taking the input list $\mathcal{L}$ and creating a word $w$ by picking a single element in every list at random. As long as the blocklength of the code (number of encoded symbols) is at least $\Omega(\frac{1}{\varepsilon})$, any word that is $2\varepsilon$-computed in $\mathcal{L}$ will be $\frac{\varepsilon}{\ell_{in}}$-computed by the resulting $w$ with overwhelming probability, at which point we can run the aLLDC. We omit the details.} Through a refined concatenation method given in \Cref{sec:alphabet}, this leads to the following binary aLLDCs:
\begin{corollary}[Polylog Rate Binary aLLDCs]\label{cor:binary-polylog-aLLDCs}
    For all large enough $N \in \mathbb{N}$ and \(\varepsilon >0\),
    there exists a strongly explicit binary 
    $\mathcal{C}: \{0,1\}^{\N} \rightarrow \{0,1\}^{\M}$  that is a $(0.01)$-approximate, $(\frac{1}{2}+\varepsilon,\ell_{out},q)$-LLDC with
    \begin{enumerate}
        \item \textbf{Rate:} \(R \geq \poly(\frac{1}{\log(N)},\varepsilon)\)
        \item \textbf{Query Complexity:} $q \leq \poly(\log(N),\frac{1}{\varepsilon})$
        \item \textbf{List Size:} $\ell_{out} \leq \frac{\polylog(\frac{1}{\varepsilon})}{\varepsilon^2}$
    \end{enumerate}
    Moreover, $\mathcal{C}$ can be locally encoded in $\Time \left(\poly(\log(N),\frac{1}{\varepsilon})\right)$ and locally decoded in logspace uniform size $\poly(\frac{1}{\varepsilon},\log N)$ and depth
    $\tilde{O}\left(\log^{2}\left(\frac{\log N}{\varepsilon}\right)\right)$.\footnote{Here $\tilde{O}$ hides $\log\log(\frac{\log N}{\varepsilon})$ factors.}
\end{corollary}
As discussed, in some cases $\poly(\frac{1}{\varepsilon})$ encoding time is insufficient. In \Cref{sec:alphabet} we also give a variant with $\poly(\log N)\cdot \log^2\frac{1}{\varepsilon}$ encoding time as well as $\tilde{O}(\frac{1}{\varepsilon^2})$-query complexity (which is the best possible for binary aLLDCs \cite{grinberg2018indistinguishability}). This comes at the cost of worse rate depending quasipolynomially on $\varepsilon$ (this trade-off is standard and comes from concatenating with a small complete complex, see below \Cref{thm:KO-locality-reduced} for further discussion).

To prove \Cref{thm:polylog-aLLDCs}, we need a hypergraph system $(V,S,G)$ with the properties discussed in \Cref{sec:subpoly} but only poly-logarithmic degree. Roughly speaking, these systems are built by replacing our direct product overing neighboring subspaces with a direct product over neighborhoods of a 3-partite \textit{high dimensional expander}.\footnote{The HDX-savvy reader might notice that the previous construction already is, in some sense, a neighborhood encoding on a high dimensional expander, albeit not a very sparse one (the Flags complex). The neighborhood encoding we use for general 3-partite HDX is a bit different, as the simplified version in \Cref{sec:subpoly} relied on special properties of subspaces.} We defer the exact construction to \Cref{sec:instantiations} where we also prove highly efficient local routing schemes for certain such families. For the moment, we just state formally the properties of the resulting hypergraph system we'll use for our code. We split this into parts: existence and circuit implementations.

\begin{proposition}[Sparse Hypergraph Systems]\torestate{\label{prop:graph-system-for-log-rate-graph}
    For every large enough $N \in \mathbb{N}$ and $\varepsilon>0$, there exists a regular hypergraph system $(V,S,G)$ with the following properties:
    \begin{enumerate}
        \item $|V| \in [N,N\poly(\log N, \varepsilon^{-1})]$
        \item $(V,S)$ is bi-regular with left and right degree at most $\poly(\log N, \varepsilon^{-1})$
        \item $G$ has a length $t \leq O(\log^{3}(N))$ unit congestion routing scheme $\mathcal{O}$
        \item For every $s \in S$, $G_s$ is a $,\frac{\varepsilon}{15^3\cdot 10^{11}t^4\log t}$-sampler
        \item The (reverse) inclusion graph $(S,V)$ is an $\frac{10^{-5}}{4t\varepsilon^2}$-sampler.
    \end{enumerate}}
\end{proposition}
We note the somewhat complicated parameters set above are set this way to obtain codes from \Cref{thm:list-recovery} with $\delta_{out}=0.01$ and input list-size generated by the Guruswami-Indyk reduction.

\begin{claim}[Circuit Implementation, Informal]\torestate{\label{claim:graph-system-for-log-rate-graph-circuit}
    $(V,S,G)$ and $\mathcal{O}$ as described in \Cref{prop:graph-system-for-log-rate-graph} have circuit implementations in logspace uniform depth $O(\log\log(N)\log(\frac{\log N}{\varepsilon}))$ and size $\poly(\frac{1}{\varepsilon},\log N)$.
    }
\end{claim}
Technically, \Cref{claim:graph-system-for-log-rate-graph-circuit} is informal only in the sense that, like our subspace circuit implementation, the circuits we construct will not \textit{exactly} encode the system $(V,S,G)$ from \Cref{prop:graph-system-for-log-rate-graph}, but do so with some vanishingly small error (in particular, the circuits we give will correspond to a very slightly larger hypergraph $(V',S')$ that includes a $\ll \poly(\varepsilon)$ fraction of extra `degenerate' nodes). As for the subspace construction, this error will be sufficiently small that it does not affect the proof in any meaningful way. This is discussed further in \Cref{sec:instantiations}.\yotamd{Not a comment for this section, but maybe we say this once clearly in prelims, instead of doing it along the way?}

We are now ready to give the proof of \Cref{thm:polylog-aLLDCs}, which largely follows the same outline as the proof of \Cref{thm:sub-poly-aLLDCs}, replacing the subspace system with \Cref{prop:graph-system-for-log-rate-graph}

\begin{proof}[Proof of \Cref{thm:polylog-aLLDCs}]
    We first describe the code, which follows exactly the same framework as \Cref{thm:sub-poly-aLLDCs}, replacing the subspace hypergraph system with \Cref{prop:graph-system-for-log-rate-graph} in each step.

    let $N$ be our target message size, and let $N'$ be the first integer greater than $N$ for which there exists a hypergraph system $(V,S,G)$ from \Cref{prop:graph-system-for-log-rate-graph} with $|V|=N'$. We are promised that $N' \in [N,N\poly(\log(N),\frac{1}{\varepsilon})]$. As before, \Cref{lem:many-to-all} shows it is enough to build a corresponding aLLDC on $N'$ message bits, and encode the original message into this space by (truncated) repetition.

    Now, our code on $N'$ message bits will be the (Guruswami-Indyk) composition of
    \begin{enumerate}
        \item The list-recoverable direct product code \(C_{(V,S)}\) corresponding to $(V,S,G)$ from \Cref{prop:graph-system-for-log-rate-graph}, and  
        \item The sampler (inclusion) graph $G'=(S,V)$ promised by \Cref{prop:graph-system-for-log-rate-graph}
    \end{enumerate}

    Toward proving the claimed code parameters for the GI-composition of these objects, let's first formally recall the list-recovery guarantees of $C_{(V,S)}$ for our parameter setting. Let $t \leq O(\log^3(N))$ be the length of the routing scheme promised in \Cref{prop:graph-system-for-log-rate-graph}. Fixing $\delta=\frac{1}{1500t}$ and $\ell_{in}=\frac{100\log t}{\varepsilon}$, we have set parameters in \Cref{prop:graph-system-for-log-rate-graph} exactly such that every intersection graph $G_s$ is a $\frac{\delta^3}{10^{3}t}$-sampler. Thus $C_{(V,S)}$ is an $\ell_{in}$-input, $(\delta, 0.01)$-approximate, $(1-\frac{10^{-5}}{t},\ell_{in},t)$-LLRC by \Cref{thm:list-recovery}.

    We can now feed the above code into the Guruswami-Indyk reduction with our sampler $G'$. This will not quite achieve the right parameters (namely the list size will be too large), but we will handle this afterwards. \Cref{prop:graph-system-for-log-rate-graph} promises the reverse inclusion graph $G'=(S,V)$ is an $\frac{10^{-5}}{4t\varepsilon^2}$-sampler, so \Cref{lem:gi-list-decoding-to-list-recovery} immediately implies that the composed code with $G'$ is a $(\frac{1}{1500^2t^2},0.01)$-approximate, $(\varepsilon,\frac{200\log t}{\varepsilon},\frac{100t\log t}{\varepsilon})$-LLDC. Plugging in $t = O(\log^3(N))$ gives the claimed parameters, with the exception of the list-size which we argue about separately below.

    Before this, however, let's confirm the rest of the parameters of the construction. The message length is $N$. The blocklength is $|V|=N' \leq N\poly(\log N, \varepsilon^{-1})$. The (log)-alphabet size is the product of the left and right degree of $(V,S)$, which is also promised to be at most $\poly(\log N, \varepsilon^{-1})$ by \Cref{prop:graph-system-for-log-rate-graph}. Together, these result in rate $\poly(\frac{1}{\log N}, \varepsilon)$ as claimed.
    
    
    Finally, the circuit complexity analysis is the same as in \Cref{thm:sub-poly-aLLDCs} given access to the routing and hypergraph implementation circuits promised by \Cref{claim:graph-system-for-log-rate-graph-circuit} and trading in the relevant parameters.

\paragraph{Improving the List Size:}

    It is left to argue that the GI-composition above in fact has list size $O(\frac{1}{\varepsilon})$, not $\frac{200\log t}{\varepsilon}$ as claimed above which grows, albeit slowly, with $N$.
    
    To do this, we need a slightly stronger variant of the list-recovery statement in \Cref{thm:list-recovery}. Recall \Cref{thm:list-recovery} proves that for any input list $\mathcal{L}$ and $f:V \to \Sigma_1$ such that a $(1-\frac{10^{-5}}{t})$ fraction of $s \in S$ have a function $g \in \mathcal{L}_s$ that $\delta$-computes $f$, the decoder $\text{Dec}(s,g,\cdot)$ from \Cref{fig:list-rec-algorithm} $15t\delta$-computes $f$ so long as
\begin{enumerate}
    \item $s$ is `good' in the sense of \Cref{eq:good-paths-distance-1}
    \item $g$ is $\delta$-close to $f$.
\end{enumerate}
    Since we prove $99\%$ of $s$ are good in this sense (and in particular also contain at least one such good starting $g$), this means we can decode $f$ with $\frac{3}{4}$-probability simply by picking a random $s$ and outputting the list of circuits $\text{Dec}(s,g,\cdot)$ for every $g $

    Instead of the above, imagine we were now in a scenario where a random $s$ is not just `good' in the above sense $99\%$ of time, but \textit{also} is promised to have \textit{$\Omega(\varepsilon)$-fraction} of the elements in its list that $\delta$-compute $f$. Since we already prove in \Cref{thm:list-recovery} that starting from any such $g \in \mathcal{L}_s$ results in a circuit that $15t\delta$-computes $f$, this means we can avoid outputting a circuit for every $g$ in the list. In particular by a Chernoff and union bound, randomly selecting $s$, then outputting $\text{Dec}(s,g,\cdot)$ for a $O(\frac{1}{\varepsilon})$ random local functions $g \in \mathcal{L}_s$ suffices to output some $\text{Dec}(s,g,\cdot)$ that $15t\delta$-computes $f$ with $98\%$ probability. Finally, note that if we are promised that our input word $\mathcal{L}$ has a $1-\frac{10^{-5}}{t}$ fraction of $s$ whose lists $\delta$-compute $f$ a $\Omega(\varepsilon)$-fraction of the time, exactly the same proof implies that a random $s$ is `good' and has such a list with $99\%$ probability.

    With the above in mind, let's turn our attention back to the Guruswami-Indyk composed code. By the above discussion, it is enough to argue that for any $f \in \Ball_{1-\varepsilon,\delta^2}(\mathcal{L})$, we output a simulated list to $C_{(V,S)}$ in the reduction such that with $99\%$ probability, a $(1-\frac{10^{-5}}{t})$ fraction of $s \in S$ have lists that $\delta$-compute $f$ for at least a $\Omega(\varepsilon)$ fraction of their local functions. Thankfully, this already follows immediately from the proof of \Cref{lem:gi-list-decoding-to-list-recovery}. In particular, it is enough to note that the Chernoff bound taken under \Cref{eq:def-B-GI} immediately implies the stronger statement that any $s \notin B$ samples an $\Omega(\varepsilon)$ fraction of neighbors that $\delta$-compute $f$) with at least $\frac{10^{-6}}{2t}$ probability. Tracing back through the Markov bound then indeed implies a $1-\frac{10^{-5}}{t}$ fraction of the simulated lists $\mathcal{L}_s$ passed to $C_{(V,S)}$ have the desired property so we are done.\yotamd{I'll say this carefully, but maybe we should be more generous in writing this out explicitly (less important for submission).}
\end{proof}

\section{Constant Rate aLLDCs}\label{sec:constant}
In this section, we cover the construction and proof of our second main result, \textit{constant rate} aLLDCs. We first state the `large-alphabet' version of the result.
\begin{theorem}[Constant Rate aLLDCs]\label{thm:constant-rate}
    For every $\varepsilon>0$ and infinitely many $N \in \mathbb{N}$, there exists an 
$(\exp(-O(\frac{1}{\varepsilon})),0.1)$-approximate, $(\varepsilon,O(\frac{1}{\varepsilon}),q)$-LLDC code
${\cal C}: \Sigma_1^{\N} \rightarrow \Sigma_2^{\M}$ 
with 
    \begin{enumerate}
        \item \textbf{Rate:} $R \geq \exp(-O(1/\varepsilon)))$
        \item \textbf{Query Complexity:} $q \leq \poly(\log(N))\exp(O(\frac{1}{\varepsilon}))$ 
        \item \textbf{Log-Alphabet and Locality:} $r \leq \log(N)^{\exp(O(\frac{1}{\varepsilon}))}$
    \end{enumerate}
    The code can be locally encoded in $\Time(\poly(\log(N),\exp(\frac{1}{\varepsilon})))$  and decoded in logspace uniform size \((\log N)^{O(\exp(\frac{1}{\varepsilon}))}\) and \Depth  $(\tilde{O}(\log \log N \exp(O(\frac{1}{\varepsilon})))$ .
\end{theorem}
When \(\varepsilon\) is a small constant, the above gives a $(10^{-10},0.1)$-approximate \((\varepsilon,O(\frac{1}{\varepsilon}),\polylog(N))\)-LLDC with \emph{constant} rate, and \(\poly\log(N)\) locality. Again, this code can easily be modified to be $\ell_{in}$-list-recoverable in the high noise setting, replacing every occurrence of $\varepsilon$ in the above with $\frac{\varepsilon}{\ell_{in}}$. Concatenating the above construction with
any sufficiently good binary list-decodable code (e.g.\ concatenated Reed-Solomon codes \cite{guruswami2000list}) immediately gives the following corollary. 

\begin{corollary}[Binary Constant Rate aLLDCs]\label{cor:constant-rate-binary}
    For every $\varepsilon>0$ and infinitely many $N \in \mathbb{N}$, there is an 
$(O(\varepsilon),0.1)$-approximate, $(\frac{1}{2}+\varepsilon,O(\frac{1}{\varepsilon^2}),q)$-LLDC ${\cal C}: \Sigma_1^{\N} \rightarrow \Sigma_2^{\M}$ with 
    \begin{enumerate}
        \item \textbf{Rate:} $R \geq \exp(-O(1/\varepsilon^3))$
        \item \textbf{Locality:} $r \leq \log(N)^{\exp(-O(1/\varepsilon^3))}$
        \item \textbf{Query Complexity:} $q \leq \log(N)^{\exp(O(\frac{1}{\varepsilon^3}))}$
    \end{enumerate}
    The code can be locally encoded in $\Time(\log(N)^{O(\exp(\frac{1}{\varepsilon^3}))})$  and decoded in logspace uniform size \((\log N)^{O(\exp(\frac{1}{\varepsilon^3}))}\) and depth \(\tilde{O}(\log^2( \log N) \exp(O(\frac{1}{\varepsilon^3})))\).
\end{corollary}
We note both codes above can actually be locally encoded in logspace uniform size $\log(N)^{O(\exp(\frac{1}{\varepsilon^3}))}$ and depth $\poly\log\log(N)\exp(O(1/\varepsilon^3))$. 
Finally, even though these theorems are stated for constant \(\varepsilon > 0\) one can also take \(\varepsilon\) slightly subconstant, e.g.\ \(\varepsilon = \frac{1}{\poly\log \log N}\).

As before, we start with an overview of the construction which requires several major changes from our previous framework:

\paragraph{Construction Overview} 

In this overview we will think of the parameters \(\varepsilon, \delta_{in},\delta_{out}\) etc.\ as constants, independent of the size of the set system.

Recall in \Cref{sec:subpoly} and \Cref{sec:polylog} we critically relied on our hypergraph system $(V,S,G)$ having intersection graphs $G_s$ that are $\frac{1}{\polylog(N)}$-samplers. Unfortunately, we do not know how to construct such systems where $(V,S)$ is additionally bounded degree in the sense that every vertex lies in at most $O(1)$ hyperedges. Indeed note that $\frac{1}{\polylog(N)}$-sampling requires the sets in $S$ to be size at least $\polylog(N)$, so such systems, if they indeed exist, must be extremely unbalanced with $|S| \ll |V|$. While we do not know how to rule out such systems formally, unbalanced objects of this type are notoriously difficult to construct.

Instead, we solve this issue by drawing inspiration from the global list decoding algorithm of \cite{dinur2019list}: in particular, we will add a \textit{third layer} $U$ to our system consisting of subsets of $V$ such that the `local view' of any $u \in U$ (more accurately, the hypergraph $(u, \set{s \cap u}_{s \in S})$) is itself a good direct product code (we'll call these the `inner codes' of $(V,S,U,G)$). Roughly speaking, we will need three key properties of $(V,S,U,G)$: 
\begin{enumerate}
    \item The graph $(V,S)$ has \textit{constant} (left-to-right) degree (and therefore the DP code $C_{(V,S)}$ is constant rate).
    \item Given any input word $w$ to $(V,S)$ and $f \in \Ball_{1-\varepsilon,\delta_{in}}(w)$, the function $f$ is in the balls of all but a \textit{sub-constant} fraction of the inner codes $u \in U$.
    \item The hypergraph system $(V,U,G)$, which we call the `outer code', has a propagation decoder for the sub-constant noise list recovery regime (in particular, we will take $(V,U)$ to be $\polylog(N)$-degree to avoid the original issue).
\end{enumerate}

Suppose we have a system that satisfies these requirements.  
Then every `inner code' positioned at $u \in U$ can decode a list of local functions $\mathcal{I}_u$ labeling the vertices in $u$. By the second property, we are guaranteed all but a $\frac{1}{\polylog(N)}$-fraction of these decoded lists contain a function $\delta'$-close to $f$, or in other words, that $f \in \Ball_{\frac{1}{\poly \log (N)},\delta'}(\mathcal{I})$. But this is exactly the low-noise list recovery setting on $(V,U)$, so we may run our decoder on $(V,U)$ to locally decode $f$ as desired, where we query these lists \(\mathcal{I}_u\) by calling of the inner decoder.

The construction of such a set system requires is non-trivial high dimensional expander machinery. (We note that these do not follow from \cite{dinur2019list}'s `double samplers' which cannot simultaneously achieve the first two properties). 

Even putting aside the difficulty of constructing such systems, there's a key issue we've brushed under the rug: every vertex in our local codes are only encoded in a constant number of sets (a property we critically rely on to achieve global constant rate), so our decoding accuracy $\delta'$ must also be constant (indeed this is inherent, as one can force a constant fraction of vertices in these codes to only see noise). This is of course a big problem for our original list-recovery scheme, which only achieves final accuracy $\delta_{out} \geq t\delta'$ for $t=\polylog(N)$.

To handle this, we present in this section a new list-recovery method for hypergraph systems that results in output accuracy \(\delta_{out} = \delta_{in} \exp(O(\frac{1}{\varepsilon}))\), allowing us to instantiate outer decoding in the constant rate regime. The technique relies on two key new ideas. The first, an adaptation of the `list pruning' method in \cite{dinur2019list}, is to ensure the lists input to our list-recovery algorithm are \textit{well-separated}. Roughly speaking, separation allows us to \textit{uniquely} match lists as we walk along the decoding graph, avoiding propagation of error (as long as we avoid a tiny set of bad edge coming from sampling error). This, of course, is slightly wishful thinking; we will not be able to reduce to the case where all lists are well-separated. However, critically we \textit{can} ensure (paying $\exp(O(\frac{1}{\varepsilon}))$ cost) that a \textit{large constant fraction} (say $99\%$) of sets in our system have this sort of lossless propagation, and moreover that this set is identifiable to the algorithm.

Now, naively, we still have a problem: since our routing scheme takes paths of length $\polylog(N)$ (and in fact we require these paths cover the graph well), it is extremely likely a random path will escape this `good' set $T$ above and ruin our error. This brings us to the second key idea: \textit{subset-internal} (`vertex-fault tolerant') routing. In particular, we will show it is possible to construct a variant of local routing which, given oracle access to any large constant fraction of the domain $T$, locally routes between almost all $u,u' \in T$ with low congestion while \textit{staying entirely inside $T$}. Together with lossless propagation, this is sufficient to ensure accurate decoding even when our original lists have constant error and (along with the existence of the hypergraph system itself) is the key to achieving constant rate.

\paragraph{Section Roadmap} We give a brief outline of the structure of the section below. We first formalize the hypergraph triple setup in \Cref{sec:constant-formal}, along with the new notion of subset internal routing schemes. In \Cref{sec:constant-graph-system} we formalize the properties of the hypergraph triple we will use. In \Cref{subsec:inner-decoder}, we give our reduction to low noise list-recovery using the inner code structure of our triple. In \Cref{subsec:outer-decoder} we introduce the well-separated condition, and give a list recovery algorithm assuming access to a subset internal routing scheme and well-separated input lists. Finally in \Cref{subsec:well-separated} we show how to reduce from generic list-recovery to the well-separated setting.

\subsection{Formal setup}\label{sec:constant-formal}

As we mentioned in the overview, we require a set system with three layers of sets, somewhat similar to the `double samplers' in \cite{dinur2019list} albeit requiring fairly different guarantees. Let \((V,S,U,G)\) be a set system where \(V\) are the vertices, \(S \subset \binom{V}{r_1}\), \(U \subset \binom{V}{r_2}\) consist of sets. (Note there is no requirement on $r_1,r_2$.) The graph \(G=(U,E)\) is a decoding graph, defined analogously to its definition in \Cref{sec:direct-products}, only with respect to \(U\). It is a regular graph whose vertices are \(U\), and for which \(u \cap u' \ne \emptyset\) whenever \(u \sim u'\).\footnote{As in the decoding graph in \cref{sec:direct-products}, not all overlapping sets have edges. The graph itself will be instance specific to our system.} The intersection graph \(G_u\) is also defined similarly: the vertices on one side are the vertices of \(u\), and on the other side we put the intersections \(\sett{u \cap u'}{\set{u,u'} \in E}\).

Along with the above, which are all analogous to graphs we analyzed on our basic hypergraph systems, we will also need several `new' local and global graphs arising from $(V,S,U,G)$. On the global side, there is only one such new graph: the \((S,U)\) set-layers graph. This is a bipartite graph whose vertices are \(S\) on one side, \(U\) on the other, and whose edges are all \(\set{s,u}\) such that \(s\cap u \ne \emptyset\).

We now move on to the definitions of the local set systems. Denote by \(In_u=(V_u,S_u)\) the internal set system with \(V_u=u\) and \(S_u = \sett{s \cap u}{s \in S, s\cap u \ne \emptyset}\). Denote by \(G_{In_u}\) the \emph{local decoding graph} of \(In_u\). That is, the vertices in this graph are \(S_u\) and, in this case unlike the above, we will connect \textit{all} \(s \sim s'\) such that \(s\cap s' \ne \emptyset\).\footnote{One could define these inner graphs more generally, but in the systems we construct connecting all overlapping sets will always give the desired inner decoder so we fix the definition this way for simplicity.}

As discussed above, our construction will only take the direct product over \((V,S)\) as before. The extra layer of sets \(U\) is used only for decoding. Given an input word $w$ to $\mathcal{C}_{(V,S)}$, we denote by \(w_{In_u}\) the induced local word on \(S_u\) obtained in the natural way. That is, for every \(s'=s \cap u\) in \(S_u\) we take \((w_{In_u})_{s'} = w_s|_{s'}\). Of course, by querying the original word at \(s\), we are also able to query word \(s'\).

Similarly, we also define the set system for \(In_s=(V_s,U_s)\) the same way. The vertices \(V_s = s\), and the sets are all \(U_s = \sett{u \cap s}{u \in U}\). We note that this set system is not necessary for the decoding itself, but its sampling properties will be important for the analysis.

\subsubsection{Subset Internal Routing}
As discussed in the overview, for our outer code to work in the constant rate setting we require a strengthened notion of routing on the decoder graph of \(U\) that only routes through \(u \in U\) that have lists with the distance properties mentioned in the overview (which we'll formalize shortly  below). Here we will treat this as an arbitrary subset $T$ of the graph $G$ (indeed we will have no control over the subset in the argument).

More formally, we now introduce our notion of `subset internal' routing, which is a low congestion routing scheme that paths between vertices \(v,v'\) in a graph \(G=(U,E)\) such that all the middle vertices between \(v\) and \(v'\) belong to some set \(T \subseteq U\) with constant size. Since \(T\) can be arbitrary, this task may of course be impossible as stated, e.g.\ a certain vertex \(v\) may have no neighbors in \(T\) so one can't reach any \(v' \ne v\) without going outside of \(T\). Nevertheless, we will show such schemes are possible on certain HDX if we allow our routing scheme to fail with \(o(1)\) probability over choice of end points \(v,v'\) and relax slightly our notion of low congestion used to avoid bad events:
\begin{definition}[Subset Internal Routing] \label{def:subset-internal-routing-scheme}
    Let \(G=(U,E)\) be a graph. A subset internal routing scheme \(\mathcal{O}^{T}(u,u')\) gets a membership oracle to some set \(T \subseteq U\) of size \(\Prob{}{T} \geq 0.9\) and outputs either \(FAIL\) or a path \(P = (u=u_0,u_1,u_2,\dots,u_m=u')\) such that \(u_1,u_2,\dots,u_{m-1} \in T\). We say that \(\mathcal{O}\) is \emph{successful} if the probability it outputs \(FAIL\) (over a uniform choice of \(u,u' \in U\)) is \(\leq \frac{1}{\log^4 |U|}\). The length of \(\mathcal{O}\) is the maximal length of a path in the support of \(\mathcal{O}\). $\mathcal{O}$ has \emph{low congestion} if for every subset $B \subset E$, the probability $\mathcal{O}$ hits $B$ is at most
    \[
        \Pr_{u,u',\mathcal{O}}[\mathcal{O}(u,u') \ne \text{FAIL} \; \wedge \;  \exists e \in \mathcal{O}(u,u'): e \in B] \leq 4t\Pr[B]^{1/4} + \frac{1}{\log^4 |U|}
    \]
\end{definition}
We comment that requiring \(\Prob{}{T} \geq 0.9\) is mainly for convenience. We will construct such schemes over bipartite graphs where \(U=L \cup R\), and the actual requirement we will need is that \(\Prob{}{T \cap L},\Prob{}{T \cap R}\) are both non-negligible. Similarly the constants \(4,1/4\) and \(\frac{1}{\log^4|V|}\) are arbitrary and fixed for simplicity to the parameters achieved by our scheme.
\subsubsection[Circuit implementation of a three layer set system]{Circuit Implementation of \((V,S,U,G)\)} \label{subsec:extended-circuit-implementation}
Now that we have a three part system, we will need a corresponding circuit implementation \((V,S,U,G)\) analogous to that defined just for $(V,S,G)$ in \Cref{sec:direct-products}.

For the outer decoder, we will need the circuit implementation of the set system \((V,U)\) as defined in \Cref{sec:direct-products}, that is \((C_{V \to U},C_{U \to V}, C_{V \to U}^{-1}, C_{U \to V}^{-1}, C_{U \cap U \to V})\).

On top of this, we will also need circuit implementations for \(In_u\) for every \(u \in U\). We will still just work with the global names of \(v \in u\) or \(s \in S_u\), rather than their labeling as \(i\)-th neighbors in \(u\). We will need the following circuits. 
\begin{enumerate}
    \item A circuit \(C_{U\to S}(u,i)\) that outputs $u$'s \(i\)-th neighbor in \(S\) (i.e.\ the \(i\)-th set \(s \in S_u\)).
    \item A circuit \(C_{U;S\to V}(u,s,i)\) that takes in \(u \in U\), \(s \in S\) that intersects \(u\), and index \(i\). It outputs the \(i\)-th neighbor of \(s' = s \cap u\).
    \item A circuit \(C_{U;V\to S}(u,v,i)\) that takes in \(u \in U\), \(v \in u\), and index \(i\). It outputs the \(i\)-th neighbor of \(v\) \emph{in \(S_u\)}.
    \item A circuit \(C_{U; S \cap S \to V}(u,s,s',i)\) that outputs the \(i\)-th neighbor of \(u \cap s \cap s'\) or \(FAIL\) if \(u \cap s \cap s' = \emptyset\).
    \item Circuits \(C^{-1}_{U; S \to V}, C^{-1}_{U; V \to S}\) that output \(i\) such that \(C_{U;S\to V}(u,s,i) = v\) or \(C_{U;V\to S}(u,v,i) = s\) respectively.
    \item A circuit \(\mathcal{C}_{\mathcal{O}}\) that implements the subset internal routing scheme for the decoding graph \(G\).
\end{enumerate}

\subsection{The Set System}\label{sec:constant-graph-system}
In the below proposition we identify the properties we use from the set system we need. This proposition is proven in \Cref{sec:instantiations}. The parameters are set so that we can plug the system directly into our decoding framework below.
\begin{proposition} \torestate{\label{prop:graph-system-for-constant-rate}
Let \(N>0\) be a large enough integer and let \(\varepsilon > \frac{1}{\log \log N}\).\footnote{In fact, no constraints on \(\varepsilon\) is necessary, but outside this range of parameters the set system becomes much larger than \(N\).} Set \(\delta_{in} = \exp(-3 \cdot 10^3 \frac{1}{\varepsilon})\), \(\eta = (\frac{\varepsilon \delta_{in}}{8})^7\) and \(\beta = (\frac{\varepsilon \delta_{in}^2}{\log N})^{28}\). There is a set system \((V,S,U,G)\) with the following properties.
\begin{enumerate}
    \item \(|V| \in [N,N \log(N)^{\exp(O(\frac{1}{\varepsilon}))}]\).
    \item Every \(s \in S\) and \(u \in U\) have size \(\leq \log(N)^{\exp(O(\frac{1}{\varepsilon}))}\).
    \item Every vertex appears in \(\exp(O(\frac{1}{\varepsilon}))\) sets.
    \item The containment graphs \((V,S), (V,U)\), the set-layers graph \((S,U)\) and the local graphs \((V_s,U_s)\), \((V_u,S_u)\) and the intersection graph \(G_u\) for every \(s \in S\) and \(u \in U\) are all bi-regular graphs. 
    \item The intersection graph \(G_u\) is a \(\beta\)-sampler.
    \item The graph \((S,U)\) is a \(\beta\)-sampler.
    \item For every \(u \in U\), \((V_u,S_u)\) is an \(\eta\)-sampler.
    \item For every \(u \in U\), local decoding graph graph \(G_{In_u}\) is \((1-O(\eta))|S_u|\) regular.
    \item For every \(u \in U\) and \(s' \in S_u\), the local intersection graph \((G_{In_u})_{s'}\) is an \(\eta\)-sampler.
    \item For every \(s \in S\) the graph \((V_s,U_s)\) is a \(\beta\)-sampler.
    \item There is a low congestion subset internal routing scheme on the decoding graph of \(G\) of length \(O(\log^3 N \poly(\eta^{-1}))\).
\end{enumerate}}
\end{proposition}

We will also implement the circuits in \Cref{sec:kms-circuit-implementation} for this set system and obtain the following guarantees.
\begin{claim}[Circuit Implementation (Informal)] \torestate{\label{claim:circuits-for-const-rate-construction}
    Let \((V,S,U,G)\) be the construction in \Cref{prop:graph-system-for-constant-rate}. Then the system is implementable in logspace uniform size \((\log N)^{\exp(O(\frac{1}{\varepsilon}))}\) and depth \(\tilde{O}(\log \log N \exp(O(\frac{1}{\varepsilon})))\).} Moreover, the set internal routing circuit \({C}_{\mathcal{O}^T}\) makes \(\exp(O(\frac{1}{\varepsilon}))\poly \log N\) queries to the oracle in parallel.
\end{claim}
As in prior sections, the above is informal only in the sense that we implement these circuits with some vanishing small error that does not effect the analysis (corresponding to encoding non-invertible matrices).
\subsection{Reduction to List-Recovery (Inner Decoding)} \label{subsec:inner-decoder}
In this subsection we present our alternate reduction to low noise list-recovery using the assumed sampling properties of $(V,S,U,G)$ and `inner codes' $In_u=(V_u,S_u)$. In particular, our main goal is to show that for any input word $w$ to $(V,S)$ and global function \(f \in \Ball_{1-\varepsilon,\delta}(w)\), we will be able to approximately decode the restriction of $f$ on $u$ as part of a small list $\mathcal I_u$ in almost every inner code \(In_u\) (roughly reducing the problem of list-recovery on $(V,U)$). Formally we show:
\begin{lemma} \label{lem:final-inner-decoding-procedure}
    Fix \(\delta_{in},\varepsilon,\eta,\beta >0\) and let \((V,S,U,G)\) be as in \Cref{prop:graph-system-for-constant-rate}. Let \(w\) be input word to $\mathcal{C}_{(V,S)}$ and \(f \in \Ball_{1-\varepsilon,\delta_{in}}(w)\). Then there exists a randomized circuit \(C_{list-u}\) that:
    \begin{enumerate}
        \item Takes as input \(u \in U\), an index \(i=1,2,\dots,\frac{10}{\varepsilon}\) and \(v \in V_u\) and outputs a symbol in \(\Sigma_1\).
        \item For \(1-\beta\) fraction of the random strings, there exists \(1-\frac{10\beta}{\varepsilon \delta_{in}^2}\)-fraction of the \(u \in U\) there exists \(i\) such that \(C_{list-u}(u,i,\cdot)\) is \(O(\sqrt[3]{\delta_{in}})\)-close to \(f\) on $u$.
        \item  \(C_{list-u}\) makes $\poly(\frac{1}{\delta_{in}},\frac{1}{\varepsilon},\log \frac{1}{\beta})$ queries to $w$
        \item \(C_{list-u}\) has size \((\log N)^{\exp(O(\frac{1}{\varepsilon}))}\) and depth \(\tilde{O}(\log^2 (\log N) \exp(O(\frac{1}{\varepsilon})))\).
    \end{enumerate} 
\end{lemma}
To be explicit, the list \(\mathcal{I}_u\) consists of the \(\frac{10}{\varepsilon}\) functions \(\set{C_{list-u}(u,i,\cdot)}_{i \in [\frac{10}{\varepsilon}]}\).

We remark the fine-grained query complexity of the above (number of bits queried from the lists rather than full symbols can be made \(\tilde{O}(\frac{1}{\varepsilon\delta_{in}^2})\), see \Cref{rem:outer-const-fine}.

The proof of \Cref{lem:final-inner-decoding-procedure} is split into three parts. The first, unsurprisingly, is to show every $In_u=(V_u,S_u)$ in our hypergraph triple is a good aLLDC. 

\begin{claim}\torestate{\label{claim:const-rate-inner-decoder}
        Let \(\varepsilon, \delta_{in} > 0\) and let \(\eta\) be as in \Cref{prop:graph-system-for-constant-rate}.   Let \(u \in U\) and \(In_u\) be a set system as in the \Cref{prop:graph-system-for-constant-rate}, i.e.\
        \begin{enumerate}
            \item the set system \((V_u,S_u)\) is an \(\eta\)-sampler.
            \item For every \(s' \in S_u\), the local intersection graph\footnote{Recall this is the graph where on one side we have the vertices in \(s\), and on the other side we have intersections \(s' \cap s\) for \(s' \in S_u\) that intersect \(s\).} \((G_{In_u})_{s'}\) is an \(\eta\)-sampler.
            \item The local decoding graph \(G_{In_u}\) is \((1-O(\eta))|S_u|\) regular.
        \end{enumerate}
        Then the direct product code on \(In_u\) is a \((\delta_{in},13\sqrt[3]{\delta_{in}})\)-approximate, \((\varepsilon, \frac{10}{\varepsilon}, q=\poly(\frac{1}{\delta_{in}},\frac{1}{\varepsilon}))\)-LLDC with decoding in logspace uniform size \((\log N)^{\exp(O(\frac{1}{\varepsilon}))}\) size and depth \(\tilde{O}(\log \log N \exp(O(\frac{1}{\varepsilon})))\).}

\end{claim}

The proof of \Cref{claim:const-rate-inner-decoder} is technical but quite similar in spirit to the original aLLDC analysis in \cite{impagliazzo2008uniform} relying on sampling and density of the inner codes (albeit in a slightly different way) so is deferred to \Cref{app:inner-code}. Nevertheless, we present the algorithm here in \Cref{fig:inner-decoder} in part so we may reference it later for clarity.
\begin{figure}[ht!]
\fbox{\parbox{\textwidth}{
\vspace{.1cm}
\underline{Inner Decoder \(InDec_{\delta_{in}}(s,v)\)}:
\begin{enumerate}
    \item[*] \textbf{Input:}
    \begin{itemize}
        \item A set \(s \in S_u\).
        \item A target vertex \(v \in V\).
        \item Oracle access to word $w$
    \end{itemize}
    \item[*] \textbf{Output:} a value in \(\Sigma\).
    \item For \(j=1,2,\dots,\frac{4 \log \delta_{in}}{\varepsilon}\):
    \begin{enumerate}
        \item Sample a set \(s'_j \ni v\).
        \item Test if:
            \begin{enumerate}
                \item The sets \(s\) and \(s'_j\) intersect, and
                \item If \(\dist_{s \cap s'_j}(w_s,w_{s'_j}) < 4\delta_{in}\).
            \end{enumerate} 
    \item If \(s'_j\) passed the test, exit the loop and output $w_{s'_j}(v)$    
    \end{enumerate}
    \item If none of the \(s'\) passed, output an arbitrary symbol.
\end{enumerate}}}
    \caption{Inner Decoder. \Cref{claim:const-rate-inner-decoder} Follows by outputting $InDec_{\delta_{in}}(s,\cdot)$ for $O(\frac{1}{\varepsilon})$ random choices of $s \in S_u$. See \Cref{app:inner-code} for formal proof.}
    \label{fig:inner-decoder}
\end{figure}

Now, note that in our definition of list decoding, the circuits we output only need to succeed in decoding a particular function \(f\) with probability \(\frac{3}{4}\). Therefore, even assuming \Cref{claim:const-rate-inner-decoder} as written, we are only guaranteed to succeed in $\delta_{out}$-computing a particular $f$ with constant probability.
This is obviously not sufficient to prove \Cref{lem:final-inner-decoding-procedure}, which requires decoding $f$ with high probability. We can amplify the success probability by simply repeating the decoder $O(\log \frac{1}{\beta})$ times, but this blows up the list size. We will see in the next section this is a real problem, as we will pay an exponential cost in the list size to reduce to the `well-separated' setting discussed in the overview, so we really need this to be constant.

Thankfully there is a simple fix to this problem (which is the second component of \Cref{lem:final-inner-decoding-procedure}). We may run a basic pruning procedure to cut the list size down to $O(\frac{1}{\varepsilon})$:

\begin{claim}[List Pruning (\Cref{claim:output-list-pruning-app})]\label{claim:output-list-pruning}
    Fix $\tau>0$ with parameters $\varepsilon,\delta_{out}$, an input word $w_{In_u}$, and $\{C_1,\ldots,C_{L}\}$ a set of output circuits from running the generating algorithm of the decoder \(InDec_{\delta_{in}}\) possibly several times. Assume that the graph \((V_u,S_u)\) is a \(\frac{\delta_{out}^2\varepsilon^2}{256}\)-sampler. There is a circuit $C_{Prune}$ which, given oracle access to the $\{C_i\}$ and input list, outputs a subset $L' \subset [L]$ of size $|L'| \leq O(\frac{1}{\varepsilon})$. With probability at least $1-\tau$ any original $C_i \in \Ball_{1-\varepsilon,\delta_{out}}(\mathcal{L})$ is $6\delta_{out}$-close to some $C_j \in L'$.
    
    Moreover $C_{Prune}$ is in logspace uniform size $\poly(|L|,\frac{1}{\varepsilon},\frac{1}{\delta_{out}},\log \frac{1}{\tau})$ and depth $O(\log^2 |L| \cdot \log \frac{1}{\delta_{out} \varepsilon \tau})$, makes at most $O(\frac{\log \frac{L}{\tau}}{\max\{\varepsilon,\delta\}})$ oracle calls to the $\{C_i\}$, word $w$, a circuit producing random $s \in S_u$ and random $v \in V_u$, and the hypergraph adjacency circuits. 
\end{claim}
Again the proof is not particularly enlightening (one simply queries the input list to discard any circuits that aren't actually computed by $w$, then finds a small cover of the rest which is promised by a bound on the combinatorial list size on the code). The proof is deferred to \Cref{app:inner-code} as well.

Now, running \Cref{claim:const-rate-inner-decoder} several times to amplify success then pruning the resulting lists, we can clearly build a circuit like in \Cref{lem:final-inner-decoding-procedure} that computes $f$ with high probability whenever $f$ is global in $In_u$. Thus the final step is to argue by sampling that any $f$ which is global in $(V,S)$ should also be global in almost every local code $In_u$ by sampling.

\begin{claim} \label{claim:global-functions-remain-global-in-inner-decoder}
    Fix any $\varepsilon,\delta,\beta>0$ with $\delta \leq 4\sqrt{\varepsilon}$, and let \((V,S,U,G)\) be such that
    \begin{enumerate}
        \item \((S,U)\) is a \(\beta\)-sampler.
        \item For every \(s \in S\) the graph \((V_s,U_s)\) is a \(\beta\)-sampler.
    \end{enumerate}
    Then for any \(f \in \Ball_{1-\varepsilon,\delta}(w)\) :
    \[\Prob{u \in U}{f|_{u} \in \mathcal{B}_{1-\frac{\varepsilon}{2},2\delta}(w_{In_u})}\geq 1-\frac{9\beta}{ \delta^2_{in}\varepsilon}\]
\end{claim}
Here one should think of \(\beta\) as \(\frac{1}{\poly \log (N)}\) and \(\delta\) as some constant multiple of \(\delta_{in}\) from above. We remark the condition on $\delta$ is just for calculational simplicity and has no effect in later arguments as we will be taking $\delta \leq \exp(-1/\varepsilon)$ anyway. We also note that if it had been the case that \(f \in \Ball_{1-\varepsilon,0}(w)\), \Cref{claim:global-functions-remain-global-in-inner-decoder} would follow immediately from sampling of \((S,U)\). However, because \(f \in \Ball_{1-\varepsilon,\delta}(w)\), a priori it could be the case that \(\dist_{s}(g,f)<\delta\) but that the distance increases upon intersecting with $u$. In this case \(g|_{s \cap u}\) would not sufficiently compute \(f|_u\) and we may fail to decode $f$. Thankfully, sampling of \((V_s,U_s)\) ensures this event is rare.

\begin{proof}[Proof of \Cref{claim:global-functions-remain-global-in-inner-decoder}]
    Fix \(f\) and let \(A \subseteq S\) be the sets where $w_s$ is \(\delta\)-close to \(f\). By definition \(\Prob{}{A} \geq \varepsilon\). If \(f\) is not global in \(In_u\) then one of the following events occurred:
    \begin{enumerate}
        \item Either \(\Prob{s \sim u}{A} < \frac{3\Prob{}{A}}{4}\), or
        \item More than \(\frac{\Prob{}{A}}{4}\) sets \(s \in A\) neighboring $u$ have $w_s$ \(\delta\)-close to \(f\) but 
        \[
        \dist_{s\cap u}(f,w|_{s \cap u}) > 2\delta.
        \]
    \end{enumerate}
    This is because if neither the first nor second items are satisfied, there is at least \(\frac{\Prob{}{A}}{2} \geq \frac{\varepsilon}{2}\) fraction of \(s \sim u\) such that \(\dist_{s\cap u}(f,w|_{s \cap u}) \leq 2\delta\) as desired.

    Thus it is enough to bound the probability of the two events. Note the first is bounded by \(\frac{16\beta}{\varepsilon^2} \leq \frac{\beta}{\delta^2\varepsilon}\) by the global sampling property between \((S,U)\). 
    
    Toward bounding the second, for any \(s \in A\) call a pair \((s,u)\) with \(s \sim u\) \emph{misleading} if \(\dist(f,w_s) < \delta\) but \(\dist_{s\cap u}(f,w_s|_{s \cap u}) > 2\delta\). By $\beta$-sampling of \((V_s,U_s)\), for every \(s \in A\) there are at most \(\frac{\beta}{\delta^2}\) neighbors \(u \sim s\) such that \((s,u)\) is misleading, so there are at most \(\frac{\beta}{\delta^2}\) misleading pairs overall. By Markov's inequality,
    \[
        \Prob{u \in U}{\Prob{s \sim u}{(u,s) \text{ is misleading}} > \frac{\Prob{}{A}}{4}} \leq \frac{4\Prob{u,s}{(u,s) \text{ is misleading}}}{\Prob{}{A}}.
    \]
    The numerator is upper bounded by \(\frac{4\beta}{ \delta^2}\) by the above, and the denominator is bounded by \(\varepsilon\). 
    
    On the other hand, if \(u\) satisfies the first item but violates the second item in the list above, then it contributes to \(\Prob{u \in U}{\Prob{s \sim u}{(u,s) \text{ is misleading}}> \frac{\Prob{}{A}}{4}}\). Hence there are at most \(\frac{8\beta}{\delta^2\varepsilon}\) such \(u\). The claim holds by adding up this probability and the probability of violating the first item.
\end{proof}

We are now ready to prove \Cref{lem:final-inner-decoding-procedure}.

\begin{proof}[Proof of \Cref{lem:final-inner-decoding-procedure}]   
Fix any input word $w$ to $(V,S)$, and $f \in \mathcal{B}_{1-\varepsilon,\delta_{in}}(w)$. The circuit \(C_{list-u}\), on input $(u,i,v)$, runs the $In_u$-inner decoder $O(\log \frac{1}{\beta})$ times\footnote{Note we must be careful about how exactly $C_{list-u}$ accesses the described inner decoder here. Namely we do not have to hardcode a separate inner code for every $u$ into the circuit (which would massively blow up its size). Instead, as described shortly below, we can simply use $L$ copies of the $InDec(s,g_i,\cdot)$ circuit in \Cref{fig:inner-decoder} combined with our circuit implementation to ensure the $s$ fed in (and queried in the procedure) are the appropriate neighbors of $u$.} and feeds the resulting $L=O(\frac{\log \frac{1}{\beta}}{\varepsilon})$ circuits $C_1,\ldots,C_L$ into $C_{prune}$ with failure probability $\tau \leq O(\beta^2)$. The circuit \(C_{prune}\) outputs the indexes of \(L' \subseteq L\) of circuits surviving the prune. Finally, \(C_{list-u}\) uses $i$ to index into the output subset $L' \subset [L]$ of $C_{prune}$ and produces $C_{L'(i)}(v)$.\footnote{If it happens to be the case that a $L'$ is smaller than the number of indices, the output may be arbitrary on larger $i$.}

We need to argue that across the internal randomness of $C_{list-u}$, a $1-\beta$ fraction of the time there exists a $1-\frac{10\beta}{\varepsilon\delta_{in}^2}$ fraction of $u \in U$ for which $C_{list-u}(u,i,\cdot)$ $O(\sqrt[3]{\delta_{in}})$-computes $f$. To see this, note that by \Cref{claim:global-functions-remain-global-in-inner-decoder} we are guaranteed that $f \in \mathcal{B}_{1-\frac{\varepsilon}{2},2\delta_{in}}(w_{In_u})$ for all except a $\frac{9\beta}{\varepsilon\delta_{in}^2}$ fraction of $u \in U$. Since we have run the inner decoder $O(\log \frac{1}{\beta})$ times, by \Cref{claim:output-list-pruning} for every such `good' $u$ we are assured that with probability at least $1-O(\beta^2)$ over the internal randomness we indeed output a deterministic circuit $O(\sqrt[3]{\delta_{in}})$-computing $f$. Thus by Markov's inequality, the probability over the internal randomness that more than a $\beta$ fraction of these good $u \in U$ fail to have such an index is at most $\beta$, leaving at most a $\frac{10\beta}{\varepsilon\delta_{in}^2}$ fraction with no good index computing $f$ as desired.

\paragraph{Circuit complexity} By \Cref{claim:const-rate-inner-decoder}, the inner decoder circuit is implementable in size \((\log N)^{\exp(O(\frac{1}{\varepsilon}))}\) and depth \(\tilde{O}(\log (\log N) \exp(O(\frac{1}{\varepsilon})))\). The list pruning procedure makes \(\poly(\log N)\exp(O(\frac{1}{\varepsilon}))\) parallel calls to the oracles of \(\set{C_i}\). Finally, the output of the inner decoder is fed to our list pruning circuit instantiated with confidence \(\tau = (\log^6 N \cdot \exp(O(\frac{1}{\varepsilon})))^{-1}\), \(L = \poly(\frac{1}{\varepsilon},\log\log N)\), and \(\delta_{out}=O(\delta_{in}^{2/3})=\exp(O(\frac{1}{\varepsilon}))\), which does not increase the asymptotic size but does increase the depth to \(\tilde{O}(\log^2 (\log N)\exp(O(\frac{1}{\varepsilon}))\).



\TP{I'm having trouble tracking $L,L'$ throughout.}

\end{proof}

\subsection{List Recovery from Constant Accuracy (Outer Decoding)}\label{subsec:outer-decoder}
The prior section gave us a new method of reducing to the low noise list-recovery regime, namely for constructing lists \(\mathcal{I}_u\) for every \(u \in U\) such that any fixed \((\delta,\varepsilon)\)-global function $f$ in \((V,S)\) will appear in almost all lists. Thus, as in \Cref{sec:subpoly}, we now just need a list recovery algorithm on $(V,U)$ that recovers functions that are \((\delta,1-\frac{1}{\poly\log(N)})\)-global in the lists \(\mathcal{I}=\set{\mathcal{I}_u}\).

As before, at a high level our list recovery algorithm for $(V,U)$ will start from a random outer set $u$ and a function $g \in \mathcal{I}_u$ that is $\delta_{in}$-close to the global function $f$. To decode $v$, we route to a random $u' \ni v$ we'd like to decode and hope to identify through this process some $g' \in \mathcal{I}_{u'}$ that is close to $f$ and output $g'(v)$. In \Cref{thm:list-recovery}, we showed by tracking the list of functions close to the initial $g$, we can ensure the final $g'$ output is
$O(\delta_{in}\polylog(N))$-close to $f$. 

Unfortunately, as we've discussed above, this is insufficient for the constant rate setting where  \(\delta_{in}\) (the decoding accuracy of our inner codes) is also constant.\footnote{This is inherent in any direct product code with constant rate (or even other codes with constant rate and locality). The rate of a direct product code is the average degree of a vertex in \((V,S)\), which we denote by \(d\). We can always find a vertex set \(V' \subseteq V\) of relative size \(\frac{1}{2d}\) whose neighborhood in \(S\) has size \(\leq \frac{1}{2}\). Denote such a set \(V' \subseteq V\). Now if we take a codeword \(Enc(f)\) and erase/replace-by-noise the sets in \(Enc(f)\) in the neighborhood of \(V'\), we still get a word that is \(\frac{1}{2}\)-close to \(Enc(f)\). However, in such a word all information on \(V'\) is lost, so we cannot hope to recover any of the bits of \(V'\) (even in a non-local algorithm). In particular, we are limited to decoding functions that are but \(\frac{1}{2d}\)-close even when \(\varepsilon = \frac{1}{2}\).} This means we cannot afford anywhere near a $\polylog(N)$ multiplicative blowup in error, and we need a new approach to propagation decoding. In this section we present our new method based on our strengthened notion of subset-internal routing that gives a bound of \(\delta_{out} = \delta_{in} \exp(O(\frac{1}{\varepsilon}))\).

\begin{theorem} \label{thm:local-outer-decoder-const-rate}
    Let \(N > 0\), \(\delta_{0}, \beta > 0\) such that \(\beta < \left ( \frac{\varepsilon \delta_{0}}{\log N} \right )^{28}\). Let \((V,U)\) be a set system satisfying:
    \begin{enumerate}
        \item \(V\) has \(\leq N^2\) vertices.\footnote{Technically we would like \(V\) to have exactly \(N\) vertices. \Cref{prop:graph-system-for-constant-rate} only promises approximately that, but the bound of \(N^2\) is more than enough.} \maxh{I'm not sure how to phrase, but this seems confusing at this point. What even is $N$, maybe we need to remind (I guess it's the original target vertex size). And where is this bound used?}
        \item \(U\) has a subset internal routing scheme of maximal length \(O(\log^3 N\exp(\frac{5}{\varepsilon}))\).
        \item For every \(u \in U\), the intersection graph \(G_u\) is a \(\beta\)-sampler.
    \end{enumerate}
    Then \(\mathcal{C}_{(V,U)}\) is a \((\delta_{0}, 8^{103/\varepsilon} \delta_0))\)-approximate, \(\frac{10}{\varepsilon}\)-input \((1-\frac{1}{\log^{10} N}, \tilde{O}(\frac{1}{\varepsilon}),\frac{\poly\log N}{\delta_0} \exp(O(\frac{1}{\varepsilon})))\)-LLRC.
\end{theorem}
Note the somewhat overcomplicated parameters are fixed so we can plug in our inner code directly to this result.

The first key idea to achieve list-recovery from constant $\delta_{in}$ is an adaptation of the `list pruning' method in \cite{dinur2019list} which ensures the lists input to our list-recovery algorithm are \emph{well-separated}. 
\begin{definition}
    Let \(\mathcal{I}_u\) be a list of functions on some domain \(u\).
    \begin{itemize}
        \item We say that a list \(L \subset \mathcal{I}_u\) is \(\delta\)-dense (with respect to \(\mathcal{I}_u\)) if every function \(g \in \mathcal{I}_u\) is $\delta$-close to some function \(g' \in L\).
        \item We say that a list \(L \subseteq \mathcal{I}_u\) is \(\delta'\)-separated if for any pair of distinct functions \(g_1,g_2 \in L\), \(\dist(g_1,g_2) > \delta'\).
        \item We say that a list \(L \subseteq \mathcal{I}_u\) is \(\delta\)-\emph{well separated} if it is \(\delta\)-dense and \(6\delta\)-separated.
    \end{itemize}
\end{definition}
We remark the constant $6$ is arbitrary (any constant larger than $2$ would suffice, the above is fixed to simplify calculations).

Roughly speaking, if every list was well separated \textit{at the same scale $\delta$}, it would allow us to \textit{uniquely} match lists as we walk along the decoding graph, avoiding additional errors in each step. Unfortunately, while it is possible to ensure every list is $\delta$-well-separated for some $\delta$, we do not know how to ensure they are separated at the same scale which is critical for this sort of lossless propagation. Instead, we will show how to ensure below that a \textit{large constant fraction} of sets in our system, say $99\%$, are well-separated at the same scale and therefore have lossless propagation (and moreover this can be done in such a way that the condition is testable). Denoting by \(T\subseteq U\) be the set of \(u \in U\) where we can get such well separated lists, the problem then reduces to subset-internal routing inside $T$ which we have by assumption (and will construct explicitly in \Cref{subsec:routing-kms}).


We are now ready to present the decoding circuit of the outer decoder in \Cref{fig:outer-decoder-constant-rate}. It is based on the following components which we abstract:
\begin{enumerate}
    \item \textbf{An implementation of the lists \(\mathcal{I} = \set{\mathcal{I}_u}_{u\in U}\):} Recall in reality we start from an input word on sets \(s \in S\), not for \(u \in U\). The circuit \(C_{list-u}\) from \Cref{lem:list-recovery-circuit}, allows us to inner-decode $w$ to lists $\mathcal{I}_u$ and simulate queries to functions in them. We stress again that the important thing about \(\set{\mathcal{I}_u}_{u \in U}\), is that functions \(f\) which are $(\varepsilon,\delta_{in})$-computed by the original word $w$ on $S$ have a close-by function in a \(1-\frac{1}{\poly\log(N)}\)-fraction of the constructed lists on $U$. We state the theorem on the outer decoder in terms of the lists \(\mathcal{I}\). We get back to the actual circuit implementation when we calculate circuit complexity.
    \item \textbf{A list well-separation circuit}: As introduced above, we need a circuit \(C_T\) that takes as input a list \(\mathcal{I}_u\) and some \(j\), and either outputs \(FAIL\) or a list \(L_u \subseteq \mathcal{I}_u\) that is \(1.16 8^{j}\delta_{in}\)-well-separated inside \(\mathcal{I}_u\). We prove the next lemma in the following subsection.
    
    \begin{lemma}\label{lem:list-sparsification}
    For every \(\delta_0 > 0\) there is a randomized circuit \(C^{\delta_0}_T\) that takes as input \(u\) (and some description of the list \(\mathcal{I}_u\)) of size \(\frac{10}{\varepsilon}\), and an index \(j \in \set{1,2,\dots,\frac{1000}{\varepsilon}}\). It either outputs \(FAIL\) or a list \(L_u \subseteq \mathcal{I}_u\) with the following guarantees.
    \begin{enumerate}
        \item For any \(\mathcal{I}_u\), for at most a \(0.01\)-fraction of the \(j \in \set{1,2,\dots,\frac{1000}{\varepsilon}}\), it outputs \(FAIL\) with probability \(> \frac{1}{\poly \log N}\).
        \item If \(C^{\delta_0}_T\) did not output \(FAIL\), then with probability \(1-\frac{1}{\poly\log N}\), the output list \(L_u \subseteq \mathcal{I}_u\) is \(\delta_j'\)-well separated in \(\mathcal{I}_u\), for \(\delta_j' = 1.16 \cdot 8^j \delta_{0}\).
    \end{enumerate}
    The \(\poly\log N\) in the guarantees can be any polynomial we choose, and in particular set to \(\frac{1}{\log^{10} N}\).
    \end{lemma}
    
    The proof of this lemma is presented in the next section, where we also analyze its circuit complexity. For now, we take it at face value and just proceed assuming it is possible to construct this circuit.
    \item \textbf{A subset internal routing scheme}: denoted \(\mathcal{O}^T\). To be explicit, this routing scheme takes as input a pair \(u,u' \in U\) and is supposed to output a path from \(u\) to \(u'\) such that all vertices (except, perhaps, the endpoints) are in the set \(T\). Such a routing scheme exists by \Cref{prop:graph-system-for-constant-rate}. Here the set \(T \subseteq U\) is the set of $u$ such that $C^{\delta_0}_T(u,j)$ outputs FAIL with probability at most $\frac{1}{\log^{10}(N)}$ (following property (a) of \Cref{lem:list-sparsification}), where $j$ will be given to the decoder as a parameter (and in particular, will be sampled at random). Formally, when we call the \(T\)-membership oracle on a set \(u'' \in U\), we apply \(C^{\delta_0}_T\) on it and check whether or not it failed.
\end{enumerate}

\begin{figure}[ht!]
\fbox{\parbox{\textwidth}{
\vspace{.1cm}
\underline{Outer decoder \(OutDec_{\delta_0,j}(u,g,v)\):}
\begin{enumerate}
    \item[*] \textbf{Input:} 
    \begin{itemize}
        \item \(u \in U\) and a function \(g \in \mathcal{I}_u\).
        \item \(v \in V\).
        \item Parameters \(j\) and \(\delta_0\).
    \end{itemize}
    \item[*] \textbf{Output:} A value in \(\Sigma_1\).
    \item Let \(T \subseteq U\) bet the set of all \(u\) such that \(C^{\delta_0}_T(u,j)\) does not fail.
    \item Sample a random \(u' \in U\) such that \(u' \ni v\).
    \item Call \(\mathcal{O}^T(u,u')\). If the routing scheme itself failed then abort and output an arbitrary symbol. Otherwise let \(P=(u=u_0,u_1,\dots,u_{t}=u')\) be its output path, and for \(k<t\), denote by \(L_{u_k}\) the list output by \(C_T(u_k,j)\).
    \item For \(k=1,\dots,t-1\):
    \begin{enumerate}
        \item Find some \(g_k \in L_{u_k}\) such that \(\dist_{u_k \cap u_{k-1}}(g_{k-1},g_k) < 3\cdot \delta_j'\).
        \item If there are several such functions, choose one of them uniformly at random.
    \end{enumerate}
    \item Find any \(g_t \in \mathcal{I}_{u_t}\) such that \(\dist_{u_t \cap u_{t-1}}(g_{t-1},g_t) < 3\cdot \delta_j'\) and output \(g_t(v)\).
\end{enumerate}}}
    \caption{Outer decoder circuit algorithm}
    \label{fig:outer-decoder-constant-rate}
\end{figure}
Our generating algorithm samples a random \(u_0\), a random $j \in [\frac{1000}{\varepsilon}]$, and outputs circuits \(\set{OutDec_{\delta_0,j_i}(u_0,g,\cdot)}_{g \in \mathcal I_u}\).

We are now ready to prove the main result of the subsection, list-recovery from subset-internal routing schemes:

\begin{proof}[Proof of \Cref{thm:local-outer-decoder-const-rate}]
    The proof is similar to \Cref{thm:list-recovery}. Fix lists \(\mathcal{I}\) and \(f \in \Ball_{\frac{1}{\log^6 N},\delta_0}(\mathcal{I})\). Our goal is to prove for most \(u_0\) and choices of $j \in [\frac{1000}{\varepsilon}]$, there exists \(g \in \mathcal I_{u_0}\) such that $OutDec_{\delta_0,j}(u_0,g,\cdot)$ approximately computes $f$:
    \[
        \Prob{u_0,j}{\exists g \in \mathcal{I}_{u_0} : \Prob{v,r}{OutDec_{\delta_0,j}(u_0,g,v) = f(v)} \geq 1-\delta_1} \geq \frac{3}{4}
    \]
    In particular, the generating algorithm will succeed with high probability by trying all \(g \in \mathcal{I}_{u_0}\).

    Toward this end, let $T=T_j$ be the set of $u$ such that $C^{\delta_0}_T(u,j)$ outputs FAIL with probability at most $\frac{1}{\log^{10}(N)}$, and recall we require $|T| \geq 0.9|U|$ for subset internal routing. By \Cref{lem:list-sparsification} property (a), the expected size of $T$ over a random $j$ is $0.99|U|$, so the probability that $|T| < 0.9|U|$ is at most $0.1$ by Markov.

    Now, condition on having sampled a good $j$ with $|T| \geq 0.9|U|$. As in \Cref{thm:list-recovery}, we now define a few further rare `bad events' such that outside these circumstances our decoder succeeds. In particular, let \(B\) denote the set of edges \(\set{u,u'}\) such that for either \(u'' \in \set{u,u'}\) one of the following conditions holds:
    \begin{enumerate}
        \item \textbf{Bad List:} There is no \(g'' \in \mathcal{I}_{u''}\) that is \(\delta_0\)-close to \(f\).
        \item \textbf{Bad Sampling:} There exists some function \(g'' \in \mathcal{I}_{u''}\) such that 
        \[ \abs{\dist_{u''}(f,g'') - \dist_{u \cap u'}(f,g'')} > \delta_0.\]
    \end{enumerate}
    We call a path \((u_0,u_1,\dots,u_m)\) good if no edge in the path is in \(B\). We claim that it is enough to prove that most \(u_0\) have mostly good oracle paths to a random endpoint \(u_t\):
    \begin{equation}\label{eq:good-paths-distance-const-rate}
        \Pr_{u_0,j}\left[\Pr_{u_t,\mathcal{O}^T}[\mathcal{O}^T(u_0,u_t) \text{ doesn't fail and the path is good}] \geq 1-\frac{\delta_1}{2}\right] \geq 99\%
    \end{equation}
    Note that the inner probability implicitly depends on \(j\) since \(T\) is defined with respect to \(C_T(\cdot,j)\).
        
    Let's first complete the proof assuming \eqref{eq:good-paths-distance-const-rate}. Fix any \(u_0,j\) such that the inner condition holds. By assumption, there exists some \(g \in \mathcal{I}_{u_0}\) that's \(\delta_0\)-close to \(f\). We claim that for any such fixed choice, the probability \(OutDec_{\delta_0,j}(u_0,g,v) = f(v)\) on a random \(v\) is at least \(1-\delta_1\). Union bounding over the choice of good $j$ and $u_0$ gives the final desired decoding guarantee.

    To see this, first recall that on input \(v\), the decoder draws a random \(u_t \ni v\), paths to it via \(\mathcal{O}^T(u_0,u_t)\) and outputs \(g_t(v)\) for some valid choice of \(g_t \in L_{u_t} \subseteq \mathcal{I}_t\). Thus our goal is to lower bound
    \[
        \Pr_{r,v}[OutDec_{j,\delta_0}(u_0,g,v) = f(v)] =\Pr_{v \in u_t,\mathcal{O}^{T},g_t}[g_t(v) = f(v)]
    \]
    where \(g_t \in L_{u_t}\) is the function obtained by the last step of the algorithm, and we recall $L_{u_t} \subset \mathcal I_{u_t}$ is the output list of the well-separation circuit $C_T(u_t,j)$. We note on both sides if \(\mathcal{O}^T\) fails or there is no valid choice of \(g_t\), we do not count this in the probability.

    Now examining the right-hand probability, notice we have drawn \(v \in V\) uniformly at random, then a random neighbor \(u_t \in v\). Since \((V,U)\) is regular, this is the same as first drawing a random \(u_t \in U\), then sub-sampling a random vertex \(v \in u_t\). This means it is enough to argue that, over the randomness of \(u_t,\mathcal{O}^T\), the final function \(g_t\) on \(u_t\) is \(\delta_1\)-close to \(f\) with high probability. Let \(L_{u_t}' \subseteq L_{u_t}\) be the possible functions that are \(3\delta_0\) close to \(g_{t-1}\). 
    \[
        \Pr_{u_t,\mathcal{O}^T}\left[\forall g_t \in L_{u_t}': \text{dist}(g_t,f) \leq \frac{\delta_1}{2}\right] \geq 1-\frac{\delta_1}{2}
    \]

    In particular, the probability we correctly output $f(v)$ over a random choice of $u_t \in U,g_t \in L_{u_t}'$ and $v \in u_t$ is at least $(1-\frac{\delta_1}{2})^2 \leq 1-\delta_1$ as desired.

    Now to show the above, we introduce one final bad event outside which our decoder succeeds: the event that our distance tester errs (recall we sub-sample points in the intersection to test distance rather than check the full symbol to ensure our decoding circuits remain small). By construction (and \Cref{claim:independent-distance-test}), this occurs with probability at most $\frac{\delta_1}{4}$.

    Now condition on: the routing oracle not failing, the $\mathcal{O}^{T}(u_0,u_t)$ being good \textit{and} the above event that our distance tester doesn't err on the path (which occurs with probability at least $1-\delta_1$ by a union bound\footnote{Here we also use the fact that \(j\) chosen above makes the failing probability \(\frac{1}{\log^4 N}\).}). We claim under this conditioning \textit{every} $g_t \in L_{u_t}'$ is $\frac{\delta_1}{2}$-close to $f$ when we start from the promised $g=g_0$ that is $\delta_0$-close to $f$. 

    This is where the proof differs from \Cref{thm:list-recovery}. By the goodness of the path, for every \(u_k\) there is a function that is \(\delta_0\)-close to \(f\) in the list \(\mathcal{I}_{u_k}\). By the \(\delta_j'\)-well separatedness of every \(L_{u_k}\) (for \(k=1,2,\dots,t-1\)), there is a \textit{unique} function \(g' \in L_{u_k}\) that is \(\delta_j' + \delta_0\)-close to \(f\). We claim that this function is the function that is chosen in every step of the algorithm. This follows by induction on \(k\). For \(k=0\), this follows because \(g=g_0\) is \(\delta_0\)-close to \(f\) by assumption. Now assume it is true for \(g_{k-1}\) and prove for \(g_k\). Our assumption is that \(\dist_{u_{k-1}}(g_{k-1},f) <(6^j+1)\delta_0\). Therefore by goodness of the edge, on the intersection
        \[\dist_{u_{k-1} \cap u_k}(g_{k-1}|_{u_{k-1} \cap u_k},f) <\delta_j' + 2\delta_0.\]
    Now on the one hand, if \(g_k\) is the function that is \(\delta_j'+\delta_0\)-close to \(f\), then also in the intersection
    \[\dist_{u_{k-1} \cap u_k}(g_{k}|_{u_{k-1} \cap u_k},f) <\delta_j'+2\delta_0\]
    and therefore by the triangle inequality,
    \[\dist_{u_{k-1} \cap u_k}(g_{k-1}|_{u_{k-1} \cap u_k},g_{k}|_{u_{k-1} \cap u_k}) <2\delta_j'+2\delta_0 \leq 3\cdot \delta_j'.\]

    On the other hand, for any other \(g' \in L_{u_k}\), by well separatedness, it is \(5\delta_j'-\delta_0 \)-far from \(f\).\footnote{This is just using the fact that \(f\) is \(\delta_j' + \delta_0\)-close to \(g_k\), that \(g_k\) and \(g'\) are \(6\delta_j'\)-far apart and the reverse triangle inequality.} By goodness of the edge, in the intersection
        \[\dist_{u_{k-1} \cap u_k}(f|_{u_{k-1} \cap u_k},g'|_{u_{k-1} \cap u_k}) >5\delta_j'-2\delta_0.\]
    By the (reverse) triangle inequality,
    \[
        \dist_{u_{k-1} \cap u_k}(g_{k-1}|_{u_{k-1} \cap u_k},g'|_{u_{k-1} \cap u_k}) >5\delta_j'-2\delta_0 - \dist_{u_{k-1} \cap u_k}(g_{k-1}|_{u_{k-1} \cap u_k},f)
    \]
    and by the bound above this is at least \(5\delta_j' - 2\delta_0 - (\delta_j'+2\delta_0) = 4\delta_j' -4\delta_0 > 3\cdot \delta_j'\). Therefore \(g_k\) and only \(g_k\) can be chosen in the \(k\)-th step.

    In the final step, we do not have the guarantee that there is a single function that will pass the test, but by a similar argument to the above, the function \(g_t\) will be \(3\delta_j'\)-close to \(f\), and this is less than \(\frac{\delta_1}{2}\).

    The rest of the proof continues to follow the same lines of \Cref{thm:list-recovery}. It is left to prove \Cref{eq:good-paths-distance-const-rate}, that many $u_0,j$ have few bad paths. To see this, first observe that by Markov's inequality it is enough to prove a random choice of $u_0,u_t$, and path $\mathcal{O}(u_0,u_t)$, the probability $\mathcal{O}^T(u_0,u_t)$ either fails or hits $B$ is very low, say:
\[
    \Pr_{u_0,u_t,\mathcal{O}^T}[\mathcal{O}^T(u_0,u_t)\text{ fails or hits } B] \leq \frac{1}{\log^2 N}.
\]
    This is the regime that is bounded by low-congestion routing. In particular, since $\mathcal{O}^T$ is low congestion, it is enough to prove the measure of $B$ is small, namely $\Pr[B] \leq \frac{1}{\log^9 N}$. This follows essentially immediately from sampling, and from the promise that all but a $\frac{1}{\log^{10} N}$-fraction of $u \in U$'s list contain a function $\delta_0$-close to $f$. The latter ensures the probability a random edge satisfies the `Bad List' condition is at most $\frac{2}{\log^{10} N}$ by a union bound. The former says that for any fixed $u$, the probability a random neighbor $u'$ mis-samples any fixed function in the list of $u$ is at most $\frac{\beta}{\delta_{in}^2}$. Union bounding over the lists of both $s$ and $s'$ in a random edge, this event therefore occurs with probability at most $\frac{2 \beta}{\varepsilon \delta_0^2}$, and a final union bound gives $\Pr[B] \leq \frac{1}{\log^9 N}$ as desired.
\end{proof}

\subsection{Reducing to Well-Separated Lists}\label{subsec:well-separated}

We now give a formal proof of \Cref{lem:list-sparsification} by constructing our list sparsification circuit \(C_T^{\delta_0}\).

Unfortunately, it is not true that any list generated by our inner codes is automatically well-spread (indeed it will almost certainly not be). It turns out, however, that it \textit{is} possible to prune the lists in such a way that at many `scales' $j$, we can find sub-lists that are \(\delta_j'\)-well-separated for \(\delta_j' \approx \delta_{in} 8^i\). The argument is an iterated variant of a similar method in \cite{dinur2019list} who proved one such scale exists.


We present our pruning procedure toward this end in \Cref{fig:distance-separator}, which is based on the following basic object we call a `distance graph'. Namely given a list of functions $L$, we prune $L$ by analyzing the (empirical) $(L,\delta_0)$-distance graphs which have vertex set \(L\), and edge set all \(\set{g,g'}\) such that \(\dist(g,g') \leq \delta_0\) measured empirically over a random sub-samples of the domain. We note that technically the circuit \(C_T^{\delta_0}\) is a randomized circuit, but as explained in the proof, it's error is $\frac{1}{\polylog N}$, and so we can safely run it without ruining the probability that the outer decoder succeeds. 
\begin{figure}[ht!]
\fbox{\parbox{\textwidth}{
\vspace{.1cm}
\underline{List sparsifier \(C_T^{\delta_0}\):}
\begin{enumerate}
    \item[*] \textbf{Input:} 
    \begin{itemize}
        \item The list \(\mathcal{I}_u\).
        \item An index \(j \in \set{1,2,\dots,\frac{1000}{\varepsilon}}\).
    \end{itemize}
    \item[*] \textbf{Output:} Either \(FAIL\) or a list \(L' \subseteq \mathcal{I}_u\).
    \item Sample $O(\frac{\log \frac{\log N}{\varepsilon}}{\delta_0})$ random points from the domain and compute the empirical distance between every pair $g,g' \in \mathcal{I}_u$.
    \item Set \(\delta_k = 8^k\delta_0\) and \(L_0 = \mathcal{I}_u\). Compute \(G_0\), the empirical \((L_0,\delta_0)\)-distance graph.
    \item For \(k=1,2,\dots,\frac{1000}{\varepsilon}\):
    \begin{enumerate}
        \item Find a maximally independent set \(L_k \subseteq L_{k-1}\) in \(G_{k-1}\) that was computed in the prior round (or before the loop in the case of \(G_0\)).
        \item Compute \(G_{k}\), the \((L_k,\delta_k)\)-distance graph.
    \end{enumerate}
    \item If the subset \(L_j\) is \(\delta_{j+1}\)-separated output \(L_j\) (according to the empirical distances), otherwise output \(FAIL\).
\end{enumerate}}}
    \caption{Outer decoder circuit algorithm}
    \label{fig:distance-separator}
\end{figure}

We comment that the algorithm runs for \(\frac{1000}{\varepsilon}\) rounds regardless of the output only depending on the first \(j\) rounds. This peculiar choice is just for ease of analysis and ease of implementation in a circuit. We can just as well stop the loop when \(k=j\).

\begin{proof}[Proof of \Cref{lem:list-sparsification}]
    The only randomized part in our algorithm is the computation of distances between every two functions in our list. Sample $O(\frac{\log \frac{\log N}{\varepsilon}}{\delta_0})$ random points from the domain and compute the empirical distance between every pair $g,g' \in L$. By a standard Chernoff and union bound, for every $\delta_{k}=\delta_0 8^k$, the probability that any pair $g,g'$ has $d(g,g') \leq 7.9\delta_k$ but has empirical distance greater than $8\delta_k$ is at most $\frac{1}{\polylog N}$, and similarly for when $d(g,g') \geq 8.1\delta_k$ but measures as less than $8\delta_k$. Condition on no such events occurring.

    We observe that in every step \(L_k\) is \(8.1\delta_{k-1}\)-dense in \(L_{k-1}\). This is because if there were a function \(f \in L_{k-1}\) that is farther than \(8.1\delta_{k-1}\) from any function in \(L_k\), then by the conditioning above, no edge between \(f\) and \(L_{k-1}\) will appear in the distance graph \(G_k\), which contradicts maximality of \(L_{k-1}\). By using the triangle inequality, for every function in \(f_0 \in L_0\) there is a function in \(f_1 \in L_1\) that is \(8.1\delta_0\)-close to it. There is a function \(f_2 \in L_2\) that is \(8.1\delta_1\)-close to \(f_1\) etc. Thus concludes via the triangle inequality that for every \(f_0 \in L_0\) there is an \(f_k \in L_k\) such that
    \[\dist(f_0,f_k) \leq \sum_{m=0}^{k-1}\dist(f_m,f_{m+1}) \leq 8.1 \delta_0 \sum_{m=0}^{k-1} 8^m \leq \frac{8.1}{7}8^{k-1} \delta_0.\]
    In particular, \(L_j\) is \(1.16 \delta_{j-1}\)-dense in \(L_0\).

    Next we observe that if \(L_{j}\) was output, then it is \(7.9\delta_{j}\)-separated. Indeed, if \(L_j\) was output, then the empirical distances of any two functions in \(L_j\) were \(\geq 8 \delta_j\), so by the event we conditioned on, their actual distances are at lesat \(7.9 \delta_j\). In particular, for \(\delta_j' = 1.16 \cdot 8^j \delta_0\), if \(L_j\) is output, then it is \(\delta_j'\)-well separated.

    Finally, note that if \(L_j\) is not output, then it means that \(L_{j+1} \ne L_j\) and in particular, that \(|L_{j+1}|\leq |L_j| - 1\). Note that after \(\frac{10}{\varepsilon}-1\) failures, the list just becomes a singleton, in which case we will always succeed. Thus there are at most a \(0.1\)-fraction of indexes where we fail for any given list.

\end{proof}

\paragraph{Circuit Complexity} Finally, we sketch the standard fact that the above algorithm can be implemented logspace uniform size and depth.
\begin{claim} \label{claim:list-pruner-circuit-complexity}
    The circuit \(C_T^{\delta_0}\) is implementable in $\poly(\frac{1}{\varepsilon},\frac{1}{\delta_0},\log\log N)$ size, $O(\frac{1}{\varepsilon}(\log^2 \frac{\log N}{\varepsilon\delta_0}))$ depth and makes at most $O(\frac{\log \frac{\log N}{\varepsilon}}{\delta_0})$ calls to the list circuit \(C_{list-u}\) and point circuits \(C_{U \to V}\).
\end{claim}

 \begin{proof}
     The number of oracle calls is immediate from the description. The first step of the algorithm calls the sample oracle in parallel. Then the algorithm runs parallel pairwise distance tests for every pair of functions in $\mathcal{I}_u$ (at every scale that will be tested in the following iterative procedure). This corresponds to calling $O(|\mathcal{I}_u|)^2$ equality circuits, and $O(|\mathcal{I}_u|)^3$ Threshold circuits (since there are \(O(|\mathcal{I}_u|)\) iterations and \(\binom{|\mathcal{I}_u|}{2}\) pairs of distinct functions).

    Now, for the iterative component of the algorithm, it is enough to have a circuit which takes as input the current indicator of the list $L_{k-1} \in \{0,1\}^{|\mathcal{I}_u|}$ and outputs the indicator for $L_{k}$, along with a bit indicating whether or not the algorithm output FAIL at that round. Note that the circuit may also be assumed to have access to the distance graph for the round, since it knows the vertex set $L_{k-1}$ and the distance threshold computations give the edges. This means it is enough to compute a maximal independent set on this input graph which is in $NC^2$ by a result of Luby \cite{luby1985simple}. To check whether FAIL is output, the algorithm simply checks if the indicator of the output maximal independent set is equal to the original input.

    All of the above procedures are in $NC^2$ over the domain of sampled points. Wiring the circuits for each of the $O(|\mathcal{I}_u|)$ scales in sequence results in the desired uniform size and depth.
\end{proof}

\begin{remark}
    If one is fine with a blowup of \(\exp(O(\frac{1}{\varepsilon^2}))\) instead of \(\exp(O(\frac{1}{\varepsilon}))\), a simpler algorithm is possible: on input \(j \leq \exp(O(\frac{1}{\varepsilon^2}))\) just find a maximal independent set in the \((\mathcal{I}_u,5^j \delta_0)\)-distance graph. The algorithm outputs a maximal independent set for every \(j\) such that no pair of functions \(f,f' \in \mathcal{I}_u\) have distance between \(5^j \delta_0\) and \(5^{j+1} \delta_0\) which is well-separated as any such set is \(6^j \delta_0\)-dense, but also \(6^{j+1} \delta_0\)-separated. As there are \(O(\frac{1}{\varepsilon^2})\) distinct possible distances between pairs of functions in \(L\), by allowing \(O(\frac{1}{\varepsilon^2})\) possible indexes \(j\), we can get large constant success rate. As we do not use this later on, we omit the details and just comment that using this procedure instead has the slight advantage of reducing the dependence on \(\varepsilon\) in the circuit depth.
\end{remark}

\subsection{Putting it All Together (Modulo Circuits)}
We are now ready to prove \Cref{thm:constant-rate} (defering the circuit complexity analysis to the next subsection). Fix \(\varepsilon > 0\) and $N \in \mathbb{N}$. We take the set system \((V,S,U,G)\) promised in \Cref{prop:graph-system-for-constant-rate}. The parameters \(\beta,\eta\) in the theorem statement are chosen so that they are small enough for using all claims and lemmas in this section. The resulting code is the direct product \(\mathcal{C}_{(V,S)}\). The hyperedge set sizes are \(\log (N)^{\exp(O(\frac{1}{\varepsilon}))}\) so locality and alphabet size follow by direct computation. Since $(V,S)$ is bi-regular, the rate of the code is just the inverse of the number of sets every vertex is contained in, that is, \(\exp(-O(\frac{1}{\varepsilon}))\). 

Let us describe the decoder:
\begin{enumerate}
    \item We set \(\delta_0 = O(\sqrt[3]{\delta_{in}})\).
    \item We use the generating algorithm of the outer decoder, where we interpret the list indexes to the starting set \(u\), as inputs to \(C_{list-u}(u,i,\cdot)\).
    \item We call the outer decoder circuits where every time an inner list \(g_j \in \mathcal{I}_{u'}\) is called, we simulate it via \(C_{list-u}(u',j,\cdot)\).
\end{enumerate}

Fix \(f \in \Ball_{1-\varepsilon,\delta_{in}}(w)\). By \Cref{claim:global-functions-remain-global-in-inner-decoder} \(f \in \Ball_{1-\frac{\varepsilon}{2},2\delta_{in}}(w_{In_u})\) for all but \(\frac{1}{2\log^{10 }(N)}\) of the \(u \in U\). Thus by \Cref{lem:final-inner-decoding-procedure} we can find a list \(\mathcal{I}_u\) of size \(<\frac{10}{\varepsilon}\) containing some function that is \(\delta_0 = O(\sqrt[3]{\delta_{in}})\)-close to \(f\). That is, \(f \in \Ball_{\frac{1}{\log^{10}(N)},\delta_0}(\mathcal{I})\). Thus we can use the outer decoder in \Cref{thm:local-outer-decoder-const-rate} to recover a function that is \(\delta_{out}=\exp(102\frac{1}{\varepsilon})\delta_0\)-close to \(f\) (which is less than \(0.01\) by our parameter setting).

Finally for our chosen parameters:
\begin{enumerate}
    \item The routing scheme queries lists of \(\log^3(N)\exp(O(\frac{1}{\varepsilon}))\) vertices.
    \item To prepare each list, the inner decoder is instantiated \(O(\frac{\log\log(N)}{\varepsilon \delta_{in}})\) times, and in each instantiation, \(O(\frac{\log\log (N)}{\varepsilon \delta_{in}})\) points are queried by \(C_{list-u}\).
    \item Every query for the inner decoder requires \(\poly(\frac{1}{\varepsilon},\frac{1}{\delta_{in}})\) points from the original lists \(\mathcal{L}\).
    \item Besides the list construction, the outer decoder tests distance between \(O(\log^3(N) \cdot \frac{1}{\varepsilon})\) pairs of functions, up to accuracy of \(\geq \frac{1}{\delta_{0}}\). This requires at most \(O(\log^3(N) \cdot \frac{1}{\varepsilon\delta_{in}})\) points in total.
\end{enumerate}
Thus the total query complexity is indeed \(\log^3(N)\exp(O(\frac{1}{\varepsilon}))\poly(\log\log(N))\).

\subsection{Circuit Implementation}
To implement a circuit of the outer decoder we perform the following.
\begin{enumerate}
    \item We need call \(C_{\mathcal{O}}^T(u,u')\) in order to route. For this we need to also call \(\poly\log(N) \exp(O(\frac{1}{\varepsilon}))\) calls to \(C_{list-u}\), in order to implement the oracle for \(T\). Fortunately we can make all calls in parallel by \Cref{claim:circuits-for-const-rate-construction}. Therefore, this step is implementable in size \((\log N)^{\exp(O(\frac{1}{\varepsilon}))}\) and depth \(\tilde{O}(\exp(O(\frac{1}{\varepsilon}))\log^2 \log N\).
    \item After that, we need to compute what is the index of a final \(g_m\) in the list of \(u'\) that is close to the function \(f\). As in the outer decoder for \Cref{thm:list-recovery}, we can do this with Savich's algorithm in size \((\log N)^{\exp(O(\frac{1}{\varepsilon}))}\) and depth \(\tilde{O}(\exp(O(\frac{1}{\varepsilon}))\log^2 \log N\). We omit the details since it is similar to those in \Cref{lem:list-recovery-circuit}.
\end{enumerate}

All in all, the circuit implementation of the decoder for \Cref{thm:constant-rate} has size \((\log N)^{\exp(O(\frac{1}{\varepsilon}))}\) and depth \(\tilde{O}(\exp(O(\frac{1}{\varepsilon}))\log^2 \log N\).
\section{The Main Construction: KMS Codes}\label{sec:instantiations}
In this section we construct the set systems in \Cref{prop:graph-system-for-log-rate-graph} and \Cref{prop:graph-system-for-constant-rate} which arise from a form of high dimensional expanders called  \textit{local spectral expanders} and in particular a variant \cite{dePeraltaVB2025high} of the elegant coset-complex HDX of Kaufman and Oppenheim \cite{kaufman2018construction}.
\TP{I am mising a high level understanding of what the essential properties of the KMS/KO HDX have that we will exploit to get a good routing scheme.Also would be good to get an preview for how we will be setting  the parameters (e.g. d, q, kappa)  that we'll be using in routing scheme. I think it may be a lot easier for reader if we focus on constant rate aLLDC theorem - Prop 5.4.}

\paragraph{Roadmap} The section is ordered as follows. In \Cref{subsec:hdx-prelims} we begin with some preliminaries on HDX. In \Cref{subsec:defining-KMS} we present a family of HDX known as Kac-Moody-Steinberg complexes (KMS complexes), that were constructed and analyzed in \cite{dePeraltaVB2025high}, and in \Cref{subsec:links-in-kms} discuss their local structure. In \Cref{subsec:routing-kms}, we use this structure to construct (subset internal) local routing schemes. In \Cref{subsec:defining-set-systems} we define the set systems in \Cref{prop:graph-system-for-log-rate-graph} and \Cref{prop:graph-system-for-constant-rate}, and in \Cref{subsec:proving-propositions-on-set-systems} prove \Cref{prop:graph-system-for-log-rate-graph} and \Cref{prop:graph-system-for-constant-rate} in full. We end with \Cref{sec:kms-circuit-implementation} constructing circuits that implement neighborhood relations in the KMS complexes and bound the circuit complexity of the implementations of the set systems.

\subsection{Preliminaries on High Dimensional Expanders}\label{subsec:hdx-prelims}
We will give only a brief primer on HDX necessary for our needs below. For a more comprehensive background we refer the reader e.g.\ to \cite{dikstein2018boolean} and the references therein.
\paragraph{Simplicial Complexes}
A \emph{$d$-dimensional simplicial complex} $X$ consists of a $(d+1)$-uniform hypergraph $X(d)$ together with its downward closure
\[
X=X(-1) \cup X(0) \cup \ldots \cup X(d),
\]
where $X(i) \subseteq \binom{[N]}{i}$, called the `$i$-faces' or `$(i+1)$-sets', are all $(i+1)$-size subsets that sit in some hyperedge in $X(d)$, and $X(-1)=\{\emptyset\}$. Given a face $t \in X(i)$, the \textbf{link} of $t$ is the sub-complex induced by localizing to faces that include $t$, that is $X_t \coloneqq \sett{ s \setminus t \in X}{s \in X, s \supseteq t}$. We say $X$ is \textbf{connected} if the base graph with vertices $X_t(0)$ and edges $X_t(1)$ of every link \(X_t\) is connected. The \textbf{inclusion graph} $(X(k),X(i))$ is the bipartite graph between $k$-faces and $i$-faces of $X$ where edges are given by inclusion. The related \emph{swap graph} (or \textit{swap walk}) between \((X(k),X(i))\) connects $k$-sets and $i$-sets of $X$ such that \(s \sim t\) if the disjoint union \(s \bigsqcup t \in X(k+i+1)\).
\TP{Mention we assume $i<k$.} For \(k < d\), The \(k\)-skeleton of a simplicial complex \(X\) is \(X^{\leq k} = X(-1) \cup X(0) \cup \ldots \cup X(k)\).

\paragraph{Partite complexes}
A $(d-1)$-dimensional complex is \(d\)-\textbf{partite} if its vertices can be partitioned into $d$ `parts' such that each top-level face has one part from each component. We denote by \(X[1],X[2],...,X[d] \subseteq X(0)\) each part (we emphasize the square brackets denote the different parts of vertices). For an index set \(C =\set{i_1,i_2,\ldots, i_j} \subseteq \set{1,2,\ldots,d}\) we also denote by \(X[C] \subseteq X(j-1)\) all \((j-1)\)-faces \(\set{v_1,v_2,\ldots,v_j} \in X(j-1)\) such that \(v_1 \in X[i_1]\), \(v_2 \in X[i_2]\),...,\(v_j \in X[i_j]\) (up to a change in their order). We call these \(C\)-colored faces. For a face \(t\) we denote by \(color(t)\) the index set \(C\) such that \(t \in X[C]\). We also denote by \(X^{C} = \bigcup_{C' \subseteq C}X[C']\). We say that a partite complex is \textbf{regular}, if the number of top-level faces containing a set \(t \in X\) only depends on the color of \(t\). We say that \(X\) is \textbf{fully regular} if \(X\) and all its links are regular.

Sometimes we also consider the colored-swap graph \((X[I],X[J])\) which is a subgraph of the swap graph above, only containing vertices in \(X[I]\) and \(X[J]\). 

\paragraph{High dimensional expanders}
Recall that a graph \(G\) is a \(\lambda\)-expander if the second largest eigenvalue of its normalized random walk operator is \(\leq \lambda\). The normalized adjacency operator is the \(V \times V\) real-valued matrix 
\[ A(v,u) = \begin{cases} \frac{1}{deg(v)} & \set{v,u} \in E \\
0 & \text{otherwise} \end{cases}.\]

For simplicity we define high dimensional expanders only for fully regular partite complexes. We say that \(X\) is a \(\lambda\)-high dimensional expander if for every \(t \in X^{\leq d-2}\) and every \(i,j \notin color(t)\), the bipartite graph between \(X_t[i]\) and \(X_t[j]\) is a \(\lambda\)-expander.\footnote{This is also the definition even if \(X_t\) is not fully regular, except that in this case we need put weights on the edges of this graph.}

\subsubsection{Theorems and Facts on HDX}
Here we mention briefly some of the facts on HDX.
\paragraph{Expansion of inclusion and swap graphs}
The inclusion and swap graphs are good expanders.
\begin{theorem}[{\cite{kaufman2020high}}] \label{thm:up-down-walk-expands}
    Let \(X\) be a \(d\)-dimensional \(\frac{1}{2d}\)-one sided HDX. Then the inclusion graph \((X(k),X(i))\) is an \(O(\sqrt{\frac{i}{k}})\)-one sided expander.
\end{theorem}

We note that in general one does not expect to do better than this bound as sets of the form \(\set{s \in X(k)}{v_0 \in s} \subseteq X(k)\) do not expand in the inclusion graph more than \(\approx \sqrt{\frac{i}{k}}\), but the swap graphs do expand better \cite{dikstein2019agreement,alev2019approximating,gur2021hypercontractivity,alev2023sequential}:
\begin{theorem} \label{thm:swap-walk}
    Let \(X\) be a \(d\)-partite \(\lambda\)-one sided HDX. Then for every \(I,J \subseteq \set{1,2,\ldots,d}\) of sizes \(i,k\), the swap graph \((X[I],X[J])\) is a \((i+k)\lambda\)-one sided expander.
\end{theorem}



\subsection{Coset and KMS Complexes} \label{subsec:defining-KMS}
In this subsection, we give an instantiation of the KMS complexes described in \cite{dePeraltaVB2025high}, which are a variant of the complexes in \cite{kaufman2018construction}. 



KMS complexes are coset complexes. That is, they are parametrized by some finite group \(G\) and subgroups \(K_1,K_2,\dots,K_d\). The vertices of the coset complex \(X=X(G;K_1,K_2,\dots,K_d)\) are cosets
\[X(0) = \bigcup_{i=1}^d g K_i\]
and the top level sets are
\[
X(d-1) = \sett{\set{gK_1,gK_2,\dots,gK_d}}{g \in G}.
\]
Equivalently these are all sets \(\set{g_1K_1, g_2K_2,\dots,g_dK_d}\) such that \(\bigcap_{i=1}^d g_iK_i \ne \emptyset\). One can verify that the intermediate level sets \(X(i)\) have a similar description within each color set. That is for $C=\set{i_1,i_2,\dots,i_j}$, \(X[C]\) is the set of \(\bigcap_{i \in C} g_iK_i \ne \emptyset\).

It is a classical fact that the mapping \[\set{g_1K_{i_1}, g_2K_{i_2},\dots,g_rK_{i_r}} \in X[C] \longmapsto \bigcap_{i \in C} g_iK_i\] is a bijection from faces to cosets of \(K_C \coloneqq \bigcap_{i\in C} K_i\). Moreover, faces $s,t \in X$ satisfy \(s \subseteq t\) if and only if \(t\cong g K_{C_1}, s \cong g' K_{C_2}\) such that \(C_1 \subseteq C_2\) and \(g K_{C_1} \supseteq g' K_{C_2}\).

\medskip

The complexes we use are based on the group \(G = SL_{d}(R)\), \(d \times d\) matrices with determinant \(1\), where \(R = \nicefrac{\F_\fsize[t]}{\phi(t)} \cong \F_{\fsize^\poldeg}\), i.e.\ \(\phi(t)\) is a degree \(\poldeg\) irreducible polynomial.

Our subgroups \(K_i\) are generated by elementary unipotent matrices.
\begin{definition}[Elementary Unipotent Matrices]
    For \(i, j \in \set{1, 2, \dots, d}\) with \(i \neq j\) and \(r \in R\), let \(e_{i,j}(r)\) denote the unipotent elementary matrix in \(G\) such that \(r\) appears in the \((i,j)\)-entry, and all other off-diagonal entries are zero, i.e., the identity matrix with the $(i,j)$th entry replaced by $r$.
\end{definition}

\begin{definition}[Subgroups \(K_i\)]
    For each \(i \in \set{1, 2, \dots,d-1}\), let \(K_i\) be
    \[
    K_i = \langle \set{e_{j,j+1}(a) : j \in [d] \setminus \set{i}, \, a \in \F_\fsize} \cup \set{e_{d,1}(at) : a \in \F_\fsize} \rangle.
    \]
    In addition, let
    \[
    K_d = \langle \set{e_{j,j+1}(a) : j \in [d-1], a \in \F_\fsize} \rangle.
    \]
\end{definition}
It is observed (see e.g.\ \cite{dePeraltaVB2025high}) that all these groups are isomorphic, and that \(K_d\) are just upper triangular matrices with \(1\)'s on the diagonal, and elements in the field \(\F_\fsize\) above the diagonal (the variable \(t\) doesn't appear). In particular, every \(|K_i| = \fsize^{\binom{d}{2}}\) and therefore \(|X(0)|=d\fsize^{-\binom{d}{2}}|SL_{d}(R)| = \fsize^{\Omega(d^2 \poldeg)}\). 
\TP{I understand $|K_i|$ but am not following the equality for $|X(0)|$.} In, \cite[Theorem 3.6]{OdonnellP2022high}, the authors prove these complexes are connected and strongly explicit. Unfortunately, the coset bit representation for which they show strong explicitness has substantial slack (i.e.\ the bit-length is much larger than $\log|X(0)|$), so cannot be used in our context (c.f.\ the discussion in \Cref{sec:circuit-subpoly}). In \Cref{sec:kms-circuit-implementation}, we show how to implement these complexes as small, low depth circuits with negligible slack.

\begin{definition}
    Let \(\fsize\) be a prime power and \(d,\poldeg > 0\) be integers. Define the complex \(\kms = \kms^{\fsize,d,\poldeg}\) to be the coset complex defined using the group and subgroups above.
\end{definition}

\subsection[Links of the KMS complex]{Links of \(\kms\)} \label{subsec:links-in-kms}
As alluded to, these complexes are high dimensional expanders.\footnote{Technically due to slight differences in the notion of local spectral expansion, this lemma uses some standard machinery from \cite{dikstein2019agreement} to translate between the two, but we omit the details. An explicit translation is given in \cite{hopkins2024hyp}.}
\begin{lemma}[\cite{OdonnellP2022high}, \cite{dePeraltaVB2025high}] \label{lem:links-of-kms-expand}
    The complex \(\kms\) is a \(\frac{10}{\sqrt{\fsize}}\)-high dimensional expander. Moreover, every vertex is connected to at most \(\leq \fsize^{\Theta(d^2)}\) top level faces, independent of \(\poldeg\). 
\end{lemma}


We can give a very explicit construction of the links using the following complex called `the opposition complex'.\footnote{In fact, \cite{Abramenko2006twin} defines opposition complexes to be a more general type of complex, but we will use the name opposition complex to denote only the complex we describe.}

Fix \(\mathcal{V}=\F_\fsize^{d}\), and let \(e_1,e_2,\dots,e_d\) be the standard basis. For \(j=1,2,\dots,d-1\), let \(E_j=span \set{e_1,...,e_j}\). The opposition complex \(P\) is a \((d-1)\)-partite complex whose vertices are
\[P[i] = \sett{W \subseteq \mathcal{V}}{W \text{ is a linear subspace, } \dim(W)=i, \text{ and } W+E_{d-i}=\mathcal{V}}.\]
Edges are pairs \(\set{W,W'}\) such that (up to reordering) \(W \subset W'\), $i$-faces are nested sequence of $i$ subspaces: $\{W_1,W_2,\ldots,W_i\}$ such that $W_1 \subset W_2 ...$ up to re-ordering, and top level faces are given by complete flags (again up to re-ordering):
\[
P(d-1) = \sett{\set{W_1,W_2,W_3,\dots,W_{d-1}}}{W_1 \subset W_2 \subset \dots \subset W_{d-1}}.
\]

\begin{claim}[\cite{Abramenko2006twin}, see discussion in \cite{dePeraltaVB2025high, OppenheimVB2025new}] \label{claim:link-structure}
    For every \(gK_i \in \kms(0)\), the link \(\kms_{gK_i}\) is isomorphic to the opposition complex \(P\). Moreover, the part \(\kms_{gK_i}[j]\) is sent to \(P[j-i]\) under this isomorphism (modulo \(d-1\)).
\end{claim}

We also note that the opposition complex can be `flipped' upside down, that is,
\begin{claim} \label{claim:flipping-opposition-complex}
    There is an automorphism of \(P\) that sends \(P[i]\) to \(P[d-i]\) and vice versa.
\end{claim}

\begin{proof}
    First note that the choice of the standard basis in the definition of \(P\) is arbitrary, and we can in fact use any ordered basis and get an isomorphic complex. 
    We will describe an automorphism \(\phi\) of \(P\) that sends subspaces of \(\mathcal{V}\) to subspaces of \(\mathcal{V}\). The automorphism will map \(W \in P[i]\) to \(\phi(W) \in P[d-i]\), and for every \(W \subseteq \mathcal{V}\), \(W+E_{d-i} = V\) iff \(\phi(W)+\tilde{E}_i=\mathcal{V}\), for appropriately defined \(\tilde{E}_i\).
    
    Let us describe the automorphism first. For a vector \(v=(v_1,v_2,\dots, v_d) \in \F_\fsize^d\) let \(v^\perp\) be the linear functional \(v^{\perp}(u_1,u_2,\dots,u_d) = v_1 u_1 + v_2 u_2 + \dots + v_d u_d\). 
    For any subspace \(W \subseteq V\), let \(\phi(W) = \sett{v \in \mathcal{V}}{\forall w \in W, w^{\perp}(v)=0}\). Note that  \(\dim(\phi(W)) = d-dim(W)\) (fixing a basis \(w_1,...w_i\) to \(W\), the equations \(w_j^{\perp}(v)=0\) are independent linear equations. The nullspace of these equations is precisely \(\phi(W)\)).
    Now let us take \(f_1=e_d,f_2=e_{d-1},\dots, f_{d-1}=e_2, f_d=e_1\) and set \(\tilde{E}_i = span(f_1,f_2,\dots,f_i)\). One can verify that \(\phi(E_i) = \tilde{E}_{d-i}\). We claim that \(W+E_{d-i}=\mathcal{V}\) if and only if \(\phi(W)+\tilde{E}_i=\mathcal{V}\). Indeed, this is equivalent to proving that \(W+E_{d-i}=\mathcal{V}\) if and only if \(\phi(W) \cap \tilde{E}_i=\set{0}\) (since \(\dim(\phi(W)+\tilde{E}_i)=d-\dim(\phi(W) \cap \tilde{E}_i)\)). Indeed, \(W+E_{d-i} = V\) if and only if \(\phi(W+E_i) = \set{0}\). By the fact that \(W \cup E_i\) spans \(W+E_i\), \(\phi(W+E_i) = \phi(W) \cap \phi(E_i) = \phi(W) + \tilde{E}_i\) so the claim is proven.
\end{proof}

The following corollary is direct from the last claim so we omit its proof.
\begin{corollary} \label{cor:link-structure-2}
    For every \(gK_i \in \kms(0)\), the link \(\kms_{gK_i}\) is isomorphic to the opposition complex \(P\). Moreover, the part \(\kms_{gK_i}[j]\) is sent to \(P[d-j+i]\) under this isomorphism (modulo \(d-1\)).
\end{corollary}

Finally, we get the following general symmetry guarantee.
\begin{claim} \label{claim:symmetry-in-links}
    Let \(C_1, C_2 \subseteq [d]\) be two sets of colors such that there is an isomorphism of the cycle \(\phi:[d] \to [d]\) that maps \(\phi(C_1) = C_2\) (that is, \(\phi(x)=\pm x + c\) modulo \(d\), for some \(c \in [d]\)). Then for every \(x \in \kms[C_1]\) and \(y \in \kms[C_2]\) there is an isomorphism such that for every color \(C \subseteq [d]\), \(\kms_x[C]\) is mapped to \(\kms_y[\phi(C)]\).
\end{claim}

\begin{proof}
    If \(C_1\) and \(C_2\) are singletons then this just follows from \Cref{claim:link-structure} and \Cref{cor:link-structure-2}. Otherwise write \(C_1 = \set{i} \cup C_1'\) and \(C_2 = \set{i'} \cup C_2'\) where \(i' = \phi(i)\). Let \(x= \set{x_1} \cup x' \in \kms[C_1]\) such that \(x \in \kms[i]\) and \(x' \in \kms[C_1']\). Let \(y_1 \in \kms[i']\) and by the claim for singletons there is an isomorphism between \(\kms_{x_1}\) and \(\kms_{y_1}\) that sends \(C_1'\) to \(\phi(C_1')=C_2'\). In particular, the set \(x' \mapsto y'\) for some \(y' \in \kms[C_2']\). Moreover, this is an isomorphism of complexes, so in particular the link \(\kms_x = (\kms_{x_1})_{x'} \cong (\kms_{y_1})_{y'} = \kms_y\) by this same isomorphism. The claim follows. 
\end{proof}
%

%
\subsection{Routing and the KMS Complexes} \label{subsec:routing-kms}
In this subsection we establish a routing scheme for this complex that will in turn give rise to the routing schemes in \Cref{prop:graph-system-for-log-rate-graph} and \Cref{prop:graph-system-for-constant-rate}. The idea is as follows. We will first find a deterministic routing scheme for swap walk graphs between two disjoint colors \(\kms[C_1]\) and \(\kms[C_2]\). Then we will use the fact that the group \(SL_d(\F_\fsize)\) acts on the complex and that the action is edge-transitive for edges between \(\kms[C_1]\) and \(\kms[C_2]\). We will use the group to randomize the routing scheme sufficiently so that it will have low congestion.\yotamd{I noticed that \cite{bafna2024quasi} also used (in their analysis) the fact that a group acts transitively on colors in their complexes. Do we want to mention this?}

In the last step we will also present a subset internal routing scheme on \(\kms[C_1],\kms[C_2]\) that is given an oracle for membership on some large \(T \subseteq \kms[C_1] \cup \kms[C_2]\), and produces a path between most \(s_1,s_2\) such that all intermediate vertices in the path are in \(T\). This scheme will also have low congestion, and is crucial for \Cref{thm:constant-rate}.

\subsubsection{A Deterministic Routing Scheme} We start by designing a deterministic routing algorithm between any two \(\kms[C_1], \kms[C_2]\) such that the colors \(C_1,C_2\) are disjoint. 

Recall for every color \(C \subseteq [d]\), the faces of \(\kms[C]\) can be identified with cosets of \(K_C = \bigcap_{i \in C} K_i\) since every \(s \in \kms[C]\) is equal to \(s=\set{gK_i}_{i \in C}\) for some \(g \in G\), and we can identify it with \(g K_C\) which is the common intersection of all vertices in \(s\) (the formal proof follows from the definitions and is therefore omitted). As \(s_1 \in \kms[C_1], s_2 \in \kms[C_2]\) are connected if and only if \(s_1 \cup s_2 \in \kms[C_1 \cup C_2]\), we can also identify edges in the swap walk graph with \(g K_{C_1 \cup C_2}\).

\begin{claim}[Neighborhood Structure] \label{claim:neighborhood-structure}
    Fix \(C_1,C_2 \subseteq [d]\) disjoint. The \(C_2\)-neighborhood of a coset \(g K_{C_1} \in \kms[C_1]\) is the set: 
    \[
    \set{gh K_{C_2} : h \in K_{C_1}}. 
    \]
\end{claim}

\begin{remark} \label{rem:paths-and-decomposition}
    Given this claim, the problem of finding a path of length \(\leq t\) from (say) \(AK_{C_1}\) to \(BK_{C_2}\) reduces to finding a decomposition of \(A^{-1}B=h_1 h_2 \dots h_{t-1}\) such that \(h_{2k+1} \in K_{C_1}\) and \(h_{2k} \in K_{C_2}\) (of course, some can be the identity), because we can take the path of the form
\[AK_{C_1},Ah_1K_{C_2},Ah_1h_2K_{C_1},\dots,Ah_1 h_2 \dots h_{t-1} K_{C_x} = BK_{C_x}\]
and then, if \(K_{C_x}=K_{C_2}\) we are done, and otherwise we just take the final step from \(BK_{C_1}\) to \(BK_{C_2}\). Let us prove the claim for completeness.
\end{remark}

\begin{proof}[Proof of \Cref{claim:neighborhood-structure}]
    The neighbors of \(gK_{C_1}\) are all \(g'K_{C_2}\) such that \(gK_{C_1} \cap g'K_{C_2} \ne \emptyset\). Clearly if \(g'=gh\) for \(h \in K_{C_1}\) then the intersection is not empty. Conversely, if the intersection contains some element \(g''\) then by definition of \(gK_{C_1}\), \(g''=gh\) for some \(h \in K_{C_1}\), and therefore \(g'K_{C_2} = g''K_{C_2}=ghK_{C_2}\).
\end{proof}

In our graphs, we will always be able to take short steps that correspond to the elementary matrices that generate the subgroups.
\begin{corollary} \label{cor:basic-steps-in-routing}
    Let \(e_{j,j+1}(r)\) be one of the generators in the definition of the subgroups \(K_1,K_2,\dots,K_d\) (i.e.\ \(r \in \F_{\fsize}\) if \(j \ne d\) and \(r=r't\) for \(r' \in \F_{\fsize}\) if \(j=d\)). Then for any disjoint \(C_1,C_2 \subseteq [d]\), and any coset \(g K_{C_1}\), there is a path of length \(\leq 2\) from \(g K_{C_1}\) to \(ge_{j,j+1}(r)K_{C_1}\).
\end{corollary}

\begin{proof}
    Recall that the colors tell us which generators \emph{are missing} from the subgroups. Observe that \(C_1 \cap C_2 = \emptyset\) and therefore \(j\) cannot be in both color sets, which means that the matrices of the form \(e_{j,j+1}(r)\) must be in one of the subgroups \(K_{C_1},K_{C_2}\). If \(j \notin C_1\) then \(e_{j,j+1}(r) \in K_{C_1}\) and then \(gK_{C_1}=ge_{j,j+1}(r)K_{C_1}\) so techincally there is a path of length \(0\) from the coset to itself.  
    Otherwise \(j \notin K_{C_2}\) and we can walk \((g K_{C_1},gK_{C_2},ge_{j,j+1}(r)K_{C_1})\) since \(e_{j,j+1}(r) \in K_{C_2}\).
\end{proof}

\begin{lemma}[Routing Scheme]\label{lem:routing}
    Let \(C_1,C_2 \subseteq [d]\) be two disjoint sets of colors. Then the swap walk graph between \(\kms[C_1] = \set{g K_{C_1}}_{g\in G}\) and \(\kms[C_2] = \set{gK_{C_2}}_{g\in G}\) admits a deterministic routing scheme \(\kms\). The maximal length of a path is \(O(\poly(d) \poldeg^{\log_2 5})\).
\end{lemma}

\begin{proof}
    To construct a path between every pair of vertices it is enough to give a path from the coset (vertex) \(K_{C_1}\) to every vertex in \(\kms[C_1]\) or \(\kms[C_2]\). To go from a generic \(AK_{C_i}\) to \(BK_{C_j}\) we may then just concatenate their paths through $K_{C_1}$.    
    Also, note that if we have a path of length \(t\) from \(K_{C_1}\) to \(AK_{C_1}\) then we also have a path of length \(t+1\) to \(AK_{C_2}\) since \(AK_{C_1} \sim AK_{C_2}\). Therefore, it suffices to construct a path from \(K_{C_1}\) to \(AK_{C_1}\) for any \(A \in SL_{d}(R)\). From now on let us fix \(A \in SL_d(R)\).
    
    Now by \Cref{rem:paths-and-decomposition} and \Cref{cor:basic-steps-in-routing}, it suffices to decompose \(A\) to \(O(\poly(d) \poldeg^5)\) generators, i.e.\ as a product of \(O(\poly(d) \poldeg^5)\) elementary matrices $e_{j,j+1}(r)$ as in the subgroup definitions.
    
    It is known that every \(A \in SL_d(R)\) is a product of \(O(d^2)\) elementary unipotent matrices\footnote{Indeed this is just by Gaussian elimination.} \(e_{i,j}(r)\) and thus it suffices to find such a decomposition assuming \(A=e_{i,j}(r)\) for some \(i,j\) and \(r \in R\). In fact, we claim it is even enough to prove this just for $r$ of the form \(r=t \cdot r'(t)\), that is, for polynomials with no constant part. The reason is the following. Suppose we know how to write \(e_{i,j}(r)\) for every polynomial of the form \(r=tr'(t)\). Then because we are working over \(\F_\fsize[t]/(\phi(t))\) and not \(\F_\fsize[t]\), this means that \(\phi(t) \equiv 0\) in our ring. As this polynomial is irreducible, this means that \(\phi(0)\ne 0\), and in particular, that it has a non-zero constant part. Thus if \(\phi(t) = \sum_{j=0}^{\poldeg} \alpha_j t^j\), then \[1 = -t(\sum_{j=1}^{\poldeg} \frac{\alpha_j}{\alpha_0} t^{j-1}).\] 
    This means that by writing \(e_{i,j}(t(\sum_{j=1}^{\poldeg} \frac{\alpha_j}{\alpha_0} t^{j-1}))\) we are effectively writing a constant. Of course, for any other polynomial with a constant part \(q(t)=a+tq'(t)\) we can just write the polynomial 
    \[t\cdot a(\sum_{j=1}^{\poldeg} \frac{\alpha_j}{\alpha_0} t^{j-1}) + tq'(t)\]
    which is equivalent to it. Thus if we can write any polynomial of the form \(t r'(t)\), we can also write all polynomials in \(R\).\footnote{This seems unnecessarily complicated since we actually have \(e_{i,j}(\alpha)\) for \(\alpha \in \F_\fsize\) constant whenever \(i<j \ne d\). However, to get to the matrices \(e_{i,d}(\alpha)\) using these generators we actually need to go to a polynomial of large degree.}
    %
    
    We now proceed by induction on \(\deg(r')\). Let \(T(m)\) be the maximal number of elements from \(K_{C_1} \cup K_{C_2}\) required to write \(tr'\) such that \(\deg(r') \leq m\).

    Recall the following formulas that can be easily verified:
    \begin{enumerate}
        \item For any \(i,j\) and \(r_1,r_2 \in R\)
        \begin{equation} \label{eq:ko-addition}
            e_{i,j}(r_1)e_{i,j}(r_2)=e_{i,j}(r_1+r_2)
        \end{equation}
        \item For any \(3\) distinct \(i,j,\ell\) and \(r_1,r_2 \in R\)
        \begin{equation} \label{eq:ko-multiplication}
            [e_{i,\ell}(r_1),e_{\ell,j}(r_2)]=e_{i,j}(r_1 \cdot r_2)
        \end{equation}
        where the brackets are the commutator \([g,h]=ghg^{-1}h^{-1}\)
    \end{enumerate}

    Let us do the induction step proving that \(T(2^m) \leq 5T(2^{m-1})\). Then, below, we prove that \(T(1)=O(d)\) by some tedious computations. Solving the recursion we get that \(T(\poldeg) \leq O(d \poldeg^{\log_2 5})\) which proves the assertion. Let \(r'\) be a polynomial of degree \(\leq 2^m\). We can write \(r' = t^{2^{m-1}}r_1 + r_2\) where \(r_1,r_2\) have degree at most \(2^{m-1}\) (and obviously \(t^{2^{m-1}}\) also has degree \(2^{m-1}\)). Therefore \(e_{i,j}(tr') = [e_{i,\ell}(tr_1),e_{\ell,j}(t^{2^{i-1}})] e_{i,\ell}(t r_2)\). This formula involves \(5\) multiplications of elementary matrices whose non-zero entry has degree \(\leq 2^{m-1}\) and therefore \(T(2^m) \leq 5T(2^{m-1})\).

    Hence it suffices to bound \(T(1)\), or equivalently to prove that every
    \(e_{i,j}(\alpha t + \beta t^2)\) can be written in \(O(d)\) elements in \(K_{C_1} \cup K_{C_2}\). By the addition formula above it suffices to be able to write \(e_{i,j}(\alpha t)\) and \(e_{i,j}(\beta t^2)\). Moreover, if we are able to write \(e_{i,j}(\alpha t)\) for every \(i,j\) then by the multiplication formula above \(e_{i,j}(\beta t^2) = [e_{i,\ell}(t),e_{\ell,j}(\beta t)]\). Therefore it really suffices just to show we can write \(e_{i,j}(\alpha t)\) for any \(i \ne j\).

    We now split to the following cases: \(i > j\), \(i < j<d\), and \(i < j=d\). By \eqref{eq:ko-multiplication} for any \(i > j\) 
    \[
    e_{i,j}(\alpha t)=[\dots [[e_{i,i+1}(p_i)e_{i+1,i+2}(p_{i+1})],e_{i+2,i+3}(p_{i+2})],\dots], e_{{j-1},j}(p_{j-1})]
    \]
    where:
    \begin{enumerate}
        \item The indexes above are to be taken modulo \(d\).
        \item \(p_{\ell} = \begin{cases}
            \alpha t & \ell=d \\
            1 & o.w.
            \end{cases}\). Note that because \(i>j\) we indeed reached \(\ell=d\) at some point.
    \end{enumerate}
    This gives us \(e_{i,j}(\alpha t)\) in \(O(d)\) number of group multiplications when \(i > j\). We note that by a similar formula, for any \(i<j\) we can write \(e_{i,j}(1)\) (including \(j=d\)). Hence for \(i<j<d\) we can write \(e_{i,j}(\alpha t) = [e_{i,d}(1),e_{d,j}(\alpha t)]\). The last case is for \(e_{i,d}(\alpha t)\) which is equal to \([e_{i,\ell}(\alpha t),e_{\ell,d}(1)]\) where \(i \ne \ell,d\) (which again, we write using the previous cases). The lemma is proven.
\end{proof}

\subsubsection{From Deterministic to Low Congestion Routing}
We note this subsection holds for any coset complex and does not use any specific properties of \(SL_d(R)\) and its subgroups.

Let \(X\) be a coset complex with respect to a group \(\Gamma\) and subgroups \(K_1,K_2,\dots,K_d\).

\begin{claim}[Transitivity of \(\Gamma\)] \label{claim:coset-complexes-have-transitive-action}
    The group \(\Gamma\) acts transitively on the top-level faces of \(X\), and more generally, on any color-set of faces. The action is given by \(h \cdot gK_i \mapsto hgK_i\). As a consequence it also acts on the swap walk graph between any two disjoint colors \(X[C_1],X[C_2]\), and it is edge transitive on this graph. The colors of the sets are preserved by this action.
\end{claim}
The proof of this claim is definition chasing, so we omit it.

Let \(G\) be a bipartite graph and let \(\Gamma\) be a group that acts on the graph that preserves the left-hand and right-hand sides. We recommend the reader think about \(\Gamma = SL_d(R)\) and \(G\) the swap walk graph between \(\kms[C_1]\) and \(\kms[C_2]\).

For a path \(P=(u_0,u_1,\dots,u_m)\) and \(g \in \Gamma\) we denote by \(g.P = (g . u_0,g. u_1,\dots,g. u_m)\).

For a routing scheme \(\mathcal{O}\) and a group \(\Gamma\) acting on a graph, we consider the following randomized routing scheme \(\mathcal{O}^{\Gamma}\) defined by:
\begin{enumerate}
    \item Input: an ordered pair of vertices \((v,v')\).
    \item Output: a path \(P\) from \(v\) to \(v'\).
    \item Algorithm:
    \begin{enumerate}
        \item Choose a uniformly random \(g \in \Gamma\).
    \item Obtain a path \(P_0 = \mathcal{O}(g^{-1}v,g^{-1}v')\) from \(g^{-1}v\) to \(g^{-1} v'\).
    \item Output \(P = g. P_0\).
    \end{enumerate}
\end{enumerate}

\begin{claim} \label{claim:routing-together-with-group-action}
    The routing scheme \(\mathcal{O}^{\Gamma}\) has unit congestion.
\end{claim}

\begin{proof}
    Let \(t\) be the length of \(\mathcal{O}^{\Gamma}\). Fix any subset of the edges $B$. We need to bound the probability \(\Prob{v,v' \mathcal{O}^{\Gamma}}{\exists e\in \mathcal{O}^{\Gamma}(v,v') : e \in B}\). This is clearly bounded by the probability \(t \Prob{v,v' \mathcal{O}^{\Gamma},e}{e \in B}\) where \(e\) is chosen uniformly at random from the output of \(\mathcal{O}^{\Gamma}(v,v')\).
    
    The randomness in this experiment is over \(v,v' \in V\), \(P_0 \sim \mathcal{O}(v,v')\) (possibly, if \(\mathcal{O}\) is itself randomized), \(g \in \Gamma\), and \(i \leq |P_0|\) (where \(e\) is the \(i\)-th edge in \(P_0\)). We claim that for any fixing of \(u=g^{-1}v,u'=g^{-1}v'\), \(P_0\) and \(i\), the element \(g \in \Gamma\) chosen by \(\mathcal{O}^{\Gamma}\) is still uniform. Note that if this holds, then by edge-transitivity any \(e\) which is the \(i\)-th edge in \(P_0\), \(g.e\) is uniformly distributed over all edges and the claim holds since \(\Prob{v,v' \mathcal{O}^{\Gamma},e}{e \in B} = \Prob{}{B}\).



    Fix any $g_0,g_1 \in \Gamma$. We'd like to show for any fixed $u,u'$ as above 
    \[
    \Pr_{g,v,v'}[g=g_0~|~g^{-1}.v=u, g^{-1}v'=u'] = \Pr_{g,v,v'}[g=g_1~|~g^{-1}.v=u, g^{-1}v'=u']
    \]
    (note the randomness of $\mathcal{O}$ and choice of $i$ are independent, so we've removed them from the probability). Looking at the left-hand side, we have
    \begin{align*}
        \Pr_{g,v,v'}[g=g_0~|~g^{-1}.v=u, g^{-1}v'=u'] &= \Pr_{g,v,v'}[g=g_0~|~v=g.u, v'=g.u']\\
        &= \Pr_{g,v,v'}[v=g.u, v'=g.u'~|~ g=g_0]\frac{\Pr_{g,v,v'}[g=g_0]}{\Pr_{g,v,v'} [v=g.u, v'=g.u']}\\
        &= \frac{\frac{1}{|V|^2}\cdot \frac{1}{\Gamma}}{\Pr_{g,v,v'} [v=g.u, v'=g.u']}
    \end{align*}
    where we've used Bayes Rule and the final equality follows from the fact that even with $g_0$ fixed in the first probability, $v$ and $v'$ are i.i.d random draws from $V$ so have a $\frac{1}{|V|}$ probability to be $g_0.u$ and $g_0.u'$ respectively, as these are simply fixed vertices in the graph. The denominator \(\Pr_{g,v,v'} [v=g.u, v'=g.u']\) is just equal to \[\sum_{g_0 \in \Gamma}\Prob{g,v,v'}{g=g_0}\Pr_{g,v,v'}[v=g.u, v'=g.u'~|~ g=g_0] = \frac{1}{|V|^2}\] by the above. Hence this is just equal to \(\frac{1}{|\Gamma|}\) and has no dependence on $g_0$.
\end{proof}

We now have our main routing scheme for \(\kms\).
\begin{corollary} \label{cor:routing-for-KMS}
    Let 
    For any \(C_1,C_2 \subseteq [d]\) that are disjoint, the swap walk graph between \(\kms[C_1]\) and \(\kms[C_2]\) has a routing scheme with length \(O(\poly(d)\poldeg^{\log_2 5})\) and unit congestion.
\end{corollary}

\subsubsection{Subset internal routing}
We now turn to showing a special setting of the swap graph on $\kms$ has a subset internal low-congestion routing scheme, which we'll eventually instantiate to give the corresponding routing scheme needed in \Cref{prop:graph-system-for-constant-rate}.

Let \(C \subseteq [d]\). We say that \(C' \subseteq [d]\) separates \(C\) if \(C=\set{i_1 < i_2 <\dots < i_j}\) are such that \(C' \cap (i_\ell < i_{\ell+1}) \ne \emptyset\) for every \(\ell=1,2,\dots,j-1\) and in addition \(C' \cap [1,i_1) \cup (i_j,d] = \emptyset\). In other words, if we embed the indexes of \(C\) and \(C'\) on the \(d\)-cycle in the obvious way, every two indexes of \(C\) are separated by an index of \(C'\). We say \(C\) is separable if there exists \(C'\) that separates \(C\). We say \(C\) is doubly separable if there exists two disjoint \(C',C''\) that separate \(C\).

\begin{claim} \label{claim:separation-implies-complete-partite-complex}
    Let \(C_1,C_2 \subseteq [d]\) and let \(C' \subseteq [d]\) that separates \(C_1 \cup C_2\). Then for any \(s \in \kms[C']\), the swap walk graph between \(\kms_s[C_1],\kms_s[C_2]\) is a complete partite graph.
\end{claim}

\begin{proof}
    Let \(s \in \kms[C']\) and write \(s = v_1 \cup s'\) where \(v_1 \notin s'\). The link \(\kms_s = (\kms_{v_1})_{s'}\). However, \(\kms_{v_1}\) is the opposition complex, and separation implies that between any two colors \(C_1 \cup C_2\) there is some color \(i \in C'\). Let us think about vertices in \(\kms_{v_1}\) as subspaces and faces in \(\kms[C_i]\) as flags of subspaces. This means that for any two faces \(t_1 \in \kms_s[C_1], t_2 \in \kms_s[C_2]\), and any two subspaces \(u_1 \in t_1, u_2 \in t_2\), there is a subspace \(w \in s'\) such that \(u_1 \subset w \subset u_2\) (or vice versa), just from the definition of a link in the opposition complex. This in particular implies that the union of every such pair \(t_1,t_2\) is a flag and hence this is a complete bipartite graph.
\end{proof}

\begin{claim} \label{claim:routing-within-a-set}
    Let \(C_1\) and \(C_2\) be such that \(C_1 \cup C_2\) is doubly separable. Then the swap walk graph $(\kms[C_1],\kms[C_2])$ has a \emph{successful} subset internal routing scheme \(\mathcal{O}^{T}(v,v')\) with the following properties:
    \begin{enumerate}
        \item It makes at most \(O(\poly(d) \poldeg \log \log N)\) queries to the membership oracle for $T$.
        \item The maximal path length is \(t = O(\poly(d) \poldeg^{\log_2 5})\).
        \item It has low congestion.
    \end{enumerate}
\end{claim}

Let us describe the routing algorithm. Let \(C_3\) and \(C_4\) be disjoint colors separating \(C_1 \cup C_2\). For simplicity we describe it only in the case where both vertices have the same color \(v,v' \in \kms[C_1]\) (the description is similar otherwise but the indexing becomes cumbersome).
\begin{figure}[ht!]
\fbox{\parbox{\textwidth}{
\vspace{.1cm}
\underline{Routing algorithm in \(T\)}:
\begin{enumerate}
    \item[*] \textbf{Input:} A pair \(v,v' \in \kms[C_1]\). 
    
    \item[*] \textbf{Output:} a path in the swap walk graph between \(\kms[C_1]\) and \(\kms[C_2]\) from \(v\) to \(v'\).
    \item Sample \(u \in \kms_{v}[C_3]\) and \(u' \in \kms_{v'}[C_3]\).
    \item Call the routing procedure on \(\kms[C_3]\) and \(\kms[C_4]\) from \(u\) and \(u'\) promised in \Cref{cor:routing-for-KMS} to get path \(P_0=(u=u_0,u_1,u_2,\dots,u_m=u')\).
    \item Denote by \(v_0=v, v_{2m}=v'\). For \(j=1,2,\dots,2m-1\) we construct the path \((v_0,v_1,v_2,\dots,v_{2m})\) as follows.
    \begin{itemize}
        \item If \(j\) is even select \(v_j \in \kms_{u_j}[C_1] \cap T\) uniformly at random.
        \item If \(j\) is odd select \(v_j \in \kms_{\set{u_{j-1},u_j}}[C_2] \cap T\) uniformly at random.
    \end{itemize}
    In both cases selection is done by rejection sampling, where for every sampled \(v_j\) we query whether it is a member of \(T\) 
    
    If no $v \in T$ is found after \(100\log \log N\) samples, abort and output \(FAIL\).
    \item Output \((v_0,v_1,\dots,v_{2m})\).
\end{enumerate}}}
    \caption{Routing in \(\kms[C_1] \cup \kms[C_2]\)}
    \label{fig:routing-in-set-T}
\end{figure}

\begin{proof}
    Let us prove first that the routing is successful, i.e.\ that the probability of outputting \(FAIL\) is bounded by \(\frac{1}{\log^4 |V|}\).
    We observe the following.
    \begin{enumerate}
        \item \(\Prob{s \in \kms[C_1]\cup \kms[C_2]}{T|col(s)=C_1} \geq 0.8\) and similarly when conditioning on \(col(s)=C_2\).
        \item The swap walk graph between \(\kms[C_3]\) and \(\kms[C_1]\) is a a \(\frac{10d}{\sqrt{\fsize}}\)-expander. Therefore the probability that \(s \in \kms[C_3]\) has that \(\Prob{s' \in \kms_s[C_1]}{s \in T} < 0.75\) is \(O(\frac{d^2}{\fsize})\).
        \item The previous item holds also for all pairs \((\kms[C_3],\kms[C_2])\), \((\kms[C_4],\kms[C_1])\), \((\kms[C_4],\kms[C_2])\), and also for pairs where one side has both colors \(C_3 \cup C_4\): \((\kms[C_3 \cup C_4],\kms[C_1])\), \((\kms[C_3 \cup C_4],\kms[C_2])\).
    \end{enumerate}
    By the above, the probability that a face \(s\) in \(\kms[C_3],\kms[C_4]\) or \(\kms[C_3 \cup C_4]\) has \(\Prob{s' \in \kms_s[C_1]}{s \in T} < 0.75\) (or \(\Prob{s' \in \kms_s[C_2]}{s \in T} < 0.75\)) is \(O(\frac{d^2}{\fsize})\). Observe that when \(v,v'\) are independent and uniformly sampled, then so are \(u,u'\) (the end points of \(P_0\) in the algorithm). Hence, by unit congestion of the routing scheme on \(\kms[C_3],\kms[C_4]\), the probability that any edge or vertex in the path \(P_0\) has this property is at most \(\frac{\poly(d) \log^{3} N}{\fsize} < \frac{1}{2\log^4 N}\). If for every such \(s=u_j\) and \(s=\set{u_j, u_{j+1}}\) the probability of \(T\) in every side is at least \(0.75\), then the probability we don't sample a vertex from \(T\) in the link of \(s\) after \(100\log \log N\) samples is \(< \frac{1}{\log^4 N}\). The routing is hence successful.
    
    We now move to proving the three claimed items. In fact the first two properties follow immediately from the definition of the algorithm and the max length of a path in the routing scheme in \Cref{cor:routing-for-KMS}. 
    As for the final property, let \(B\) be a subset of the edges. 
    
    Before bounding the probability of \(B\), we define three `bad events' for possible edges \(e = \set{u_j,u_{j+1}}\). After we bound these bad events, we will bound the probability of sampling \(B\) when these events did not occur:
    \begin{enumerate}
        \item The density of \(T\) inside \(\kms_{u_j}[C_1]\) or \(\kms_{\set{u_j,u_{j+1}}}[C_2]\) is less than \(0.75\).
        \item The link of \(\kms_{u_j}\) has more than \(\Prob{}{B}^{1/2}\) edges in \(B\).
        \item The link of \(\kms_{u_j}\) has less than \(\Prob{}{B}^{1/2}\) edges in \(B\) but the graph between \(\kms_{u_j}[C_1]\) and \(\kms_{\set{u_j,u_{j+1}}}[C_2]\) has more than \(\Prob{}{B}^{1/4}\) edges in \(B\).
    \end{enumerate}
    When an edge is sampled uniformly at random, the first item happens with probability \(< O(\frac{d}{\fsize})\) and in particular it occurs with probability less than \(\frac{1}{2t \log^4 |V|}\) as we set \(\fsize\) to be sufficiently large. 

    The second event occurs with probability \(\Prob{}{B}^{1/2}\) by Markov's inequality. Moreover, for any \(\kms_{u_j}\) whose link has less than \(\Prob{}{B}^{1/2}\) edges in \(B\), the fraction of \(u_{j+1}\) such that the third event occurs is also at most \(\Prob{}{B}^{1/4}\) by Markov's inequality (again, all of this is when edges are sampled uniformly at random).

    Denote these bad edge \(R\). Up until now, we only bounded these events when sampling an edge uniformly at random. Note however that the internal routing we used has unit congestion, therefore the probability we hit such a bad edge is bounded by \(t\Prob{e}{R}\) (where \(e\) is a random edge in the swap walk). This is at most \(2t\Prob{}{B}^{1/4} + \frac{1}{2\log^4 |V|}\).
     
     Observe as before that the probability of hitting the edge set \(B\) by one of the edges in the output path is \(\leq t \Prob{v,v', e \in \mathcal{O}^T(v,v')}{B}\). The same holds if we also assume on the path not being in \(R\). Therefore we need to prove that \(\Prob{v,v', e \in \mathcal{O}^T(v,v')}{B \text{ and } \mathcal{O}^T(v,v') \text{ doesn't hit } R} \leq \Prob{}{B}^{1/4}\) to get the required congestion since:
     \begin{align*}
         \Prob{v,v', \mathcal{O}^T(v,v')}{\exists e \in \mathcal{O}^T(v,v')\; e \in B} &\\
         &\leq \Prob{}{\mathcal{O}^T(v,v') \text{ hits }R} + t \Prob{v,v', e \in \mathcal{O}^T(v,v')}{B \text{ and } \mathcal{O}^T(v,v') \text{ doesn't hit } R}\\
         &\leq \frac{1}{2\log^4 |V|} + 2t\Prob{}{B}^{1/4} + t \Prob{v,v', e \in \mathcal{O}^T(v,v')}{B \text{ and } \mathcal{O}^T(v,v') \text{ doesn't hit } R}.
     \end{align*}
     
     Note that we can choose a random edge in \(\mathcal{O}^T(v,v')\) by first choosing a random edge \(\set{u_j,u_{j+1}} \in P_0\) and then setting the edge \(e=\set{v_j,v_{j+1}}\). We also observe that this is a random edge between \(\kms_{u_j}[C_1]\cap T\) and \(\kms_{\set{u_j,u_{j+1}}}[C_2] \cap T\).

    \maxh{All the above are based on the fact that $(u_j,u_{j+1})$ is marginally random in the prior scheme, right? But this is not explained, so it's confusing where the above bounds are coming from.}\yotamd{I hope now it is clearer.}

    
    If \(e \notin R\), then by the fact that a choice of edge is a choice of a random pair of vertices (one of which is in the link of the vertex \(u_j\) and the other is in the link of \(\set{u_j,u_{j+1}}\)), the fraction of edges in \(T \times T\) between \(\kms_{u_j}[C_1]\) and \(\kms_{\set{u_j,u_{j+1}}}[C_2]\) is bounded by \(0.75^2 \geq 0.5\). Moreover, the probability of an edge in \(B\) is at most \(\Prob{}{B}^{1/4}\) so even conditioning on an edge going between end points in \(T\), the probability is at most \(2\Prob{}{B}^{1/4}\). Thus the probability of choosing an edge in \(\Prob{v,v', e \in \mathcal{O}^T(v,v')}{B \text{ and } \mathcal{O}^T(v,v') \text{ doesn't hit } R}\) is bounded by
    \(2t\Prob{}{B}^{1/4}.\)
\end{proof}

\begin{remark}
    It is reasonable to ask why in even steps we go straight from a face in \(\kms_{\set{u_{j-1},u_j}}\) to a face in \(\kms_{u_{j}}\) rather than a face in \(\kms_{\set{u_j,u_{j+1}}}\). If we were to do so, then the third `bad event' would have depended on the three vertices \(u_{j-1},u_j,u_{j+1}\) and not only on a single edge. While our routing scheme has unit congestion we did not prove any guarantee on events on pairs of edges.
\end{remark}

\subsection[The set systems arising from the KMS complex]{The Set Systems Arising from \(\kms\)}\label{subsec:defining-set-systems}
We first describe the set systems for \Cref{prop:graph-system-for-log-rate-graph} and \Cref{prop:graph-system-for-constant-rate}

\paragraph{The set system \((V,S,G)\) for \Cref{prop:graph-system-for-log-rate-graph}} For this construction set \(d=3\) and \(\fsize > C \frac{1}{\varepsilon} \log^{10} N \log \log N\) to be a prime power\footnote{For reasons of implementations we take \(\fsize = 2^z\) for some large enough integer \(z\).} for some absolute constant \(C > 0\). We take \(\kms\) for these parameters \(\fsize,d\) and \(\poldeg \to \infty\). The set system in \Cref{prop:graph-system-for-log-rate-graph} is:
\begin{itemize}
    \item \textbf{The vertex set \(V\).}
        \[V=\kms[1].\]
    \item \textbf{The sets \(S\).}
        \[ S = \sett{\kms_{s}[1]}{s \in \kms[2] \cup \kms[3]}.\]
        In words, we take as sets the neighborhoods in \(\kms[1]\) of every vertex \(s \in \kms[2] \cup \kms[3]\). For brevity when we write \(v \in s\) we actually mean \(v \in \kms_s[1]\).
    \item \textbf{The decoding graph \(G\).} This is the graph between \(\kms[2]\) and \(\kms[3]\).
\end{itemize}

Let us also spell out some of the graphs we need to analyze in \Cref{prop:graph-system-for-log-rate-graph}:
\begin{itemize}
    \item \textbf{The graph \((V,S)\).} This is the graph between \(\kms[1]\) and \(\kms[2] \cup \kms[3]\).
    \item \textbf{The graph \(G_s\).} For \(s \in \kms[2]\) (resp.\ \(s \in \kms[3]\)), the intersection graph is the link of \(s\), i.e.\ graph between \(\kms_s[3]\) (resp. \(\kms_s[2]\)) and \(\kms_s[1]\). This just follows from the fact that if \(s \sim s'\) in \(G\), then their intersection is \(\kms_{s}[1] \cap \kms_{s'}[1]\). One can verify from the definitions that this is equal to \(\kms_{s,s'}[1]\) which is the neighborhood of \(s'\) in the link of \(s\).
\end{itemize}

\paragraph{The set system for \Cref{prop:graph-system-for-constant-rate}}
Let \(\eta, \beta > 0\) be as in \Cref{prop:graph-system-for-constant-rate}. Set \(d= \frac{14}{\eta}\) and without loss of generality let us assume it is divisible by \(7\). Set \(\fsize > 4\eta^2\beta^2\) to be a prime power. Let \(\kms\) be the KMS complex with parameters \(d,\fsize\) and \(\poldeg \to \infty\).\footnote{Again, for circuit reasons it is simplest to assume \(\fsize\) is a power of \(2\).} Let \(C = \set{7,14,\dots,d}\) and for \(i\in \set{1,2,\dots,6}\) we denote by \(C_{+i} = \set{i,7+i,\dots,d-7+i}\).
\begin{enumerate}
    \item \textbf{The vertex set \(V\).}
            \[V=\kms[C],\]
    \item \textbf{The sets \(S\).} For a vertex \(s \in \bigcup_{i \in C} \kms[i]\) define \(N_{\kms}(s) = \sett{v \in \kms[C]}{s \in v}\). The sets
        \[S = \sett{N_{\kms}(s)}{s \in \bigcup_{i \in C} \kms[i]}.\]
    This may be slightly confusing, so we spell it out explicitly. For every \emph{vertex $s$ of $\kms$} whose color is in \(C\), we define \emph{the set} \(N(s)\) which is all the faces that contain the vertex \(s\) (which are elements in \(V\)). This is the \textit{reversal} of the typical encoding, which would put message bits on vertices of $\kms$, and take the direct product over faces. We use the reverse graph because every vertex (in \(V\)) is contained in exactly \(d = O(\frac{1}{\eta})\)-sets (in \(S\)). This implies that the rate of the direct product code is \(\frac{1}{d} = \Omega(\eta)\). On the other hand, our complex is not bounded degree (to ensure strong enough sampling), so the normal encoding would not be constant rate. Henceforth, when we write \(s \in S\) we actually mean \(N_{\kms}(s) \in S\).

    \item \textbf{The sets \(U\).} These are links \(\kms_u[C]\) for \(u \in \kms[C_{+2}] \cup \kms[C_{+5}]\). That is,
    \[
    U = \sett{\kms_u[C]}{u \in \kms[C_{+2}] \cup \kms[C_{+5}]}.
    \]
    As before, we write \(v \in u\) instead of \(v \in \kms_u[C]\) for brevity. The set system \(In_u=(V_u,S_u)\) consists of \(V_u=X_u[C]=u\). The sets are indexed by vertices \(s \in \bigcup_{u \in C}\kms_u[i]\). The set \(s\) is actually 
    \[N_{\kms_u}(s) \coloneqq \sett{v \in \kms_u[C]}{v \in s}.\]
    In words, these are all faces of color \(C\) that contain \(s\) in the link of \(u\). We note for the observant reader that another way of defining these sets is just intersecting a set \(s\) with \(u\), i.e.\ 
    \[N_{\kms}(s) \cap \kms_u[C] = N_{\kms_u}(s).\]
    \item \textbf{The decoding graph \(G\).} This is the swap walk graph between \(\kms[C_{+2}]\) and \(\kms[C_{+5}]\).
\end{enumerate}

Again, let us spell out the graphs in \Cref{prop:graph-system-for-constant-rate}. 
\begin{itemize}
    \item \textbf{The graph \((V,S)\).}
    \[V = \kms[C]\]
    \[S = \kms^{C}(0) = \bigcup_{i \in C}\kms[i]\]
    and \(v \sim s\) if and only if if the face in \(\kms[C]\) corresponding to \(v\) contains the vertex in \(\kms^{C}(0)\) that corresponds to \(s\) (i.e.\ containment is reversed).
    \item \textbf{The graph \((V,U)\).} 
    \[U = \kms[C_{+2}] \cup \kms[C_{+5}]\]
    \[V=\kms[C]\] 
    and \(v \sim u\) if and only if \(\set{v,u}\) is an edge in the swap walk graph (between the \(C\) and \(C_{+i}\) which is the color of \(u\)).
    \item \textbf{The graph \((S,U)\).}
    \[U = \kms[C_{+2}] \cup \kms[C_{+5}]\]
    \[S=\kms^{C}(0)\]
    and \(s \sim u\) if and only if \(\set{s,u}\) is an edge in the swap walk graph (between the color of \(u\) and the color of \(s\)).
\end{itemize}
We also have the local graphs.
\begin{itemize}
    \item \textbf{The graph \(G_u\).} Let \(u \in \kms[C_{+i}]\) for \(i \in \set{2,5}\) and let \(j \in \set{2,5}, j\ne i\) be the other color. Similar to the graph in \Cref{prop:graph-system-for-log-rate-graph} this graph is the following bipartite graph \((L,R,E)\).
    \[ L = u = \kms_u^{C}(0)\]
    \[ R = \kms_u[C_{+j}]\]
    and \(\set{u',v} \in E\) if and only if \(u' \sim v\) in the swap walk graph. The explanation is the same as in \Cref{prop:graph-system-for-log-rate-graph}.
    \item \textbf{The graph \((V_s,U_s)\).} This graph has \(N_{\kms}(s)\), the faces \(\kms[C]\) that contain \(s\) on one side. On the other side it has \(\kms_s[C_{+2}] \cup \kms_s[C_{+5}]\). An edge is between \(u \sim v\) if and only if \(\set{v,u}\) is a swap walk edge. Observe that we can identify \(N_{\kms}(s) \cong \kms_s[C \setminus \set{i}]\) where \(s \in \kms[i]\) (and the identification is \(v \in N_{\kms}(s) \mapsto v\setminus \set{s}\)). Thus by definition, \(v \sim u\) if and only if \(v \setminus \set{s} \sim u\) \emph{in the swap walk in the link of \(\kms_s\)}.  Hence this graph is isomorphic to the graph \((L,R,E)\) where
    \[L = \kms_s[C_{+2}] \cup \kms_s[C_{+5}]\]
    \[R=\kms_s[C \setminus \set{i}]\] 
    and \(u \sim v\) are the swap walk edges in \(\kms_s\).

    \item \textbf{The graph \((V_u,S_u)\).} This is the (reverse) containment graph in the link, i.e.\ 
    \[V_u = \kms_u[C]\]
    \[S_u = \kms_u^{C}(0)\]
    and \(v \sim s\) if and only if if the face in \(\kms_u[C]\) corresponding to \(v\) contains the vertex in \(\kms_u^{C}(0)\) that corresponds to \(s\).

    \item \text{The graph \((G_{In_u})_{s}\).} Fix \(u \in U \cong \kms[C_{+2}] \cup \kms[C_{+5}]\) and \(s \in S_u\cong \kms^{C}(0)\). The local graph of \((G_{In_u})_{s}\) is a graph \((L,R,E)\) with
        \[L = s= \sett{v \in \kms_u[C]}{v \text{ contain \(s\) as a face}}\]
        \[R = \sett{s' \in \kms^{C}(0)}{s \cap s' \ne \emptyset}\]
    and edges are by containment.
\end{itemize}

\subsection[Proofs of set system propositions]{Proofs of \Cref{prop:graph-system-for-log-rate-graph} and \Cref{prop:graph-system-for-constant-rate}}\label{subsec:proving-propositions-on-set-systems}
\restatetheorem{prop:graph-system-for-log-rate-graph}
\begin{proof}[Proof of \Cref{prop:graph-system-for-log-rate-graph}]
    Let us go through the items one by one.
    
    \begin{enumerate}
        \item Recall that \(\fsize = \poly \log (N,\frac{1}{\varepsilon})\). The number of vertices in \(V=\kms[1]\) is the number of cosets, which is \(\frac{|SL_d(\F_{\fsize^{\poldeg}})|}{|K_1|}\). The size of \(|K_1|=\fsize^3\) and the size of \(|SL_3(\F_{\fsize^{\poldeg}})| = c \fsize^{8}\) for some \(c \in (0.99,1)\). This is equal to \(c\fsize^{8\poldeg -3}\) for some \(c \in (0.99,1)\). Set \(\poldeg\) to be the integer value of \( \frac{3}{8} + \frac{\log N}{8\log \fsize}\), rounded up. Then obviously \(|V| \in [N,N\fsize^8]\) or in other words \([N,N\poly(\log N, \frac{1}{\varepsilon})]\).
        \item Recall that the graph between \((V,S)\) is just the graph between \(\kms[1]\) and \(\kms[2]\cup \kms[3]\). The graph is indeed bi-regular. To see this, note the complex has a transitive action on the top-level faces so it is clear neighborhoods of all vertices of the same color have the same sizes. In particular, every \(v \in V=\kms[1]\) appears in the same number neighborhoods of sets in \(\kms[2]\cup \kms[3]\). In the other direction, recall that by \Cref{subsec:links-in-kms} we can identify the vertices of color \(1\) in a link of a vertex in color \(2\) with lines in the opposite complex, and the same is true for color $3$ by duality so in particular the degree is the same for any \(s \in \kms[2] \cup \kms[3]\).
        \item The routing scheme in \Cref{cor:routing-for-KMS} between \(\kms[2]\) and \(\kms[3]\) is a routing scheme in length \(t = O(\poldeg^{\log_2 5}) \leq  O(\log^3 N)\) with unit congestion.
        \item Recall that for every \(s \in S\), the graph \(G_s\) is the link of \(\kms_s\). By \Cref{lem:links-of-kms-expand} this is a \(\frac{10}{\sqrt{\fsize}}\)-expander. By \Cref{claim:expander-sampler} these are also \(\frac{100}{\fsize}\)-samplers. Thus the link is a sufficient sampler by choice of \(\fsize\).

        \item The following is a standard claim for expander graphs we will also use a few times in the proof of \Cref{prop:graph-system-for-constant-rate}.
    \begin{claim}\label{claim:convex-combination-of-graphs}
        Let \(G=(L,R,E)\) be a bi-regular graph. Assume there is a partition \(L=L_1 \cup L_2 \cup \dots \cup L_{m}\) and \(R=R_1 \cup R_2 \cup \dots \cup R_{m'}\) such that the induced bipartite graphs between every \(L_i\) and \(R_j\) are bi-regular and isomorphic to one another. Then if the graph between \(L_1\) and \(R_1\) is a \(\beta'\)-expander, it implies that \(G\) is a \(\beta'\)-expander.
    \end{claim}
    We can use this claim by identifying \(L=L_1 = \kms[1]\) and \(R=R_2 \cup R_3\) where \(R_2=\kms[2]\) and \(R_3=\kms[3]\). The graph between \(\kms[1]\) and \(\kms[i]\) is a \(\frac{10}{\sqrt{\fsize}}\)-expander for \(i \in \set{2,3}\) separately. Following the definitions, one also observes that these are also biregular with the same regularity. Thus the graph \((V,S)\) is a \(\frac{10}{\sqrt{\fsize}}\)-expander. Sampling follows by choice of \(\fsize\) and \Cref{claim:expander-sampler}.
    \end{enumerate}
\end{proof}

\restateproposition{prop:graph-system-for-constant-rate}

\begin{proof}[Proof of \Cref{prop:graph-system-for-constant-rate}]
Again, we go over the grocery list of items one-by-one.
\begin{enumerate}
    \item The number of vertices is \(|\kms[C]|\) where \(C = \set{7,14,21,\dots,d}\). This is the number of cosets, or \(\frac{|SL_d(\F_{\fsize^\poldeg})|}{|\kms[C]|}\). In other words \(\fsize\) to the power \(\poldeg(d^2-1) - 3d\). By taking \(\poldeg = \frac{\frac{\log N}{\log \fsize}+3d}{d^2-1}\) rounded up to the nearest integer value we have \(|V| \in [N,N\fsize^{O(d^2)}]\). As \(\fsize^{O(d^2)} = \log (N)^{\exp(O(\frac{1}{\varepsilon}))}\) the first item is clear.
    \item Let us bound the size of \(s \in S\) (and analogously $u \in U$). Recall that sets \(s\) are indexed by vertices \(s \in \kms(0)\) whose colors are in \(C\). The set itself is actually \(N_{\kms}(s)\), faces of color \(C\) that contains \(s\). Note that for \(s \in \kms[i]\), the set \(N_{\kms}(s) = \sett{v \cup \set{s}}{v \in \kms_s[C \setminus \set{i}]}\). Thus to bound the size of the set \(s\), it is sufficient to bound the size of \(\kms_s[C']\) for any color \(C'\). Recall that this is the number of cosets of \(K_{C'}\) inside the subgroup \(K_i\). Obviously this is bounded by \(|K_i|\) itself which is \(\fsize^{O(d^2)}\). The correct bound then follows from our choice of \(d=\exp(O(\frac{1}{\varepsilon}))\) and \(\fsize = \poly(\log(N), \exp(O(\frac{1}{\varepsilon}))\). 
    
    A similar bound holds for \(|\kms_u[C]| \leq |K_{C_{+i}}|\) (for \(i \in \set{1,6}\)). This is at most the size of \(K_i\) since adding vertices to a set only makes the link smaller, so the bound from above still holds.
    \item Every vertex is actually a face \(v \in \kms[C]\). It is contained in \(\frac{d}{7} = \exp(O(\frac{1}{\varepsilon}))\) sets since it contains exactly \(\frac{d}{7}\) vertices of $\kms$.

    \item Let us examine each of the five graphs separately, making many uses of \Cref{claim:symmetry-in-links}:
    \begin{enumerate}
        \item The graph \((V,S)\) is the containment graph between \(\kms[C]\) and \(\kms^{C}(0) = \bigcup_{i \in C}\kms[i]\). We already observed that any face in \(\kms[C]\) contains the same number of vertices. In the other direction, we again use the fact that for \(s \in \kms[i]\), \(|N_{\kms}(s)|=|\kms_s[C \setminus \set{i}]|\). Thus we prove that \(|\kms_s[C \setminus \set{i}]|\) is independent of \(i\). For this we use \Cref{claim:symmetry-in-links}. For any \(i_1,i_2 \in C\) the isomorphism \(j \overset{\psi}{\mapsto} j + (i_2-i_1) \text{ (mod \(d\))}\) is an isomorphism of the cycle. Thus by \Cref{claim:symmetry-in-links}, for any \(s \in \kms[i_1], s' \in \kms[i_2]\) the links \(\kms_{s} \cong \kms_{s'}\) by an isomorphism that maps \(\kms_s[C] \mapsto \kms_{s'}[\psi(C)]\) As \(C=\psi(C)\) the regularity follows.
        \item \yotamd{oh nice can I see another one of those?} The graph \((V,U)\) is bi-regular by similar considerations. The vertices all have the same color and \(v \in u\) if and only if \(u \in \kms_v[C_{+i}]\). As all \(v \in V\) have the same color set, their links are all isomorphic (recall there is a  group action on \(\kms\) that is transitive on every \(\kms[C']\)). As for \(u \in U\), recall that these are indexed by \(\kms[C_{+2}] \cup \kms[C_{+5}]\) and the vertex set of \(u\) is actually \(\kms_u[C]\). So we need to show that the size of \(|\kms_u[C]|\) is independent of whether \(u \in \kms[C_{+2}]\) or \(u \in \kms[C_{+5}]\). Similarly to the item above, this follows from the fact that the isomorphism \(j \overset{\psi}{\mapsto} -j \text{ (mod \(d\))}\) of the cycle that maps \(\psi(C_{+2}) = C_{+5}\) and \(\psi(C)=C\). Thus by \Cref{claim:symmetry-in-links}, the link of any \(u \in \kms[C_{+2}]\) is isomorphic to a link of any \(u' \in \kms[C_{+5}]\) by an isomorphism sending \(\kms_{u}[C] \mapsto \kms_{u'}[C]\) and regularity follows.
        \item \yotamd{and another one I guess.} The set-layers graph is actually the graph between \(U \cong \kms[C_{+2}] \cup \kms[C_{+5}]\) and \(S \cong \kms^{C}(0)\) where there is an edge \(\set{u,s}\) if and only if \(u \cup s \in \kms\). The isomorphism in \((b)\) will also show that any two \(u, u' \in U\) have the same degree (regardless of their color). The isomorphism in \((a)\) proves the same for \(s,s' \in S\).
        \item \yotamd{OK Yotam, I think I understand...} Fix \(s \in S\). Identifying \(s \in \kms[i]\), the local graph \((V_s,U_s)\) has sets in \(\kms[C]\) that contain \(s\), and \(u \in \kms[C_{+2}] \cup \kms[C_{+5}]\) that are connected to \(s\) via the swap walk. The edges are \(u \sim v\) if and only if \(v \in u\) if and only if \(u \sim v\) in the swap walk. This is isomorphic to the graph between \(\kms_s[C \setminus \set{i}]\) and \(\kms_s[C_{+2}] \cup \kms_s[C_{+5}]\) with swap-walk edges. Regularity on the \(V_s\) side follows because there is just one color. To show regularity for \(U_s\) we note that for every \(i \in C\) there is an isomorphism of the cycle mapping \(C_{+2}\) to \(C_{+5}\) and \(C\) and \(i\) to themselves, this is \(j \mapsto -j + 2i \text{ (mod \(d\))}\). This proves there is an isomorphism of this graph mapping \(\kms_s[C_{+2}]\) to \(\kms_{s}[C_{+5}]\).
        \item \yotamd{oh god please no not another one} For \((V_u,S_u)\), we note that this is the containment graph between sets \(V_u \cong \kms_u[C]\) and \(S_u \cong \kms_u^{C}(0)\). Regularity on the \(V_u\) side is because every face in \(\kms_u[C]\) contains the same number of vertices. Regularity on the \(S_u\) side follows from the same isomorphism of the cycle in item \((a)\). We omit the details.
        \item Let us describe the case for \(G_u\) for \(u \in \kms[C_{+2}]\) (the other case is the same except for switching \(2\leftrightarrow 5\)). The graph has sets indexed by \(u' \in \kms_u[C_{+5}]\) and vertices that are \(\kms_u^{C}(0)\). Every \(u'\) intersects \(u\) on vertices \(\kms_{u \cup u'}^{C}(0)\). So this is just the graph with swap walk edges between the two sides. This graph is bi regular by a similar argument to the above, which we skip.
    \end{enumerate}
    \item We saw above in our description of the set system that the graph \(G_u\) has \(\kms_u[C_{+j}]\) on one side (for \(j \in \set{2,5}\)) and \(\kms^{C}_u(0)\) on the other side, and the edges are swap walk graph edges in the link of \(u\). For every \(i \in C\), the graph between \(\kms_u[C_{+j}]\) and \(\kms_u[i]\) is a \(\frac{10d}{\sqrt{\fsize}}\)-expander by \Cref{thm:swap-walk}. The different parts also have the same regularity by another use of \Cref{claim:symmetry-in-links}. Thus by \Cref{claim:convex-combination-of-graphs} the graph is a \(\frac{10d}{\sqrt{\fsize}}\)-expander. It is a \(\beta\)-sampler by \Cref{claim:expander-sampler} and choice of \(\fsize\).
    \item Recall that the graph \((S,U)\) is the swap walk graph where one side is \(\kms^{C}(0)\) and the other side is \(\kms[C_{+2}] \cup \kms[C_{+5}]\). This item is similar to the above. Note that for any choice of parts \(i \in C\) and \(C' \in \set{C_{+2},C_{+5}}\), the swap walk graphs between \(\kms[i]\) and \(\kms[C']\) are isomorphic by \Cref{claim:symmetry-in-links}. They are also \(\beta'=\frac{O(d)}{\sqrt{\fsize}}\)-expanders. \Cref{claim:convex-combination-of-graphs} together with \Cref{claim:expander-sampler} implies that the graph \((S,U)\) is a \(\frac{O(d^2)}{\fsize}=\beta\)-sampler.
    \item As we saw above the graph \((V_u,S_u)\) is the containment graph between \(\kms_u[C]\) and \(\kms_u^{C}(0)\) in the link of \(u \in U \cong \kms[C_{+1}] \cup \kms[C_{+6}]\). This is a \(O(\frac{1}{|C|})\)-expander by \Cref{thm:up-down-walk-expands}. As \(\frac{1}{|C|} = \Omega(\eta)\) the item follows.
    \item Recall the local decoding graph $G_{In_u}$ is the graph where we connect \(s \sim s'\) if and only if \(s \cap s' \ne \emptyset\). Recall that \(S_u \cong \kms_u^{C}(0)\). Observe that the set \(C_{+i}\) \emph{separates} \(C\). Therefore, by \Cref{claim:separation-implies-complete-partite-complex} the link \(X_u^C\) is a complete \(|C|\)-partite graph. In particular, \(s \cap s' \ne \set{0}\) (i.e.\ there is a \(C\)-face containing \(\set{s,s'}\)) if and only if \(s\) and \(s'\) have a different color. Hence, \(s\) is connected to all but a \(\frac{1}{|C|}\)-fraction of the vertices and therefore is \((1-O(\eta))|S_u|\)-regular.
    \item Let \(i\) be the color of \(s\). Similar to the above, by the description of this graph, it is just the containment graph between \(\kms_{u \cup \set{s}}[C \setminus \set{i}]\) and \(\kms^{C \setminus \set{i}}(0)\). This is a \(O(\frac{1}{|C|-1}) \leq \eta\)-sampler by \Cref{thm:up-down-walk-expands} and \Cref{claim:expander-sampler}.
    \item As we saw above, the graph \((V_s,U_s)\) is isomorphic to the graph between \(\kms_s[C]\) and \(\kms_s[C_{+2}] \cup \kms_s[C_{+5}]\) where there is an edge if and only if \(v \sim u\) in the swap walk. The swap walk graph is a \(\frac{10d}{\sqrt{\fsize}}\)-expander and therefore by \Cref{claim:convex-combination-of-graphs} this is also a \(\frac{10d}{\sqrt{\fsize}}\)-expander. Sufficient sampling follows from \Cref{claim:expander-sampler}.
    \item Recall that the decoding graph \(G\) is the swap walk graph between \(\kms[C_{+2}]\) and \(\kms[C_{+5}]\). The set of colors \(C_{+2}\cup C_{+5}\) is doubly separated by the sets \(D_1 = C_{+1} \cup C_{+4}\) and \(D_2 = C_{+3} \cup C_{+6}\). Hence there is a subset internal routing scheme with the desired properties by \Cref{claim:routing-within-a-set} where we note that the length is bounded by \(O(\kappa^{\log_2 5} \poly(\frac{1}{\eta})) \leq  O(\log^3 (N) \exp(O(\frac{1}{\varepsilon})))\).
\end{enumerate}

\end{proof}
\subsection{HDX Circuit Implementation}\label{sec:kms-circuit-implementation}
\maxh{Remember to include some comment about encoding time being improved to $\log(\fsize)$ dependence by having repeated neighbor indexes.}
We now present a number of circuits that are necessary to construct encoders and decoders for the HDX based codes. At a high level, we need to encode certain cosets of the special linear group $SL_d(R)$. The natural approach is to simply encode using the set of all matrices, and map a given input matrix to its corresponding coset. Unfortunately, as in the subspace setting in \Cref{sec:circuit-subpoly}, this encoding has too much redundancy, and we need a way to encode our cosets such that the bit representation is truly log the total number of cosets (up to a vanishing error). Below, we will see how to do this essentially via (unfortunately notationally involved) variants of Gaussian elimination. Using the coset structure, we will define a canonical matrix form for each coset given by sequentially zero-ing out all `free' coefficients via allowed column operations in the coset, resulting in an exact representation up to vanishing error coming from non-invertible matrices.

For a matrix \(A\) we write \(col(A,j)\) to denote its \(j\)-th column. 

\subsubsection{Another point of view on cosets}
The encoder and decoder circuits of our code need to take as input a vertex in $\kms[C]$, which is a coset of some \(K_i\), or a face which is some coset of \(K_S\) for a set \(S \subseteq [d]\). Internally, our circuits will just use a matrix and an index \(i\) to represent a coset. Unfortunately, like in the subspace setting in \Cref{sec:circuit-subpoly}, this cannot be the way our circuits take vertices or faces of the complex as input. The reason is that representations of matrices contain too many repetitions and invalid matrices (repetitions from many matrices representing the same coset, invalid matrices are those that are not in \(SL_d(R)\)). This fact adversely affects the rate of the actual code obtained if one tries to use the naive representation. Therefore we must construct a format for representing a coset of \(K_i\) or \(K_S\) in roughly \(\log ([SL_d(R):K_i])\) bits (resp. \(\log ([SL_d(R):K_S])\) bits).

This is what we do next.

The subgroups \(K_i\) can be described as follows. The subgroup \(K_d\) is the set of all upper triangular matrices with entries in \(\F_\fsize\). The subgroup \(K_i\) for \(i<d\) is the set of matrices with:
\begin{enumerate}
    \item Ones on the diagonal.
    \item Elements in \(\F_\fsize\) in indexes \(I_{i}^u = \sett{(j,k)}{i+1 \leq j<k\leq d} \cup \sett{(j,k)}{1\leq j < k \leq i} \).
    \item Elements in \(t \cdot \F_\fsize\) in indexes \(I_i^{\ell} = \sett{(j,k)}{1\leq k<i+1 \leq j\leq d}\).
    \item The rest of the entries are zero.
\end{enumerate}
\begin{figure}[ht]
    \centering
    \[K_2 : \begin{pmatrix}
1 & a_4 & 0 &0 &0\\
0 & 1 & 0 &0 &0\\
a_5 t & a_6 t & 1 &a_1 &a_2 \\
a_7 t & a_8 t & 0 &1 &a_3 \\
a_9 t & a_{10} t & 0 &0 &1\\
\end{pmatrix}\]
    \caption{\(K_2\) in \(SL_5(R)\). \(I_2^u = \set{(3,4),(3,5),(4,5)} \cup \set{1,2}\). \(I_2^{\ell} = \set{3,4,5} \times \set{1,2} \).}
    \label{fig:k2-example}
\end{figure}
See \Cref{fig:k2-example} for an example. In fact, for \(K_d\) we get the same description plugging in $i=0$ in the definition of \(I_i^u, I_i^{\ell}\). Staring at \Cref{fig:k24-intersect-example}, one notices that it easy to find the intersection \(K_{J} = \bigcap_{j \in J}K_j\) for non-empty \(J \subseteq [d]\). This is just the subgroup that has:
\begin{figure}[ht]
    \centering
    \[K_4 : \begin{pmatrix}
1 & a_1 & a_2 &a_3 &0\\
0 & 1 & a_4 &a_5 &0\\
0 & 0 & 1 &a_6 &0 \\
0 & 0 & 0 &1 &0 \\
a_7 t & a_8 t & a_9 t &a_{10}t &1\\
\end{pmatrix},\; K_{\set{2,4}} : 
\begin{pmatrix}
1 & a_1 & 0 &0 &0\\
0 & 1 & 0&0 &0\\
0 & 0 & 1 &a_2 &0 \\
0 & 0 & 0 &1 &0 \\
a_3 t & a_4 t & 0 &0 &1\\
\end{pmatrix}\]
    \caption{\(K_4\) and \(K_{\set{2,3}}\) in \(SL_5(R)\).}
    \label{fig:k24-intersect-example}
\end{figure}

\begin{enumerate}
    \item Ones on the diagonal.
    \item Elements in \(\F_\fsize\) in indexes \(I_J^u = \bigcap_{j \in J}I_j^u\).
    \item Elements in \(t \cdot \F_{\fsize}\) in indexes \(I_J^{\ell} = \bigcap_{j \in J}I_j^{\ell}\).
    \item The rest of the entries are zero.
\end{enumerate}
Finally, we notice that these groups are also generated by 
\[Gen(J) = \sett{e_{j,k}(\alpha)}{(j,k) \in I_J^{u}, \alpha \in \F_\fsize} \cup \sett{e_{j,k}(\alpha t)}{(j,k) \in I_J^{\ell}, \alpha \in \F_\fsize}.\]
With this in mind, multiplying a matrix \(A\) with a matrix \(B \in Gen(J)\) on the right corresponds to performing one of the elementary column operations
\begin{equation} \label{eq:col-op-up}
    col(A,k) \leftarrow col(A,k) + \alpha \cdot col(A,j)
\end{equation}
if \((j,k) \in I_J^u\), or

\begin{equation} \label{eq:col-op-low}
    col(A,k) \leftarrow col(A,k) + \alpha t \cdot col(A,j)
\end{equation}
if \((j,k) \in I_J^{\ell}\).

Therefore \(A' \in AK_J\) if and only if one can reach \(A'\) from \(A\) via the above elementary column operations.

\subsubsection{Canonical representations for the constant rate construction}
Let \((V,S,U)\) be a set system in the family constructed in \Cref{prop:graph-system-for-constant-rate}. We now construct canonical representatives for \(v \in V\). Recall that in this construction we take some \(d=7d'\), and \(V=\kms[C]\) for \(C=\set{7,14,\dots,d}\) in a KMS complex of dimension \(d\). A vertex \(v \in V\) is a coset of \(K_C\) so let us describe \(K_C\) first using the above observations. Matrices in \(K_C\) are upper triangular matrices (i.e. \(I_C^{\ell} = \emptyset\)) and 
\[
I_C^u = \cup_{k=0}^{d'-1} \sett{(i,j)}{7k+1 \leq i < j \leq 7k+7}.
\]

Thus, by the characterization given by the operations in \eqref{eq:col-op-up}, two matrices \(A,A' \in SL_d(R)\) belong to the same coset of \(K_C\) if and only if for every \(j=7k+r\) (where \(r \in [6]\)), 
\[
col(A',j) = col(A,j) + \sum_{1\leq r' < r} \alpha_{r'} \cdot col(A,7k+r'),
\]
where \(\alpha_{r'} \in \F_{\fsize}\).
In particular, every column indexed by \(j \equiv_7 1\) is identical between \(A\) and \(A'\). Every column that is indexed by \(j \equiv_7 2\) has that the difference \(col(A,j)-col(A',j)\) is in \(\F_{\fsize} \cdot col(A,j-1)\), and so on.

With this in mind, we will build a canonical representation of \(A K_C\) as a particular matrix \(A' \in AK_C\) constructed by zero-ing as many coefficients as possible via a sort of Gaussian elimination based on the above structure. This allows us to remove extra degrees of freedom such that the bit-length of the encoding is indeed roughly $\log([SL_d(R): K_C])$.

Toward this end, for a matrix \(A\) let \(\ell_j(A)\) be the first non-zero row in the \(j\)-th column. Let \(z_{j}(A)\) be the degree of the polynomial \(A_{\ell_j(A),j}\). At a high level, the canonical representation is a form where we zero-out the coefficient of \(t^{z_j(A)}\) at \(A_{\ell_j(A),j'}\) for any \(7k+1\leq j<j'<7(k+1)\), by subtracting multiples of \(col(A,j)\) from \(col(A,j')\). The definition is as follows.\begin{enumerate}
    \item Columns indexed by \(j \equiv_7 1\) are \(col(A',j)=col(A,j)\).
    \item For columns indexed by \(j \equiv_7 2\) we require the following: \(col(A',j)=col(A,j) + \alpha \cdot col(A,j-1)\) with the constraint that the coefficient in the polynomial in \(A'_{\ell_{j-1}(A'),j}\) of the monomial \(t^{z_{j-1}(A')}\) is \(0\). Note that we can always get such a matrix from \(A\) by adding the appropriate multiple of \(col(A,j-1)\).
    \item For columns indexed by \(j \equiv_7 3\), we require the following: \(col(A',j)=col(A,j) + \alpha \cdot col(A,j-2) + \beta \cdot col(A,j-1)\) with the constraints that (1) the coefficient in the polynomial \(A'_{\ell_{j-2}(A'),j}\) of the monomial \(t^{z_{j-2}(A')}\) is \(0\), and that (2) the coefficient in the polynomial \(A'_{\ell_{j-1}(A'),j}\) of the monomial \(t^{z_{j-1}(A')}\) is \(0\). Note that given a matrix \(A'' \in A K_C\) that already satisfies the first two items in this list we can get to a matrix satisfying this third item as well. We just add a multiple of \(col(A'',j-2)\) to zero out the coefficient of \(t^{z_{j-2}(A'')}\) in \(A'_{\ell_{j-2}(A'),j}\), and then we add a multiple of \(col(A'',j-1)\) to zero out the coefficient of \(t^{z_{j-1}(A'')}\) in \(A'_{\ell_{j-1}(A'),j}\). Note that as \(A''\) satisfies the second item in this list, adding multiples of \(col(A'',j-1)\) does not affect the coefficient of \(t^{z_{j-2}(A'')}\) in \(A'_{\ell_{j-2}(A'),j}\).
    \item In general, the definition of a canonical representative is the following.
    \begin{definition}
        We say that \(A' \in A K_C\) is \emph{canonical} if for every \(r \in \set{2,\dots,7}\), column \(j \equiv_r 7\) and every \(r' \in [1,r-1]\) the coefficient of \(t^{z_{j-r'}(A')}\) in \(A'_{\ell_{j-r'}(A'),j}\) is \(0\).

        We say a column in \(A'' \in AK_C\) is canonical if it satisfies the above (even if other columns do not).
    \end{definition}
\end{enumerate}

\begin{claim} \label{claim:canonical-representative-for-KC}
    For every coset \(A K_C\) there exists a unique canonical \(A' \in AK_C\).
\end{claim}

The proof of this claim is analogous to Gaussian elimination and RREF. We go over the columns one-by-one and add multiples of previous columns to zero-out specific entries. As in Gaussian elimination, we order the column operations so that later operations do not interfere with the zeroing out done by previous operations.

\begin{proof}[Proof of \Cref{claim:canonical-representative-for-KC}]
    Let us prove separately existence and uniqueness, starting with existence. Set \(A=A^1\). We construct the matrix \(A'=A^7\) by constructing a sequence of matrices \(A^2,A^3,\dots,A^7\) such that the matrix \(A^r\) will satisfy the canonicity requirement for every \(r' \leq r\). Every \(A^r\) will be obtained by performing column operations on columns that are \(j \equiv_r 7\) (the rest of the columns will be untouched in the \(r\)-th step).

    Given \(A^{r-1}\) and a column \(j=7k+r\), for every \(r'=1,2,\dots,r-1\) we:
    \begin{enumerate}
        \item Set \(b\) to be the coefficient of \(t^{\ell_{7k+r'}(A^{r-1})}\) in \(A^{r-1}_{\ell_{7k+r'}(A^{r-1}),7k+r'}\).
        \item Set \(a\) to be the coefficient of \(t^{\ell_{7k+r'}(A^{r-1})}\) in \(A^{r-1}_{\ell_{7k+r'}(A^{r-1}),j}\).
        \item If \(a \ne 0\) then do \(col(A^r,j) \leftarrow col(A^{r-1},j) - \frac{a}{b} col(A^{r-1},7k+r')\).
    \end{enumerate}

    Note that if \(r' < r''\), then the coefficient of \(t^{\ell_{7k+r'}(A^{r-1})}\) in \(A^{r-1}_{\ell_{7k+r'}(A^{r-1}),7k+r''}\) is already set to \(0\), therefore once the coefficient of \(t^{\ell_{7k+r'}(A^{r-1})}\) is set to zero in the \(j\)-th column when considering the \(7k+r'\)-column, it does not change by subsequent column operations. Thus \(A^r\) satisfies the canonicity requirements necessary.
    
    As for uniqueness, assume towards contradiction that there exists two distinct \(A',A'' \in AK_C\) that have canonical form. Let \(j=7k+r\) be the first column on which they differ. By the above, this means that \(col(A',j) - col(A'',j)\) is in the span of \(col(A',7k+r')\) for \(r' \in \set{1,2,\dots, r-1}\). Let \(7k+r''\) be the minimal column with a non-zero coefficient in the linear combination of \(col(A',j) - col(A'',j)\) (recall that the columns are independent, so there is indeed a single such combination). Then observe that the coefficient of \(t^{\ell_{7k+r''}(A')}\) in \(A'_{\ell_{7k+r''},7k+r''}\) is not zero, but it is zero for all \(A'_{\ell_{7k+r''},7k+r'}\) where \(r''<r'<r\). Thus, \(col(A',j) - col(A'',j)\) has a non-zero coefficient of \(t^{\ell_{7k+r''}(A')}\), which contradicts canonicity of one of the \(A'\) or \(A''\).
\end{proof}

Now that we have established what a canonical representation is, we can construct a circuit that gets a short bit description of a canonical representation of a coset, and constructs the canonical representation of the coset. The circuit does the following:
\begin{enumerate}
    \item Every column \(j \equiv_1 7\) is just encoded as a column of \(d\) elements in \(\F_{\fsize}[t]/\phi(t) \cong \F_\fsize^{\poldeg}\). The circuit just reconstructs these columns by reading their full description.
    \item Columns of the form \(j=7k+r < d\) are encoded using \(\poldeg d  - (r-1)\) elements in \(\F_{\fsize}\), where there are \(r\) entries that correspond to the coefficients of \(t^{\ell_{7k+r'}(A)}\) (for \(r'<r\)) that are taken to be zero, so they are not encoded. The circuit reconstructs these columns one by one, and uses its knowledge of the leading coefficients in \(7k+r'<7k+r\) to understand which coefficients to leave blank.
    \item Finally the last column is encoded using \(\poldeg d  - \poldeg -7\) elements in \(\F_{\fsize}\). Instead of giving exact descriptions of the coefficients, this last column encodes a solution to the linear equations that force the leading coefficients from the previous \(6\) rows to be zero \emph{and} that the determinant of the whole matrix to be \(1\). Observe that indeed the equation that the determinant is \(1\) is a linear equation in the entries of the last column (for example, by computing the determinant using the last column and the minors from the rest of the columns).
\end{enumerate}
We note that if the first \(d-1\) columns given to the circuit are linearly independent, then the linear equations in the last circuit indeed have a solution space of size \(\F_{\fsize}^{\poldeg d -\poldeg - 7}\). This is true because we can take any column independent of the first \(d-1\) columns to get a matrix with whatever determinant we want, and use the above algorithm to `canonize' it without making changes to the determinant (this shows that indeed there are \(\fsize^{\poldeg}\) cosets for constraints regarding the determinant).

Let us compute the length of the representation vs.\ log the number of cosets. The size \(|K_C| = \fsize^{21d'} = \fsize^{3d}\). The size of \(SL_d(R) = \fsize^{\poldeg (d^2-1)}(1 - O(\frac{1}{\fsize^{\poldeg}}))\). Therefore \(\log [SL_d(R):K_C] = (\poldeg d^2 -3d - \poldeg )\log \fsize - O(\fsize^{-\poldeg})\). On the other hand our description has length \(\frac{d}{7} \sum_{r=0}^6 (\poldeg d-r) - \poldeg = \poldeg d^2 -3d - \poldeg\), matching up to the vanishing $o(1)$ term that stems from our encoding of a tiny set of non-invertible matrices (which we may safely ignore in our arguments in the prior sections).

\subsubsection[Canonical representations of cosets of Ki]{Canonical representations of cosets \(K_i\)}
Our constructions will also require us to construct canonical representations of cosets of vertices, \(AK_i\). For cosets of \(K_d\), the canonical matrices are similar to those of \(K_C\) above, only that we do not partition the indexes to intervals of length \(7\), but rather we add to the \(j\)-th column multiples of \emph{all} columns \(j'<j\) to zero out leading coefficients.

For cosets of \(K_i\) for \(i<d\), however, a problem arises. In these cosets, the \((i+1)\)-st column is the one that remains unchanged (not the first as in \(K_d\)). The \((i+2)\)-nd column can have multiples of \(col(A,i+1)\) added to it, and so on. However, once we circle back to columns \(1\) through \(i\), the rule changes. The reason is that now we also have column operations as in \eqref{eq:col-op-low}, where we add \(t\)-multiples of certain columns. To deal with this, we use a different basis to write out the polynomials in columns \(1,2,\dots,i-1\), that plays nicely with adding \(t\)-multiples. Details follow.

We denote by \(E=(1,t,t^2,\dots,t^{\poldeg-1})\) which is a basis to the polynomial ring we are working in. We denote by \(\tilde{E}=(t,t^2,\dots,t^{\poldeg})\) which is also a basis. We interpret \(t^{\poldeg}\) to be the polynomial \(t^{\poldeg}-\phi(t)\).\footnote{Assuming the polynomial we are quotienting with, \(\phi(t)\), is monic, this is indeed a polynomial of degree \(\leq \poldeg-1\). In addition, as \(\phi(t)\) is irreducible, this is indeed a basis: \(\phi(t)\) has no root at \(0\), thus \(t^{\poldeg} - \phi(t)\) is the only polynomial with a constant part.} We note that the map \(f(t) \mapsto tf(t)\) maps a polynomial \(\sum_{j=0}^{\poldeg -1} \beta_j t^j\) (i.e.\ \(\beta_j\) is the coefficient of the \((j+1)\)-st vector in the basis \(E\)) to a polynomial \(\sum_{j=1}^{\poldeg} \beta_j t^j\). That is, now \(\beta_j\) is the coefficient of the \(j\)-th element of the basis \(\tilde{E}\).

The trick that allows us to draw an analogy between the canonical form of \(AK_C\) to that of \(AK_i\), is to think about the columns of \(i+1 \leq j \leq d\) as polynomials written in basis \(E\), and of the columns of \(1 \leq j \leq i\) as written in basis \(\tilde{E}\). Then when we zero out coefficients in a column where \(j\leq i\), we zero out coefficients with respect to \(\tilde{E}\), which plays nicely with the fact that we multiply by \(t\) columns \(j \geq i+1\).

With this in mind, consider the following definition of a canonical representation. For a matrix \(A\) and \(j \in [d]\) let \(\ell_j(A)\) be the first non-zero row on the \(j\)-th column. Let \(z_j(A)\) be the degree of the polynomial in \(A_{\ell_j,j}\), and let \(\tilde{z}_j(A)\) be the degree of the polynomial at \(\ell_j(A)\) with respect to the basis \(t,t^2,\dots,t^{\poldeg}\). Unfortunately while the idea here is simple, it is notationally cumbersome and we will need to separately give the definition for $1 \leq i \leq d-2$ and `edge-cases' $i=d-1$ and $i=d$ for clarity. We start with the former:
\begin{definition}[canonical form]
    Let \(i \leq d-2\) and let \(AK_i\) be a coset. We say that a matrix \(A' \in AK_i\) is \emph{canonical} if
    \begin{enumerate}
        \item For every \(j \in [i+2,d]\), and \(i+1\leq j' < j\), the polynomial in \(A'_{\ell_{j'}(A'),j}\) has no coefficient for \(t^{z_{j'}(A')}\).
        \item For every \(j \in [1,i]\), and \(1\leq j' < j\), the polynomial in \(A'_{\ell_{j'}(A'),j}\) has no coefficient for \(t^{\tilde{z}_{j'}(A')}\) when written in the basis \(\tilde{E}\).
        \item For every \(j \in [1,i]\), and \(i+1 \leq j' \leq d\), the polynomial in \(A'_{\ell_{j'}(A'),j}\) has no coefficient for \(t^{z_{j'}(A')+1}\) when written in the basis \(\tilde{E}\).
    \end{enumerate}
\end{definition}
Unfortunately, this definition can be interpreted in different ways when \(i\geq d-1\), since \(i+2\) that appears in the definition is greater than \(d\). The correct definition in these two cases is:
\begin{definition}[canonical form - edge cases modulo \(d\)]
    For \(i=d-1\), only items \(2\) and \(3\) cases should appear, that is, we only have constraints on columns \(j \in [1,d-1]\). So it should read:
    \begin{enumerate}
        \item No item \(1\).
        \item For every \(j \in [1,d-1]\), and \(1\leq j' < j\), the polynomial in \(A'_{\ell_{j'}(A'),j}\) has no coefficient for \(t^{\tilde{z}_{j'}(A')}\) when written in the basis \(\tilde{E}\).
    \item For every \(j \in [1,d-1]\), and \(i+1 \leq j' \leq d\), the polynomial in \(A'_{\ell_{j'}(A'),j}\) has no coefficient for \(t^{z_{j'}(A')+1}\) when written in the basis \(\tilde{E}\).
    \end{enumerate}
    For \(i=d\) only the first item should appear, instead of \(j \in [i+2,d]\) we should have \(j\in [2,d]\). It should read:
    \begin{enumerate}
        \item For every \(j \in [2,d]\), and \(1\leq j' < j\), the polynomial in \(A'_{\ell_{j'}(A'),j}\) has no coefficient for \(t^{z_{j'}(A')}\).
        \item No item \(2\).
        \item No item \(3\).
    \end{enumerate}
\end{definition}

\begin{claim} \label{claim:canonical-representative-exists}
      There exists a unique canonical matrix \(A' \in AK_i\).
\end{claim}

\begin{proof}[Proof of \Cref{claim:canonical-representative-exists}]
    We prove existence. The proof of uniqueness is the same as in \Cref{claim:canonical-representative-for-KC} so we leave it to the reader. The proof of existence is just applying an algorithm analogous to Gaussian elimination. We observe that the operations we can do on columns \(i+2 \leq j\leq d\), are adding \(\F_{\fsize}\) multiples of columns \(i+1 \leq j' < j\). For columns \(1 \leq j < i\) we can either add multiples of columns \(1 \leq j' <j\) or add \(t\F_{\fsize}\) multiples of columns \(i \leq j' \leq d\).

    Thus our algorithm will go from column to column, starting from \(j=i+1\), up to \(j=d\), then back to \(j=1\) and up to \(j=i\). For ease of notation we denote this as an order \(i+1 \prec i+2 \prec \dots \prec d \prec 1 \prec 2 \prec \dots \prec i\) which we will use.
    
    In each step, the algorithm has matrix \(A^j\), and it performs column operations on \(col(A^j,j)\) using only the columns of index \(j'\prec j\), i.e.\ that were already modified in previous rounds (of course, in addition to the \(i\)-th column that does not get modified in the algorithm). That is, \(A^{j+1}\) is equal to \(A^j\) in all columns except the \(j\)-th column, and 
    \[
    col(A^{j+1},j) = col(A^j,j) +\sum_{j' \prec j} \alpha_{j'} \cdot col(A^j,j')
    \]
    where \(\alpha_{j'} \in \F_\fsize\) when \(j \geq i+2\) or when \(j,j' \leq i\) and otherwise \(\alpha_{j'}=\alpha t\) for some \(\alpha_{j'} \in \F_\fsize\).

    We prove by induction on the ordering \(\prec\) that we can find a linear combination that will zero out the coefficients specified in the claim for the \(j\)-th column. Let us begin with the base case:
    \begin{enumerate}
        \item If \(i<d-1\): in the first round at index \(A^{i+2}_{i+1,\ell_{i+1}(A^{i+2})}\) there is a polynomial of degree \(z_{i+1}(A^{i+2})\), therefore if \(col(A^{i+2},i+2)\) has some non-zero coefficient \(t^{z_{i+1}(A^{i+2})}\) we can add a field multiple of \(col(A^{i+2},i+1)\) to zero-out that coefficient.
        \item If  \(i=d\): then the same argument as the previous case holds here, only replacing \(i+1,i+2\) with \(1,2\) resp.
        \item If \(i=d-1\): then we do something similar. At index \(A^{1}_{d,\ell_d(A^{1})}\) there is a polynomial of degree \(z_d(A^{1})\). Thus if we multiply \(col(A^{1},d)\) by \(t\) then the resulting polynomial is of degree \(z_d(A^{1}) + 1\) when writing it in basis \(\tilde{E}\). 
        Thus if \(col(A^{1},1)\) has some non-zero coefficient \(t^{z_d(A^{d}+1)}\) (according to the basis \(\tilde{E}\)), we can add a field multiple of \(col(A^{1},d)\) to zero-out the coefficient of \(t^{z_d(A^{d})}\) in \(col(A^{d-1},1)\).
    \end{enumerate}
    Now for the inductive step, we also split to the case where \(i+2 \leq j \leq d\)  and \(1 \leq j \leq i\). Starting with the former. By the inductive assumption, in \(A^j\), for every \(k \prec k' \prec j\), the \(\ell_k(A^j)\) row in \(col(A^j,k')\) is a polynomial where the coefficient of \(t^{z_k(A^j)}\) is zero. Therefore, we can go over all \(k=i,i+1,\dots,j-1\), and add multiples of \(col(A^j,k)\) to \(col(A^j,j)\) in order to zero out \(t^{z_k(A^j)}\) in \(A^j_{\ell_k(A^j),j}\). Once we zero out \(t^{z_k(A^j)}\) in \(A^j_{\ell_k(A^j),j}\), adding any \(col(A^j,k')\) for \(k'\succ k\) will maintain that the coefficient of \(t^{z_k(A^j)}\) is zero. We note also that because the columns are independent, then after these operations \(col(A^{j+1},j) \ne 0\) so \(\ell_j(A^{j+1}),z_j(A^{j+1})\) are still well defined.

    Now let us also do the case for \(1\leq j \leq i\). Similar to the above, we claim that if we consider the vectors \(t \cdot col(A^j,i), t \cdot col(A^j,i+1),\dots, t \cdot col(A^j,d)\) and \(col(A^j,1),col(A^j,2),\dots,col(A^j,j-1)\) then for any \(k \prec k' \prec j\) if one writes out \(col(A^j,k')\) with respect to \(\tilde{E}\) then:
    \begin{enumerate}
        \item if \(i \geq k,k'\) then the \(\ell_k(A^j)\) row in \(t\cdot col(A^j,k')\) is a polynomial where the coefficient of \(t^{z_k(A^j)+1}\) is zero.
        \item if \(i \geq k\) and \(k' < i\) then the \(\ell_k(A^j)\) row in \(col(A^j,k')\) is a polynomial where the coefficient of \(t^{z_k(A^j)+1}\) is zero.
        \item if \(k,k' < i\) then the \(\ell_k(A^j)\) row in \(col(A^j,k')\) is a polynomial where the coefficient of \(t^{\tilde{z}_k(A^j)}\) is zero.
    \end{enumerate}
    Therefore, we can go over all \(k=i,i+1,\dots,j-1\), and add multiples of \(t \cdot col(A^j,k)\) (if \(k\geq i\)) and \(col(A^j,k)\) (if \(k<i\)) to \(col(A^j,j)\) in order to zero out \(t^{z_k(A^j)+1}\) (if \(k\geq i\)) or \(t^{\tilde{z}_k(A^j)}\) (if \(k< i\)) in \(A^j_{\ell_k(A^j),j}\). Once we have zeroed out the relevant coefficient for \(k\) in \(A^j_{\ell_k(A^j),j}\), adding any \(col(A^j,k')\) (or \(t \cdot col(A^j,k')\)) for \(k'\succ k\) again maintains that the coefficient of \(k\) is zero.
\end{proof}
We can now encode a coset \(A K_i\) in the same manner we encoded cosets of \(AK_C\).
\begin{enumerate}
    \item The \((i+1)\)-th column is encoded explicitly.
    \item For \(j=i+2,i+3,\dots,d\) we encode the \(j\)-th column using \(\poldeg d - (j-i-1)\) elements of \(\F_{\fsize}\). We decode the \(j\)-the column given the columns \(i+1,i+2,\dots,j-1\), where we zero out the coefficients of \(t^{z_k(A)}\) in \(A_{\ell_k(A),j}\) for all \(i+1 \leq k < j\). We note that these coefficients are only known after we construct the previous columns.
    \item For \(j=1,2,\dots,i-1\) we encode using \(\poldeg d - (d-i+j-1)\) elements in \(\F_{\fsize}\). Our elements now specify coefficients with respect to the basis \(\tilde{E}\). We zero out the coefficients of \(t^{z_k(A)+1}\) in \(A_{\ell_k(A),j}\) for all \(i+1 \leq k \leq d\), and of \(t^{\tilde{z}_k(A)}\) for all \(1\leq k < j\).
    \item For the final column, we use \(\poldeg d - d\) elements in \(\F_{\fsize}\). Our elements now specify a solution to the linear equation where all appropriate coefficients are zeroed out \emph{and} that the determinant is \(1\).
\end{enumerate}
We note that for \(i=d-1,d\) the same encoding operations hold: in \(i=d-1\) we skip step \(2\) and in \(i=d\) we skip step \(3\). Finally, a similar calculation to the $K_C$ case shows that the length the bit representation generated above is within $O(\fsize^{-\poldeg})$ of \(\log ([SL_d(R):K_i])\). Below we will be slightly informal and simply say the circuits take in \(\log (|SL_d(R)|/|K_i|)\) bits, ignoring this slight discrepancy which we note is inverse polynomially small in $|SL_d(R)|/|K_i|$ for constant $d$, and therefore acceptable in all our applications.

\begin{corollary}[Circuit complexity of the canonical representative]\label{cor:finding-canonical-rep}~
\begin{enumerate}
    \item A circuit \(C_{input\to mat}\) that takes as input \(\log (|SL_d(R)|/|K_C|)\) (resp. \(\log (|SL_d(R)|/|K_i|))\) bits, and outputs the canonical representative \(AK_C\) (resp.\ \(A K_i\)) associated with these bits.
    \item A circuit \(C_{mat\to can}\) that takes as input a matrix \(A\) and outputs the canonical matrix \(A' \in AK_C\) (resp.\ \(AK_i\)).
    \item A circuit \(C_{mat\to input}\) that takes a canonical matrix \(A' \in AK_C\) (resp.\ \(AK_i\)) and outputs it in input form.
\end{enumerate}
All three circuits are implementable in logspace uniform size \(O(d^3 \poldeg^3 \log^2 \fsize)\) and depth \(O(d^2 \log^2(d\poldeg) \log \log \fsize)\).
\end{corollary}

\begin{proof}
    Throughout we do addition, multiplication, finding inverses and comparisons of elements in \(\F_\fsize\). These are always implementable in logspace uniform depth \(O(\log \log \fsize)\) and size \(O(\log^2 \fsize)\). Therefore we count size and depth assuming we can do this by circuits with gates that do field operations, and in the end just multiply depth by \(O(\log \log \fsize)\) and size \(O(\log^2 \fsize)\).
    \begin{enumerate}
        \item The algorithm above computes the first column in constant depth and \(O(d \poldeg)\) size since it just copies the input. Then there is a loop over \(d-2\) columns, where in each iteration we find the first non-zero element in \(d \poldeg\) elements (possibly after some change of basis from \(E\) to \(\tilde{E}\), in \(\poldeg^2\) operations and \(\log^2 \poldeg\) depth) and flag the next columns to not use this register when copying output. For the last column we need to solve a set of linear equations in \(d \poldeg\) variables (and \(d \poldeg\) equations). This is done in size \((d \poldeg)^3\) and depth \(\log^2(d \poldeg)\). In total we have size \(O(d^3 \poldeg^3 \log^2 \fsize)\) and depth \(O(d^2 \log^2(d\poldeg) \log \log \fsize)\).
        \item This is done via the described sequential algorithm transforming any $A' \in AK_i$ (or $AK_{C}$) into canonical form. We note the basis change from \(E\) to \(\tilde{E}\) can be performed in depth \(O(\log \poldeg + \log \log \fsize)\).\footnote{This is a convenient basis change since the only different monomial is the change from \(1\) to \(t^d\). Thus in one direction if the constant coefficient of the polynomial is \(a\), we add to the polynomial \(\frac{-a}{\phi(0)} \phi(t)\). In the other direction, we replace \(at^d\) by \(a(t^d-\phi(t))\) in the polynomial and collect terms.} All this can then also be done in size \(O(d^3 \poldeg^2 \log^2 \fsize)\) and \(O(d^2 \log ^2(d\poldeg)\log \log \fsize)\) depth.
        \item This is just the same as the second circuit, plus a layer that outputs all coefficients of the matrix (except the zeroed out ones). Depth and size stay the same asymptotically.
    \end{enumerate}
\end{proof}
\begin{remark}
    We note that in both our code constructions, the sets $S$ are indexed by cosets of \textbf{multiple} distinct $K_i$, namely all cosets of $K_i$ for all $i \in C$ in the constant rate case, and by all cosets of $K_2$ and $K_3$ in the polylog rate case. To encode this family of sets, we may take as input an index $i$ along with the bit-representation of a coset of $K_i$, then use $i$ to index into the appropriate $K_i$-specific circuit $C_{input \to mat}$ defined above. Picking $d$ appropriately, we can ensure the number of different coset classes we need to encode is always a power of $2$, so this does not add any slack to the bit-representation of the sets. This increases the circuit size by at most a factor of $d$ and has no asymptotic effect on depth.
\end{remark}

\subsubsection{Circuit complexity of KMS complex elements}
We turn to the implementations of the circuits in \Cref{sec:direct-products} and \Cref{sec:constant-formal}, towards proofs of \Cref{claim:graph-system-for-log-rate-graph-circuit} and \Cref{claim:circuits-for-const-rate-construction}.

For a coset \(AK_J\) we define \(lex(AK_J)\) to be the first matrix in \(AK_J\) in lexicographical order.
\begin{claim}\label{claim:finding-lex-A}
    There exists a circuit that takes as input \(A' \in AK_J\) (as a matrix), and outputs \(lex(AK_J)\) in size \(O(|K_J|d^3\poldeg \log \fsize) = \fsize^{O(d^2)}d^3 \poldeg \log \fsize\) and depth \(O(  (\log \log \fsize + \log d + \log \poldeg)\log|K_J|)=O((\log \log \fsize + \log d + \log \poldeg)d^2\log \fsize)\).
\end{claim}

\begin{proof}
    The circuit multiplies \(A'\) with all matrices in \(K_J\) in parallel (finding the whole coset \(A'K_J\)). Then compares them via a tournament tree. The first step is done in parallel, so the depth is the depth of matrix multiplication over a field \(\F_{\fsize^\poldeg}\), i.e.\ size \(|K_J|d^2 \poldeg \log \fsize\) and depth \(\log^2 (d) (\log \poldeg + \log \log \fsize)\). The tournament tree has size \(|K_J| \cdot (d^2 \poldeg \log \fsize)\) (i.e.\ number of matrices times length of matrix as a bit string). Its depth is \(\log |K_J| \cdot \log (d^2 \poldeg \log \fsize)\). The claim follows. 
\end{proof}

\maxh{If it's already proved, what's the point in not doing this? If it makes things more annoying notationally somehow that's fine, but maybe say this? True that if we don't have to use this for the polylog construction it doesn't matter much, but we do need the fast encoding time for polylog one scaling with $\log(\fsize)$.}\yotamd{I commented this out since it is obvious that what I meant isn't clear. So far the barrier to work with those, is that in order to make them useful in finding neighbors and such, you need to (1) define canonical forms of \(K_{i,j}\) and (2) observe that canonical forms of \(K_{i,j}\) inside \(K_i\) are extremely simple to construct. We did not do this since it required to either be more general/do another case. If one were to do this, then \Cref{cor:finding-canonical-rep} would have guaranteed us better circuit complexity.}

\begin{claim} \label{claim:neighborhood-circuits}
    Fix \(\fsize,\poldeg,d\) and let \(\kms\) be the complex determined by these parameters. Then
    \begin{enumerate}
        \item For every pair of colors \(C_1 \supseteq C_2\) there is a circuit \(C_{C_1\searrow C_2}\) whose input is a representative \(A'\) of a face \(s=A K_{C_1}\) (as a matrix). It outputs a representative of the face \( t \subseteq s\) of color \(C_2\). This circuit is implementable in constant size and depth.
        \item For every pair of colors \(C_1 \subseteq C_2\) there is a circuit \(C_{C_1\nearrow C_2}\) whose input is a representative \(A'\) of a face \(s=A K_{C_1}\) (as a matrix) and an index \(i\). It outputs a representative of the \(i\)-th face \(t \supseteq s\) of color \(C_2\). This circuit is implementable in size \(\fsize^{O(d^2)}d^3 \poldeg \log \fsize\) and depth \(O(  (\log \log \fsize + \log d + \log \poldeg)d^2 \log \fsize)\).
        \item For every pair of disjoint colors \(C_1, C_2\) there is a circuit \(C_{C_1 \overset{\cdot}{\to} C_2}\) whose input is a representative \(A'\) of a face \(s=A K_{C_1}\) (as a matrix) and an index \(i\). It outputs a representative of the i-th neighbor \(t \sim s\) in the \(\kms[C_1]\) vs.\ \(\kms[C_2]\) swap walk. This circuit is implementable in size \(\fsize^{O(d^2)}d^3 \poldeg \log \fsize\) and depth \(O(  (\log \log \fsize + \log d + \log \poldeg)d^2 \log \fsize)\).
    \end{enumerate}
    The same is true for links of faces in \(\kms\).
\end{claim}

\begin{proof}
    The first circuit is trivial: it just outputs the input. Recall that if \(C_1 \supseteq C_2\) then \(K_{C_1} \subseteq K_{C_2}\). Moreover, \(t \subseteq s\) if and only if the coset of \(t\) \emph{contains} the coset of \(s\). Thus the only possible coset of \(K_{C_2}\) that contains \(AK_{C_1}=A'K_{C_1}\) is \(A'K_{C_2}\).

    The second circuit requires more care. Let \(B_1,B_2,\dots,B_m\) be representatives of distinct cosets of \(K_{C_2}\) inside \(K_{C_1}\). We do the following steps:
    \begin{enumerate}
        \item We find \(A'' = lex(AK_{C_1})\).
        \item We multiply \(A''B_i\) and output the result.  
    \end{enumerate}
    The reason this works is as follows. First, as we find \(lex(AK_{C_1})\) the result is independent of the representative \(A'\). Second, we note that \(t \supseteq s\) if and only if the coset \(B'K_{C_2}\) is contained in \(AK_{C_1}\). Thus we need to output one such coset. We note that \(A''B_iK_{C_2}\) is contained in \(AK_{C_1}\) if and only if \(B_iK_{C_2} \subseteq A''^{-1} A_{C_1}\). As \(A\) and \(A''\) are in the same coset, \(A''^{-1}A \in K_{C_1}\) and so we give the correct answer. We note that the fact that all \(B_i\)'s represent different cosets also implies the \(A''B_i\) represent different cosets. Finally, the size of the circuit is the size of the circuit that finds \(lex(AK_{C_1})\), plus the number of representatives times the size of a matrix multiplication circuit. The first step costs the most in size, \(\fsize^{O(d^2)}d^3 \poldeg \log \fsize\) by \Cref{claim:finding-lex-A}. The depth is similarly dominated by the first step, i.e.\ \(O(  (\log \log \fsize + \log d + \log \poldeg)d^2 \log \fsize)\).
    
    The third circuit is again trivial, given the second circuit. Starting from \(s \in \kms[C_1]\) we use the second circuit to get the face \(s \cup t \in \kms[C_1 \cup C_2]\), and then output \(t \in \kms[C_2]\) (using the first circuit). Size and depth are the same as the second circuit.

    Finally, we comment that being in \(\kms_x\) instead of \(\kms\) does not make a real difference, since if \(x \in \kms[C_3]\) we can always work with \(s \cup x \in \kms[C_1 \cup C_3]\) instead of \(s \in \kms_x[C_1]\), and take the other color to be \(C_2 \cup C_3\) in the first two circuits, and \(C_2\) in the third one.
\end{proof}
We are now ready to implement the main circuits in \Cref{sec:direct-products} and \Cref{sec:constant-formal}.

\subsubsection[Circuit for polylog rate construction]{Circuits of \Cref{prop:graph-system-for-log-rate-graph}}
The explanation below proves \Cref{claim:graph-system-for-log-rate-graph-circuit}.
\paragraph{The circuits \(C_{V\to S}\) and \(C_{S\to V}\)} These are just circuits where we input a representative from \(V = \kms[1]\) (resp. \(S = \kms[2] \cup \kms[3]\)) and output a neighbor of color \(2\) or \(3\) (resp. \(1\)). Other than specifying where the neighbor should be of color \(2\) or \(3\) (in \(C_{V \to S}\)), this is just the third circuit in \Cref{claim:neighborhood-circuits}. In particular they are implementable in size \(\fsize^{8}\poldeg \log \fsize\) and depth \(O(\log \fsize(\log \log \fsize + \log \poldeg))\). Instantiating parameters this is size \(\poly(\frac{1}{\varepsilon},\log N)\) size and \(O(\log \log N \log \frac{\log N}{\varepsilon})\) depth.

\paragraph{The circuits \(C^{-1}_{V \to S}, C^{-1}_{S \to V}\)} In this circuit, we get representatives of \(A' \in A K_i\) and a neighboring coset \(B K_j\) and we need to find \(i\) such that \(C_{V \to S}(A',i) \in B K_j\). We start by finding \(lex(AK_i)\). Then we know that there is a distinct \(i\) such that \(lex(A K_i) \cdot B_i \in BK_j\) which holds if and only if \(lex(A K_i) B_i B^{-1}\) in \(K_j\). We find this by testing all representatives \(B_i\). This is costly but also dominated by the cost of finding \(lex(A K_i)\) and so it is implementable in size \(\fsize^{8}d^3\poldeg \log \fsize\) and depth \(O(d^2 \log \fsize(\log \log \fsize + \log d + \log \poldeg))\). Instantiating parameters this is size \(\poly(\frac{1}{\varepsilon},\log N)\) size and \(O(\log \log N \log \frac{\log N}{\varepsilon})\) depth.

\paragraph{The circuit \(C_{S \cap S \to V}\).} This circuit gets \(i\) and two neighboring \(s, s'\) represented as \(A_2 \in AK_2\) and \(A_3 \in A'K_3\). If we know some \(A'' \in AK_2 \cap A'K_3\), then we can just use the third circuit to compute a neighbor of \(A''\). To find this representative we can just exhaustively search in \(AK_2\), which as in lexicographical search is implementable in size \(\fsize^{8}\poldeg \log \fsize\) and depth \(O(\log \fsize(\log \log \fsize +  \log \poldeg))\). Instantiating parameters this is size \(\poly(\frac{1}{\varepsilon},\log N)\) size and \(O(\log \log N \log \frac{\log N}{\varepsilon})\) depth.

\paragraph{The routing circuit \(C_{\mathcal{O}}\).} The circuit gets \(AK_i, A' K_j\) and needs to calculate the path between \(A\) and \(A'\). We instead calculate a path between the identity and \(B=A^{-1} A'\) and then just multiply by \(A\). We begin by performing Gaussian elimination and find how to write \(B\) as \(O(1)\)-unipotent elementary matrices \[B=e_{i_1,j_1}(r_1)e_{i_2,j_2}(r_2)\dots e_{i_k,j_k}(r_k)\]
and then decompose each of these separately via the routing scheme in \Cref{subsec:routing-kms}. Note that the algorithm in \Cref{subsec:routing-kms} is recursive. The size necessary to implement it is the size of the recursion tree, and the depth is the depth of the tree. Thus the size necessary is the size of \(O(\poldeg^{\log_2 5})\) matrix multiplications, plus \(O(\poldeg^{\log_2 5})\) computations in the recursion. Taking into account the field operations, this is size \(O(\poldeg^3 \log^2 \fsize)\) and depth \(O(\log \poldeg  \log \log \fsize)\). Instantiating parameters this is size \(\poly(\log \frac{1}{\varepsilon},\log N)\) size and \(O(\log \log N \log \frac{\log N}{\varepsilon})\) depth.

\paragraph{Accessing decoder} To do decoding, we need to query a noisy codeword or list. For this we need to convert the relevant matrix representation we want to query to input bits, which can be done via \Cref{cor:finding-canonical-rep}. This takes \(O(\poldeg^2 \log ^2 \fsize)\) size and \(O(\log ^2 \poldeg + \log \log \fsize)\). Instantiating parameters this is \(O((\log \log N)^2)\).

\paragraph{Encoding} We note that a circuit for encoding just accepts a pair \(s,i\) (as input bits, not a matrix) and oracle access to a function \(f\), and must output \(f(v)\) where \(v\) is the \(i\)-th neighbor of \(s\). If we were to do this as explained above, we would calculate the canonical representation of \(s\) via \Cref{cor:finding-canonical-rep}, \(A\), call \(C_{S \to V}(A,i)=B\), and output \(f(B)\). This is done in time \(\poly(\frac{1}{\varepsilon},\log N)\) (since this is a bound on the circuit size).

However, this is not totally sufficient for our purposes, as we have claimed our codes have encoding time $\polylog(N)\log \frac{1}{\varepsilon}$ (and indeed encoding locality of this type is necessary for our strongly locally computable codes used in low-complexity hardness amplification). To move to this regime, we develop an alternative encoding (and decoder access) circuits in \Cref{sec:alternative-encode-decode}. These give the necessary circuit bounds.

\subsubsection[Circuit for constant rate construction]{Circuits of \Cref{prop:graph-system-for-constant-rate}}
We move to \Cref{claim:circuits-for-const-rate-construction}.
\paragraph{The Circuits \(C_{V\to U},C_{U\to V},C^{-1}_{V\to U}, C^{-1}_{U\to V}, C_{U \cap U \to V}, C_{U\to S}\)}
Are implemented the same as those in the subsection above for \Cref{claim:graph-system-for-log-rate-graph-circuit}. The size and depth of these operations are dominated by finding the lexicographic minimum. This is done in size \(\fsize^{O(d^2)}d^4 \poldeg \log \fsize\) and depth \(O(d^2 \log \fsize(\log \log \fsize + \log d + \log \poldeg)\) by \Cref{claim:finding-lex-A}. Instantiating parameters this is size \((\log N)^{\exp(O(\frac{1}{\varepsilon}))}\) and \(\tilde{O}(\log \log N \exp(O(\frac{1}{\varepsilon})))\) depth.

\paragraph{The circuits \(C_{U; S \to V},C_{U; V \to S},C^{-1}_{U; S \to V},C^{-1}_{U; V \to S},C_{U; S\cap S \to V}\).} These circuits are the `internal circuits', where we first specify \(u\), and then we need to do the set system operations in \(In_u\). Fortunately, for any \(u \in U\), this just means that all the operations are done in \(\kms_u\) and therefore, these inherit the same circuit complexity as the corresponding global operations. For example, for the global circuit \(C_{S \to V}\) (which only needs to be implemented for the purposes of encoding, and is not needed by our decoding procedure), we would take some \(s \in \kms^{C}(0)\) and output some \(v \in s\) (or in other words, a matrix representing a \(C\)-face that contains the \(s\) as a vertex in \(\kms^{C}(0)\)). For the local circuit \(C_{U; S\to V}\) we do the same only for the link \(\kms_u\). This can be done via the same methods using \Cref{claim:neighborhood-circuits}, only in the link of \(u\). Thus these circuits can be implemented also in size \((\log N)^{\exp(O(\frac{1}{\varepsilon}))}\) and \(\tilde{O}(\log \log N \exp(O(\frac{1}{\varepsilon})))\) depth.

\paragraph{The circuit \(C_{\mathcal{O}}^T\).} The complexity is essentially the complexity of \(C_{\mathcal{O}}\) in the previous section, plus an additive depth and size of \(\poly (\log N)\exp(\frac{1}{\varepsilon})\) of the trial and errors in sampling \(T\). Instantiating parameters this remains in size \((\log N)^{\exp(O(\frac{1}{\varepsilon}))}\) and \(\tilde{O}(\log \log N \exp(O(\frac{1}{\varepsilon})))\) depth. Moreover, by following the algorithm, one observes it makes \(\exp(O(\frac{1}{\varepsilon}))\poly (\log N)\) queries to the \(T\)-oracle, and that one can do all of them in parallel after having sampled the $u$-path $P_0$.

\paragraph{Encoding} Encoding is done exactly the same way as in the subsection above. Finding the canonical form is done in time \(\poly (\log N) \exp(\frac{1}{\varepsilon})\) and likewise, applying the circuit \(C_{S\to V}\) is done in \(\poly (\log N) \exp(\frac{1}{\varepsilon})\). The depth of the circuits is \(\exp(O(\frac{1}{\varepsilon})) \log^2( \log N)\) by \Cref{cor:finding-canonical-rep} and \Cref{claim:neighborhood-circuits}.
\yotamd{Added depth}

\paragraph{Accessing decoder (Decoding)} As above, we do this via finding the canonical form and then converting it back to input bits. This is also done in size \((\log N)^{\exp(O(\frac{1}{\varepsilon}))}\) and \(\tilde{O}(\log \log N \exp(O(\frac{1}{\varepsilon})))\) depth.

\subsubsection[Improved encoding and decoding circuits for inverse polylog-rate codes]{Improved encoding and decoding for \Cref{prop:graph-system-for-log-rate-graph}} \label{sec:alternative-encode-decode}
To end this section we construct an improved encoding and decoding (accessing decoder) circuits for \Cref{prop:graph-system-for-log-rate-graph}. Toward this end, let's write down explicitly the subgroups used in this case where \(d=3\). Instead of set notation, we just write the format of matrices in the respective subgroups with \(\alpha,\beta,\gamma \in \F_{\fsize}\). 

\[K_1 : \begin{pmatrix}
1 & 0 & 0\\
\beta t & 1 & \alpha\\
\gamma t & 0 & 1
\end{pmatrix}, \; K_2 : \begin{pmatrix}
1 & \alpha & 0\\
0 & 1 & 0\\
\beta t & \gamma t & 1
\end{pmatrix}, \; K_3: \begin{pmatrix}
1 & \alpha & \beta\\
0 & 1 & \gamma\\
0 & 0 & 1\\
\end{pmatrix}\]
and the intersections
\[K_{\set{1,2}}: \begin{pmatrix}
1 & 0 & 0\\
0 & 1 & 0\\
\alpha t & 0 & 1
\end{pmatrix}, \; K_{\set{1,3}} : \begin{pmatrix}
1 & 0 & 0\\
0 & 1 & \alpha\\
0 & 0  & 1
\end{pmatrix}, \; K_{\set{2,3}}: \begin{pmatrix}
1 & \alpha & 0\\
0 & 1 & 0\\
0 & 0 & 1\\
\end{pmatrix}.\]

Using these formats and the fact that \(x \in YK\) if and only if \(X Y^{-1} \in K\), it is easy to verify that the following matrices are distinct representatives of \(K_{\set{1,3}}\) inside \(K_3\). 
\[A_{13\subseteq 3}(\beta,\gamma) \coloneqq
\begin{pmatrix}
1 & \beta & \gamma\\
0 & 1 & 0\\
0 & 0  & 1
\end{pmatrix}.
\]

As there are \(\fsize^2 = \frac{|K_3|}{|K_{\set{1,3}}|}\) of these, we conclude that for every every coset \(A K_{\set{1,3}}\) inside \(K_3\) has a representative of this form.\footnote{In fact, if we were to define canonical forms for these subgroups, this would have been the canonical form.}

Similarly for every distinct \(k,j \in \set{1,2,3}\) we define \(A_{kj \subseteq j}\) such that every coset of \(K_k \cap K_j \) that is contained in the subgroup \(K_j\) has a unique matrix of such form. These are,
\[A_{23 \subseteq 3} \coloneqq \begin{pmatrix}
1 & 0 & \beta\\
0 & 1 & \gamma\\
0 & 0 & 1
\end{pmatrix}, \; A_{23 \subseteq 2} \coloneqq \begin{pmatrix}
1 & 0 & 0\\
0 & 1 & 0\\
\beta t & \gamma t & 1
\end{pmatrix} , \; A_{13 \subseteq 1} \coloneqq \begin{pmatrix}
1 & 0 & 0\\
\beta t & 1 & 0\\
\gamma t & 0 & 1
\end{pmatrix}\]
and
\[\; A_{12 \subseteq 2} \coloneqq \begin{pmatrix}
1 & \beta & 0\\
0 & 1 & 0\\
0 & \gamma t & 1
\end{pmatrix}, \; A_{12 \subseteq 1} \coloneqq \begin{pmatrix}
1 & 0 & 0\\
\beta t & 1 & \gamma\\
0 & 0 & 1
\end{pmatrix}\]

Therefore, for \(d=3\) and colors \(C_1=\set{j}, C_{2}=\set{k,j}\) we can implement \(C_{C_1 \nearrow C_2}(A,i)\) where \(i=(\beta,\gamma)\) where the circuit:
\begin{enumerate}
    \item Finds \emph{the canonical representation} \(A'=can(AK_j)\).
    \item Outputs \(A'\cdot A_{kj\subseteq j}(\beta,\gamma)\).
\end{enumerate}
Let \(s = AK_j\). As before, because these are distinct unique representations of the cosets of \(K_{\set{k,j}} \leq K_j\), this indeed gives us all neighboring cosets \(t\supseteq s\) of \(K_{\set{k,j}}\) (since it outputs all cosets \(A'B_i K_{\set{k,j}}\) where \(B_i \in K_j\)). The size of this circuit is \(\tilde{O}(\log^2 N \log^2 \frac{1}{\varepsilon})\) and the depth is \(O(\log^2 N + \log \frac{1}{\varepsilon})\) by \Cref{cor:finding-canonical-rep}.\footnote{Assuming \(\poldeg \leq \log N\) and \(\fsize = \frac{\poly \log N}{\poly(\varepsilon)}\).} The same holds for the circuit \(C_{C_1 \overset{\cdot}{\to} C_2}\) for \(C_1 = \set{k}, C_2=\set{j}\), and we may therefore implement \(C_{V\to S},C_{S\to V}\) in size \(\tilde{O}(\log^2 N \log^2 \frac{1}{\varepsilon})\) and \(O(\log^2 N + \log \frac{1}{\varepsilon})\), which in turn implies that \textbf{Encoding} can be done in this time and circuit complexity.

Let us also depict how to implement the \textbf{Decoding} circuit (i.e.\ accessing the decoder), or in other words, the circuit that implements \(C^{-1}_{S\to V}\). Inverting \(C_{S \to V}\) boils down the following problem. Given a `set' representative \(A' \in AK_{j}\) and neighboring `vertex' representative \(B' \in BK_1\), find the matrix \(A_{1j \subseteq j}(\beta,\gamma)\) such that \(can(AK_j) \cdot A_{1j \subseteq j}(\beta,\gamma) \in BK_1\). The reason is this: \(C_{S\to V}\) takes in a representative of \(AK_j\) (for \(j \in \set{2,3}\) indicated to the circuit by a bit), and some \(i=(\beta,\gamma)\) and outputs the representative \(can(AK_j) \cdot A_{1j\subseteq 1}(\beta,\gamma)\) of a neighboring \(K_1\) coset. 

Therefore, to find such \(\beta,\gamma\) we may simply:
\begin{enumerate}
    \item Compute \(A_{c}=can(AK_j)\) and \(B_{c}=can(BK_1)\)
    \item Multiply \(B_{c}^{-1}A_{c}\).
    \item Find \(\beta,\gamma\) such that \((B_{c}^{-1}A_{c})\cdot A_{1j \subset j}(\beta,\gamma) \in K_1\), and output \(i=(\beta,\gamma)\).
\end{enumerate}
The final step gives the correct \(i\) since \(A_c\cdot A_{1j \subset j}(\beta,\gamma) \in B_cK_1=BK_1\) if and only if \(B_c^{-1}A_c \cdot A_{1j \subset j}(\beta,\gamma) \in K_1\). 

The final step can be done just by solving a set of linear equalities: Obviously, the entries of \(B_c^{-1}A_c \cdot A_{1j \subset j}(\beta,\gamma)\) are linear in \(\beta, \gamma\). Moreover, \(K_1\) is an affine subspace over \(\F_{\fsize}\),\footnote{that is, subtracting the identity from the matrices of \(K_1\) results in a linear subspace.} so this easily translated to linear equations with \(O(\poldeg)\) variables and equations over \(\F_\fsize\).

Thus, setting \(\fsize = \poly(\frac{\log N}{\varepsilon})\) and \(\poldeg \leq \log N\) (and \(d=3\)), the first step is in size \(\poly(\log N,\log \frac{1}{\varepsilon})\), depth \(O(\log N (\log \log N + \log \log \frac{1}{\varepsilon}))\). The second and third steps are in size \(\poly(\log N, \log \frac{1}{\varepsilon})\) and depth \(O(\log^2 N (\log \log N + \log \log \frac{1}{\varepsilon}))\). So in total we get size \(\poly(\log N, \log \frac{1}{\varepsilon})\) and depth \(O(\log^2 N (\log \log N + \log \log \frac{1}{\varepsilon}))\)
\section{Lower Bound}\label{sec:lower}
\newcommand{\Nbr}{\rm{Nbr}}
\newcommand{\m}{\bf x}
\newcommand{\code}{\bf c}
\newcommand{\Blk}{\textsf{Blk}}
In this section, we prove any high rate (weakly locally computable) aLLDC must make logarithmically many queries in $N$. In particular, we prove the following tight bound for any such code in terms of the rate $R$, error tolerance $\varepsilon$, and domain size $N$:
\begin{theorem}\label{thm:lowerbound}
There exist universal constants $c_1,c_2>0$ such that the following holds. For any $N \in \mathbb{N}$, $r \leq N^{0.99}$, and $S \subset \binom{[N]}{r}$, if ${\cal C}_{([\N],S)}$ is a $.01$-approximate, $(\epsilon,\ell_{out},q)$-LLDC with list size $\ell_{out} \leq \exp(N^{c_1})$, then for any $N^{-c_2} \leq \varepsilon \leq \frac{1}{4}$, the query complexity of ${\cal C}_{([N],S)}$ is at least
\[
q \geq \Omega\left(\frac{\log N}{\varepsilon\log\frac{2}{R}}\right)
\]
where we recall $R=\frac{N}{|S|r}$ is the rate of the code.
\end{theorem}
We note the assumption $\varepsilon \leq \frac{1}{4}$ is not necessary for our argument, and can be replaced with any constant (strictly) less than $\frac{1}{2}$. Similarly, $r \leq N^{0.99}$ may be replaced with $r \leq N^{1-\alpha}$ for any $\alpha>0$ at the cost of decreasing the constant $c_1$ in the list size accordingly. The bound is trivial (or not well-defined) when $R \geq 2$, but in this case no such codes exist (see discussion before \Cref{defn-dist}). Finally, we note the bound holds for any message alphabet.

Since any code with encoding locality $r$ can be made into a direct product code with analogous parameters, replacing $R$ with $\frac{N}{|S|r}$ the above holds for any $r$-locally encodable code.\footnote{Given an $r$-locally encodable code $C$, the corresponding DP code $C_{DP}$ records at any block the $r$ values of the message the original code $C$ depends on. To decode a word $w$, the decoder simulates a word $\tilde{w}$ to the original code $C$ by reading the $r$-bits at any code block $w_s$ and computing the corresponding encoded symbol at $C$ based on them. If $w$ $\varepsilon$-computes a message $f$, then so does $\tilde w$, so we may locally decode simply by simulating $C$'s decoder in this way.} As immediate corollaries, we get that any code with $\poly(\log N, \frac{1}{\varepsilon})$ rate and locality must make at least $\tilde{\Omega}(\frac{\log N}{\varepsilon})$ queries, matching our information theoretic upper bound in \Cref{sec:polylog} up to a $\log\log N$ factor.\footnote{Recall that while our main result in \Cref{sec:polylog} uses $\frac{\polylog N}{\varepsilon}$ queries, if one is not concerned with circuit size this can easily be reduced to $\tilde{O}(\frac{\log(N)}{\varepsilon})$ by taking $\log(N)$-length random paths found by brute force instead of via our longer strongly explicit routing scheme.} Similarly, any $2^{\sqrt{\log N}}$ rate and locality code must make $\Omega(\frac{\sqrt{\log N}}{\varepsilon})$ queries, matching our simple subspace construction.





\paragraph{Proof Intuition} We briefly give some intuition behind the lower bound. At its core, the key difficulty in (approximate) local list decoding lies in the fact that there may be two (or more) global messages $f_1$ and $f_2$ correlated with an input word $w$, yet we are only given a \textit{small local view} (via queries) around the target by which to distinguish them. In particular, since we may have $f_1(v) \neq f_2(v)$ for any given target vertex $v$, we need some globally consistent method in determining which encoded symbols in $w$ surrounding $v$ correspond to $f_1$, and which correspond to $f_2$. We argue, in essence, that this is only possible if the decoder makes enough queries to `link' $v$ to a good fraction of other vertices in the code to enforce this consistency globally, which requires roughly $q \geq \frac{\log N}{\log \frac{1}{R}}$ queries (since the rate $R$ controls the degree of the inclusion graph we use to connect $v$ to any other $v'$). In our upper bound, this corresponds to the fact that we link each $v$ to some global starting advice $s_0 \in S$ (and therefore to each other $v'$ via $s_0$).

The additional $\frac{1}{\varepsilon}$ in the lower bound simply stems from the fact that the adversary may also make all but an $O(\varepsilon)$ fraction of the input word $w$ noise (then distribute the rest randomly between $f_1$ and $f_2$). Thus, on average, it takes $\frac{1}{\varepsilon}$ queries for the decoder to get a single `informative' query about $f_1$, giving the additional $\frac{1}{\varepsilon}$ multiplicative factor.

\subsection{Proof of \texorpdfstring{\Cref{thm:lowerbound}}{Lemma~\ref{thm:lowerbound}}}

Before jumping into the proof, we need to set up some relevant definitions and objects. We start by recalling the standard inclusion graph of a direct product code and some corresponding notation.

\begin{definition}
Let ${\cal C}_{([N],S)}$ be a direct product code, where $S \subset \binom{[\N]}{r}$.
We define the $[N] \times S$ bipartite graph $G^{\Enc}$ where each left vertex $v \in [N]$ corresponds to a message coordinate, and each right vertex $s \in S$ corresponds to an encoded symbol (we'll refer to these as `code blocks'). Each $s \in S$ is connected to the $r$ left vertices that make up that code block.
Let $d_{\rm avg}=|S|/\N$ be the average left-degree in $G^{\Enc}$.
For $s \in S$, we abuse notation slightly as in prior sections and write $s$ both to mean a right-vertex in this graph, and to denote ${\rm Nbr}(s)$, the set of left vertices $v \in [N]$ that are incident with $s$ when clear from context. For $v \in [\N]$, we write ${\rm Nbr}(v)$ to denote the set of right neighbors of $u$, i.e.\ all sets $s \in S$ that contain $v$.
\end{definition}

Before moving on, we make a few simplifications without loss of generality. First, we note we may assume the inclusion graph has maximum left-degree at most $100d_{avg}$. In particular, given a direct product encoding by the above parameters, by Markov there are at most $\N/100$ left vertices in $G^{\Enc}$ of degree more than $100d_{\rm avg}$. For the rest of this section, we will slightly abuse notation and let $[\N]$ denote the set of $\Omega(N)$ left vertices in the original graph $G^{\Enc}$ of degree at most $d = 100 d_{\rm avg}$. The distinction has essentially no effect on the below proof.

We may also assume that $q \geq \frac{1}{100\varepsilon}$, as a lower bound of this form was already proven in the stated regime in \cite{artemenko2014lower} for the appropriate choices of $c_1,c_2$, and that $|S| \geq \frac{C}{\varepsilon}$ for any fixed constant $C>0$. Assuming otherwise, consider a word $w$ given by selecting an arbitrary $\varepsilon$ fraction of $S$ on which to encode some fixed message $f:[N] \to \{0,1\}$. Since $r \leq N^{0.99}$, this means there are $\exp(O(N))$ many $0.01$-far messages $f$ that lead to the same encoding (i.e.\ for which $f \in \Ball_{1-\varepsilon}(w)$), so no such code is possible with the claimed parameters. By similar reasoning, we may also assume the rate $R \leq \frac{3}{2}$.

With these modifications out of the way, we are ready to introduce the distribution over corrupted codewords $w \in (\{0,1\}^{r})^{S}$ that witnesses our lower bound:\footnote{As mentioned, our bound holds for any input alphabet. For generic $\Sigma$, simply identify two of the symbols with $\{0,1\}$ and ignore the remainder in the construction.}

\begin{definition}
\label{defn-dist}
Let ${\cal D}^{\Enc}$ be the distribution over corrupted codewords $w \in (\{0,1\}^{r})^{|S|}$ generated by the following sampling procedure: \\

\noindent {\bf Sampling Procedure for ${\cal D}^{\Enc}$}
\begin{enumerate}
\item Pick random messages ${\cal F} = \{f_1,f_2\}$, $f_i \in \{0,1\}^{\N}$.
\item Pick a random subset $D \subset S$ of measure $4\varepsilon$
\item Pick random ${\overline b} \in \{1,2\}^{|S|}$.
\item For each $s \in S$
\begin{enumerate}
    \item For each $s \in D$: let $w_s = f_{b_s}|_s$
    \item For each $s \notin D$: let $w_s=1^{r}$
\end{enumerate}
\end{enumerate}
\end{definition}

In other words, we pick two random functions $f_1,f_2$ over the vertex set, and set $w$ to encode a random one of these functions on a random $4\varepsilon$ fraction of the code blocks (and otherwise simply give the all $1$'s function, which we view as containing no information about the $f_i$). Note that since we may assume $|S|$ is not too small, the resulting word $w$ will $\varepsilon$-compute both $f_1$ and $f_2$ with extremely high probability. On the other hand, we will show any aLLDC which locally decodes these functions with good probability must make many queries (or else have exponentially large list size).

We formalize this via a multi-round game between two players, Alice and Bob. We will prove (Lemma \ref{lemma:ub-for-lb}) that on the one hand,
if a too-good-to-be true direct product code ${\cal C}_{([\N],S)}$ exists, then
Bob can win most rounds of the game with non-negligible probability (in particular, with roughly probability $\Omega(\frac{1}{\ell_{out}})$). On the other hand,
we will prove (Lemma \ref{lemma:mainlb}) that if Bob is restricted by the locality conditions of ${\cal C}_{([\N],S)}$, then he loses most rounds of the game except with negligible $\exp(-\Omega(N^{0.005}))$ probability.

\begin{figure}[ht!]
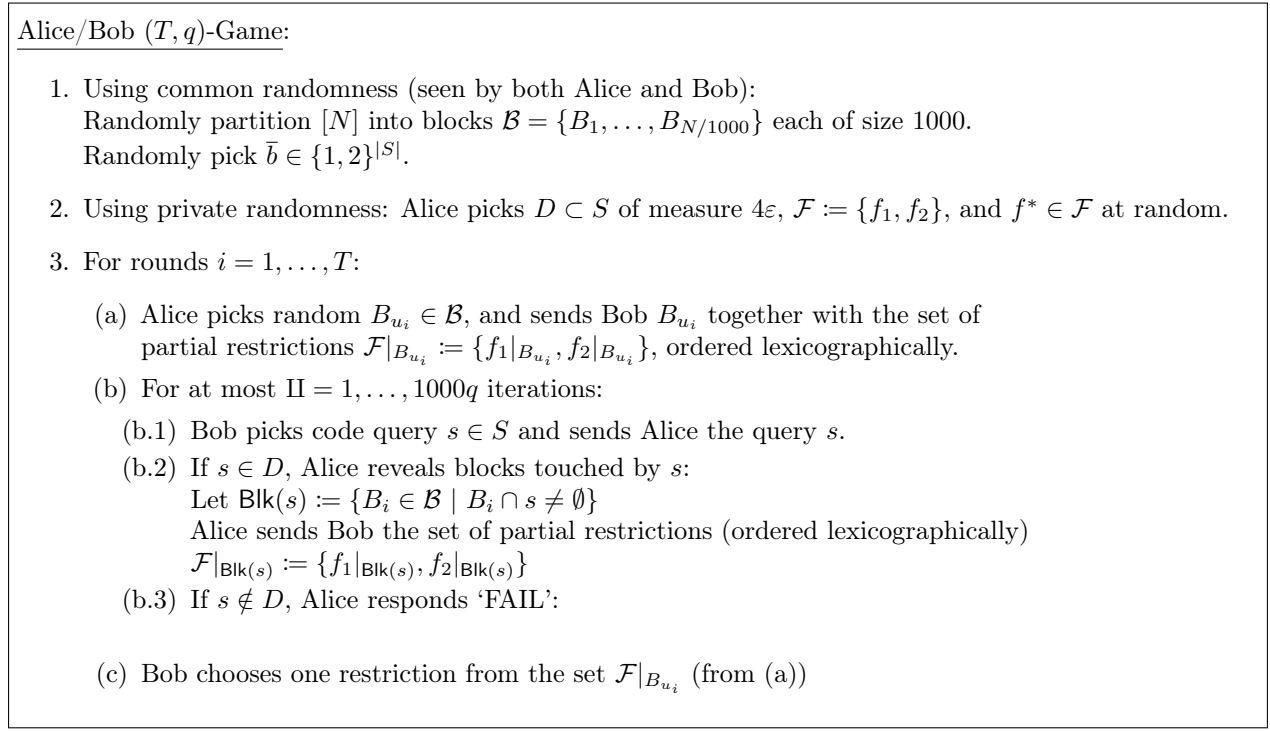

\fbox{\parbox{\textwidth}{
\vspace{.1cm}
\underline{Alice/Bob $(T,q)$-Game}:

\begin{enumerate}
    \item[1.] Using common randomness (seen by both Alice and Bob): \\
	      Randomly partition $[\N]$ into blocks ${\cal B} = \{B_1,\ldots,B_{N/1000}\}$ each of size $1000$. \\
	      Randomly pick ${\overline b} \in \{1,2\}^{|S|}$.
    \item[2.] Using private randomness: Alice picks $D \subset S$ of measure $4\varepsilon$, ${\cal F} \coloneqq \{f_1,f_2 \}$, and $f^* \in {\cal F}$ at random.
    \item[3.] For rounds $i=1,\ldots,T$:
    \begin{enumerate}
        \item[(a)] Alice picks random $B_{u_i} \in {\cal B}$, and sends Bob $B_{u_i}$ together with the set of \\
	partial restrictions ${\cal F}|_{B_{u_i}} \coloneqq \{f_1|_{B_{u_i}},f_2|_{B_{u_i}}\}$, ordered lexicographically. 
         \item[(b)] For at most ${\cal q} = 1,\ldots, 1000q$ iterations:
		\begin{itemize}
		\item[(b.1)] Bob picks code query $s \in S$ and sends Alice the query $s$.
		\item[(b.2)] If $s \in D$, Alice reveals blocks touched by $s$: \\
			Let ${\Blk}(s) \coloneqq \{B_i \in {\cal B} ~|~ B_i \cap s \neq \emptyset\}$ \\
			Alice sends Bob the set of partial restrictions (ordered lexicographically) \\
			${\cal F}|_{\Blk(s)} \coloneqq \{f_1|_{\Blk(s)}, f_2|_{\Blk(s)} \}$ 
        \item[(b.3)] If $s \notin D$, Alice responds `FAIL': \\
		\end{itemize}
	\item[(c)] Bob chooses one restriction from the set ${\cal F}|_{B_{u_i}}$ (from (a))
     
     \end{enumerate}
\end{enumerate}}}
			
 \caption{Game between Alice and Bob}
    \label{fig:alice-and-bob}
\end{figure}

Figure \ref{fig:alice-and-bob} below describes the game between Alice (the adversary) and Bob (the guesser). The game has two parameters, the number of rounds $T=N^{0.005}$, and the query complexity $q$ which controls the number of queries Bob makes per round (mimicking the aLLDC). We will assume without loss of generality $\N$ is divisible by $1000$.\footnote{We may discard vertices to reach the next factor of $1000$ which has negligible effect on the argument for large enough $N$.}

We say that Bob {\it wins} round $i$ of the game if his choice of restriction in the last step (c) of  round $i$ is equal to $f^*|_{B_{u_i}}$.

A random play of the game in \Cref{fig:alice-and-bob} induces a 
distribution over pairs $({\cal F},{\overline b})$ where both ${\cal F}$ and ${\overline b}$ are
chosen randomly. 
In turn, each random pair $({\cal F},{\overline b})$ induces a
codeword vector $w \in (\{0,1\}^r)^{|S|}$, by setting $w_s \coloneqq f_{b_s}|_{s}$ whenever $s \in D$, and $w_s=1^r$ when $s \notin D$. The distribution over $w$ induced by the random choice of $({\cal F},{\overline b})$ is then exactly the distribution ${\cal D}^{\Enc}$ defined above, so from now on we will write $w \sim {\cal D}^{\Enc}$ to denote 
a random $w$ induced by a play of the game.

The proof of \Cref{thm:lowerbound} is broken into two parts. We first show that if a good aLLDC exists, Bob can win $75\%$ of rounds in the game with non-negligible probability, scaling with the list-size of the code.

\begin{lemma}
\label{lemma:ub-for-lb}
Assume ${\cal C}_{([N],S)}$ is a $0.01$-approximate $(\varepsilon,\ell_{out},q)$-aLLDC. For every $T \in \mathbb{N}$, there exists a strategy for Bob such with probability at least $\Omega(\frac{1}{\ell_{out}})$, Bob wins at least $\frac{3}{4}T$ rounds of the $(T,q)$-game.
\end{lemma}

On the other hand, we show that when $q$ is small, no randomized strategy for Bob wins more than $0.6 T$ rounds except with exponentially small probability.

\begin{lemma}
\label{lemma:mainlb}
There exists a universal constants $C'>0$ such that for $T=N^{0.005}$ and $ \frac{1}{100\varepsilon} \leq q \leq C'\frac{\log N}{\varepsilon\log\frac{2}{R}}$, the probability any (possibly randomized) strategy for Bob wins more than $0.6 T$ rounds of the $(T,q)$-game is at most $3\exp(-N^{0.005}/100)$.
\end{lemma}

Together, these lemmas immediately imply \Cref{thm:lowerbound}.

\begin{proof}[Proof of \Cref{thm:lowerbound}]
Assume for the sake of contradiction there exists a $0.01$-approximate, $(\varepsilon,q,\ell_{out})$-aLLDC with $q \leq C'\frac{\log N}{\varepsilon\log\frac{2}{R}}$, $\ell_{out} \leq \exp(N^{c_1})$, and $N^{c_2} \leq \varepsilon \leq \frac{1}{4}$ for any pair $c_1<c_2$ sufficiently small as to satisfy the statement of \cite[Theorem 1.4]{artemenko2014lower} (the aforementioned $q \geq \frac{1}{100\varepsilon}$ lower bound), and $C'$ the constant in \Cref{lemma:mainlb}. By \Cref{lemma:ub-for-lb}, for any $T$, Bob therefore has a strategy to win $\frac{3}{4}T$ rounds of the $(T,q)$-game with probability $P=\Omega(\exp(-N^{c_1}))$. If $q < \frac{1}{100\varepsilon}$, we are done by \cite{artemenko2014lower}. Otherwise, setting $T=N^{0.005}$, \Cref{lemma:mainlb} states no strategy for Bob wins the $(T,q)$-game on more than $0.6T$ rounds with probability greater than $3\exp(-N^{0.005}/100) < P$ for $N$ sufficiently large and $c_1 < 0.005$, giving the desired contradiction.
\end{proof}

\subsubsection{Proof of \texorpdfstring{\Cref{lemma:ub-for-lb}}{Lemma~\ref{lemma:ub-for-lb}}}

We start with the proof lower bounding the probability Bob wins given access to the claimed aLLDC.

\begin{proof}[Proof of \Cref{lemma:ub-for-lb}]
Let ${\cal C}_{([N],S)}$ be the direct product code assumed to exist,
and let ${\cal A}$ be the randomized aLLDC decoding algorithm.  Then Bob's strategy in the game is as follows. Recall by standard amplification (see \Cref{sec:apx-local-list}), we may assume without modifying the code's parameters asymptotically that the decoder outputs a list of circuits $C_1,\ldots,C_{\ell_{out}}$ such that for any word $w$ and $f \in \Ball_{1-\varepsilon}(w)$ with $99\%$ probability, some $C_i$ deterministically $0.1$-computes $f$ (in particular, $\cal A$ itself randomly fixes the internal randomness of the original $C_i$ after majority amplification).
\\

\noindent{\bf Bob's Strategy:} First, Bob runs ${\cal A}$ to obtain 
randomized circuits $C_1,\ldots,C_{\ell_{out}}$, and randomly selects one of the circuits, $C$.

At the start of the game, Alice and Bob pick a random partition of $[N]$ into $\frac{N}{1000}$ blocks, ${\cal B} = \{B_1,\ldots,B_{N/1000}\}$ each of size $1000$, and a random vector ${\overline b} = b_1,\ldots,b_{|S|}$ where each $b_i \in \{1,2\}$.
For each round Bob's strategy is as follows. 
In round $i \in [T]$,
in step 3(a), Alice sends Bob a challenge block $B_{u_i} \in {\cal B}$ 
to decode, and sends Bob the set of partial restrictions (ordered lexicographically)
$\{\rho_1=f_1|_{B_{u_i}}, \rho_2=f_2|_{B_{u_i}}\}$. 
Now for each $v \in B_{u_i}$, Bob simulates the circuit $C^{\rm}$ on input $v$.
Whenever $C^{\rm}(v)$ makes a codeword coordinate query $s \in S$, he sends Alice the query $s$ in step (b.1)
and she returns the set of partial restrictions (ordered lexicographically)
$\{\sigma_1 = f_1|_{\Blk(s)}, \sigma_2 = f_2|_{\Blk(s)}\}$ in step (b.2), or `FAIL' if $s \notin D$.
These restrictions correspond to the values of each $f_i$ restricted to the
blocks touched by $s$. Bob then uses ${\overline b}$ to compute the answer $w_s$ to the query by setting
$w_s = f_{b_s}|_{s}$ in the former case, and $w_s=1^r$ when Alice responds FAIL.
Bob proceeds in this fashion until he has simulated the entire computation of $C$ on input $v$. 
Note that by the locality of the direct product code, $C(v)$ makes at most $q$ queries, so he can simulate the computation of  
$C(v)$ using $q$ iterations of the inner loop (b). Since he simulates $C$ on every $v \in B_{u_i}$, altogether he makes at most $1000q$ iterations of the inner loop in round $i$.
 
After completing the (b)-loop in round $i$, Bob has simulated $C(v)$ for all $v \in B_{u_i}$ which produces 
a string $\rho' \in \{0,1\}^{1000}$. In the final step (c) of round $i$, he compares this string $\rho'$ with the
strings $\rho_1$ and $\rho_2$ given by Alice in step 3(a),
and outputs the string $\rho_i$ closest to $\rho'$ in hamming distance (where ties are broken arbitrarily).

\bigskip

We now prove Bob's strategy defined above wins most rounds of the game with non-negligible probability. Toward this end, we first define a few useful properties that will help Bob win the game.

\begin{definition}[Atypical Rounds]
    We say that a round $i \in [T]$ of a play of the game is {\it atypical} if $f_1|_{B_{u_i}}$ and $f_{2}|_{B_{u_i}}$ have normalized hamming distance less than $0.4$.
\end{definition}


\begin{definition}[Lucky] We say a run of the game is lucky for Bob if the following events all hold:
\begin{itemize}
\item[(1)] The random choice of $w \sim {\cal D}^{\Enc}$ induced by $({\cal F},{\overline b})$ satisfies $f^* \in \Ball_{1-\epsilon}(w)$. 
\item[(2)] The number of atypical rounds is less than $T/100$.
\item[(3)] The random circuit Bob selects from the aLLDC is $0.1$-close to Alice's chosen $f^* \in {\cal F}$.
\end{itemize}
\end{definition}

\begin{lemma} 
\label{lemma-prob-good}
Assuming the existence of an aLLDC with the pre-specified parameters, a run of the game is lucky for Bob with probability $\Omega(\frac{1}{\ell_{out}})$, 
\end{lemma}

The proof of the above Lemma uses the following claims.

\begin{claim}
\label{claim-w-closeto-f}
The probability over $\mathcal D^{Enc}$ that $f^* \in \Ball_{1-\epsilon}(w)$ is at least $\frac{1}{200}$.
\end{claim}
\begin{proof}
    Fix Alice's choice of $f^*$ and the sub-sampled domain $D \subset S$ of measure $4\varepsilon$. For each $s \in D$, there is an independent $50\%$ probability that $w_s=f^*|_s$, so the expected agreement of $f^*$ with $w$ is $2\varepsilon$. A Chernoff bound promises the probability that $f^* \notin \Ball_{1-\varepsilon}(w)$ is at most $\exp(-\Omega(\varepsilon |S|)) \leq \frac{1}{200}$ for $|S| \geq \Omega(\frac{1}{\varepsilon})$ sufficiently large.
    \end{proof}

\begin{claim}
\label{claim-atypical}
For any $T \in \mathbb{N}$ sufficiently large and $\epsilon$ sufficiently small, the probability that there are more than $T/100$ atypical rounds is at most $\frac{1}{200}$.
\end{claim}

\begin{proof}
    We argue the probability any particular round $i$ of the game is atypical is small, and therefore that by a Chernoff bound over the $T$ independent blocks that almost all rounds are typical with high probability.

    In any fixed round $i$, the probability that $dist(f_1|_{B_{u_i}},f_2|_{B_{u_i}}) < 0.4$ is at most $e^{-10}$ by a Chernoff bound since the expected distance between $f_1,f_2$ is at most $\frac{1}{2}$ (potentially larger for larger alphabets) and the $|B_{u_i}|=1000$ bits are chosen uniformly at random. Now, since each of the $T$ blocks in the game are independent and the expected number of atypical rounds is at most $e^{-10}T$, a second Chernoff bound promises the probability there are more than $T/100$ atypical rounds is at most $\frac{1}{200}$ for $N$ sufficiently large.
\end{proof}

We are now ready to prove Lemma \ref{lemma-prob-good}.
\begin{proof}[Proof of \Cref{lemma-prob-good}]

By \Cref{claim-w-closeto-f} and \Cref{claim-atypical}, events (1) and (2) occur with $99\%$ probability.

Conditioned on these events, Bob's run of the (amplified) aLLDC outputs a list of circuits $\{C_1,\ldots,C_{\ell_{out}}\}$ such that some $C_i$ is $0.1$-close to $f^*$ with $99\%$ probability. Since Bob selects one of these circuits uniformly at random, the probability he selects $C_i$ is $\frac{1}{\ell_{out}}$, so the probability the game is lucky for Bob is at least $\frac{1}{2\ell_{out}}$.
\end{proof}

We can now complete the proof of \Cref{lemma:ub-for-lb}. First, we argue that when the game is lucky for Bob, Bob wins any fixed typical round with $90\%$ probability. Assuming this is the case we are done by Markov's inequality: since $99\%$ of the rounds are typical when the game is lucky, Bob's expected number of wins is at least $0.8T$ and the probability he wins fewer than $\frac{3}{4}T$ rounds is at most $0.8$. Since Bob is lucky with probability $\Omega(\frac{1}{\ell_{out}})$, he therefore wins $\frac{3}{4}T$ rounds with probability at least $\Omega(\frac{1}{\ell_{out}})$ as desired.

It's left to argue Bob wins a fixed typical round in a lucky run with $90\%$ probability. To see this, note that Bob automatically wins so long as his choice of circuit is $0.15$-close to Alice's choice of $f^*$ on the block, since by assumption $\{f_1,f_2\}$ are $0.4$-far, so $f^*$ is then the unique closer function. Without conditioning on typicality (but still assuming Bob's circuit is globally $0.1$-close to $f^*$), the probability it is $0.15$-close to $f^*$ on a given block $B_{u_i}$ is at least $1-e^{-8}$ by a Chernoff bound. On the other hand, we proved above the probability $B_{u_i}$ is typical is at least $1-e^{-10}$ (and this event is independent of Bob's choice of circuit, so this remains true under such conditioning). Thus even conditioned on typicality, Bob's circuit remains $0.15$-close to $f^*$ on $B_{u_i}$ with at least $90\%$ probability, completing the proof.
\end{proof}

\subsubsection{Proof of \texorpdfstring{\Cref{lemma:mainlb}}{Lemma~\ref{lemma:mainlb}}}

We now need to show toward contradiction that when $q$ is smaller than the specified parameter, Bob cannot win more than $0.6 T$ rounds except with negligible probability.

\begin{figure}[ht!]
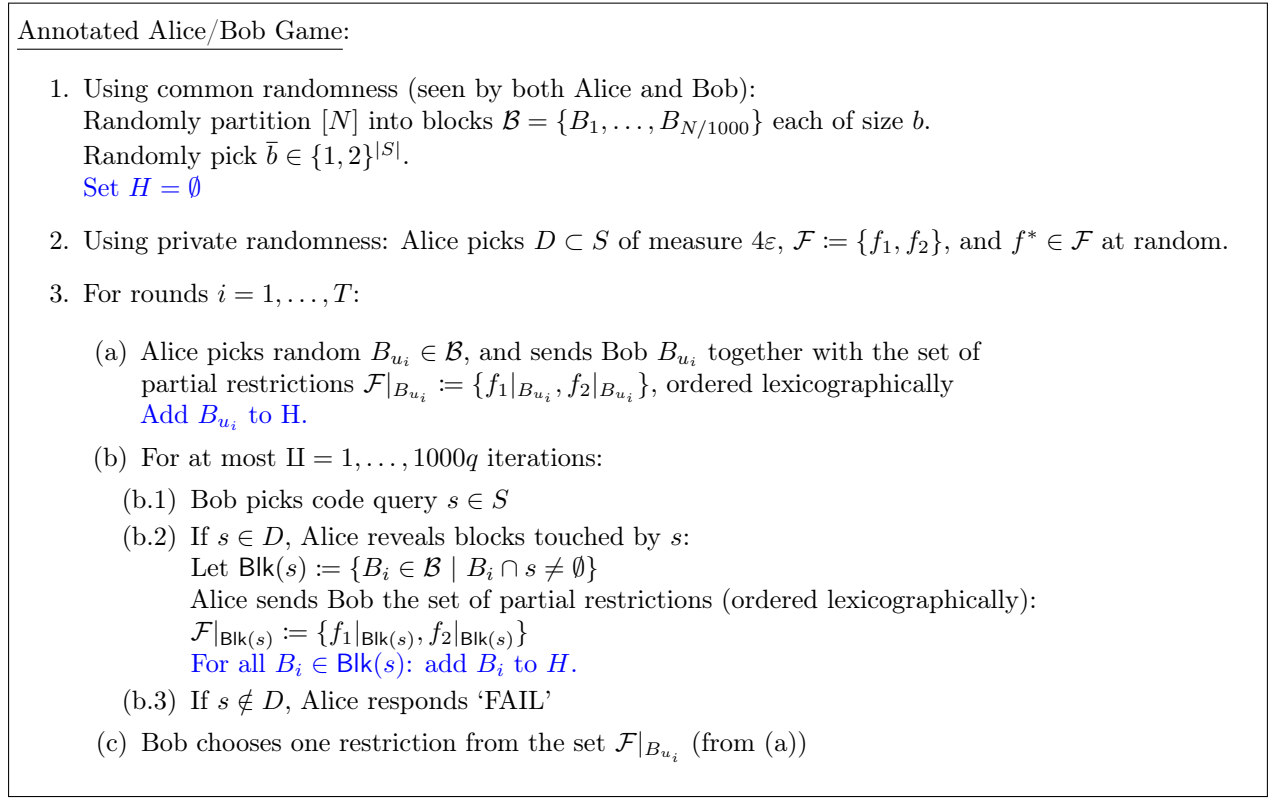

\fbox{\parbox{\textwidth}{
\vspace{.1cm}
\underline{Annotated Alice/Bob Game}:

\begin{enumerate}
    \item[1.] Using common randomness (seen by both Alice and Bob): \\
        Randomly partition $[\N]$ into blocks ${\cal B} = \{B_1,\ldots,B_{N/1000}\}$ each of size $b$. \\
        Randomly pick ${\overline b} \in \{1,2\}^{|S|}$. \\
        \textcolor{blue}{ Set $H=\emptyset$}
    \item[2.] Using private randomness: Alice picks $D \subset S$ of measure $4\varepsilon$, ${\cal F} \coloneqq \{f_1,f_2 \}$, and $f^* \in {\cal F}$ at random.

    \item[3.] For rounds $i=1,\ldots,T$:

    \begin{enumerate}
        \item[(a)] Alice picks random $B_{u_i} \in {\cal B}$, and sends Bob $B_{u_i}$ together with the set of \\
        partial restrictions ${\cal F}|_{B_{u_i}} \coloneqq \{f_1|_{B_{u_i}}, f_2|_{B_{u_i}}\}$, ordered lexicographically \\
                 \textcolor{blue}{Add $B_{u_i}$ to H.}

         \item[(b)] For at most ${\cal q} = 1,\ldots, 1000q$ iterations:
                \begin{itemize}
                \item[(b.1)] Bob picks code query $s \in S$
                \item[(b.2)] If $s \in D$, Alice reveals blocks touched by $s$: \\
                        Let $\Blk(s) \coloneqq \{B_i \in {\cal B} ~|~ B_i \cap s \neq \emptyset\}$ \\
                        Alice sends Bob the set of partial restrictions (ordered lexicographically): \\
                        ${\cal F}|_{\Blk(s)} \coloneqq \{f_1|_{\Blk(s)}, f_2|_{\Blk(s)} \}$ \\
                        \textcolor{blue}{For all $B_i \in \Blk(s)$: add $B_i$ to $H$.}
                \item[(b.3)] If $s \notin D$, Alice responds `FAIL'
                \end{itemize}
        \item[(c)] Bob chooses one restriction from the set ${\cal F}|_{B_{u_i}}$ (from (a))
     \end{enumerate}
\end{enumerate}}}
\caption{Annotated Game between Lazy Alice and Bob}
    \label{fig:annotated-alice-and-bob}
\end{figure}

The high level intuition behind the lower bound is that each round of the game is local in the sense that the number of message blocks seen by Bob in any given round is small since Bob only makes $1000q$ queries per challenge, and each query only touches $r$ message bits. Setting $T=N^{0.005}$, as long as $q$ and $r$ are small, we will argue it is likely that the challenge block $B_{u_i}$ in the $i$-th round is `far' from any previously queried information -- i.e.\ that even making $1000q$ queries, Bob cannot `link' the challenge block $B_{u_i}$ to any previous information. As a result, Bob cannot correlate his answer with prior rounds, and can at best guess whether $f_1$ or $f_2$ on $B_{u_i}$ is $f^*$, giving an independent $\frac{1}{2}$ chance of losing the round. As long as almost all rounds satisfy this property, basic concentration implies Bob cannot win significantly more than half the rounds except with $\exp(-O(T))$ probability, giving the desired upper bound on Bob's success.

\begin{proof}[Proof of \Cref{lemma:mainlb}]
We will prove that for any strategy for Bob, the probability he wins more than $0.6T$ rounds of the game is at most $3\exp(-N^{0.005}/100)$. To analyze the dependencies between the challenge blocks throughout the game,
we will annotate the game with an \textit{associated graph $G=(V,E)$}, which has a vertex $v_i$ for each challenge block $B_i \in \mathcal B$, and an edge between $B_i$ and $B_j$ iff there exists a code block $s \in S$ whose vertices intersect both blocks, and an \textit{evolving vertex-induced subgraph $H \subset G$} throughout each round of the game which tracks which blocks have been revealed through Bob's queries. We will typically refer to $H$ just by the vertices/blocks that define it as a vertex-induced subgraph.

Roughly speaking, we will argue that disconnected components of the subgraph $H$ are independent in the game (in the sense that the value of Alice's secret $f^*$ on any fixed connected component of $H$ does not depend on its value on the other components). We then argue that in most rounds, because the graph $G$ is low-degree, Alice's random challenge block $B_{u_i}$ will be very far from $H$, and in particular Bob will not have enough queries to connect $B_{u_i}$ to any component of $H$. As a result, the value of $f^*$ on $B_{u_i}$ will be independent of all prior rounds, so Bob can do no better than simply guessing, and loses with probability $\frac{1}{2}$ \textit{independent} of his wins/losses in prior rounds. Standard concentration then promises that Bob cannot win much more than $\frac{T}{2}$ rounds except with $\exp(-O(T))$ probability. See Figure \ref{fig:annotated-alice-and-bob} for 
a description of the annotated game.

More formally, at the start of the game, the vertex-induced subgraph $H$ is empty. In each round, Alice picks a random challenge block $B_{u_i}$, which is added to $H$, and Bob makes up to $1000q$ queries $s \in S$. For each queried $s \in D$, we add every block $B_{j}$ that intersects $s$ (denoted $\Blk(s)$) to $H$. For convenience, we write $H_i$ to denote the current vertex-induced subgraph $H$ at the end of round $i$, and for a block $B \in \mathcal B$ write $H_i(B)$ to denote the connected component of $H_i$ containing $B$.

We will show in most rounds the challenge block $\mathcal{B}_{u_i}$ has distance at least $8000\varepsilon q$ from $H_{i-1}$, and that Bob is unlikely to make this many `informative' queries (i.e.\ queries in $D$). Thus Bob cannot link the block with any prior challenges and has an independent $\frac{1}{2}$ chance of loss for the round.
\begin{definition}[Typical and Lucky for Alice]
    We call round $i$ of the game typical if
    \begin{itemize}
        \item Bob queries at most $8000\varepsilon q$ sets in $D$, and
        \item $B_{u_i}$ has distance less than $7000\varepsilon q$ from $H_{i-1}$,
    \end{itemize}
    We say a run of the game is lucky for Alice if no more than $\frac{T}{50}$ rounds are atypical.
\end{definition}

We call the sequence of Bob's wins and losses throughout the game his win-loss transcript. We now argue the probability Bob wins a typical round is $\frac{1}{2}$, independent of his transcript up to that point:

\begin{claim}
The probability Bob wins a typical round is $1/2$, independent of his prior win-loss transcript.
\end{claim}
\begin{proof}
By definition, each query $s \in D$ adds a clique to the subgraph $H$, and each query $s \not\in D$ adds nothing. As a result, if $B_{u_i}$ has distance at least $8000q\varepsilon$ from $H_{i-1}$ and Bob makes at most $7000q\varepsilon$ queries in $D$, $\mathcal{B}_{u_i}$ must be disconnected in $H_i$ from every prior challenge $\mathcal{B}_{u_j}$ (recall $H_{i-1}$ contains every prior challenge block by construction).

It is left to argue that if $\mathcal{B}_{u_i}$ is disconnected from every prior challenge block in the subgraph $H_i$, the probability Bob wins is $\frac{1}{2}$ independent of his prior win-loss record. Roughly speaking, this is immediate from the fact that the value of $f^*$ (Alice's secret function) restricted to the message bits in the connected component $H_i(B_{u_i})$ is completely independent of its value on all other connected components.

Formally, fix Alice's choice of $f^*=f_1$ or $f^*=f_2$ (say $f_1$), and fix any possible 
\[
\mathcal{F}|_{H_i(B_{u_i})}=\{f_1|_{H_i(B_{u_i})},f_2|_{H_i(B_{u_i})}\}
\] 
Alice gives Bob\footnote{More accurately, Alice gives the values for each Blk$(s)$ separately, but it is only easier for Bob if he is given the aggregated values.} in round $i$. Note this does \textit{not} fix the values of $f_1,f_2$ themselves, which are equally likely to be either of the functions in $\mathcal{F}|_{H_i(B_{u_i})}$). Bob must decide whether $f^*$ is the first or second function in this lexicographically ordered set. 

The key is to observe that conditioning on any prior win-loss transcript for Bob has no further effect on Alice's choice of $f_1,f_2$ on $H_i(B_{u_i})$ --- in other words, over the randomness of Alice, the probability $f^*$ is either of the fixed functions in $\mathcal{F}|_{H_i(B_{u_i})}$ remains exactly $\frac{1}{2}$, so any strategy for Bob wins with probability $\frac{1}{2}$ independent of his prior win-loss transcript. This simply follows from the fact that fixing a win-loss transcript by definition fixes the values of $f_1,f_2$ on every prior $B_{u_j}$ challenge block. This information may effect which function in $\mathcal{F}|_B$ is $f_1$ in any block connected to $B_{u_j}$ using Bob's queries, but has no effect on the value of $f_1$ on message bits in $H_i(B_{u_j})$ which is disconnected from every other component (namely Bob never queries any $s$ that contains a message bit in both $H_i(B_{u_j})$ and $H_i(B_{u_i})$, so fixing the value of $f_1$ on the former has no effect on the value of $f_1$ on the latter). Thus either possible value of $f_1$ remains valid under any such conditioning, and is chosen independently with probability $\frac{1}{2}$ over Alice's randomness as desired.

\end{proof}

We now argue that, except with negligible probability, the game is lucky for Alice.
\begin{claim}\label{claim:Alice-Lucky}
The game is lucky for Alice except with probability $2\exp(-T/3)$.
\end{claim}
\begin{proof}
We need to prove the probability more than $\frac{T}{50}$ rounds are atypical is at most $2\exp(-T/3)$. First note that by a Chernoff bound (without replacement), the probability over Alice's choice of $D \subset S$ that Bob queries at most $7000\varepsilon q$ $s \in D$ is at least $1-\exp(-T/3)$. Condition on this event. We claim that every round is then typical except with probability 
\begin{equation}\label{eq:typical-alice}
\frac{1000T(1+8000\varepsilon qr)(10^5/R)^{8000\varepsilon q}}{N} + \exp(-1000\varepsilon q),
\end{equation}
even conditioned on Alice and Bob's choices in prior rounds. Since we assume $q \geq \frac{1}{100\varepsilon}$, the latter term is at most $\frac{1}{200}$. On the other hand, assuming $q \leq C'\frac{\log N}{\varepsilon\log \frac{2}{R}}$ for small enough $C'$, and plugging in $T=N^{.005}$ and $r \leq N^{0.99}$ the former term is bounded by
\begin{align*}
    \frac{1000T(1+8000\varepsilon qr)(10^5/R)^{8000\varepsilon q}}{N} &\leq O\left(N^{-0.005}\log N(10^5R^{-1})^{8000C'\frac{\log(N)}{\log \frac{2}{R}}}\right)\\
    &\leq O(N^{C''-0.005}\log N)
\end{align*}
where we may take $C''>0$ as small as desired by taking $C'$ sufficiently small. Taking $C'' < 0.005$, we get the entire expression is bounded by $\poly(1/N)$, and in particular by $\frac{1}{200}$ for large enough $N$. Thus for large enough $N$, \Cref{eq:typical-alice} is always upper bounded by $\frac{1}{100}$. The total number of atypical rounds is therefore stochastically dominated by $\text{Bin}(T,\frac{1}{100})$, and a Chernoff (and union) bound implies the desired result.

To prove \Cref{eq:typical-alice}, observe that by our conditioning, at any point in the game the number of vertices in $H$ is at most $T(1+8000\varepsilon qr)$, i.e.\ the queries in $D$ (each of which may add up to $r$ blocks to $H$) and challenge blocks themselves. On the other hand, our graph has maximum degree at most $1000dr=10^5R^{-1}$, since each block $B$ has $1000$ message bits, each of which touches at most $d$ sets $s \in S$, and each $s$ respectively touches at most $r$ blocks. Thus the total number of vertices in $G$ at distance $8000\varepsilon q$ from $H_i$ is at most
\[
T(1+8000\varepsilon qr)(10^{5}R^{-1})^{8000 \varepsilon q}.
\]
The challenge block $B_i$ is chosen uniformly at random, giving the desired probability of landing near $H_{i-1}$. Similar to above, a Chernoff bound also promises that in each round, the probability Bob selects more than $7000\varepsilon q$ sets in $D$ is at most $\exp(-1000\varepsilon q)$ as desired.
\end{proof}

The proof is now essentially immediate. Alice is lucky except with $2\exp(-T/3)$ probability, in which case all but a $\frac{1}{50}$ fraction of rounds in the game are typical. Conditioned on this event, Bob's probability of winning any typical round is $\frac{1}{2}$, independent of prior rounds. Thus the total number of rounds won by Bob is stochastically dominated by $\frac{T}{50}+\frac{49}{50}\text{Bin}(\frac{1}{2},\frac{49}{50}T)$, and a Chernoff bound implies the probability Bob wins more than $0.6T$ rounds is at most $\exp(-T/100)$. Union bounding over these events completes the proof.
\end{proof}

\section*{Acknowledgements}

We are very grateful to Shashank Srivastava for several helpful discussions regarding list decoding and list recovery, including pointing us to the works of Guruswami and Indyk, and to Swastik Kopparty, Noga Ron-Zewi, and Mary Wootters for discussion of modern parallel and local decoding algorithms. We are also thankful to Irit Dinur, Avi Wigderson, and Ronen Shaltiel for helpful discussions regarding HDX and hardness amplification, and to Izhar Oppenheim for pointing us to the KMS-variant of the Kaufman-Oppenheim HDX construction. Finally, we thank Gil Melnik for help creating various figures throughout the work.

\bibliographystyle{unsrt}  
\bibliography{references} 

\appendix
\section{Alphabet and Locality Reduction}\label{sec:alphabet}
In the main body we constructed codes with good rate and query complexity but large (log)-alphabet and local encoding time scaling with $\poly(\varepsilon^{-1},\log(N))$. In this section, we use variants of standard concatenation machinery to turn these into \textit{binary}, \textit{locally computable} codes. We study two regimes. First, if we just wish to achieve binary codes and are willing to accept $\poly(\varepsilon^{-1},\log(N))$ locality, it suffices to concatenate with essentially any good list-decodable code with RNC$^2$ decoding (e.g.\ those classically following from concatenated Reed-Solomon codes \cite{guruswami2000list,alekhnovich2002linear}). This gives the following corollary from \Cref{sec:polylog} which we restate here for convenience and prove at the end of the section:

\begin{corollary}[Binary aLLDCs]\label{cor:binary-polylog-aLLDCs-restated}
    For all large enough $N \in \mathbb{N}$ and \(\varepsilon >0\),
    there exists an explicit binary code $\mathcal{C}: \{0,1\}^N \to \{0,1\}^{N'}$ that is $(0.01)$-approximately $(\frac{1}{2}+\varepsilon,\ell_{out},q)$-list decodable with
    \begin{enumerate}
        \item \textbf{Rate:} \(R \geq \poly(\frac{1}{\log(N)},\varepsilon)\)
        \item \textbf{Query Complexity:} $q \leq \poly(\log(N),\frac{1}{\varepsilon})$
        \item \textbf{List Size:} $\ell_{out} \leq \frac{\polylog(\frac{1}{\varepsilon})}{\varepsilon^2}$
    \end{enumerate}
    Moreover, $\mathcal{C}$ can be locally encoded in Time$(\poly(\log(N),\frac{1}{\varepsilon}))$ and decoded by logspace uniform circuits of size $\poly(\frac{1}{\varepsilon},\log N)$ and depth
    $\tilde{O}\left(\log^{2}\left(\frac{\log N}{\varepsilon}\right)\right)$.\footnote{Here $\tilde{O}$ hides $\log\log(\frac{\log N}{\varepsilon})$ factors.}
\end{corollary}

The proof of the binary codes in the sub-polynomial (\Cref{cor:sub-poly-aLLDCs-binary}) and constant (\Cref{cor:constant-rate-binary}) regime are essentially exactly the same (though for the former one first concatenates with \cite{impagliazzo2008uniform}'s derandomized subspace code to remove the $2^{\sqrt{\log(N)}}$ term from the encoding locality.), so we omit them.





Our second goal in the section is to construct binary codes via concatenation that are locally computable in time \textit{polylogarithmic} in $\frac{1}{\varepsilon}$. This is critical to perform exponential hardness amplification while remaining in natural classes like $P^{NP}$. While the ideas needed to do this are still largely standard, a few tweaks are needed due to our strong requirement on $\delta_{in}$ and to achieve optimal query complexity and list-size:
\begin{corollary}[Locally-Computable Binary aLLDCs]\label{thm:KO-locality-reduced}
    For all sufficiently large $N \in \mathbb{N}$ and $\varepsilon>0$, there exists an explicit binary code $\mathcal{C}: \{0,1\}^N \to \{0,1\}^{N'}$ that is $0.01$-approximate $(\frac{1}{2}+\varepsilon,\ell_{out},q)$-list-decodable with
    \begin{enumerate}
        \item \textbf{Rate:} $R \geq \frac{1}{\polylog N} \cdot2^{-O(\log^2(1/\varepsilon))}$
        \item \textbf{Query Complexity}: $q \leq \frac{\polylog(N,\frac{1}{\varepsilon})}{\varepsilon^{2}}$
        \item \textbf{List size}: $\ell \leq \frac{\polylog \frac{1}{\varepsilon}}{\varepsilon^2}$
    \end{enumerate}
    Moreover, $\mathcal{C}$ can be locally encoded in Time$(\poly\log(N)\cdot\log^2\frac{1}{\varepsilon}))$ and list decoded by logspace uniform circuits of size $\poly(\frac{1}{\varepsilon},\log N)$ and depth $O\left(\log\left(\frac{\log N}{\varepsilon}\right)\log\log(N)\right)$.
\end{corollary}
A few remarks are in order before moving on. First, on the positive side, the query complexity of \Cref{thm:KO-locality-reduced} has essentially optimal dependence on $\varepsilon$ by \cite{grinberg2018indistinguishability}. The \textit{rate}, however, has decayed fairly substantially, from polynomial to quasi-polynomial. This is a regrettable but standard barrier in uniform hardness amplification (see \cite{impagliazzo2008uniform}), and not one we address in this work. Indeed the question is essentially orthogonal to achieving near-linear dependence on $N$, as if uniform aLLDCs with better $\varepsilon$-dependence and even sub-exponential dependence on $N$ were known, we could use them directly in our framework to analogously improve the $\varepsilon$-dependence while maintaining near-linearity in $N$.

\paragraph{Proof Overview} We briefly outline our general proof strategy for the locally computable setting before moving on. At a high level, the basic idea is the same as in standard code concatenation: we'd simply like to replace each big symbol in the original code with its own inner aLLDC such that whenever the `outer code' wants to query (part of) a symbol, it calls the (small alphabet) inner code to decode the (probably correct) value. This strategy is employed in \cite{impagliazzo2008uniform} to reduce the alphabet and locality of their subspace construction by composing with a small complete complex.

Unfortunately, this method won't work directly to prove \Cref{thm:KO-locality-reduced} for two reasons. First, recall from \Cref{sec:outer} that for technical reasons, we need our inner code to decode to accuracy $\delta_{out} \ll \frac{1}{\log(N)}$, and in particular thus must also have $\delta_{in} \ll \frac{1}{\log(N)}$. To do this with a direct product inner code would require using sets of size at least $\log(N)$ which, in effect, defeats the point of alphabet reduction (and more concretely is too large to then concatenate with Hadamard). With this in mind, we'll need an inner code that keeps the alphabet size independent of this $\delta$.

Thankfully, the construction of such a code isn't too difficult once one moves away from the direct product paradigm. In particular, brushing some details under the rug, this can be achieved by composing variants of two known codes: a binary direct-sum code that decodes from $\varepsilon$-agreement to $99\%$-agreement, and a second binary \textit{non-DP} code based on methods from hardness amplification that decodes from $99\%$-agreement to $1-\delta_{out}$ agreement. While the latter still requires encoding locality $O(\frac{1}{\delta_{out}})$, critically we are able to maintain binary alphabet throughout the whole procedure.

The second main issue is that known concatenation theorems for aLLDCs are lossy and would result in a large polynomial loss in queries and list size. We handle this by introducing a new strengthened model of aLLRCs we call the `single circuit' model (which in essence allows for query re-use between circuits), and additionally tightening the standard concatenation argument to allow decoding at multiple potentail error scales to optimize the list size and query complexity.

\subsection{Single Circuit (High Noise) List Recovery}\label{sec:single-circuit}

In the main body, we typically thought of list recovery in the \textit{low noise regime}, where we reduced list-decoding from high noise to list-recovery where almost every index contains the desired function $f$. In concatenation, we additionally need to handle the \textit{high noise} list recovery regime because only a small fraction of inner codes will have a valid list-decoding. Thus running these inner codes generates a high noise list recovery input to the other code. Before we wrote list recovery in the notation $(1-\eta,\ell_{out},q)$-LLRC to emphasize the low-noise regime. Now, we will switch to writing simply $(\varepsilon,\ell_{out},q)$-LLRC to match the notation of our high noise LLDC definition, where the notation still means that we should decode any $f \in \Ball_{1-\varepsilon,\delta_{in}}(\mathcal{L})$.

While the above is standard in the coding literature, we will also need a new strengthened formulation of list decoding/recovery to achieve optimal parameters under concatenation that we call \textit{single-circuit} decoding. Informally, a \textit{single-circuit} aLLRC is an aLLRC which instead of outputting a \textit{list} of $\ell_{out}$ circuits (one of which is promised to decode the target message $f$ with good probability), outputs a \textit{single} circuit with an output in $\Sigma_1^{\ell_{out}}$ such that one index of the output is promised to compute $f$ with good probability.

\begin{definition}[Single-Circuit aLLRC]\label{def:single-circuit}
Let \(\varepsilon,\delta, > 0\), and \(\ell_{in},\ell_{out},q \in \mathbb{N}\). Let \(\mathcal{C}:\Sigma_1^m \to \Sigma_2^n\) be a code. $\mathcal{C}$ is called a $(\delta_{in},\delta_{out})$-approximate, $\ell_{in}$-input, $(\varepsilon,\ell_{out},q)$-single-circuit {\rm aLLRC} if there is a probabilistic TM $\mathcal{A}$ which on input $1^k$ outputs a (probabilistic) circuit $C^{\mathcal{L}}:[m] \to \Sigma_1^{\ell_{out}}$ such that for any fixed \(f \in \Ball_{1-\varepsilon,\delta_{in}}(\mathcal{L})\) there is an index $i \in [\ell_{out}]$ such that the $i$-th output of $C^{\mathcal{L}}$ $\delta_{out}$-computes $f$ with good probability:
\[
\forall f \in \Ball_{1-\varepsilon,\delta}(\mathcal{L}): \Pr_{\mathcal{A}}\Big[ \exists i \in [\ell_{out}]:\Pr_{r, x \in [k]}\left[ C^{\mathcal{L}}(x)_i = f(x)\right] \geq 1-\delta_{out}\Big] \geq \frac{3}{4}
\]
\end{definition}
A few remarks are in order. First, why is this a useful definition? The key idea is the following: in standard concatenation, an outer code calls the inner code to produce a list of circuits, on which it typically needs to check every circuit for one that agrees with the target function (or else picks a random one and simply hopes it is good). If each of these circuits uses separate queries, this means the outer code must pay $q_{inner}$ queries for every circuit in the list, blowing up the complexity by a factor of the list size. On the other hand, if there is only a single circuit (that in some sense \textit{re-uses} the queries to compute all the list elements), the outer algorithm only pays $q_{inner}$, saving the factor in the list size.

Finally, when the input list size is $1$, we call the above a single-circuit aLLDC, and when $\delta_{in},\delta_{out}=0$ just a single-circuit LLDC. 

\subsection{Single-Circuit aLLRC Concatenation}

We now present a near-optimal concatenation theorem for (single-circuit) LLRCs. Recall the \textit{concatenation} of two codes $\mathcal{C}_1$ and $\mathcal{C}_2$ encodes a message $x$ by first applying $\mathcal{C}_1$, then encoding each symbol with $\mathcal{C}_2$. It is well known that LLRC's can be concatenated up to losing polynomial factors in query complexity \cite{impagliazzo2008uniform}. Below, we show how this loss can be avoided.

To state the result, we need to use \textit{fine-grained query complexity}, discussed in \Cref{rem:fine-grained}, which we recall for a direct product code is the total number of $\Sigma_1$-symbols queried by the decoder in a single $(\Sigma_1)^r$ encoded symbol (recall our decoder frequently sub-samples symbols and only reads a few of their bits). We will write the fine-grained query complexity as capital $Q$ rather than lowercase $q$ to distinguish from prior sections.

\begin{proposition}[Single-Circuit LLRC Concatenation]\label{prop:basic-concat}
    Fix a constant $c \in [0,1)$, error parameter $\varepsilon>0$, input list size $\ell'_{in} \in \mathbb{N}$, and for every $t \in [1+\log \frac{1}{\varepsilon}]$ fine-grained query complexities $Q^{(t)},Q'^{(t)}$, and output list sizes $\ell'^{(t)}_{in},\ell^{(t)}_{out} \in \mathbb{N}$. Let $\mathcal{C}_1$ and $\mathcal{C}_2$ be codes such that for every $t$:
    \begin{enumerate}
        \item \textbf{Outer Code:} $\mathcal{C}_1:\Sigma_1^k \to \Sigma_2^{n_1}$ is a $(\delta_{in},\delta_{out})$-approximate, $(\ell_{out}'^{(t)}\log\frac{1}{\varepsilon})$-input, $(\frac{2^{-t}}{\log\frac{1}{\varepsilon}},\ell_{out}^{(t)},Q^{(t)})$-LLRC
        \item \textbf{Inner Code:} $\mathcal{C}_2: \{0,1\}^{\log|\Sigma_2|} \to \Sigma_3^{n_2}$ is a
        $(\delta_{in}',\delta_{in}/2)$-approximate, $(\ell'_{in})$-input $(c+2^{t-2}\varepsilon,\ell_{out}'^{(t)},Q'^{(t)})$-LLRC
    \end{enumerate}
    Then $\mathcal{C}_1 \circ \mathcal{C}_2: \Sigma_1^{k} \to \Sigma_3^{[n_1] \times [n_2]}$ is a $(\delta'_{in},\delta_{out})$-approximate, $\ell'_{in}$-input, $(c+\varepsilon,q,\ell_{out})$-LLRC with
    \begin{enumerate}
        \item \textbf{Query Complexity:} $q \leq O\left(\log(Q^{(T)})\sum\limits_{t \in [\log \frac{1}{\varepsilon}+1]}Q'^{(t)}Q^{(t)}\right)$
        \item \textbf{List Size:} $\ell_{out} \leq \sum\limits_{t \in [\log \frac{1}{\varepsilon}+1]}\ell_{out}^{(t)}$
    \end{enumerate}
    where $T=1+\log \frac{1}{\varepsilon}$.\footnote{For simplicity, we assume $1+\log\frac{1}{\varepsilon}$ is integral and $Q^{(t)}$ is increasing in $t$ as is typically the case. One can replace the below $\log Q^{(1+\log \frac{1}{\varepsilon})}$ with $\max_t \log Q^{(t)}$ otherwise.} 
\end{proposition}
\begin{proof}
Fix an input word $\mathcal{L} \in ((\Sigma_2)^{\ell'_{in}})^{[n_1] \times [n_2]}$ and a function $f \in \Ball_{1-\varepsilon,\delta'_{in}}(\mathcal{L})$ of the concatenated code. For each $i \in [n_1]$, let $\mu(i)$ denote the fraction of $j \in [n_2]$ such that $\mathcal{L}_{i,j}$ $\delta_{in}'$-computes $f$. In \cite{impagliazzo2008uniform}, the authors observe by Markov there must be an $\Omega(\varepsilon)$ fraction of indices $i \in [n_1]$ with $\mu(i) \geq c+\Omega(\varepsilon)$. We prove a refinement of this observation that allows us to sharpen the bound: up to log factors, there exists some $t$ such that at least a $2^{-t}$ fraction of $i$ have $\mu(i) \geq c+ 2^t\varepsilon$. As a result, either the inner or outer code (or both), can be run assuming much larger input $\varepsilon^{(t)}$.
\begin{claim}\label{claim:bucketed-Markov}
    There exists $ t \in [1+\log\frac{1}{\varepsilon}]$ such that at least a $\frac{2^{-t+1}}{3(1+\log\frac{1}{\varepsilon})}$ fraction of $i \in [n_1]$ have $\mu(i) \geq c+2^{t-2}\varepsilon$.
\end{claim}
We defer the proof of the claim which is a simple bucketed Markov argument. Now, note that while we  cannot explicitly find this $t$ without knowledge of $f$, since a good $t$ must exist, we will simply run the inner and outer decoders for every $t$. Our final output circuit will wire these in parallel (or more accurately success-amplified versions of the circuits), meaning we pay the sum of query costs and list sizes.


Formally, let $T=\lceil\log \frac{1}{\varepsilon} +1 \rceil$ and fix $t'$ satisfying the above guarantee, and call an index $i \in [n_1]$ `good' if at least a $c+\frac{2^t\varepsilon}{4}$-fraction of symbols $w_{i,j}$ $\delta_{in}'$-compute $\mathcal{C}_1 \circ \mathcal{C}_2(f)$. By assumption, on good indices our inner code $\mathcal{C}_2$ (run at level $t'$) outputs a single-circuit $\mathcal{C}_i$, one of whose $\ell'^{(t)}_{out}$ output indices (say the $\ell$-th) $\delta_{in}/2$-computes $f$. By taking a majority over each $\mathcal{C}_i$ $O(\log Q^{(T)})$ times, we can ensure the $\ell$-th index deterministically $\delta_{in}$-computes $f$ for a random fixed choice of $r$ with probability at least $1-O(\frac{1}{Q^{(T)}})$.

We now run the outer code for every $t$ using these inner circuits. Pick a random string $r$ to use for the internal circuits. For the `good' $t$, our outer code queries at most $Q^{(t)} \leq Q^{(T)}$ inner circuits, so by a union bound, the probability every (amplified) `good' inner circuit $\mathcal{C}_i(\cdot,r)$ has an output index $O(\delta_{in})$-computing $f$ is at least (say) $99\%$.

Conditioned on this event, we are exactly in the approximate list-recovery setting. In other words, our outer code $\mathcal{C}_1$ has (simulated) query access to a list of inner words on each index (given by the output indices of each $\mathcal{C}_i$), a $c+2^{t-2}\varepsilon$ fraction of which contain a function which $\delta_{in}$-computes $f$. Thus our outer code makes $Q^{(t)}$ total queries to the inner $\mathcal{C}_i$ and produces a single-circuit $\ell_{out}^{(t)}$-dimensional output one of whose indices $\delta_{out}$-computes $f$ with probability at least $1/2$. Repeating the process independently $O(1)$ times (wiring the output circuits in parallel) we can reach $3/4$ success as desired. The query complexity and list size are immediate from construction.


It is left to prove \Cref{claim:bucketed-Markov}. The key observation, as in prior work, is simply that by assumption we have global expectation $\mathbb{E}_{i \in [n_1]}[\mu(i)]=c+\varepsilon$, since $\mu$ averages over a random $j \in [n_2]$. Assume towards contradiction that for every $t$, the fraction of $i \in [n_1]$ with $\mu(i) \in [c+2^{t-3}\varepsilon,c+2^{t-2}\varepsilon]$ is less than $\frac{2^{-t+1}}{3\log\frac{1}{\varepsilon}}$. Then:
\begin{align*}
    \mathbb{E}[\mu(i)] &= \sum_{a \in [0,1]}\Pr[\mu(i)=a]a\\
    &\leq c+\frac{\varepsilon}{2}+\sum\limits_{t \in [1+\log \frac{1}{\varepsilon}]} \Pr\Big[\mu(i) \in [c+2^{t-2}\varepsilon,c+2^{t-1}\varepsilon]\Big]2^{t-1}\varepsilon\\
    &\leq c+\frac{\varepsilon}{2}+\sum\limits_{t \in [1+\log \frac{1}{\varepsilon}]}\frac{2^{-t+1}}{3(1+\log\frac{1}{\varepsilon})}2^{t-1}\varepsilon\\
    &< c+\varepsilon
\end{align*}
giving the desired contradiction. In particular, there must exist some $t$ with $> \frac{2^{-t+2}}{3(1+\log \frac{1}{\varepsilon})}$ fraction of $i \in [n_1]$ satisfying $\mu(i) \geq c+2^{t-2}\varepsilon$ as desired.
\end{proof}
We remark the $\log Q^{(t)}$ term in the final query complexity can be removed assuming $\varepsilon \geq \poly(\frac{1}{n_1})$ is sufficiently large.

\subsection{A Poly-Size, Low Error aLLDC}
In this subsection, we construct the full inner code of our locally computable aLLDCs:
\begin{proposition}[The Inner Code]\label{prop:alph-red-inner}
        For all large enough $N \in \mathbb{N}$, $\delta_{out} \geq \Omega(\frac{1}{\poly(N)})$, and $\varepsilon>0$, there exists a binary code $\mathcal{C}: \{0,1\}^N \to \{0,1\}^{N'}$ that is a $\delta_{out}$-approximate, $(\varepsilon,\ell_{out},q)$-single-circuit LLDC with
    \begin{enumerate}
        \item \textbf{Rate:} $R \geq \poly(N^{-1})\cdot2^{-O(\log^2 \frac{1}{\varepsilon})}$
        \item \textbf{Query Complexity:} $q \leq \frac{\polylog (\frac{1}{\delta_{out}},\frac{1}{\varepsilon})}{\varepsilon^2}$
        \item \textbf{List Size:} $\ell_{out} \leq \frac{\polylog (\frac{1}{\varepsilon})}{\varepsilon^2}$
        \item \textbf{Locality:} $\poly(\delta_{out}^{-1})\log(\frac{1}{\varepsilon})$
    \end{enumerate}
    Moreover $\mathcal{C}$ is locally encodable in TIME$(\polylog(N)\log^2(1/\varepsilon))$ and decodable by logspace uniform circuits of size $\poly(\log(N),\frac{1}{\varepsilon})$ and depth $O(\log^2(\frac{\log N}{\delta_{out}})+\log \frac{1}{\varepsilon})$
\end{proposition}
\Cref{prop:alph-red-inner} is the composition of two codes. A locality-efficient binary aLDC:
\begin{proposition}[Amplification in the unique-decoding regime]\label{lem:unique-amp}
    For all large enough $N \in \mathbb{N}$, $\delta_{out} \geq \Omega(\frac{1}{\poly(N)})$, and every $N' \geq \frac{1}{\poly(\delta_{out})}N$ there exists a binary code $\mathcal{C}: \{0,1\}^N \to \{0,1\}^{N'}$ that is a $\delta_{out}$-approximate, $(0.9,q)$-query LDC  for $q \leq \polylog (\frac{\log N}{\delta_{out}})$.
    Moreover, $\mathcal{C}$ is locally encodable in Time$(\poly(\log(N),\frac{1}{\delta_{out}}))$ and can be decoded by logspace uniform circuits of size $\poly(\log(N),\frac{1}{\delta_{out}})$ and depth $O(\log^2(\frac{\log N}{\delta_{out}}))$.
\end{proposition}
and a binary direct product code in the list decoding regime from \cite{impagliazzo2008uniform,ben2010note}:
\begin{proposition}[Polynomial Rate Binary aLLDC]\label{thm:poly-rate-aLLRC}
    Let $p$ be a prime power, $\varepsilon \geq \Omega(1/p)$, and $N=\mathbb{F}_p^m$ for any $m \geq 10$. There exists a binary code $\mathcal{C}: \{0,1\}^N \to \{0,1\}^{N'}$ that is a $0.001$-approximate, $(\varepsilon,\ell_{out},q)$-single-circuit LLDC with
    \begin{enumerate}
        \item \textbf{Rate:} $R \geq \poly(N^{-1})\cdot2^{-O(\log^2 \frac{1}{\varepsilon})}$
        \item \textbf{Query Complexity:} $q \leq \tilde{O}\left(\frac{1}{\varepsilon^2}\right)$
        \item \textbf{List Size:} $\ell_{out} \leq \tilde{O}(\frac{1}{\varepsilon^2})$
        \item \textbf{Locality:} $O(\log \frac{1}{\varepsilon})$
    \end{enumerate}
    Moreover, $\mathcal{C}$ is locally encodable in Time$(\polylog(N)\cdot\log^2(1/\varepsilon))$ and can be decoded by logspace uniform circuits of size $\poly(\log(N),\frac{1}{\varepsilon})$ and depth $O(\log(\frac{1}{\varepsilon}))$.    
\end{proposition}

We cover these in separate subsections below. \Cref{prop:alph-red-inner} is immediate from combining the two.
\begin{proof}[Proof of \Cref{prop:alph-red-inner}]
    The stated code is obtained by composing appropriate settings of \Cref{thm:poly-rate-aLLRC} ($\mathcal{C}_2$) and \Cref{lem:unique-amp} ($\mathcal{C}_1$). Namely, let $p$ be the smallest prime power greater than $\Omega(\varepsilon^{-3/4})$. We may instantiate \Cref{lem:unique-amp} as an $[N'=p^m,N]$ code for $m \geq 10$ and $N' \leq N\cdot \poly(\log N, \frac{1}{\delta_{out}},\frac{1}{\varepsilon})$. This gives the right setting to compose with (a constantly success amplified) \Cref{thm:poly-rate-aLLRC} which requires such a prime power message length (i.e. take the encoding $\mathcal{C}_2(\mathcal{C}_1(f))$).

    Now, by success amplification (see discussion after \Cref{def:aLLDC}), given a word $w$ to the composed code, we may assume $\mathcal{C}_2$'s list-decoder (over the choice of randomness) produces with $98\%$ probability a deterministic circuit one of whose outputs $0.1$-computes $\mathcal{C}_1(f)$. Conditioned on this event, we may view this decoding as the input to our unique decoder to $\mathcal{C}_1$. This (again, suitably amplified), produces a function that is $1-\delta_{out}$ close to $f$ with $99\%$ probability as desired (union bounding the success is clearly greater than the desired $\frac{3}{4}$).

    The query complexity, rate, encoding time, and circuit complexity are immediate from construction. Namely in composition the queries and rate are the product of the constituent code, the circuit size grows by at most the product of the circuit sizes and query complexity, the depth is additive, the encoding locality is the product, and the encoding time is at most the time of the inner code plus the product of the outer codes time and locality.
\end{proof}

\subsubsection[The list decoding regime]{The List Decoding Regime (\Cref{thm:poly-rate-aLLRC})}
In this subsection, we prove \Cref{thm:poly-rate-aLLRC}. As discussed above, this code is simply given by concatenating three standard aLLDCs modified to the single-circuit list-recovery setting: \cite{impagliazzo2008uniform}'s subspace code, the complete complex, and the Hadamard code. Indeed this code is already implicit in \cite{impagliazzo2008uniform} and explicitly used in \cite{ben2010note} -- the only thing new is giving an improved analysis of the query complexity and list size via \Cref{prop:basic-concat}.

We first state the guarantees of \cite{impagliazzo2008uniform}'s subspace code in the single-circuit aLLRC model:

\begin{theorem}[Subspace Code \cite{impagliazzo2008uniform}]\label{lem:subspace-ss}
    For any $\varepsilon>0$, let $p \geq \Omega(\frac{1}{\varepsilon^{3/4}})$ be a prime power, $m \geq 10$, $V=\mathbb{F}_p^m$, and $S$ be the set of $8$-dimensional subspaces of $V$. Then the direct product code $C_{(V,S)}$ 
    is an $(0.0005,0.001)$-approximate, $\ell_{in}$-input, $(\varepsilon,\ell_{out},q)$-single-circuit LLRC with
    \begin{enumerate}
        \item \textbf{Rate:} $R \geq \frac{1}{\poly(|V|)}$
        \item \textbf{Query Complexity:} $q \leq O(\frac{1}{\varepsilon^2})$
        \item \textbf{List Size:} $\ell_{out} \leq O(\frac{\ell_{in}}{\varepsilon})$
        \item \textbf{Locality:} $r \leq O(p^8)$
    \end{enumerate}
    Moreover $C_{(V,S)}$ can be locally encoded in Time$(O(\log \frac{N}{\varepsilon}))$ and decoded in logspace uniform size $O(\poly(\log(N),\frac{1}{\varepsilon}))$ and depth $O(\log (\frac{1}{\varepsilon}))$.
\end{theorem}
The only difference in the above result and \cite{impagliazzo2008uniform} as stated is that one must run their decoding algorithm starting from every function in the input list.\footnote{Technically, this also then requires setting the sampling parameters so that we can union bound over the list, but this is a fairly low order factor.} Queries can be re-used across the list-elements, but naively cannot be re-used across the $O(1/\varepsilon)$ different starting points they use as advice, which results in the $O(1/\varepsilon^2)$ complexity. We remark this actually can be improved to $q \leq O(\frac{1}{\varepsilon})$ by modifying their algorithm further, but this will not help the final query complexity of the concatenated code so we omit this fact.


The subspace code still has reasonably large log-alphabet (on the order of $\poly(\varepsilon^{-1})$), but it is critically \textit{independent of the number of vertices $N=q^m$}. This allows us to concatenate with the complete complex without incurring large blowup in $N$:
\begin{theorem}[Complete Complex Code {\cite{impagliazzo2008uniform}}]\label{lem:complete-ss}
    There exists $C>0$ such that for any $\varepsilon,\delta>0$, $\ell_{in} \in \mathbb{N}$, $d \geq C\ln \frac{\ell_{in}}{\varepsilon}$, and $N \geq d$ the direct product code $\mathcal{C}$ on the complete complex $([N],\binom{[N]}{d})$ is a $0.0005$-approximate, $\ell_{in}$-input, $(\varepsilon,\ell_{out},q)$-single-circuit aLLRC with
    \begin{enumerate}
        \item \textbf{Query Complexity:} $q \leq \tilde{O}(\frac{1}{\varepsilon^2})$
        \item \textbf{List Size:} $\ell_{out} \leq \tilde{O}(\frac{\ell_{in}}{\varepsilon})$
        \item \textbf{Locality:} $O(\log \frac{\ell_{in}}{\varepsilon})$
    \end{enumerate}
    Moreover $\mathcal{C}$ can be locally encoded in Time$(O(\log N\log \frac{1}{\varepsilon}))$ and locally decoded in logspace uniform size $\poly(\log(N),\frac{1}{\varepsilon})$ and depth $O(\log \frac{1}{\varepsilon})$.
\end{theorem}
The above requires the exact same modifications as discussed for the subspace code.

Finally, we need to move from the small alphabet setting afforded by the complete complex to the binary setting. This is classically done via the \textit{Hadamard code}, which we recall maps a string $x \in \{0,1\}^k$ to $Had(x)_\gamma \coloneq \langle x, \gamma\rangle$. Goldreich and Levin \cite{goldreich1989hard} gave a low-complexity local list decoding algorithm for the Hadamard code arbitrarily close to distance $1/2$. In fact, Goldreich and Levin's analysis already gives a single-circuit LLDC (i.e.\ the query locations do not depend on the advice, and so can be re-used) of the following form:\footnote{We refer the reader to \cite{Trevisan2005GoldreichLevin} for an exposition of the result in which the claimed version below is particularly clear.}
\begin{theorem}[Hadamard Code \cite{goldreich1989hard}]\label{lem:Hadamard-ss}
    For every $N \in \mathbb{N}$ and $\varepsilon>0$, the Hadamard code $\mathcal{C}_{Had}: \{0,1\}^N \to \{0,1\}^{2^N}$ is a $(\frac{1}{2}+\varepsilon,\ell_{out},q)$-single-circuit LLDC with 
    \begin{enumerate}
        \item \textbf{Query Complexity:} $q \leq O(\frac{1}{\varepsilon^2})$
        \item \textbf{List Size:} $\ell \leq O(\frac{1}{\varepsilon^2})$
    \end{enumerate}
    Moreover $\mathcal{C}_{Had}$ can be locally encoded in Time$(O(N))$, and locally decoded by logspace uniform circuits of size $\poly(N,\frac{1}{\varepsilon})$ and depth $O(\log \frac{1}{\varepsilon})$.
\end{theorem}

The main disadvantage of the Hadamard code is of course its exponential blocklength, but this is not an issue since we will be applying it to blocks of size $N=O(\log \frac{1}{\varepsilon})$ after concatenating with the complete complex. Concatenating these three codes gives the desired polynomial rate binary aLLRC from \Cref{thm:poly-rate-aLLRC}:

\begin{proof}[Proof of \Cref{thm:poly-rate-aLLRC}]
    The proof is essentially immediate from combining \Cref{prop:basic-concat}, \Cref{lem:subspace-ss}, \Cref{lem:complete-ss}, and \Cref{lem:Hadamard-ss}. Namely we take the three-wise concatenation starting with the subspace code, then the complete complex, then Hadamard. More explicitly, by \Cref{prop:basic-concat}, the concatenation of the subspace code on $N$ vertices and the complete complex code of dimension $O(\log \frac{1}{\varepsilon})$ is a $0.001$-approximate, $\ell_{in}$-input $(\varepsilon,\ell_{out},q)$-LLRC with 
    \[
    q \leq \tilde{O}(\frac{1}{\varepsilon^2}) \quad \quad \text{and} \quad \quad \ell_{out} \leq \tilde{O}(\frac{\ell_{in}}{\varepsilon})
    \]
    Applying \Cref{prop:basic-concat} again, concatenating this with the Hadamard code gives a $0.001$-approximate, $(\varepsilon,\ell_{out},q)$-LLDC with 
    \[
    q \leq \tilde{O}(\frac{1}{\varepsilon^2}) \quad \quad \text{and} \quad \quad \ell_{out} \leq \tilde{O}(\frac{1}{\varepsilon^2})
    \]
    The locality, encoding, and decoding complexity follows from directly from construction (i.e.\ from code concatenation). The locality is inherited from the complete complex step. The bound on encoding time follows from tracing back encoding time and locality of each constituent code noting that our message sizes for the subspace complex, complete complex, and Hadamard codes are $N_{subspace}=\poly(\log N,\frac{1}{\varepsilon})$, $N_{complete} = \poly(1/\varepsilon)$, and $N_{Had}=O(\log \frac{1}{\varepsilon})$ respectively. The resulting circuit complexity follows similarly.
\end{proof}

\subsubsection[The unique decoding regime]{The Unique Decoding Regime (\Cref{lem:unique-amp})}
In this subsection, we prove \Cref{lem:unique-amp}, our locality and query efficient unique decoding procedure that amplifies a circuit computing $f$ $99\%$ percent of the time to one that computes $f$ except with error $\delta_{out}$. This is the easier `unique decoding regime', and is well studied in the hardness amplification literature (e.g.\ \cite{guruswami2006hardness}) and in coding theory \cite{mon2024approximate}. We will use a variant of a simple construction of Guruswami and Kabanets \cite{guruswami2006hardness} from the former, which encodes $f$ using a direct-product over an expander-walk, then `flattens' with an error correcting code.

To avoid losing extra $\poly\log(N)$ factors in the locality and query complexity, instead of using expander walks we'll take a direct product over neighborhoods of the KMS expander from \Cref{sec:instantiations} and `flatten' using a locally decodable code. To handle arbitrary blocklengths we repeat the output as in \Cref{lem:many-to-all}, the same approach we used previously to build codes of all message lengths.

Before giving the proof, let's formally state these two main components. First we recall our logspace uniform construction of expanders from \Cref{sec:instantiations} in the desired parameter regime:
\begin{corollary}[Log-Space Samplers]
    There exists $C>0$ such that the following holds. Fix any $N \in \mathbb{N}$, $\beta \in (0,1)$. There exists $N' \in [N,\frac{C}{\poly(\beta)}N]$ and a strongly explicit, $\poly(\beta^{-1})$-regular $\beta$-sampler graph $G_{KMS}=([N'],[N'],E)$. Moreover, $G_{KMS}$ has a logspace uniform circuit implementation of size $\poly(\log(N),\beta^{-1})$ and depth $O(\log\log(N)\cdot \log \frac{1}{\beta})$.
\end{corollary}
In particular, the above follows from taking the colored swap walk on ($\kms[1]$,$\kms[2]$) in the HDX underlying the hypergraph system in \Cref{prop:graph-system-for-log-rate-graph} with $\fsize=\poly(\beta)$, i.e.\ the $2$-dimensional KMS complex.

Our second component is a standard construction of polylog query binary locally decodable codes given by concatenating Reed-Muller with any good error correcting code (see e.g.\ \cite{arora1998probabilistic,yekhanin2012locally}).
\begin{theorem}[Polylog Query LDCs]\label{thm:subpoly-lldc}
    For every large enough $N \in \mathbb{N}$ and $N' \geq N^{2+o(1)}$ a multiple of $2^{O(\log^{2/3}(N))}$, there is an explicit binary code $\mathcal{C}_N:\{0,1\}^N \to \{0,1\}^{N'}$ that are $(0.99,O(\log^2(N)))$-locally decodable and can be locally encoded in Time$(\poly(N))$ and locally decoded in logspace uniform size $O(\poly(N))$ and depth $O(\log^2(N))$.
\end{theorem}
We now briefly overview the construction itself. As discussed, our encoding consists of two main parts, sandwiched with repetition codes to ensure the block and message lengths of each part match up. Namely, starting with a message in $x \in \{0,1\}^k$, we first repeat $x$ to match the size of our sampler graph $G_{KMS}$, concatenate each symbol with the desired LDC, and repeat the result to reach the final desired blocklength.

\begin{proof}[Proof of \Cref{lem:unique-amp}]
    It is sufficient to prove that the concatenation of the first $G_{KMS}$-direct-product encoding with $\beta=O(\delta_{out})$ (and therefore degree $\poly(1/\delta_{out})$) and vertex size $N' \in [N,\poly(\delta_{out}^{-1})N]$, with the LDCs from \Cref{thm:subpoly-lldc} is a $\delta_{out}$-approximate $(0.99,q)$-query LDC. \Cref{lem:many-to-all} then implies the full code (with initial and final repetitions) is the desired aLDC. Denote the former code as $\mathcal{C}$.

    Fix a word $w$ and target function $f \in \Ball_{0.01}(w)$. We decode $f$ layer by layer. By Markov's inequality the fraction of LDC `blocks' (that is encodings of symbols in the first layer DP code by the LDC) that are flipped on more than $20\%$ of their coordinates is at most $1/10$. As a result, the LDC decoding on $90\%$ of the blocks (and therefore on $90\%$ of the message symbols into that layer) are correct with probability $2/3$ over the randomness of the LDC.

    Now, conditioned on the success of the above decoders, it is immediate from sampling that all except a $O(\delta_{out})$-fraction of message bits see at least $80\%$ `good' blocks (i.e.\ blocks with fewer than $20\%$ corruptions above). Querying such a block results in a correct bit at least $2/3$ of the time, so a Chernoff bound promises that taking the majority decoding across $O(\log(\frac{1}{\delta_{out}}))$ random encodings containing the bit $x$ decodes the correct bit except with probability $1-O(\delta_{out})$.

    Altogether, union bounding over our failure events and $O(\delta_{out})$ `bad' vertices that fail to see sufficiently many good LDC blocks, our decoding succeeds with probability at least $1-O(\delta_{out})$ over a random message bit and the internal randomness of our algorithm, as desired. It is left to analyze the various parameters of the algorithm, which we do separately below:    

    \paragraph{Blocklength} Repeating the initial symbol via \Cref{lem:many-to-all} from $N$ to the next smallest vertex size of $G_{KMS}$ increases blocklength by at worst a factor of $\poly(\delta_{out})$. The direct product encoding increases the log-alphabet by $\poly(\delta_{out})$, then encoding each symbol with the LDC then squares the length of each block, increasing by another $\poly(\delta_{out})$ factor. Finally repetition brings the blocklength to any desired $N' \geq N\poly(\delta_{out}^{-1})$.    

    \paragraph{Query Complexity}
    
    The base decoder makes $O(\log(\frac{1}{\delta_{out}}))$ queries, each of which are a call to the inner LDC, which makes $O(\log^2 \frac{1}{\delta_{out}})$ queries. The repetition codes make $O(1)$ queries each, so the total query complexity is $O(\log^3\frac{1}{\delta_{out}})$.

    \paragraph{Encoding Time}

    Each encoded bit depends only on a single `block' of the $G_{KMS}$ DP code (which has degree $\poly(\delta_{out}^{-1})$) further encoded by the LDC. The LDC encoding takes time $\poly(\delta_{out})$. Computing each message bit in the LDC layer simply corresponds computing the original message on the neighbors of that block, which can be done in time $\poly(\log(N),\delta_{out}^{-1})$ by our implementation of KMS. The repetition encodings only take $O(\log(N))$ time. Note the locality of the code is upper bounded by the encoding time (of course it is in fact better, namely it is just the degree of $G_{KMS}$, but we will not use this).

    \paragraph{Decoding Complexity}

    Our decoding procedure requires 1) sampling $O(\log \frac{1}{\delta})$ random neighbors in $G_{KMS}$ and taking a majority 2) decoding these neighbors by running the appropriate LDC, and 3) decoding repetition codes. Each of these procedures can be performed in logspace uniform size $\poly(\log(N),\delta_{out}^{-1})$ and depth $O(\log^2(\frac{\log(N)}{\delta_{out}}))$ as desired.
\end{proof}

\subsection{Putting it All Together}
We are essentially ready to prove the main result of the section. All that is missing is the following standard construction of list-decodable codes used in our large locality result. The construction itself is simply Reed-Solomon concatenated with any good binary inner code below the Johnson list-decoding radius, padded to reach any message length:
\begin{theorem}[Binary List-Decodable Codes (see e.g.\ \cite{sudan1997decoding,guruswami2000list,vadhan2012pseudorandomness})]\label{thm:binary-list-decode}
    For every large enough $N \in \mathbb{N}$ and $\varepsilon>\frac{1}{\poly(N)}$, there exists a binary linear code $\mathcal{C}:\{0,1\}^N \to \{0,1\}^{N'}$ with rate $\poly(\varepsilon)$ that is $(\varepsilon,O(\frac{1}{\varepsilon^2}))$-list-decodable. Moreover the code is encodable in TIME$(\poly(N,\frac{1}{\varepsilon}))$ and decodable in logspace uniform size $O(\poly(N,\frac{1}{\varepsilon}))$ and depth $\tilde{O}(\log^2 N)$.\footnote{Here $\tilde{O}$ hides $\log\log(N)$ factors. We remark the depth bound is folklore. It is not explicitly stated in \cite{sudan1997decoding} but the decoding algorithms uses only root finding and linear system solving over a field size $O(N)$, which are in logspace uniform $\tilde{O}(\log^2(N))$-depth.}
\end{theorem}

We are now ready to prove both of the main statements of this section.

\begin{proof}[Proofs \Cref{cor:binary-polylog-aLLDCs-restated} and \Cref{thm:KO-locality-reduced}]
    We start with the former. Recall \Cref{thm:polylog-aLLDCs} (setting $\delta \leq O(\frac{1}{\log^5(N)})$ sufficiently small) is an $(O(\frac{1}{\log^5(N)}),0.01)$-approximate, $\ell_{in}$-input, $(\varepsilon,O(\frac{\ell_{in}}{\varepsilon}),\tilde{O}(\frac{\log^5(N)}{\varepsilon^2}))$-LLRC for any $\varepsilon>0$ over a log-alphabet size of $\poly(\frac{\log(N)}{\varepsilon})$ (note we may also take this polynomial to be as large as desired just by increasing $\fsize$ polynomially). We may therefore concatenate with \Cref{thm:binary-list-decode} and \Cref{prop:basic-concat} implies the final code has the desired list-decoding parameters. Rate, encoding, and decoding complexity are immediate from concatenation (noting that the message length `$N$' in the inner code \Cref{thm:binary-list-decode} is set to the log-alphabet size $\poly(\frac{\log(N)}{\varepsilon})$).

    The proof of the latter result is similar. The only change is we concatenate with \Cref{prop:alph-red-inner} with the same setting of parameters. The list decoding parameters follow similarly. The decoding circuit depths are additive, and the original outer codes depth $O(\log(\frac{\log N}{\varepsilon})\log\log N)$ dominates. The circuit sizes are at worst bounded by the product of all the circuit sizes and number of queries, which remains $\poly(\log(N),\frac{1}{\varepsilon})$ as well. Finally the encoding time is at most the time of the inner code, plus the locality of the inner code times the encoding time of the outer code which is at most $\polylog(N)\cdot \log^2\frac{1}{\varepsilon}$ as desired.
\end{proof}
Finally, we remark that \Cref{cor:constant-rate-binary} follows from exactly the same argument, since concatenating with \Cref{thm:binary-list-decode} only loses constant factors in the rate.

\section{Hardness Amplification}\label{app:hardness}
In this section, we will give some consequences of our
results for hardness amplification, constructions of 
reliably difficult problems from occasionally difficult
problems.  Hardness amplification is important for a quantitative
understanding of complexity, but also for cryptography and
pseudorandomness, where reliable difficulty is useful.

The different versions of codes we give will correspond to 
hardness amplification for different purposes, in different
settings. The connection is to view the somewhat hard function as a message to be sent.  The reliably hard function is the coding of
that message.  From a highly noisy version of this code (i.e., an algorithm that solves the problem just slightly better than random guessing), we could decode an approximate version of the message, i.e., an algorithm that solves almost all instances of the original problem.  From this contradiction with the assumed difficulty of the original problem,  the constructed function must not even be solvable slightly better than guessing. 

Often, we want to look at hardness within a complexity
class, so we want the reliably hard problem
to have essentially the same complexity as the sometimes difficult problem.  This will correspond to codes with very efficient locally encoding algorithms.  In other settings, we assume the worst-case difficulty of the somewhat hard problem is exponential, and wish the reliably hard problem to still be exponential with close to the same exponent.  This will correspond to codes whose decoding algorithms are sub-linear in the length of the message, but still a power of the message.   If we want to preserve the exact quantitative difficulty of problems (as a function of the input size), we need the input size for the constructed function to be very close to that of the original problem.  This corresponds to codes with (relatively) high rates.  For non-uniform models, exact list size is usually less important, but for uniform models, it is key.  Our codes have different trade-offs between rate, locality, list size, and the complexity of decoding, so lead to different hardness amplification results.  In the following, we will only give a few consequences of these codes, and are highlighting simplicity over optimality.

\subsection{Hardness amplification within complexity classes}

We will first consider hardness amplification that preserves the
complexity of a sometimes difficult problem.   This will give us fine-grained time/error tradeoffs possible within relatively small complexity classes such as $NP \cap co-NP$ or $P^{NP}$, closed under polynomial time Turing reductions. 
We will give a version for uniform models and for non-uniform models, but they both will follow directly from the same coding and decoding algorithms.

\begin{theorem}[Uniform complexity-preserving hardness amplification]\label{thm:ucpha}
Let $T(n)$ be a time complexity,   $1> \varepsilon(n) \in 2^{-o(\sqrt{n})} , $ and $A$ a non-negative integer.  Let 
\[
f: \{0,1\}^n \rightarrow \{0,1\},
\]
such that
\[
f \not\in Heur_{\frac{1}{n^A}}-BPTIME(T(n))/O(\log (n/\varepsilon))) .
\]
Then there is a
length $n' = n(1+o(1))$ and a $F: \{0,1\}^{n'} \rightarrow \{0,1\}$ so that $F \in P^f $ and $F \not\in Heur_{\frac{1}{2} - \varepsilon(n)}-BPTIME(T(n) \poly (\frac{\varepsilon(n)}{n}))$
\end{theorem}

\begin{theorem}[Non-uniform complexity preserving hardness amplification]\label{thm:nucpha}
Let $S(n)$ be a circuit size function, $1> \varepsilon(n) \in 2^{-o(\sqrt{n})}$$, A$ a non-negative integer.  Let 
\[f: \{0,1\}^n \rightarrow \{0,1\}\]
such that
\[f \not\in Heur_{\frac{1}{n^A}}-Size(S(n),n).\]  
Then there is a
length $n' = n(1+o(1))$ and a $F: \{0,1\}^{n'} \rightarrow \{0,1\}$ so that $F \in P^f $ and $F \not\in Heur_{\frac{1}{2} - \varepsilon(n)}-Size(S(n) \poly(\frac{\varepsilon(n)}{n}))$
\end{theorem}

We prove these two theorems via an identical construction.
First, let $N=2^n$ and view $f$ as an $N$ bit message, where
the $x$-th bit is $f(x)$.  
We use the code $C_1$ from \Cref{lem:unique-amp} (using $\delta_{out}= n^{-A}$ and $N=2^n$) to construct a new function
$f_1$ on $n_1= n+ O( \log n)$ bit inputs, where $f_1(y)$ is the $y$-th output bit of $C_1(f)$. 
$C_1$ has the following properties:
\begin{enumerate}
    \item $|C_1(f)|= N_1= N/\poly(1/\delta_{out}),$ so $n_1=\log N_1= n +O( \log n)$
    \item Each bit of $C_1(f)$ can be computed with queries to $f$ in time $\poly(\log N, 1/\delta_{out})= \poly(n)$.  Thus, $f_1 \in P^f$.  
    \item $f$ can be $\delta_{out}$-approximately uniquely decoded from any corrupted message with agreement $.9N'$ with $C_1(f)$ and the decoded message is represented as a circuit of size 
    $$\poly(\log N, 1/\delta_{out}) =\poly(n)$$ 
    with queries to the corrupted code word.
\end{enumerate}
Then if 
\[
f_1 \in Heur_{.1}-Size(S_1(n), n_1),
\]  
there is a circuit
$Circ_1$ of size $S_1(n)$ that agrees with $C(f)$ on $.9 N_1$ bit
positions $y$.   Using the decoding algorithm above yields a circuit $Circ$ that uses $Circ_1$ as a subroutine and otherwise has size
$\poly(n)$ that agrees with $f$ on all but $\delta_{out}N$ inputs $x$.
Since $Circ_1$ has size $S_1(n)$, $Circ$ will have size $S_1(n)\poly(n)$
So assuming $f \not\in Heur_{\delta_{out}}-Size(S_1 \poly(n), n)$, it follows that $f_1 \not\in Heur_{.1}-Size(S_1(n),n_1)$. 

We then apply the list-decodable code $C_2$ from Corollary A.2 to get a function $F$, where $F(z)$ is the $z$-th bit of $C_2(f_1)$.
This code has the following properties: 

\begin{enumerate}
\item The length of the encoded message is 
$$|C_2(f_1)|=N_2 = N_1\poly(\log N_1)2^{O(\log^2 1/\varepsilon)}.$$ Thus the input size to $F$ is $n_2=\log N_2= n_1+ O(\log n_1) + O(\log^2 1/\varepsilon)= n+O(\log n)+O(\log^2 1/\varepsilon)$.  Since $\varepsilon \in  2^{-o(n^{1/2})}$, $n_2=n+o(n)$.
\item Each output bit of $C_2(f_1)$ is computable in time $\poly\log(N_1) \cdot \log^2 1/\varepsilon =
\poly(n)$ with queries to the bits of $f_1$.
Thus, $F \in P^{f_1} \subseteq P^f$.  
\item From any circuit $Circ_2$ that differs from $F$ on fewer than $\frac{1}{2}- \varepsilon$ fraction of bit positions, we can compute in time $\poly\log(N_2) \cdot \poly(1/\varepsilon)$ a list of $\poly(1/\varepsilon)$ circuits $Circ_{1,i}$, one of which has $.9$ agreement with $f_1$.  
\end{enumerate}

If $F \in Heur_{\frac{1}{2}-\varepsilon}-Size (S_2(n), n_2)$ via circuit $C_2$, then the best circuit output on the list output by the decoder using $Circ_2$ as the corrupted message would be a $Circ_1 \in Size (S_2(n) \poly(n/\varepsilon))$ computing $f_1$ on all but $.1$ fraction of inputs.  So $f_1 \in Heur_{.1}-Size(S_2(n) \poly(n/\varepsilon), n_1)$. From what we said above, this implies
$f \in Heur_{\delta(n)} - Size( S_2(n) \poly(n) \cdot \poly(n/\varepsilon), n)$, a contradiction  for some  $S_2(n) < S(n) poly(\varepsilon /n)$.  

For \Cref{thm:ucpha}, the following partial characterization
of $Heur_{\delta(n)}-BPTIME(T(n))/a(n)$ will be convenient:
\begin{lemma}
If $f \in Heur_{\delta(n)}-BPTIME(T(n))/a(n)$, there is a randomized algorithm that given $n$, in time $T(n+a(n))\poly(n)$ returns a circuit $Circ_{guess}$ of size $S(n)= T(n+a(n)) \poly(n)$ so that with probability
at least $1/2^{-a(n)+1}$, $Circ_{guess}$  agrees with $f$ on all but $2 \delta(n)$ fraction of inputs
\end{lemma}
\begin{proof}
Let $A(n,r,advice)$ be the randomized algorithm witnessing $f \in Heur_{\delta(n)}-BPTIME(T(n))/a(n)$.
The randomized algorithm picks a random $r_0$ and a random
string $advice_0$ of length $a(n)$, and returns a circuit simulting $Circ_{guess}(x)= A(x, r_0, advice_0)$.  With probability $1/2^{a(n)}$, $advice_0=advice(r_0)$, and conditioned on this occuring, by Markov's inequality, $A(x,r_0, advice_0)$  differs from $f$ on more than twice the expected number of inputs with probability at most $1/2$.  
\end{proof}
The following gives a partial converse:
\begin{lemma}
If there is a time $T(n)$ randomized algorithm that produces circuits $Circ$ so that with probability $\alpha(n)$, $Circ(x)$ differs from $f(x)$ on at most a $\delta(n)$ fraction of inputs, then $f \in Heur_{2 \delta(n)}-BPTIME(T(n)+\poly(n))/O(\log 1/\alpha+\log 1/\delta)$.  
\end{lemma}
\begin{proof}
Consider a pairwise independent sequence of random strings $r_1,r_2,...r_K$ so that given $i$, $r_i$ can be computed in polynomial time.  (For example, if we let $r_i= a*i+b$ for elements $a$ and $b$ chosen from a finite field of size $2^{|r|}$, where we
interpret each $i$ as a different field element.)   The probability that no $r_i$ generates a circuit $Circ_i$ with the desired property is at most $\frac{1}{\alpha K}$ by Chebyshev's inequality.  Picking $K = \frac{1}{\alpha \delta}$ makes this probability at most $\delta$.  If we pick the randomness used to generate the sequence at random, and use as advice an $i$ for which this holds, except with
probability $\delta$, the conditional error of the generated circuit is at most $\delta$.  With probability $\delta$, it could be 1, meaning the overall expected error is at most $2 \delta$.  This shows $f \in Heur_{2 \delta}-BPTIME(T(n)\poly(n))/(\log 1/\alpha(n)+ \log 1/\delta(n))$
\end{proof}
Returning to \Cref{thm:ucpha}, we use the same construction, with $f_1= C_1(f)$
and $F= C_2(f_1)$. Since this is the same construction, $F \in P^f$ and $n_2=n(1+o(1))$ just as before.

Assume $F \in Heur_{\frac{1}{2}-2\varepsilon}-BPTIME(T(n))$.  We want to show
$$f \in Heur_{2 \delta(n)} -BPTIME(T(n)\poly(n/\varepsilon)/ O(\log 1/\varepsilon+ \log n).$$  
By the lemma, it suffices to give an algorithm running in this time that with probability $\poly(\varepsilon)$ produces a circuit $Circ$ computing $f$ on all but $\delta(n)$-fraction of inputs.  
We only need to change a few steps from the non-uniform case.  

First, we pick a random tape for the randomized heuristic $A(x,r)$ for $f$ and fix it.
With probability at least $\varepsilon$, the heuristic with this fixed tape computes $F$ on all but a $\frac{1}{2} -\varepsilon$ fraction of inputs.   We call this algorithm $A_2$.  
Then we run the decoder for $C_2$ on $A_2$ as a corrupted message.  This gives a family of $\poly(1/\varepsilon)$ circuits of size $T(n) \poly(n/\varepsilon)$, one of which computes $f_1$ with at most $.1$ errors.  We randomly guess which of this list of circuits to keep as $Circ_1$.  We then use the decoder for $C_1(f)$ on $Circ_1$ to get a circuit $Circ$.  If $Circ_1$ is correct, then $Circ$ agrees with $f$ on all but a $\delta_{out}$ fraction of inputs. 

The probability of success is $\varepsilon$ for fixing the correct random tape, times $\poly(\varepsilon)$ for picking the correct list element, which make $\poly (\varepsilon)$ probability overall.  
The lemma characterizing randomized heuristic classes with advice then completes
the proof.  

\subsection{Hardness amplification preserving approximate time to compute}

When hard problems are used in pseudorandom generators, we typically do not care about ensuring that the new reliably hard function is in some small complexity class, only that we are able to compute its truth table
relatively efficiently.  In fact, constructions of reliably hard problems from worst-case hard problems are known (e.g., \cite{aszl1993bpp, impagliazzo1997p, sudan1999pseudorandom}), which implicitly or explicitly use exactly decodable LLDCs.  The catch is that these codes cannot be locally computable.  
Here, we give another tradeoff point between constructions based on average-case difficulty that preserve complexity exactly and constructions based on worst-case difficulty that do not preserve complexity except with respect to large classes such as $E$. The advantage over the results in the previous subsection is that we can start from a very weak average-case assumption and amplify
to difficulty $1/2- \varepsilon$ for $\varepsilon=2^{-\Omega(n)}$ while keeping input sizes small, where the previous results required $\varepsilon= 2^{-o(n^{1/2})}$.  

We will later use this result to give pseudo-random generators that are efficiently computable, following \cite{chen2021simple}. 
This uses non-uniform lower bounds, so we only present the hardness amplification for non-uniform complexity here.  

\begin{theorem}
There exists an $\alpha > 0$ so that the following holds.  Let $n \le S < 2^n$,  let $\frac{1}{n}> \varepsilon > 2^{- \alpha n}$ and $f \not\in Heur_{\varepsilon}-Size(S,n)$.  Then there exists Boolean function $F \in TIME^f(\poly (n , 1/\varepsilon))$
on inputs of size $n'=n+O(\log 1/\varepsilon)$ so that 
$F \not\in Heur_{1/2-\varepsilon}-Size(S \poly(\varepsilon), n')$
\end{theorem}

Proof:  The proof is similar to the ones in the previous subsection, except we compose the code $C_1$ in \Cref{lem:unique-amp} with $\delta_{out}=\varepsilon$ with the code $C_3$ from Theorem 1.1 and Corollary 4.2 to get $f_1$ on $n_1=n+O(\log n)$ bit inputs and $F$ on $n'$ bit inputs, respectively.

The properties of the code $C_1$ are:
\begin{enumerate}
    \item $|C_1(f)|= N_1= N/\poly(1/\delta_{out}),$ so $n_1=\log N_1= n +O( \log 1/\varepsilon)$
    \item Each bit of $C_1(f)$ can be computed with queries to $f$ in time $\poly(\log N), 1/\delta_{out})= \poly(n/\varepsilon)=\poly(1/\varepsilon)$.  Thus, $f_1 \in TIME^f (\poly(n/\varepsilon))$.  
    \item $f$ can be $\delta_{out}=\varepsilon$ uniquely decoded from any corrupted message with agreement $.9N_1$ with $C_1(f)$ and the decoded message is represented as a circuit of size $$\poly(\log N, 1/\delta_{out}) =\poly(1/\varepsilon)$$ with queries to the corrupted code word.
\end{enumerate}
So it follows that $f_1 \in TIME^f(\poly(1/\varepsilon))$ and $f_1 \not\in Heur_{.1}-Size(S\cdot \poly(\varepsilon), n_1)$.

We then construct $F$ from $f_1$ using code $C_3$. The code $C_3$ has the following properties:
\begin{enumerate}
    \item Rate $\poly (1/\log N_1, \varepsilon)$.
    Thus, the domain size $N_3 = N_1 \poly(n, 1/\varepsilon)= N_1 \poly(1/\varepsilon)$, so
    $n_3=\log N_3 = n_1+ O(\log 1/\varepsilon)=n+O(\log 1/\varepsilon)$.  
    \item Encoding time $\poly(\log N_1, 1/\varepsilon)= \poly(1/\varepsilon)$, using query access to the message bits.  Thus, $F \in TIME^{f_1}(\poly(1/\varepsilon)) \subseteq TIME^f (\poly(1/\varepsilon))$
    \item List decoding circuit size $\poly(\log N_3, 1/\varepsilon)= \poly (1/\varepsilon)$.   Thus, given a circuit of
    size $S_3$ computing $F$ with error rate $1/2-\varepsilon$, there exists a circuit on this list of total size $S_3 \poly(1/\varepsilon)$ that computes $f_1$ with error rate at most $.1$
\end{enumerate}

So assuming $F \in Heur_{1/2-\varepsilon} -Size(S_3, n_3)$ via some Circuit $Circ_3$, there is a circuit $Circ_1$ on the decoding list using $Circ_3$ as the corrupted message that computes $f_1$ on all but $.1 $ fraction of inputs, and has size $S_1=S_3 \poly(1/\varepsilon)$.  The approximate decoding algorithm from \Cref{lem:unique-amp} run on $Circ_1$ will then return a circuit $Circ$ of size $S_1 \poly(1/\varepsilon)$ computing $f$ on all but $ \varepsilon$ fraction of instances.
Unless $S_3 < S \poly(\varepsilon)$, this contradicts the assumed hardness for $f$.

One particularly interesting set of values to apply this
theorem is when $\varepsilon=2^{-\alpha n}$ and $S = 2^{(1-\alpha)n}$ for some small enough $\alpha>0$. In this case, the average-case assumption is equivalent to the worst case assumption:
\begin{corollary}\label{cor:hard-for-PRG}
There exists an $\alpha > 0$ so that the following holds for all $\alpha < \alpha_0$.  Let  $f \not\in Size(2^{(1-\alpha)n},n)$.  Then there exists Boolean function $F \in TIME^f(2^{O(\alpha)n})$ 
on inputs of size $n'=n (1+ O(\alpha)) $ so that 
$F \not\in Heur_{1/2-\varepsilon}-Size( 2^{(1-O(\alpha))n'}, n')$
\end{corollary}

Proof: Let $\varepsilon = 2^{\alpha n}/n^2$.  If $f \in Heur_{\varepsilon}-Size (2^{(1-\alpha)n}/ 2 , n) $, then the circuit that witnesseses this
makes at most $2^{(1-\alpha)n}/n^2$ mistakes, which can be corrected by a circuit of size $2^{(1-\alpha)n}/2$ to make a circuit of size $2^{(1-\alpha)n} $ that exactly computes the function. (We can actually do better using more sophisticated circuit constructions, but it will not make much of a difference here.)

Now applying the previous theorem to $f$ with this $\varepsilon$
and $S= 2^{(1-\alpha)n}/2$
gives a function 
\[
F \in TIME^f (\poly (1/\varepsilon))= TIME^f(2^{O(\alpha n)})
\]
on input size $n'= n+O(\log 1/\varepsilon)=n(1+O(\alpha))$  so that $F \not\in Heur_{1/2-\varepsilon} -Size (S 2^{-O(\alpha n)}, n')= Heur_{1/2-2^{-\alpha n }}- Size(2^{(1-O(\alpha))n'},n')$. 

\subsection{Application to Pseudorandom Generators}
Here, we use the hardness amplification lemma in a pseudorandom generator construction due to \cite{chen2021simple}, which leverages the PRG construction from \cite{nisan1994hardness} with that of cryptographic pseudorandom generators \cite{blum2019generate, yao1982theory, haastad1999pseudorandom}. 

A pseudorandom generator is a function $G:\{0,1\}^r \rightarrow \{0,1\}^m$ with $r < m$.   It {$\gamma$-\em fools} size $S$ if for any circuit $Circ \in Size(S,m)$, $|\Pr_z [Circ(G(z))=1] -\Pr_R[Circ(R)=1]|< \gamma$ where $R$ is the uniform distribution on $\{0,1\}^m$.

The advantage of using our codes over those used in \cite{chen2021simple} is that we get a very fast generator $G$.

\begin{theorem}\label{thm:PRG}
There exists an $\alpha_1$ so that for every $\alpha_0 < \alpha_1$ the following holds. Assume 
\begin{enumerate}
    \item There are one-way functions secure against polynomial sized circuits
    \item $f \in TIME(2^n)$ is not
in $i.o.-Size(2^{n(1- \alpha_0)}, n) $
\end{enumerate}
Then for all sufficiently large $S$ there is an $(S,1/4)$-pseudorandom generator 
\[
G: \{0,1\}^r \to \{0,1\}^S
\]
with seed length $r=(1+O(\alpha_0))\log S$ bits that is computable in time $S^{1+O(\alpha_0)}$
\end{theorem}
\begin{proof}
    The proof is standard and follows exactly as in \cite{chen2021simple}, replacing their initial amplification step with \Cref{cor:hard-for-PRG} (equivalently, replacing their use of Reed-Muller + Hadamard with our codes). We sketch the proof below.

     We first apply \Cref{cor:hard-for-PRG} to get a function computable in time $TIME(2^{(1+O(\alpha_0))n})$
so that $F \not\in Heur_{1/2- 2^{-\alpha_0 n}}-Size (2^{(1-O(\alpha_0))n'}, n')$ for $n' = n(1+O(\alpha))$.

A set system $S_1,\ldots,S_m$ of $S_i \subset \{1,...R\} $ is a $(R, n', I,m)$-{\em NW design} if $|S_i|=n'$ for all $i$, and for all $j \neq i$, $|S_I \cap S_J| \leq I$. \cite{nisan1994hardness} give constructions of NW -designs for any  with $R=O(n')$ and  $I =  n'/4$ with $m= 2^{\Omega( n')}$.  Implicitly, \cite{tell2018quantified} describes how to convert such a design into a $R= n(1+O(\beta)), I=n' (1- \beta)$ design.  Simply divide the $R$ elements into $n(1-2\beta)$ fixed elements $Fixed$  and a set of $O(\beta n) $ variable elements.  Take an $n''= 2 \beta n, R'=O( n'')=O(\beta n), I' = 1/4 n'' < \beta n $ design over the variable elements, $S'_1,..S'_m$. 
Define $S_i= S'_i \cup Fixed$.   Then $S_i \cap S_j = Fixed \cup (S'_i \cap S'_j)$, so has size at most $|Fixed|+I' = n(1-2\beta n)+\beta n= n(1-\beta)$

Our first generator chooses $n' = (1+O(\alpha_0)\log(S) $ and uses the NW generator with this design to produce a series
of $m= 2^{\Omega(\alpha_0)n}$ inputs to $F$ of size $n'$, and concatenates the results. This generator can be computed in time
$m*2^{n' (1+\alpha_0)} = S^{1+O(\alpha_0)} $ time.
From \cite{nisan1994hardness}, any distinguishing circuit for
this generator of size $S$ yields an approximating $1/2-\Omega(1/m)$ circuit of size $S+ O(2^{I} m)$ for $F$, contradicting the hardness of $F$.  Following \cite{chen2021simple}, we then apply an almost linear time cryptographic PRG to increase the stretch to $S$.  
\end{proof}

\section{Applications in Coding Theory}\label{app:list-decoding}

Following Trevisan \cite{trevisan2003list}, in this section we give applications in the view of aLLDCs as a tool for complexity-preserving distance amplification. This is by now a very standard method in the literature and is the core of many list-decoding algorithms (see e.g.\ \cite{ben2010note,hemenway2019local,jeronimo2021near,srivastava2025list}). We use this strategy to give the first family of good codes with list-decoding in \textit{logarithmic depth} (RNC$^1$), as well as new families of constant rate LLDCs with state-of-the-art time and query complexity.

We remark the first of these codes can also be made LDPC, or to be \textit{encodable} in NC$^1$. Achieving all three properties (log depth encoding and decoding for LDPC codes) remains open even in the unique decoding regime --- if such codes existed, we would immediately obtain the same for list-decoding by the same method.

\subsection[List Decoding in RNC1]{List Decoding in RNC$^1$}

Parallel computation dominates modern computing. Error correcting codes are no exception, and have an extensive history of parallelization dating back to Gallager's breakthrough 1960 thesis \cite{gallager2003low} introducing \textit{low-density parity check} (LDPC) codes, and Zyablov and Pinsker's \cite{zyablov1975estimation} corresponding low-depth decoders. We briefly recall some background. A linear code $C$ may always be represented as the kernel of a corresponding `parity check' matrix $H$. A code is called LDPC if the rows of this matrix have constant hamming weight. 

In early influential work, Zyablov and Pinsker \cite{zyablov1975estimation} proved Gallager's random LDPC can be decoded in parallel by near-linear size circuits. In particular, they gave a simple logarithmic depth belief propagation decoder that worked for `most' random LDPC codes, and remains a very standard method in coding today. In a landmark result, Sipser and Spielman \cite{sipser2002expander} famously constructed \textit{explicit} examples of Gallager's codes called \textit{expander codes} with corresponding logarithmic depth and linear time decoders. Their codes and simple belief-propagation scheme (itself based on Zyablov-Pinsker \cite{zyablov1974decoding}) have since been hugely influential in coding theory, and are in some sense the `gold standard' in combinatorial decoding algorithms.

Surprisingly, parallelization is much less understood in the high noise regime. In particular, while a long line of fruitful work has developed list-decodable codes with linear \cite{guruswami2003linear} or near-linear time global list-decoding \cite{alekhnovich2002linear,goyal2024fast} and sub-linear time local list decoding \cite{kopparty2015list,kopparty2015list,guo2016list,hemenway2019local,kopparty2020list}, there are \textit{no} examples of good codes (even existentially!) that are list-decodable in better than $\log^2(N)$ depth. Indeed all list-decoding algorithms for known codes are either algebraic, spectral, or rely on semi-definite programs, putting them in at best NC$^2$ in our current understanding.

Like Zyablov-Pinsker \cite{zyablov1975estimation} and Sipser-Spielman \cite{sipser2002expander}, our aLLDC is based purely on message passing and has a low depth decoder. As a result, composing our codes with any log depth unique decoder immediately gives the first constructions of good list-decodable codes with (randomized) logarithmic depth decoders.

There is a little bit of subtelty in defining the list decoding model here. We will use the following strong notion of (probabilistic) list decoding. We say $\mathcal{C}$ is $(\frac{1}{2}+\varepsilon,\ell_{out})$-list decodable in RNC$^1$ if there is a list $C_1,\ldots,C_{\ell_{out}}$ of logspace uniform, logarithmic depth randomized circuits such that for any $w$, the probability over the internal randomness $r$ of the $C_i$ that the list $\{C^w_1(\cdot,r_1),\ldots,C^w_{\ell_{out}}(\cdot,r_{\ell_{out}})\}$ contains every function in $\mathcal{B}_{\frac{1}{2}-\varepsilon}(w)$ is at least $\frac{3}{4}$. Note that given a \textit{fixed} $w$, one can easily amplify success by running the circuit family several times on independent randomness, and prune the above list in NC$^1$ down to its combinatorial list size (typically $O(\frac{1}{\varepsilon^2}))$. Nevertheless, we state the `pre-pruned' list-size below that is independent of the word $w$. The advantage of this model is that one only has to generate this slightly larger list of randomized circuits \textit{once}, not every time a new word $w$ is received.

\begin{theorem}[List Decoding in RNC$^1$ (Constant Rate Regime)]\label{thm:constant-rate-NC}
There exists a universal constant $C>0$ such that for any $\varepsilon>0$ and infinitely many $N \in \mathbb{N}$ there is an explicit linear code $\mathcal{C}_N:\{0,1\}^N \to \{0,1\}^{N'}$ that is $(\frac{1}{2}+\varepsilon,\ell_{out})$-list-decodable with
\begin{enumerate}
    \item \textbf{Rate:} $R \geq \exp(-O(1/\varepsilon^3))$
    \item \textbf{List Size:} $\ell_{out} \leq \frac{\polylog(\frac{1}{\varepsilon})}{\varepsilon^2}$
\end{enumerate}
Moreover $\mathcal{C}_N$ is list decodable in RNC$^1$ with $N\log(N)^{\exp(O(1/\varepsilon^3))}$-size circuits.
\end{theorem}
Depending on the instantiation, the codes can be made either be made LDPC (Siper-Spielman \cite{sipser2002expander}) or encodable in NC$^1$ (Spielman \cite{spielman1995linear}).

For smaller $\varepsilon \leq \frac{1}{\log N}$, we can appeal to our poly-logarithmic rate codes to give the following improved construction:
\begin{theorem}[List Decoding in NC$^1$ (Sub-constant Error Regime)]\label{thm:log-rate-NC}
For infinitely many $N \in \mathbb{N}$ and any $\varepsilon \leq \frac{1}{\log N}$, there exists a explicit linear code $\mathcal{C}_N:\{0,1\}^N \to \{0,1\}^{N'}$ that is $(\frac{1}{2}+\varepsilon,\ell_{out})$-list-decodable with
\begin{enumerate}
    \item \textbf{Rate:} $R \geq \poly(\varepsilon)$
    \item \textbf{List Size:} $\ell_{out} \leq \frac{\polylog(\frac{1}{\varepsilon})}{\varepsilon^2}$
\end{enumerate}
Moreover the codes can be decoded in RNC$^1$ with $N\poly(\log(N),\frac{1}{\varepsilon})$-size circuits whenever $\varepsilon \geq \exp(-\tilde{O}(\sqrt{\log N}))$.
\end{theorem}

We note we can also achieve list decoding from $1-\varepsilon$ errors with larger alphabet simply by concatenating our original large alphabet aLLDC with any inner code that is list-decodable from $1-\varepsilon$ errors. We give only the binary case here for simplicity. 

As discussed, the proof of these results is essentially immediate from combining our codes with the appropriate base code, namely either Sipser-Spielman \cite{sipser2002expander} (for the LDPC property), or Spielman \cite{spielman1995linear} (for fast encoding):
\begin{proof}[Proofs of \Cref{thm:constant-rate-NC} and \Cref{thm:log-rate-NC}]
    We prove only the former. The latter follows similarly. Our construction will be the composition of two codes
    \begin{enumerate}
        \item A linear binary \textit{base code} $\mathcal{C}_{unique}$ that has
        \begin{itemize}
            \item Constant rate and distance
            \item An NC$^1$, size $N\polylog(N)$ decoder from a constant $c>0$ fraction of errors, and
            \item Is either encodable in NC$^1$ \cite{spielman1995linear}, or LDPC \cite{sipser2002expander}
        \end{itemize}
        \item Our constant rate linear binary aLLDC $C_{apx}$ from \Cref{cor:constant-rate-binary}
    \end{enumerate}

    In more detail, fix any $N \in \mathbb{N}$ for which there exists a code in our constant rate family \Cref{cor:constant-rate-binary} on $N$ message bits that is $\frac{c}{100}$-approximately $(\frac{1}{2}+\varepsilon,\tilde{O}(1/\varepsilon^2))$-list-decodable (recall it is always easy to constantly amplify $\delta_{out}$, which formally is set to just $\frac{1}{10}$ in \Cref{cor:constant-rate-binary} for simplicity). Note that since the decoding circuits are promised to be in logspace uniform size $\log(N)^{2^{O(1/\varepsilon^3)}}$ and depth $O_{\varepsilon}(\poly\log\log(N))$, simply running the decoders in parallel for every message bit gives an NC$^1$ algorithm for global (approximate) decoding in size $N\log(N)^{2^{O(1/\varepsilon^3)}}$ as claimed in the statement. Finally, note that the generating algorithm of our codes is independent of $w$. In the main body its only job is to select a random starting hyperedge (and potentially list element) for propagation. This can be done internally by the circuits themselves, so they indeed only need to be generated one time then work for all in put words $w$.\footnote{In hardness amplification this does not make sense, as then there is no single fixed circuit which computes the original function. In list-decoding, however, we do not care if on different runs the same circuit outputs a different valid message.}
    
    Now, using any logspace constructions of constant degree expander graphs (e.g.\ \cite{lubotzky1988ramanujan,morgenstern1994existence,viola2018local} or the 1-skeleton of the KMS complexes we construct), the expander code frameworks of \cite{sipser2002expander} and \cite{spielman1995linear} respectively imply the existence of binary linear base codes $\mathcal{C}$ with blocklength $N$ and either the LDPC property or NC$^1$ encoding.\footnote{Note that while such expanders might not exist for all $N$, they exist within a constant factor of any target $N$ so we can simply pad the blocklength to the desired $N$ while maintaining constant rate and decoding radius.} Our code will be given by first encoding a message with $\mathcal{C}_{unique}$, then encoding the output by our constant rate aLLDC $C_{apx}$. Denote this composed code by $\mathcal{C}=C_{apx}(\mathcal{C}_{unique})$. Note that the claimed rate and linearity are immediate from construction.

    We now prove this composed code is indeed list-decodable from a $\frac{1}{2}-\varepsilon$ fraction of errors in NC$^1$. First, by standard direct product and concatenation analysis, note the composed code can be taken to have distance much larger than $\geq \frac{1}{2} - O(\varepsilon^2)$. Thus the Johnson bound (see e.g.\ \cite{goldreich2000learning,guruswami2007algorithmic}) promises the combinatorial list size of the code is at most $O(\frac{1}{\varepsilon^2})$.

    Now, fix an input word $w$ and any $f \in \mathcal{B}_{\frac{1}{2}-\varepsilon}(w)$. Running our aLLDC decoder for $\mathcal{C}_{apx}$ outputs with probability at least $3/4$ (possibly randomized) NC$^1$ circuits $C_1,\ldots, C_{\ell_{out}}$ one of which $\frac{c}{100}$-computes $\mathcal{C}_{unique}(f)$. Fixing for each $C_i$ a random setting of internal randomness, we have that one of the output circuits deterministically $c$-computes $\mathcal{C}_{unique}(f)$ with at least $\frac{2}{3}$ probability. Finally, since the combinatorial list size of $\mathcal{C}$ is at most $O(1/\varepsilon^2)$, repeating the above process $O(\log(1/\varepsilon))$ times ensures \textit{every} $\varepsilon$-computed function $f$ is $c$-computed by some circuit in the list with high constant (say $99\%$) probability.

    We may now feed these circuits in parallel to the decoder for $\mathcal{C}_{unique}$, which is also in NC$^1$. By the unique decoding guarantee, presuming the previous step succeeded the decoder will output a slightly larger list of $\tilde{O}(1/\varepsilon^2)$ functions guaranteed to contain every $f \in \Ball_{\frac{1}{2}-\varepsilon}(w)$. If desired, one can then prune this list in NC$^1$ to the correct $O(1/\varepsilon^2)$ size by removing duplicates and checking agreement with the word $w$ (though this is a `$w$-specific' operation, so as explained our list size in the formal model remains $\frac{\polylog\frac{1}{\varepsilon}}{\varepsilon^2}$.

    Finally, we note that if we take our base code to be the LDPC expander code of Sipser-Spielman \cite{sipser2002expander}, then the entire construction is LDPC as long as the inner code we concatenate with in \Cref{cor:constant-rate-binary} is LDPC and systematic (the latter allows for short parity checks for the direct product equality constraints). Similarly, if we take Spielman's code, encoding the base code and the aLLDC are both in NC$^1$, so encoding the composed code is in NC$^1$ as well as desired.
\end{proof}
\subsection{New LLCCs with sub-Polynomial Time Decoding}\label{app:LLCC}

There is a huge breadth of literature on the construction of locally (list) decodable and correctable codes ranging from classical algebraic methods like Reed-Muller \cite{rubinfeld1996robust,arora1997improved,sudan1999pseudorandom}, combined algebraic and combinatorial methods like matching vector codes \cite{yekhanin2008towards,efremenko20093}, and even purely combinatorial/random constructions (e.g.\ sparse random linear codes) \cite{kaufman2007sparse,kopparty2010local}. Traditionally, most constructions of LDCs only work in the sub-constant rate regime. While this is in part due to the regime's powerful use as a primitive in complexity and cryptography, it is also due to a substantial lack of tools in the high rate regime which is of great interest within modern coding theory, data storage, and algorithm design. Indeed this is evidenced by a growing body of work over the past 20 years constructing \textit{relaxed} versions of LDCs with high rate \cite{ben2004robust,blocki2018brief,gur2020relaxed,gur2021power,cohen2022relaxed,goldberg2024linear} as well as the related quest towards constant rate locally \textit{testable} codes \cite{dinur2021,panteleev2021asymptotically}, culminating in recent work achieving weaker relaxed local decoding with constant rate and only polylog queries \cite{kumar2024relaxed,cohen2024asymptotically}.

Turning our attention back to standard LDCs, the past 15 years have also seen a renaissance in the high rate regime. While for many years the only known sublinear time decodable codes were Reed-Muller codes (which can achieve $k^{\alpha}$-time decoding at rate $\alpha^{\Omega(1/\alpha)}$ for certain parameter settings), several works starting around 2010 managed to achieve $k^\alpha$ time unique decoding with much higher (even close to 1) rate including multiplicity codes \cite{kopparty2014high}, lifted codes \cite{guo2013new, guo2015high}, expander codes \cite{hemenway2015local}, and tensor codes \cite{hemenway2019local}.

In applications like modern data storage, sub-linear time decoding is far from sufficient, and one would indeed hope for the sub-polynomial or even polylogarithmic time decoders known for sub-constant rate codes. Such codes were achieved in the constant rate regime for the first time by Kopparty, Meir, Ron-Zewi, and Saraf \cite{kopparty2017high}, who combined multiplicity codes with classic combinatorial amplification techniques \cite{alon1995linear} to achieve sub-polynomial time local decoders. In \cite{gopi2018locally,kopparty2018improved}, the authors argued the multiplicity codes are further list-recoverable, allowing them to extend sub-polynomial time decoding to the list-decoding regime up to capacity at the cost of non-trivially increasing the query complexity from $\exp(\sqrt{\log N})$ to $\exp(\log^{3/4}(N))$. Using our codes, we give new constant rate LLDCs achieving the former, though our codes are far from capacity. 

More generally, the method we apply below works for any base LDC, and in particular implies the claimed statement in \Cref{sec:intro-applications} regarding complexity-preserving amplification of LDCs to LLDCs with at worst polynomial blowup in time and query complexity (ignoring dependence on $\varepsilon$). The only barrier to constructing $\polylog(N)$-time LLDCs is therefore in the unique decoding regime, where the problem is a longstanding open question.


Before giving the main result, we briefly state the stronger notion of local correction our codes will satisfy, since it has not yet appeared formally in this work. To better match the coding literature on this topic, we will also move away from the notion of decoding \textit{circuits} in this section in favor of classical time-bounded decoding using oracle (Turing) machines with query access to the codeword $w$. We will also move back to the more standard guarantee that the output list should cover \textit{every} function near the input word $w$, rather than covering each fixed such function with high probability. We now state the standard notion of locally list correctable codes. As discussed, these are codes which can recover the value of the closest \textit{codewords} at a particular index, rather than the value of the message:
\begin{definition}[Locally List Correctable Codes (LLCC)]
        A code $\mathcal{C}: \{0,1\}^k \to \{0,1\}^n$ is $(\frac{1}{2}+\varepsilon,q,\ell_{out})$-locally list correctable in time $T$ if there exists a probabilistic algorithm $A$ producing randomized algorithms $M_1,\ldots,M_{\ell_{out}}$ such that for all $w \in \{0,1\}^{n}$:
    \[
    \Pr_{A}\left[\forall f \in \Ball_{\frac{1}{2}-\varepsilon}(w) ~ \exists i, \forall j \in [n]:~ \Pr_{}[{M_i}^{w,r}(j) = Enc(f)_j] \geq \frac{3}{4} \right] \geq \frac{3}{4}.
    \]
    Moreover, every $M_i$ queries $w$ in at most $q$ locations, and both the $M_i$, and the algorithm $A$ generating the description of the $M_i$ run in time $T$.
\end{definition}
An LLCC is automatically locally (list) decodable if it is systematic (or otherwise encodes the message directly in its encoding). This is true for our codes, so we will focus only on local correction. Our main result of the section is the following LLCC with sub-polynomial time decoding

\begin{theorem}[Binary LLCCs with sub-Polynomial Time Decoders]\label{thm:sub-Poly-LLDCs}
For every \(\varepsilon >0\) and infinitely many $N \in \mathbb{N}$ there exists a explicit infinite family of binary linear $(\frac{1}{2}+\varepsilon,q,\ell_{out})$-LLCCs with
\begin{enumerate}
  \item \textbf{Rate}:  \(R \geq \exp(-O(1/\varepsilon^3))\)
  \item \textbf{Query and Time Complexity:} $q,T \leq \exp\left(O\left(\sqrt{\log N \cdot \log\log N}\right)\right)$
  \item \textbf{List Size:} $\ell_{out} \leq \frac{\polylog(\frac{1}{\varepsilon})}{\varepsilon^2}$
\end{enumerate}
\end{theorem}
Note we focus on the binary case above for simplicity. We can also easily achieve large alphabet codes with the same parameters that decode from radius $1-\varepsilon$ just by concatenating our large-alphabet aLLDC with any sufficiently good list-decodable code in this regime (rather than with a binary one as we have here). The above codes are also locally list decodable with the same parameters.

\Cref{thm:sub-Poly-LLDCs} is a fairly immediate corollary of \Cref{cor:constant-rate-binary} and \cite[Theorem 1.1]{kopparty2017high}:
\begin{theorem}[{\cite[Theorem 1.1]{kopparty2017high}}]\label{thm:LLDC-Unique}
For every \(r \in (0,1)\), there exist \(\delta > 0\) and sufficiently large $N \in \mathbb{N}$ there is an explicit binary linear code $C_N$ satisfying:
\begin{enumerate}
  \item \(C_N\) has block length \(N' \in [N,N2^{O(\sqrt{\log N \cdot \log\log N})}]\), rate at least \(r\), and relative distance at least \(\delta\),
  \item \(C_N\) is locally correctable from a \(\delta/2\)-fraction of errors with running time at most $\exp\bigl(O(\sqrt{\log N \cdot \log\log N})\bigr)$
\end{enumerate}
\end{theorem}
We remark the statement about the size of $N'$ does not appear explicitly in \cite{kopparty2014high}, but is immediate from examining the parameters of their construction.

We're almost ready for the proof of \Cref{thm:sub-Poly-LLDCs}, which follows essentially as one would suspect with one wrinkle: in both cases, we have codes for \textit{infinitely many} but not \textit{all} $N$, so we must be careful to ensure the message and blocklengths line up for composition. In fact, in order to handle this, we need the following variant of \Cref{cor:constant-rate-binary} handling super constant size message alphabet. 
\begin{corollary}[Binary aLLDCs with Large Input Alphabet]\label{cor:constant-rate-large-in}
    There is a universal constant $C>0$ such that for all large enough $M \in \mathbb{N}$ and $\frac{C}{\log\log M} < \varepsilon<0$, there is an $N \in [M,M\log(M)^{2^{O(1/\varepsilon^3)}}]$ and an explicit binary linear code $\mathcal{C}:\Sigma_{in}^{N} \to \{0,1\}^{N'}$ that is a $0.1$-approximate, $(\frac{1}{2}+\varepsilon,\tilde{O}(\frac{1}{\varepsilon^2}),q)$-LLDC with 
    \begin{enumerate}
        \item \textbf{Rate:} $R \geq 2^{-O(1/\varepsilon^3)}$
        \item \textbf{Time and Query Complexity:} $q \leq \log(\Sigma_{in}) \log(N)^{2^{O(1/\varepsilon^3)}}$
    \end{enumerate}
\end{corollary}
The proof is exactly the same, modulo the time increase to read and compute equality over the input symbols, and to concatenate over the eventual output log-alphabet size $\log \Sigma_{in} \cdot \log(N)^{2^{O(1/\varepsilon^3)}}$ (note now that we are not in the circuit model, we just require the inner list-decodable code used in concatenation has polynomial time encoding and decoding). We are now ready to prove the existence of good sub-polynomial time LLDCs:

\begin{proof}[Proof of \Cref{thm:sub-Poly-LLDCs}]
Fix some constant $r \in (0,1)$ and corresponding $\delta_{out}=\frac{\delta}{2}$ corresponding to the unique local decoding radius of the code family from \Cref{thm:LLDC-Unique}. Let $N \in \mathbb{N}$ be any integer such that there exists an $\frac{\delta}{200}$-approximate $(\frac{1}{2}+\varepsilon,\tilde{O}(\frac{1}{\varepsilon^2}),\poly(\frac{\log N}{\varepsilon}))$-LLDC $\mathcal{C}_{apx}$ with $\exp(-O(1/\varepsilon^3))$-rate as in \Cref{cor:constant-rate-binary}.\footnote{Again, note it is easy to achieve any constant $\delta/200$ approximate decoding just by shifting the constants in our construction, or by composing with an appropriate expander. We set $\delta_{out} = 0.01$ in our Theorem statements for just for concreteness.}

Now, \Cref{thm:sub-Poly-LLDCs} promises the existence of a binary code $\mathcal{C}_{unique}$ with blocklength $N' \in [N,N2^{O(\sqrt{\log N\log\log N})}]$. Naively, we cannot compose $\mathcal{C}_{unique}$ with $\mathcal{C}_{apx}$, because the blocklength $N'$ of the former does not match the message length $N$ of the latter. Thankfully, there is a standard trick to construct, based on $\mathcal{C}_{unique}$, a \textit{larger alphabet} code $\mathcal{C}^{fold}_{unique}$ with blocklength exactly $N'$. In particular, we may simply fold the code by viewing each block of $\lceil \frac{N'}{N} \rceil$ bits as a single alphabet symbol (`bundling' the symbols), and pad the last block with 0s to reach the same alphabet. Thus our final code will simply be the composition $\mathcal{C}=\mathcal{C}_{apx}(\mathcal{C}_{unique}^{fold})$. 

We now give our local correction algorithm for $\mathcal{C}$. As always, fix an input word $w$ and a function $f \in \Ball_{\frac{1}{2}-\varepsilon}(w)$. Running our local list decoding algorithm for $\mathcal{C}_{apx}$, we get (randomized) oracle machines $M_1,\ldots,M_{\ell_{out}}$ (for $\ell_{out} \leq \tilde{O}(1/\varepsilon^2)$) such that with probability at least $3/4$, at least one of the machines $\frac{\delta}{200}$-computes $\mathcal{C}^{fold}_{unique}(f)$ with oracle access to $w$ in $2^{O(\sqrt{\log(N)\log\log(N)})}$ time. By Markov, a random choice of internal randomness for these machines therefore $\frac{\delta}{2}$-computes $\mathcal{C}^{fold}_{unique}(f)$ with $99\%$ probability. Note that since the combinatorial list size bound on our composed code is at most $O(1/\varepsilon^2)$ by the Johnson bound (as in the proof of \Cref{thm:constant-rate-NC}), repeating this process $O(\log(1/\varepsilon))$ times ensures that with high probability \textit{every} $f \in \Ball_{\frac{1}{2}-\varepsilon}(w)$ has an $M_i$ that $\frac{\delta}{2}$-computes it.

Now, say we are given a codeword index $(s,i)$ we'd like to correct, where $s$ stands for the hyperedge in the original code direct product $\mathcal{C}_{(V,S)}$, and $i$ is the index of the concatenated inner code. To correct $(s,i)$, it suffices to correctly compute the (pre-concatenated) large alphabet encoding at $s$, then encode this with the inner code and output the $i$th bit. Note the latter takes at most $2^{O(\sqrt{\log(N)\log\log(N)})}$-time by our assumption that the inner code has a polynomial time decoder.

Now since $C_{(V,S)}$ is a direct product code, to compute the full symbol at $s$ it of course suffices to correctly compute the encoded value $\mathcal{C}_{unique}^{fold}(f)$ at every $v \in s$ (which itself is a `block' of $2^{O(\sqrt{\log(N)\log\log(N)}}$ encoded bits from $\mathcal{C}_{unique}^{fold}(f)$). Thus, if $M_i$ is the oracle machine that $\frac{\delta}{2}$-computes $\mathcal{C}_{unique}(f)$, we may run the unique correction algorithm for $\mathcal{C}_{unique}$ on every encoded bit in the block corresponding to every $v \in s$. The local corrector is guaranteed to output the correct bit with $\frac{3}{4}$ probability, so taking the majority answer over $O(\log(N))$ repetitions suffices to ensure we encode the entire block for every $v \in s$ correctly with (say) $99\%$ probability. Putting everything together, our final list decoder simply outputs the list of decoders applying $\mathcal{C}_{unique}$ to each oracle machine $M_i$ produced above, each of which makes at most $2^{O(\sqrt{\log(N)\log\log(N)})}$ queries. Repeating the processes above constantly many times, this indeed outputs $M_i$ locally correcting every $f \in \Ball_{\frac{1}{2}-\varepsilon}(w)$ with probability at least $3/4$ as desired.

\end{proof}

\section{Inner Code Analysis}\label{app:inner-code}
\begin{figure}[ht!]
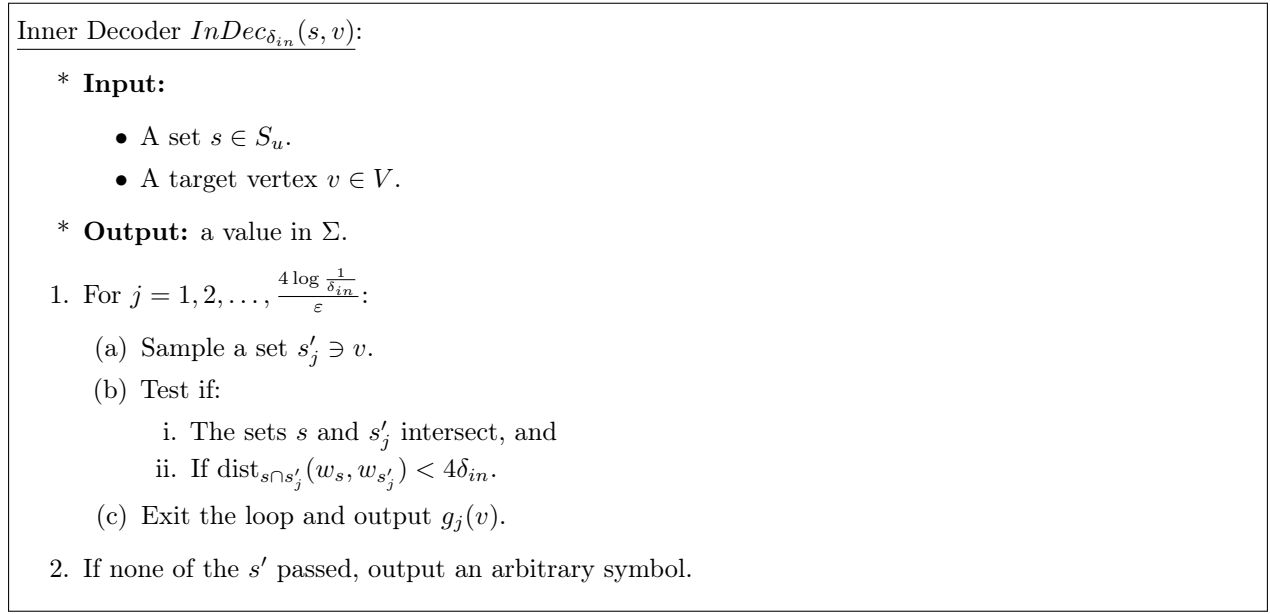

\fbox{\parbox{\textwidth}{
\vspace{.1cm}
\underline{Inner Decoder \(InDec_{\delta_{in}}(s,v)\)}:
\begin{enumerate}
    \item[*] \textbf{Input:}
    \begin{itemize}
        \item A set \(s \in S_u\).
        \item A target vertex \(v \in V\).
    \end{itemize}
    \item[*] \textbf{Output:} a value in \(\Sigma\).
    \item For \(j=1,2,\dots,\frac{4 \log \frac{1}{\delta_{in}}}{\varepsilon}\):
    \begin{enumerate}
        \item Sample a set \(s'_j \ni v\).
        \item Test if:
            \begin{enumerate}
                \item The sets \(s\) and \(s'_j\) intersect, and
                \item If \(\dist_{s \cap s'_j}(w_{s},w_{s'_j}) < 4\delta_{in}\).
            \end{enumerate} 
    \item Exit the loop and output \(g_j(v)\).
    \end{enumerate}
    \item If none of the \(s'\) passed, output an arbitrary symbol.
\end{enumerate}}}
    \caption{Inner Decoder}
    \label{fig:inner-decoder-app}
\end{figure}

In this appendix we prove \Cref{claim:const-rate-inner-decoder} and \Cref{claim:output-list-pruning}, our inner decoder and pruning process for constant rate codes. We restate the algorithm and claims here for convenience. As noted in the main body, the algorithm presented is very similar to the original aLLDCs for dense systems in \cite{impagliazzo2008uniform}. For the moment, we fix \(In_u\), input word \(w_{In_u} = \set{w_s}_{s \in S_u}\), and parameters \(\varepsilon, \delta_{in} > 0\). The decoding circuit is defined in \Cref{fig:inner-decoder-app}.

As one might expect, the generating algorithm samples \(\frac{10}{\varepsilon}\) sets \(s \in S_u\) and outputs circuits that compute \(InDec_{\delta_{in}}(s, \cdot)\). 

\restateclaim{claim:const-rate-inner-decoder}


\begin{remark}[Testing distance and fine-grained query complexity]\label{rem:outer-const-fine}
    We implement the distance testing in step \((b)\) via the distance tester in \cref{app:distance-tester} without querying the whole intersection. In particular, we may do this probabilistically by querying \(m=\log \left (\frac{100 \log \frac{1}{\delta_{in}}}{\delta_{in} \varepsilon}\right ) \frac{1}{\delta_{in}^2}\) points in \(s \cap s'\) and calculating the empirical distance up to accuracy \(\delta_{in}^2\) with confidence \(1-\delta_{in}\) \emph{over all functions checked}. Hence the expression that should appear in the query complexity of \cref{claim:const-rate-inner-decoder} is \(\frac{4 \log \frac{1}{\delta_{in}}}{\varepsilon} \cdot \log \left (\frac{100 \log \frac{1}{\delta_{in}}}{\delta_{in} \varepsilon}\right ) \frac{1}{\delta_{in}^2}\). 
    
    In the proof of \cref{claim:const-rate-inner-decoder}, we will just assume that this test passes deterministically for simplicity. We point the meticulous reader to the proof of \cref{thm:list-recovery} to see how the general case is dealt with.
\end{remark}

We comment that we did not try to optimize the query complexity and the output approximation parameter \(13\sqrt[3]{\delta_{in}}\), since it is set to roughly \(\approx \exp(-\frac{1}{\varepsilon})\) and gets hits by a factor of \(\approx \exp(\frac{1}{\varepsilon})\) in the outer decoder anyway. A linear dependence between \(\delta_{in}\) and \(\delta_{out}\) is likely possible with more care, see e.g.\ \cite{impagliazzo2008uniform}. In addition, we remark the output circuits can be derandomized by having the generating algorithm fix their randomness in advance if desired.

\begin{proof}[Proof of \Cref{claim:const-rate-inner-decoder}]
    In the proof we will use the fact that \(In_u=(V_u,S_u)\) itself and the intersection graphs \((G_{In_u})_s\) are \(\eta\)-samplers. In particular, in the statement of the claim we choose \(\eta > 0\) sufficiently small so that, in every inequality where \(\eta\) appears as a multiplicative factor, the resulting term is strictly smaller than the term on the other side of the inequality. This remains true even after taking square roots. E.g.\ if we write  an expression of the form \(\sqrt{\eta}{\delta^2_{in}} \leq \varepsilon^3\), this will be because \(\eta\) was set to be small enough with respect to \(\delta_{in},\varepsilon\) so that this holds true. We note this here once for clarity, rather than verifying it repeatedly in the text.
    
    The query complexity and list size are immediate from the construction. We must show that for any function \(f \in \Ball_{1-\varepsilon,\delta}(w_{In_u})\), we output a circuit that is \(13\sqrt[3]{\delta_{in}}\)-close to \(f\) with probability \(\frac{3}{4}\).

    Let \(A = \sett{s \in S_u}{\dist(w_s,f) \leq \delta_{in}}\) and note by assumption that \(\Prob{}{A} \geq \varepsilon\). We would like to argue that there is a subset \(A' \subseteq A\) of probability \(\Prob{}{A'} \geq \frac{\varepsilon}{2}\) such that for every \(s \in A'\) decoding from $s$ succeeds with high probability
    \begin{equation}\label{eq:good-s-decode-most-vertices-in-inner-decoder}
        \Prob{v \in V_u,r}{InDec_{\delta_{in}}(s,v) = f(v)} \geq 1-13\sqrt[3]{\delta_{in}}   
    \end{equation}
    Here \(r\) is the random string the decoding circuit uses. If this holds then we have proven the desiderata, since the generating algorithm samples enough $s$ to hit \(s \in A'\) with probability greater than \(\frac{3}{4}\).

    Like in \Cref{thm:list-recovery}, towards defining \(A'\) we call an edge \(\set{s,s'}\) `bad' if for either \(s'' \in \set{s,s'}\)
    \[
        \abs{\dist_{s''} (f,w_{s''}) -  \dist_{s \cap s'}(f,w_{s''}|_{s\cap s'}) } \geq \delta_{in}.
    \] 
    Recall that the graph \((G_{In_u})_{s''}\) is an \(\eta\)-sampler. Bad edges are exactly intersections \(s \cap s'\) which \(\delta_{in}\)-missample the fraction of vertices in \(s''\) where \(f \ne w_s\), so sampling promises for every \(s'\) there are at most \(\frac{\eta}{\delta_{in}^2}\) such intersections. Hence there are at most a \(2 \cdot \frac{\eta}{\delta_{in}^2}\) fraction of `bad edges' overall, and Markov's inequality implies there are at most \(\sqrt{\frac{2\eta }{\delta_{in}^2}}\) sets \(s\) that are connected to more than a \(\sqrt{\frac{2 \eta }{\delta_{in}^2}}\)-fraction of bad edges. By choice of \(\eta\), this is less than an \(\varepsilon^3\)-fraction of the sets. So we define 
    \[A' = \sett{s \in A}{s \text{ is connected to less than \(\sqrt{\frac{2\eta}{\delta_{in}^2}}\) bad edges}}.\]
    Of course \(\Prob{}{A'} \geq \frac{\varepsilon}{2}\), and in particular our generating algorithm hits \(A'\) with probability \(\geq 0.9\). 
    
    It remains to prove that for any \(s \in A'\), \(InDec_{\delta_{in}}(s,\cdot)\) computes \(f(v)\) with probability \(1-13\sqrt[3]{\delta_{in}}\).

    Fix \(s \in A'\). We will define a set \(V' \subseteq V\) of vertices and prove that (1) \(\Prob{}{V'} < 5\sqrt[3]{\delta_{in}}\) and (2) that if \(v \notin V'\) then
    \begin{equation}\label{eq:good-vertices-get-decoded-in-inner-decoder}
        \Prob{r}{InDec_{\delta_{in}}(s,v)=f(v)} \geq 1-8\sqrt[3]{\delta_{in}}.
    \end{equation}
    This will prove \eqref{eq:good-s-decode-most-vertices-in-inner-decoder} since the probability of \emph{not} correctly decoding is bounded by \(\Prob{}{V'} + 8\sqrt[3]{\delta_{in}} \leq 13\sqrt[3]{\delta_{in}}\).
    
    \paragraph{Defining \(V'\):}Let \(M\) be the sets \(s' \in S_u\) which pass the distance test in step \((2)\) (for some fixed \(s \in A'\) defined above; in particular this is a subset of the neighbors of \(s\)). The set \(V'\) will be all vertices \(v \in V\) such that (at least) one of the following holds:
    \begin{enumerate}
        \item \label{item:M-is-small}\(\Prob{s' \in S_u}{M| s' \ni v} < \frac{\varepsilon}{2}\).
        \item \label{item:v-is-misleading-in-M}\(\Prob{s' \in S_u}{w_{s'}(v) \ne f(v)| s' \in M, s' \ni v} > 7\sqrt[3]{\delta_{in}}\). Spelled out explicitly, this is the probability of the output to be incorrect on \(v\) in the decoding algorithm (assuming the algorithm didn't fail to find a set in \(M\)).
    \end{enumerate}
    Let's first confirm that, as claimed, if $v \notin V'$ then \eqref{eq:good-vertices-get-decoded-in-inner-decoder} indeed holds.
    \paragraph{\(v \notin V' \Rightarrow \eqref{eq:good-vertices-get-decoded-in-inner-decoder}\) holds:} If \[\Prob{s' \in S_u}{M| s' \ni v} \geq \frac{\varepsilon}{2},\] 
    then the probability we selected a set \(s' \in M\) is \(\geq 1-(1-\frac{\varepsilon}{2})^{\frac{4 \log \frac{1}{\delta_{in}}}{\varepsilon}} \geq 1-\sqrt[3]{\delta_{in}}\). When this occurs, the set \(s' \in M\) is uniform conditioned on \(s' \ni v\). Hence by the second item the \(s'\) selected when the loop is exited satisfies \(w_{s'}(v) \ne f(v)\) with probability \(\leq 7\sqrt[3]{\delta_{in}}\), so
    \[
        \Prob{r}{InDec_{\delta_{in}}(s,v)=f(v)} \geq (1-\sqrt[3]{\delta_{in}})(1-7\sqrt[3]{\delta_{in}}) \geq 1-8\sqrt[3]{\delta_{in}}.
    \]

    \paragraph{Bounding \(\Prob{}{V'} < 5\sqrt{\delta_{in}}\):} We will prove that each of the two items that define \(V'\) are only violated with probability \(O(\sqrt{\delta_{in}})\). Loosely speaking, our proof strategy is to show the set $M$ defined above is roughly the set of $s'$ that $O(\delta_in)$-compute $f$:
    \[
    M \approx \sett{s'}{\dist_{s'}(w_{s'},f) \leq O(\delta_{in})}.
    \] 
    Then we will argue that \(\Prob{}{M} \approx \Prob{}{A}\geq \frac{3\varepsilon}{4}\) and therefore by sampling most vertices have \(\Prob{s' \in S_u}{M| s' \ni v} \geq \frac{\varepsilon}{2}\), proving that the fraction of vertices that violate the first item is \(\sqrt{\delta_{in}}\). To deal with the second item, we will argue that because most \(s' \in M\) have \(\dist(w_{s'},f) = O(\delta_{in})\) decoding succeeds on average, so a Markov argument implies all except a \(\sqrt[3]{\delta_{in}}\)-fraction of the vertices must violate the second item in the definition of \(V'\). Details follow.

    We begin by arguing that that most sets \(s'\) with \(w_s\approx f\) are in \(M\) and conversely, that \(s' \in M\) implies $w_s$ is close to \(f\). Formally,
    \begin{enumerate}
        \item \textbf{Every \(s' \in A\) neighboring $s$ such that \(\set{s,s'}\) is not a bad edge is in \(M\):} by assumption, both \(\dist_s(w_s,f), \dist_{s'}(w_{s'},f) \leq \delta_{in}\). The edge \(\set{s,s'}\) is not bad, so
        \[
        \dist_{s\cap s'}(w_{s'}|_{s\cap s'},f), \dist_{s\cap s'}(w_{s}|_{s\cap s'},f) < 2\delta_{in},
        \]
        but then by the triangle inequality
        \[
        \dist_{s\cap s'}(g|_{s\cap s'},g'|_{s\cap s'}) < 4\delta_{in}
        \]
        so \(s'\) passes the test in step \((2)\).
    
        \item \textbf{If \(s' \in M\) and \(\set{s,s'}\) is not a bad edge, then \(\dist(w_{s'},f) < 6\delta_{in}\):} Since \(\dist_{s}(f,w_s) < \delta_{in}\) and $\{s,s'\}$ is not a bad edge, we have \(\dist_{s \cap s'}(f,w_s|_{s \cap s'}) < 2\delta_{in}\). On the other hand, since $s'$ passed the test, by the triangle inequality:
        \[
        \dist_{s \cap s'}(f,g'|_{s \cap s'}) < \dist_{s \cap s'}(f,g|_{s \cap s'}) + \dist_{s \cap s'}(g,g') \leq 6\delta_{in},
        \]
        and since the edge is not bad we have \(\dist_{s'}(f,g') < 7\delta_{in}\).
    \end{enumerate}
    We will use these assertions to bound the probabilities in \cref{item:M-is-small} and \cref{item:v-is-misleading-in-M} in the definition of $V'$.

    \paragraph{Proof of \Cref{item:M-is-small}}
    The first assertion above implies that \[\Prob{}{M} \geq \Prob{}{A} - \Prob{s'}{s' \text{ is not a neighbor of } s} - \Prob{s'}{\set{s',s} \text{ is bad}}.\]
    Recall that \(\Prob{}{A} \geq \varepsilon\). From the assumption on density, \(\Prob{s'}{s' \text{ is not a neighbor of } s} \leq \eta\) and by assumption on \(s\), \(\Prob{s'}{\set{s',s} \text{ is bad}} \leq \sqrt{\frac{2\eta }{\delta_{in}^2}}\). Hence by assumption on the degree \(d\), \(\Prob{}{M} > \frac{3\varepsilon}{4}\). By sampling between \(V_u\) and \(S_u\), there are less than \(\frac{4\eta}{\varepsilon^2} <\sqrt[3]{\delta_{in}}\) vertices such that \(\Prob{}{M| s' \ni v}  < \frac{\varepsilon}{2}\); this proves that only \(\sqrt[3]{\delta_{in}}\) vertices violate \cref{item:M-is-small} in the definition of \(V'\).

    \paragraph{Proof of \Cref{item:v-is-misleading-in-M}:} We now argue the fraction of $v \in V$ satisfying \Cref{item:v-is-misleading-in-M} is at most $4\sqrt[3]{\delta_{in}}$. Toward this end, call a pair \((v,s')\) misleading if \(s' \in M\), but $w_{s'}(v) \neq f(v)$. We claim the fraction of misleading pairs is at most \(8\delta_{in} \Prob{}{M}\). To see this, observe
    \[
    \Pr_{v \in s'}{[(v,s')~\text{is misleading}]}= \Pr[M]\cdot \Pr_{v \in s'}[w_{s'}(v) \neq f(v)~|~s' \in M] = \Pr[M]\cdot \Pr_{v \in s'}[dist(w_{s'},f)~|~s' \in M].
    \]
    On the other hand, if $(s,s')$ is not a bad edge, we have $dist(w_{s'},f) \leq 6\delta_{in}$, so
    \[
    \Pr_{v \in s'}{[(v,s')~\text{is misleading}]} \leq \Pr[M]\cdot (6\delta_{in} + \Pr_{s'}[(s,s')~\text{is bad}~|~s' \in M]) \leq 8\delta_{in}\Pr[M]
    \]
    where in the last inequality we have used the fact that $\Pr[M] \geq \frac{3\varepsilon}{4}$, by assumption at most a $\sqrt{\frac{2\eta}{\delta_{in}^2}}$ fraction of $(s',s)$ are bad, and $\eta$ is sufficiently small.
    

    Now, recall that if the second item is violated, rephrasing the inequality we have
    \[
    \Pr_{s'}[w_{s'}(v) \neq f(v) \land s' \in M~| s' \ni v] \geq 7\sqrt[3]{\delta_{in}}\Pr_{s'}[M~|~s' \ni v].
    \]
    Then so long as $\Pr_{s'}[M~|~s' \ni v] \geq \frac{\Pr[M]}{2}$ (which happens except with probability \(\frac{4\eta }{\Prob{}{M}^2} < \sqrt{\delta_{in}}\) by sampling and the fact that \(\Prob{}{M} \geq \frac{3\varepsilon}{4}\)) we have
    \[
    \Pr_{s'}[(v,s')~\text{is misleading}~| s' \ni v] \geq \frac{7}{2}\sqrt[3]{\delta_{in}}\Pr[M].
    \]    
    but the fraction of $v$ which can satisfy the above is at most
    \[
    \frac{\Prob{v \in s'}{(v,s') \text{ is misleading}}}{\frac{7}{2}\sqrt[3]{\delta_{in}}\Prob{}{M}} \leq 3\delta_{in}^{2/3}
    \]
    by Markov's inequality. Thus altogether, \Cref{item:v-is-misleading-in-M} occurs with probability at most $4\sqrt[3]{\delta}$ as desired.
\end{proof}

\paragraph{Circuit implementation}
Carefully examining the inner decoder, we observe that it makes \(\poly(\frac{1}{\delta_{in}},\frac{1}{\varepsilon})\) calls to:
\begin{enumerate}
    \item The circuit \(C_{U;V\to S}\) in step \((a)\) with a random index \(i\), to sample a random \(s' \ni v\).
    \item The circuit \(C_{U;S\cap S \to V}\) (sampling random points to assess distances) in step \((c)\). This circuit also tells whether the sets \(s,s'\) actually intersect.
    \item The circuit \(C^{-1}_{U;S \to V}\) so that we can find the correct index on which to query the word $w_{In_u}$.
    \item A decoder circuit that maps the inner representation of a set \(s\) to input bits for the function list circuit (or noisy codeword in the list decoding scenario).
\end{enumerate}
Each of these circuits are implementable in size \((\log N)^{ \exp(O(\frac{1}{\varepsilon}))}\) and \(\tilde{O}(\log \log N \exp(O(\frac{1}{\varepsilon})))\) in our regime of parameters by \Cref{claim:circuits-for-const-rate-construction}.

Thus the size of the circuit remains \((\log N)^{ \exp(O(\frac{1}{\varepsilon}))}\) and the depth of the circuit is \(\tilde{O}(\log \log N \exp(O(\frac{1}{\varepsilon})))\).

\subsection{List Pruning}

We now move to turning our vanilla decoder above to the desired high confidence decoder. Namely by calling the generator algorithm for the inner code \(In_u\) multiple times, we can modify the inner decoders to decode a function close to \(f\) with probability \(1-\tau\) for any \(\tau > 0\) (namely it is enough to make \(\log \frac{1}{\tau}\) calls for this). This increases the list size from \(O(\frac{1}{\varepsilon})\) to \(O(\log \frac{1}{\tau} \cdot \frac{1}{\varepsilon})\) which seems minor, but would actually break our reduction to `well-separated' lists which has parameters depending on the input list size. Thus in order to maintain `constant' list size in the inner codes, we implement the simple circuit \(C_{Prune}\) described in the main body that prunes a large list to a smaller list of size \(O(\frac{1}{\varepsilon})\) while ensuring we do not remove the circuit that approximately computes $f$ (or more accurately, any function that is global within $In_u$). We re-state the claim here for convenience and give its formal proof.
\begin{claim}[List Pruning]\label{claim:output-list-pruning-app}
    Fix $\tau>0$ with parameters $\varepsilon,\delta_{out}$, an input word $w_{In_u}$, and $\{C_1,\ldots,C_{L}\}$ a set of output circuits from running the generating algorithm of the decoder \(InDec_{\delta_{in}}\) possibly several times. Assume that the graph \((V_u,S_u)\) is a \(\frac{\delta_{out}^2\varepsilon^2}{256}\)-sampler. There is a circuit $C_{Prune}$ which, given oracle access to the $\{C_i\}$ and input list, outputs a subset $L' \subset [L]$ of size $|L'| \leq O(\frac{1}{\varepsilon})$. With probability at least $1-\tau$ any original $C_i \in \Ball_{1-\varepsilon,\delta_{out}}(\mathcal{L})$ is $6\delta_{out}$-close to some $C_j \in L'$.
    
    Moreover $C_{Prune}$ is in logspace uniform size $\poly(|L|,\frac{1}{\varepsilon},\frac{1}{\delta_{out}},\log \frac{1}{\tau})$ and depth $O(\log^2 |L| \cdot \log \frac{1}{\delta_{out} \varepsilon \tau})$, makes at most $O(\frac{\log \frac{L}{\tau}}{\max\{\varepsilon,\delta\}})$ oracle calls to the $\{C_i\}$, word $w$, a circuit producing random $s \in S_u$ and random $v \in V_u$, and the hypergraph adjacency circuits. 
\end{claim}
We note the \(\delta_{out}\) actually used later on is \(O(\delta_{in}^{2/3})\), the output approximation parameter we get from our inner decoder in the previous subsection.

\begin{proof}
    We proceed in two steps. First, we remove any decoded circuits that do not agree with the global input list by empirically checking their distance from $w_{In_u}$. In particular, draw $O(\frac{\log \frac{L}{\tau}}{\varepsilon})$ hyperedges in the domain, and for each hyperedge $s$, sub-sample $O(\frac{\log \frac{L}{\tau}}{\delta_{out}})$ random $v \in s$. For each circuit $C_i$, we remove $C_i$ from the list if it is $2\delta_{out}$-far from $w_{In_u}$ on more than a $1-\varepsilon/2$ fraction of the domain. By a standard Chernoff and Union bound, the remaining set of circuits, which we denote by $L''$, has the following properties with probability at least $1-\tau/2$:
    \begin{enumerate}
        \item Any $C_i$ which is $(\varepsilon,\delta_{out})$-computed by $w_{In_u}$ remains in the list
        \item Every function in the list is in $\Ball_{1-\varepsilon/4,4\delta_{out}}(w_{In_u})$.
    \end{enumerate}

    Now by \Cref{claim:list-recovery-size-bound-on-samplers}, the second property, and the fact that \(G_{In_u}=(V_u,S_u)\) is a \(\frac{\delta_{out}^2 \varepsilon^2}{256}\)-sampler, we are promised that any set of pairwise $\delta$-separated circuits in $L''$ has size at most $\frac{2}{\varepsilon}$. To find the desired cover of $L''$,
    consider the following natural `empirical distance graphs'. Given a subset of functions $L' \subset L''$ and distance $\delta'$, the $(L',\delta')$-distance graph is simply the graph on vertex set $L'$ where $g,g' \in L'$ have an edge iff the empirical distance between them is at most $\delta'$ (measured with respect to the samples taken in the first step). Note that with probability at least $1-\tau/2$, all empirical distances are within $\delta_{out}$ of the true distances between functions.

    Now, consider a maximal independent set $L'$ of the $(5\delta_{out},L'')$-distance graph based on the empirical distances computed from $O(\frac{\log \frac{L}{\tau}}{\delta_{out}})$ sampled points from $V$. Assuming the empirical estimates are good in the above sense, $L'$ is a $4\delta_{out}$-separated subset of $\Ball_{1-\varepsilon/4,4\delta_{out}}(w_{In_u})$ so $|L'| \leq \frac{2}{\varepsilon}$. On the other hand, every $C' \in L''$ is within true distance $6\delta_{out}$ of $L''$. Since $L''$ contains every original $C_i \in \Ball_{1-\varepsilon,\delta_{in}}$ in the list, we are done.

    \paragraph{Circuit Complexity} Finally, we sketch the standard fact that the above algorithm can be implemented in the claimed logspace uniform size and depth. The number of oracle calls is immediate from the description. The first samples the sets in parallel, then points in them all in parallel. It then tests the fraction of sets on which we are close, and uses a threshold function to decide which indexes go into \(L''\). The size of such a circuit is $\poly(|L|,\frac{1}{\varepsilon},\frac{1}{\delta_{out}},\log \frac{1}{\tau})$ and the depth is \(O(\log \frac{\log \frac{L}{\tau}}{\varepsilon})\).
    
    The second step of the algorithm calls the sample oracle in parallel, then runs parallel pairwise distance tests for every pair of functions in $L''$. The depth and size of this step is also similar to the previous step.

    The additional step necessary for computing \(L'\) is to compute a maximal independent set on the distance graph which is well known to be in $NC^2$ \cite{luby1985simple}. In particular, all the above procedures are in $NC^2$ over the domain of sampled points.
\end{proof}
\section{Some Approximate List-Decoding Lemmas}
In this subsection, we cover several basic facts used in approximate local list decoding/recovery. Most of these are at least in some sense implicit in the prior literature or well known to the community, but to the best of our knowledge are not written anywhere so we include them for completeness.

\subsection{Combinatorial List Size on Samplers}

The following is a variant of a standard argument showing that the direct product encoding on sampler graphs has small combinatorial list size.

\begin{claim}\label{claim:list-recovery-size-bound-on-samplers}
    Fix \(\delta_{in}, \varepsilon > 0\), \(\ell_{in} \in \mathbb{N}\), \(\beta \leq \frac{\delta_{in}^2\varepsilon^2}{4\ell_{in}}\), and $\delta \geq \delta_{in}$. Let \((V,S)\) be a hypergraph whose inclusion graph is a $\beta$-sampler. Let $\mathcal{C}_{(V,S)}$ be the corresponding direct product code and \(\mathcal{L} = \set{L_s}\) be a set of lists of cardinality \(\leq \ell_{in}\). Let \(W \subseteq \Ball_{1-\varepsilon,\delta_{in}}(\mathcal{L})\) be a maximal set such that every \(f_1,f_2 \in W\) are \(\delta\)-far from one another. Then 
    \begin{enumerate}
        \item \(|W| \leq 2\frac{\ell_{in}}{\varepsilon}\), and
        \item For every \(f \in \Ball_{1-\varepsilon,\delta_{in}}(\mathcal{L})\) there exists \(f' \in W\) such that \(\dist(f,f') \leq \delta\).
    \end{enumerate}
\end{claim}

\begin{proof}[Proof of \cref{claim:list-recovery-size-bound-on-samplers}]
    The second item is automatic from maximality and the assumption that $\delta \geq \delta_{in}$. If there existed $f' \in \Ball_{1-\varepsilon,\delta_{in}}(\mathcal{L})$ that is $\delta$-far from every $f \in W$, then $f'$ can be added to $W$ maintaining pairwise $\delta$-distance, violating maximality.


    For the first item, it is sufficient to consider the case $\delta=\delta_{in}$. Assume towards contradiction there exists \(2\frac{\ell_{in}}{\varepsilon}+1\) distinct $\delta_{in}$-separated functions \(f_1,f_2,\dots,f_{2\frac{\ell}{\varepsilon}+1} \in W\). Let \(A_i \subseteq S\) be the \(s \in S\) such that \(f_i\) is \(\delta_{in}\)-close to a function in \(L_{s}\) and such that for all \(j \ne i\) \(\dist_s(f_i|_s,f_j|_s) > 2\delta_{in}\). By our sampling guarantee we have \(\Prob{}{A_i} \geq \varepsilon - 2\frac{\ell_{in}}{\varepsilon}\frac{\beta}{\delta_{in}^2} \geq \frac{\varepsilon}{2}\).

    Now, we claim that by an averaging argument, there must exist some $s$ which is at least $\ell_{in}+1$ of the $A_i$'s since
    \[
    \Ex{}{\sum_{i=1}^{2\frac{\ell_{in}}{\varepsilon} + 1} \one_{A_i}} \geq \left (2\frac{\ell_{in}}{\varepsilon} + 1 \right )\cdot \frac{\varepsilon}{2} > \ell_{in}
    \]

    This is of course impossible, since for every \(s \in A_i\), if \(\dist(f_i,g) \leq \delta_{in}\) then for any \(f_j\), \(\dist(f_j,g) > \delta_{in}\) since otherwise by the inequality \(\dist_s(f_i|_s,f_j|_s) \leq 2\delta_{in}\). Thus \(f_j,f_i\) cannot be close to the same elements of the input list at $s \in A_i,A_j$, and in particular no $s$ can be in more than $\ell_{in}$ of the $A_i$ reaching the desired contradiction.
    

\end{proof}

\subsection{From Many to All Message Lengths}
In this subsection, we describe a simple repetition procedure which takes any construction of an approximate locally list-recoverable code that works for many message sizes and transforms it to one that works for \textit{all} message sizes up to a loss in the rate.

In particular, say we have a construction of $(\delta_{in},\delta_{out})$-approximate $\ell_{in}$-input, $(\varepsilon,\ell_{out},q)$-aLLRCs that can be instantiated on message sizes $s(k)$ for some increasing size function $s(k) \in \mathbb{N}$. We claim this immediately implies a family of aLLRC's for \textit{all} message lengths, such that the rate for message lengths between $s(k)$ and $s(k+1)$ loses a factor of $\frac{s(k+1)}{s(k)}$. The code family achieving this is extremely simple, one simply repeats the input message until reaching size $s(k+1)$ (truncating the final repetition). In other words, the final code is just a truncated repetition code composed with the original code. To decode a message bit, we simply pick one of the repetitions at random and decode that via the original code. This means the resulting code maintains its locality and query complexity (and is in fact still a direct product code if the original code is, albeit possibly with multisets).

The above is captured formally by the following simple lemma, showing how to take a code $C_1: \Sigma_1^k \to \Sigma_2^{n}$ to a code $C_2: \Sigma_1^{k'} \to \Sigma_2^{n}$ over fewer $k' \leq k$ message bits. Let $Rep_{k',k}$ be the truncated repitition code above. Then:
\begin{lemma}\label{lem:many-to-all}
    If $C:\Sigma_1^{k} \to \Sigma_2^n$ is a $(\delta_{in},\delta_{out})$-approximate, $\ell_{in}$-input, $(\varepsilon,q,\ell_{out})$-aLLRC, then for any $k' \leq k$, the composition $C(Rep_{k',k}): \Sigma_1^{k'} \to \Sigma_2^n$ is a $(\delta_{in},O(\delta_{out}))$-approximate, $\ell_{in}$-input, $(\varepsilon,O(q),O(\ell_{out}))$-aLLRC
\end{lemma}
\begin{proof}
        Let $C'$ denote $C(Rep_{k',k})$. Fix an input list $\mathcal{L}$ and function $f \in \Ball_{1-\varepsilon,\delta_{in}}(\mathcal{L})$. By assumption running the original code $C$ outputs a list of (possibly randomized) circuits $C_1,\ldots,C_{\ell_{out}}$ on $k$ bit inputs such that one of these one of the circuits $\delta_{out}$-computes $Rep_{k',k}(f)$ with probability at least $3/4$. As discussed in \Cref{sec:preliminaries}, by repeating this process $O(1)$ times and moving to the $O(1)$-majority versions of each $C_i$, we can ensure that with $99\%$ probability, there is a circuit $C_j$ in the list such that $C_j$ deterministically $O(\delta_{out})$-computes $f$ for $99\%$ of choices of its random string.

        Now, condition on the existence of such a $C_j$, and choice of random string $r$ such that $C_j^r$ deterministically $O(\delta_{out})$-computes $Rep_{k',k}(f)$. We now use the following elementary decoding: sample a random message bit $x \in [k']$, a random encoded position of $x \in [k]$, and output the encoded bit. If every message bit were encoded the same number of times, this would clearly succeed except with $O(\delta_{out})$ probability. Because of truncation, however, the resulting distribution over message bits may be slightly skewed, but even in the worst case the algorithm is at most twice as likely to query any particular encoded bit. This means the error can at worst double, so the final probability of failure is still at most $O(\delta_{out})$ as desired even under the above conditioning.
\end{proof}

\section{Testing distance between functions} \label{app:distance-tester}
We will frequently need to test the distance between pairs of functions \(f,g\) on some domain \(s\). We will use the following basic query-efficient procedure:
\begin{figure}[ht!]
\fbox{\parbox{\textwidth}{
\vspace{.1cm}
\underline{\((p,\delta)\)-distance tester}:
\begin{enumerate}
    \item[*] \textbf{Input:} Functions \(f,g:s\to \Sigma\).
    \item[*] \textbf{Output:} True/False.
    \item Sample \(p\) i.i.d random points \(v_1,v_2,\dots,v_p \in V\). 
    \item If \(\frac{\abs{i: f(v_i)=g(v_i)}}{p} \leq \delta\) output True, otherwise output False.
\end{enumerate}}}
    \caption{Distance tester}
    \label{fig:distance-tester}
\end{figure}
The following claim follows from a standard Chernoff bound, we omit its proof.
\begin{claim} \label{claim:independent-distance-test}
    Let \(\alpha, \gamma,\delta > 0\) and let \(p \geq \frac{5\log \frac{1}{\gamma}}{\alpha}\). Then the \((p,\delta)\)-distance tester has the following guarantees.
    \begin{enumerate}
        \item If \(\dist(f,g) \leq \delta - \alpha\) then \(\Prob{v_1,v_2\dots,v_p}{\text{\((p,\delta)\)-distance(f,g)=True} }\geq 1-\gamma\).
        \item If \(\dist(f,g) \geq \delta + \alpha\) then \(\Prob{v_1,v_2\dots,v_p}{\text{\((p,\delta)\)-distance(f,g)=False} }\geq 1-\gamma\).
    \end{enumerate}
\end{claim}
As for circuit complexity, we have the following guarantees:
\begin{claim} \label{claim:distance-tester-circuit}
    Assuming oracle access to \(f,g:s \to \Sigma\), and oracle access to a circuit that outputs random points in \(s\), a circuit \((p,\delta)\)-distance tester can be implemented using \(O(\log p)\)-depth and \(O(p \log |V|)\)-size. It makes \(2p\)-calls to \(f,g\).
\end{claim}

\begin{proof}
    We sample all \(p\) points. For each point we test equality \(f(v)=g(v)\) separately by a constant size circuit\footnote{Technically we are assuming the alphabet is constant. If it is not, then the depth includes another additive factor of \(\log |\Sigma|\), and size also scales multiplicatively with \(\log |\Sigma|\).}. We note that the size is \(p \log |V|\) mainly because the identifier for every point \(v\) has size \(\log |V|\). After this, we implement a \(\delta p \)-threshold circuit over the \(p\)-points which has size \(O(p)\) and depth \(O(\log p)\) by standard calculations.
\end{proof}

Finally, the following straightforward generalization will be necessary later on. Let \(L_1,L_2\) be two lists of functions on the same domain \(s\). The following algorithm tells us which function \(g \in L_2\) are close to \emph{some} function in \(L_1\).
\begin{figure}[ht!]
\fbox{\parbox{\textwidth}{
\vspace{.1cm}
\underline{\((p,\delta)\)-distance tester}:
\begin{enumerate}
    \item[*] \textbf{Input:} Function lists $L_1=\{f_i\}$ and $L_2=\{g_i\}$ on a set \(s\).
    \item[*] \textbf{Output:} Function list $L_{out} \subset L_2$
    \item Sample \(p\) i.i.d random points \(v_1,v_2,\dots,v_p \in V\). 
    \item Set $L_{out} \gets \emptyset$
    \item For each $g_i \in L_2$
    \begin{itemize}
        \item If $\exists f_j \in L_1$ s.t.\ \(\frac{\abs{i: f(v_i)=g(v_i)}}{d} \leq 2\delta\), $L_{out} \gets L_{out} \cup g_i$
    \end{itemize}
    \item Return $L_{out}$
\end{enumerate}}}
    \caption{Distance tester}
    \label{fig:list-distance-tester}
\end{figure}

The following claim is straightforward from \cref{claim:independent-distance-test} and \cref{claim:distance-tester-circuit}.
\begin{claim}\label{claim:list-distance-tester}
    Let \(\delta > 0\) and \(\gamma >0\) and let \(p \geq \frac{12 \log\frac{|L_1||L_2}{\gamma}}{\delta}\).
    The algorithm in \cref{fig:list-distance-tester} has the following guarantees.
    \begin{enumerate}
        \item If \(\dist(g,L_1) \leq \delta\) then \(\Prob{v_1,v_2\dots,v_p}{g \in L_{out} }\geq 1-\gamma\).
        \item If \(\dist(g,L_1) \geq \delta + \alpha\) then \(\Prob{v_1,v_2\dots,v_p}{g \notin L_{out} }\geq 1-\gamma\).
    \end{enumerate}
    \Cref{fig:list-distance-tester} can be implemented in logspace uniform \(O(\log p + \log |L_1||L_2|)\)-depth, \(O(p(|L_1||L_2| +(|L_1|+|L_2|)\log |V|))\)-size and \(p(|L_1|+|L_2|)\) calls to the functions.
\end{claim}

\end{document}